\documentclass[12pt]{book}
\usepackage[activeacute,spanish]{babel}
\usepackage{fancyhdr,epsf,layout,times,indentfirst,eucal,amsmath,amsfonts}
\usepackage[spanish]{babel}
\usepackage{epstopdf}
\usepackage{times}
\usepackage{fancybox}
\usepackage{graphicx}

\newcommand{\be}{\begin{equation}}
\newcommand{\ee}{\end{equation}}

\input{epsf}
\newtheorem{thm}{Teorema}[section]
\newtheorem{cor}[thm]{Corolario}
\newtheorem{lem}[thm]{Lema}

\newtheorem{defn}[thm]{Definici{\'o}n}

\numberwithin{section}{part}

\newcommand{\eps}{\varepsilon}

\newcommand{\n}{\~n}
\newcommand{\bs}{\bigskip}
\newcommand{\fin}{\begin{flushright}$\Box$\end{flushright}}

\begin{document}

\pagenumbering{arabic}


\title{\vspace{-30mm}
{\large Universidad Nacional de La Plata}\\
\vspace{20mm}
{\bf Funciones Espectrales de Operadores Singulares}\\
\vspace{10mm}
{\Large Pablo Pisani}
\vspace{15mm}
\begin{figure}[h]
\center
  \includegraphics[scale=0.5]{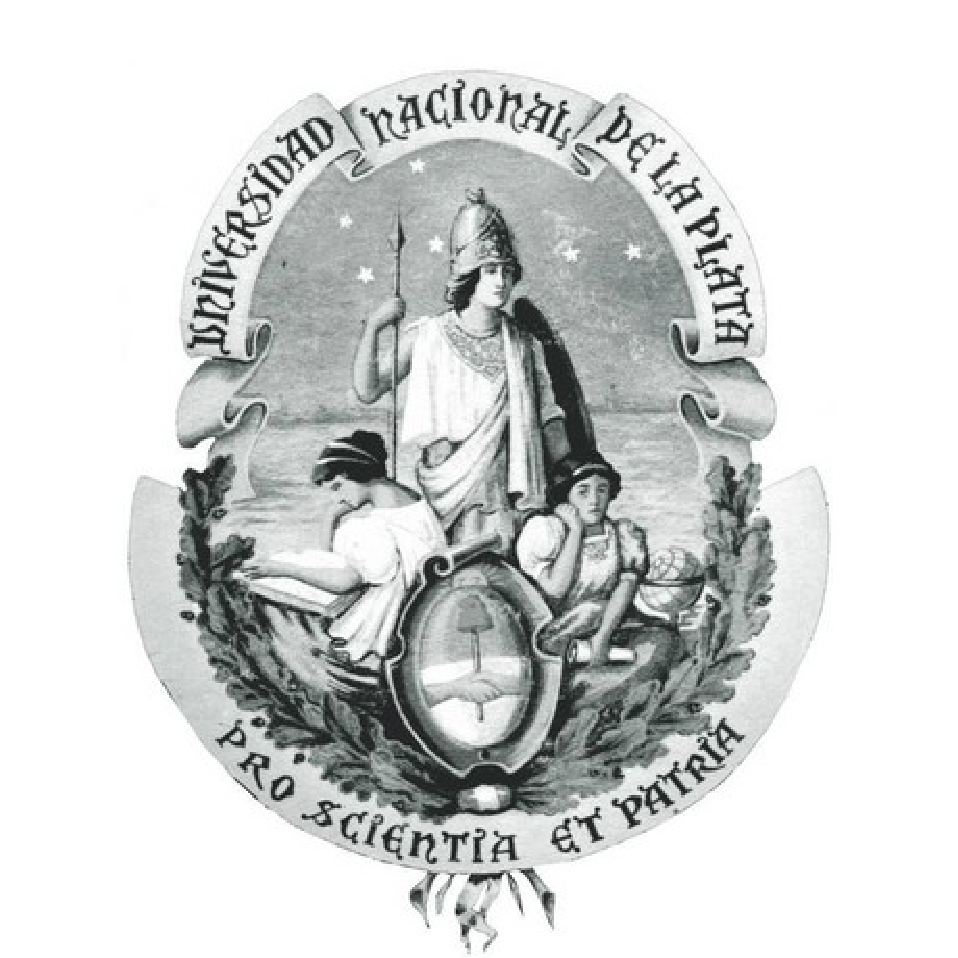}
\end{figure}}

\date{17 dec 2004}

\maketitle


\pagebreak

\mbox{}

\pagebreak

\begin{large}
\noindent
\\ \\ \\ \\ \\Tesis Doctoral del Departamento de F{\'\i}sica\\
\\de la Facultad de Ciencias Exactas\\
\\de la Universidad Nacional de La Plata.\\ \\ \\ \\ \\ \\
\noindent Director: Dr.\ Horacio A.\ Falomir.\\ \\ \\ \\ \\ \\
\end{large}

\pagebreak

\mbox{}

\pagebreak

\mbox{}

\vspace{30mm}

\begin{flushright}{\it Dedicado al inalterable recuerdo de mi viejo, que est{\'a}
presente en cada paso.}\end{flushright}

\pagebreak

\thanks{{\it Desocupado lector, sin juramento me podr{\'a}s creer que quisiera que
 este libro, como hijo
del entendimiento, fuera el m{\'a}s hermoso, el m{\'a}s gallardo y
m{\'a}s discreto que pudiera imaginarse. Pero no he podido yo
contravenir al orden de naturaleza; que en ella cada cosa engendra
su semejante.\\
\begin{flushright}
Miguel de Cervantes\\
(Primeras palabras del pr{\'o}logo al\\
Ingenioso Hidalgo Don Quijote de la Mancha.)
\end{flushright}
\vspace{80mm} For the most wild yet most homely narrative which I
am about to pen, I neither expect nor solicit belief. Mad indeed
would I be to expect it, in a case where my very senses reject
their own evidence. Yet, mad am I not -and very surely do I not
dream. But tomorrow I die, and today I would unburden my soul. My
immediate purpose is to place before the world, plainly,
succinctly, and without comment, a series of mere household
events. In their consequences, these events have terrified -have
tortured- have destroyed me. Yet I will not attempt to expound
them. To me, they have presented little but horror -to many they
will seem less terrible than baroques. Hereafter, perhaps, some
intellect may be found which will reduce my phantasm to the
commonplace -some intellect more calm, more logical, and far less
excitable than my own, which will perceive, in the circumstances I
detail with awe,
nothing more than an ordinary succession of very natural causes and effects.\\
\begin{flushright}Edgar Allan Poe (The Black Cat.)\end{flushright}}}


%


\tableofcontents



\part{Introducci{\'o}n}

\vspace{5mm}\begin{flushright}
{\it No other question has ever moved so profoundly the spirit of man;\\
no other idea has so fruitfully stimulated his intellect;\\
yet no other concept stands in greater need of clarification\\
than that of the infinite.\\
(David Hilbert.)}
\end{flushright}

\vspace{25mm}

\section{Funciones Espectrales y Operadores Singulares}

En una serie de trabajos publicados entre 1967 y 1969, R.T.\
Seeley \cite{Seeley1,Seeley3,Seeley4} estudi{\'o} la existencia y
propiedades de la resolvente $(A-\lambda)^{-1}$ de un operador
diferencial $A$ con coeficientes infinitamente derivables definido
sobre secciones de un fibrado vectorial sobre una variedad de base
compacta $M$ con borde suave $\partial M$. El
pro\-ce\-di\-mien\-to utilizado consiste en construir una
aproximaci{\'o}n al n\'ucleo de la resolvente $(A-\lambda)^{-1}$,
con $\lambda\in\mathbb{C}$, para grandes valores de $|\lambda|$ a
partir de una aproximaci{\'o}n al s\'\i mbolo de la resolvente.
Esto permite, por su parte, definir e investigar las propiedades
del operador pseudodiferencial $A^{-s}$, para una variable
$s\in\mathbb{C}$.

\bigskip

Los trabajos \cite{Seeley1,Seeley3,Seeley4} contienen un resultado
fundamental en la teor{\'\i}a de las funciones espectrales: la
traza de $A^{-s}$, tambi{\'e}n denominada funci{\'o}n-$\zeta$ del
operador $A$, es una funci{\'o}n meromorfa de la variable $s$
cuyas {\'u}nicas singularidades consisten en una sucesi{\'o}n de
polos simples $s_n$ ubicados en puntos del eje real dados por,
\begin{equation}\label{resu}
\begin{picture}(300,40)(-50,0)
    \put(-5,0){\line(0,1){40}}
    \put(225,0){\line(0,1){40}}
    \put(-5,0){\line(1,0){230}}
    \put(-5,40){\line(1,0){230}}
    \put(15,18){$\displaystyle{s_n=\frac{m-n}{d}}\qquad{\rm con}
    \qquad\ n=0,1,2,\ldots$}
\end{picture}
\end{equation}
donde la cantidad $m$ designa la dimensi{\'o}n de la variedad $M$
y $d$ el orden del operador di\-fe\-ren\-cial $A$.

\bigskip

Es notable que los polos de la funci{\'o}n espectral
$\zeta_A(s):={\rm Tr}\,A^{-s}$, s{\'o}lo dependan del orden del
operador $A$ y de la dimensi{\'o}n de la variedad $M$. No
dependen, por ejemplo, de par{\'a}metros externos que aparecieren
en los coeficientes del operador diferencial ni de la forma
funcional de estos coeficientes.

\bigskip

No obstante, el resultado (\ref{resu}) es v{\'a}lido bajo las
hip{\'o}tesis que hemos mencionado. La variedad $M$ debe ser
compacta y su borde $\partial M$ suficientemente suave. Por otra
parte, existen restricciones sobre las condiciones de contorno del
problema, esto es, sobre el comportamiento de las funciones del
dominio $\mathcal{D}(A)$ del operador en el borde $\partial M$ de
la variedad. Las condiciones de contorno deben estar definidas por
o\-pe\-ra\-do\-res de borde locales que son combinaciones lineales
de las derivadas normales al borde (v{\'e}ase la ecuaci{\'o}n
{\ref{opbo}}.) El operador diferencial $A$ y los
o\-pe\-ra\-do\-res de borde deben, adem{\'a}s, definir un sistema
el{\'\i}ptico para el cual se sa\-tis\-fa\-ga la condici{\'o}n de
Agmon (v{\'e}anse las definiciones (\ref{deuno}), (\ref{dedos}) y (\ref{decuatro}).) Asimismo, el operador $A$ debe
ser regular, {\it i.e.}, sus coeficientes deben pertenecer a la
clase $\mathcal{C}^{\infty}(M)$ de funciones infinitamente
derivables sobre la variedad.

\bigskip

Antes de mencionar las aplicaciones de este resultado en
Teor{\'\i}a Cu{\'a}ntica de Campos hemos de decir que, a pesar de
la gran actividad de\-sa\-rro\-lla\-da en el estudio de las
funciones espectrales, no existe informaci{\'o}n suficiente acerca
de la validez del resultado (\ref{resu}) en el caso de operadores
diferenciales con coeficientes singulares.

\bigskip

El objetivo central de esta Tesis es estudiar la estructura de
polos de la funci{\'o}n $\zeta_A(s)$ de un operador diferencial
$A$ con coeficientes singulares definido sobre una variedad con
borde. En el cap{\'\i}tulo \ref{fe} presentaremos sucintamente la
derivaci{\'o}n del resultado (\ref{resu}) pero mostraremos que
este resultado pierde validez en presencia de cierto tipo de
singularidades. Deduciremos entonces el Teorema \ref{elthm11} que
muestra la ubicaci{\'o}n de los polos de la funci{\'o}n-$\zeta$ de
operadores de Schr\"odinger unidimensionales cuyos potenciales
poseen ese tipo de sin\-gu\-la\-ri\-dad. Veremos que, en este
caso, la posici{\'o}n de los polos puede depender de algunos otros
par{\'a}metros que caracterizan la singularidad del
o\-pe\-ra\-dor.

\bs

Algunos ejemplos ser{\'a}n tratados en el cap{\'\i}tulo
\ref{apli}, en el que calcularemos los polos y residuos de las
funciones-$\zeta$ correspondientes a algunos operadores
di\-fe\-ren\-cia\-les con coeficientes singulares que pueden
resolverse en forma expl{\'\i}cita\,\footnote{En los casos que
hemos estudiado, las autofunciones pueden expresarse en
t{\'e}rminos de funciones especiales y los autovalores est{\'a}n
determinados por las soluciones de ecuaciones trascendentes.}.

\bigskip

Consideraremos, en particular, hamiltonianos de Schr\"odinger,
\begin{equation}\label{sch}
    A=-\partial^2_x+U_\nu(x)\,,
\end{equation}
en variedades de base unidimensionales $M$ compactas y no
compactas, cuyos potenciales $U_\nu(x)$ presentan una singularidad
en el borde $\partial M$. Tambi{\'e}n resolveremos un pro\-ble\-ma
asociado al hamiltonianos de Dirac,
\begin{equation}\label{dir}
    A=i\gamma_0\left(\not{\!\partial_x}-i \not{\!\!{\mathcal{A}_\nu
    (x)}}\right)\, ,
\end{equation}
definidos sobre una variedad no compacta $M$ de dos dimensiones en
la que el campo de gauge $\mathcal{A}_\nu(x)$ presenta una
singularidad aislada.

\bs

Es importante se{\~n}alar que, para ambos operadores, el ``grado''
de la sin\-gu\-la\-ri\-dad presente en los coeficientes $U_\nu(x)$
y $\mathcal{A}_\nu(x)$ coincide con el orden $d$ del
o\-pe\-ra\-dor diferencial; {\it id est}, el potencial del
operador de Schr\"odinger, para el que $d=2$, posee un t{\'e}rmino
singular proporcional a $x^{-2}$, en tanto que el t{\'e}rmino
singular en el campo de gauge del operador de Dirac, para el cual
$d=1$, es proporcional a $x^{-1}$. Esta propiedad permite estudiar
las funciones espectrales en t{\'e}rminos de las transformaciones
de escala en las proximidades de la singularidad. El
sub{\'\i}ndice $\nu$, por su parte, est{\'a} relacionado, tanto en
uno como en otro caso, con el coeficiente de los t{\'e}rminos
singulares, caracterizando as\'\i\ la ``intensidad" de la
singularidad.

\bigskip

De acuerdo con el resultado (\ref{resu}), los polos de la
funci{\'o}n-$\zeta$ de los operadores regulares de primer y
segundo orden definidos sobre variedades de base de una y dos
dimensiones est{\'a}n ubicados en valores enteros o semienteros
del eje real. No obstante, los operadores diferenciales
(\ref{sch}) y (\ref{dir}) que hemos estudiado en el cap{\'\i}tulo
\ref{apli} poseen un coeficiente\,\footnote{Nos referimos al
potencial $U_\nu(x)$ y al campo de gauge $\mathcal{A}_\nu(x)$ como
los coeficientes del t{\'e}rmino de orden cero en las derivadas.}
con una singularidad regular de la forma $x^{-2}$ y $x^{-1}$,
respectivamente. En consecuencia, no podemos afirmar {\it a
priori} que se verifique (\ref{resu}) pues los operadores no
satisfacen las hip{\'o}tesis que lo rigen.

\bigskip

La resoluci{\'o}n expl{\'\i}cita para los operadores (\ref{sch}) y
(\ref{dir}) que daremos en el cap{\'\i}tulo \ref{apli} muestra
que, por el contrario, se verifica la estructura de polos de las
funciones-$\zeta$ que probaremos en el cap{\'\i}tulo \ref{fe} para
operadores con coeficientes singulares. Como dijimos, esta
estructura no responde a la ecuaci{\'o}n (\ref{resu}) seg{\'u}n la
cual, en los casos $d=1,2$ y $m=1,2$, los polos est{\'a}n ubicados
en enteros o semienteros del eje real. En presencia de las
singularidades re\-gu\-la\-res mencionadas, las funciones-$\zeta$
correspondientes tienen polos simples en el eje real cuyas
posiciones dependen con continuidad del par{\'a}metro $\nu$ que,
como hemos dicho, caracteriza la intensidad de la singularidad.
Por consiguiente, como el par{\'a}metro $\nu$ toma va\-lo\-res en
un intervalo del eje real, las posiciones de los polos de las
funciones-$\zeta$ pueden tomar, incluso, valores irracionales.
Asimismo, se observa que, aunque las singularidades de las
funciones-$\zeta$ son aisladas, la distancia entre polos sucesivos
disminuye en la direcci{\'o}n del semieje real negativo y su
densidad aumenta conforme la variable $s$ tiende a $-\infty$ sobre
el eje real. En el caso de operadores regulares, por el contrario,
las singularidades de la funci{\'o}n-$\zeta$ est{\'a}n, como
indica (\ref{resu}), igualmente espaciadas.

\bigskip

Daremos a continuaci{\'o}n expresiones para las funciones
espectrales que estudiaremos en esta Tesis y veremos algunas
consecuencias del resultado (\ref{resu}). Las definiciones y
propiedades de las funciones espectrales se presentan con mayor
detalle en la secci{\'o}n \ref{funcionesespectrales}.

\bs

Si el operador diferencial $A$ tiene un espectro de autovalores
dado por $\{\lambda_n\}_{n\in\mathbb{N}}$ entonces,
\begin{equation}
    \zeta_A(s)={\rm Tr}\,A^{-s}=\sum_{n\in\mathbb{N}}\lambda_n^{-s}\,,
\end{equation}
que converge para $\mathcal{R}(s)$ suficientemente grande. Por
otra parte, si el valor absoluto de los autovalores crece
suficientemente r{\'a}pido conforme $n\rightarrow\infty$, entonces
la traza de la resolvente existe y est{\'a} dada por,
\begin{equation}
    {\rm Tr}\,(A-\lambda)^{-1}=\sum_{n\in\mathbb{N}}
    \frac{1}{\lambda_n-\lambda}\,.
\end{equation}
Asimismo, si la parte real de los autovalores est{\'a} acotada
inferiormente, la traza del heat-kernel ${\rm Tr}\,e^{-tA}$
verifica,
\begin{equation}
    {\rm Tr}\,e^{-tA}=\sum_{n\in\mathbb{N}}e^{-t\lambda_n}\,,\\
\end{equation}
siendo $t\in\mathbb{R}^+$.

\bs

En la secci{\'o}n \ref{funcionesespectrales} mostraremos que la
transformada de Mellin de la traza del heat-kernel es la
funci{\'o}n-$\zeta$ y que su transformada de Laplace es la traza
de la resolvente (v{\'e}anse las ecuaciones (\ref{reshea}) y
(\ref{zethea}).) Esto implica que las singularidades de
$\zeta_A(s)$ aportan informaci{\'o}n acerca de los desarrollos
asint{\'o}ticos de ${\rm Tr}\,(A-\lambda)^{-1}$ y de ${\rm
Tr}\,e^{-tA}$ para grandes valores de $|\lambda|$ y para
peque\~nos valores de $t$, res\-pec\-ti\-va\-men\-te. Si la
funci{\'o}n $\zeta_A(s)$ posee polos simples en $s=s_n$, con
$n\in\mathbb{N}$, entonces ${\rm Tr}\,(A-\lambda)^{-1}$ admite un
desarrollo asint{\'o}tico para grandes valores de $|\lambda|$ en
potencias de la forma $\lambda^{-s_n-1}$ en tanto que ${\rm
Tr}\,e^{-tA}$ admite un desarrollo asint{\'o}tico para
peque{\~n}os valores de $t$ en potencias de la forma $t^{-s_n}$.
Los coeficientes de estas potencias en ambos desarrollos
asint{\'o}ticos est{\'a}n determinados por los residuos de la
funci{\'o}n $\zeta_A(s)$ en el polo co\-rres\-pon\-dien\-te. En
consecuencia, si se cumplen las hip{\'o}tesis del resultado
(\ref{resu}), los exponentes de las potencias de $\lambda$ y $t$
en los desarrollos asint{\'o}ticos de ${\rm Tr}\,(A-\lambda)^{-1}$
y de ${\rm Tr}\,e^{-tA}$ est{\'a}n determinados por el orden del
operador y la dimensi{\'o}n de la variedad.

\bigskip

Consideremos, por ejemplo, un operador diferencial de
Schr\"odinger regular,
\begin{equation}
    A=-\triangle + V(x)\,,
\end{equation}
definido sobre secciones $\phi$ de un fibrado vectorial $E$ de
rango $k$ y conexi{\'o}n $\omega$ sobre una variedad de base
compacta $M$ que satisfacen la condici{\'o}n de contorno local,
\begin{equation}
    \left.\left(\partial_m+S\right)\phi\right|_{\partial M}=0
\end{equation}
donde $\partial_m$ es la derivada respecto de la coordenada
$x^{m}$ normal al borde $\partial M$ y $S:\partial
M\rightarrow\mathbb{C}^{k\times k}$. Se demuestra entonces
\cite{Gilkey} que la traza del heat-kernel $e^{-tA}$ satisface el
siguiente de\-sa\-rro\-llo asint{\'o}tico para peque{\~n}os
valores de $t$,
\begin{equation}\label{teohea}
    {\rm Tr}\,e^{-t A}\sim \sum_{n=0}^\infty c_n(A)\cdot
    \,t^{ \frac{-m+n}{2}}\,,
\end{equation}
donde los coeficientes $c_n(A)$ est{\'a}n dados por,
\begin{equation}\label{666}\begin{array}{l}
    {\displaystyle c_{2k}(A)=\frac{1}{(4\pi)^{m/2}}
    \int_M c_{2k}(A,x)\,d\mu(x)+\frac{1}{(4\pi)^{m/2}}
    \int_{\partial M} c^b_{2k}(A,x)\,d\mu^b(x)}\,,\\ \\
    {\displaystyle c_{2k+1}(A)=\frac{1}{(4\pi)^{(m-1)/2}}
    \int_{\partial M} c^b_{2k+1}(A,x)\,d\mu^b(x)}\,,\end{array}
\end{equation}
siendo $d\mu(x)$ y $d\mu^b(x)$ medidas de integraci{\'o}n sobre la
variedad y sobre su borde, respectivamente. Como puede verse de la
expresi{\'o}n (\ref{teohea}), los exponentes de las potencias de
$t$ est{\'a}n de acuerdo con la ubicaci{\'o}n de los polos de la
funci{\'o}n-$\zeta$ correspondiente, dada por (\ref{resu}).

\bs

Los coeficientes locales
$c_{2k}(A,x),c^b_{2k}(A,x),c^b_{2k+1}(A,x)$ de las expresiones
(\ref{666}) son funciones del potencial $V(x)$, de la conexi{\'o}n
$\omega$, de la matriz $S$, del tensor de curvatura
$\mathcal{R}_{\mu\nu\rho\sigma}$ de la variedad $M$ y del tensor
de curvatura extr{\'\i}nseca $K_{\mu\nu}$ del borde $\partial M$.
Los valores calculados para los primeros de estos coeficientes son
\cite{Gilkey,Klaus1,Vass1},
\begin{eqnarray}
    c_0(A,x)=k\,,\label{coefi1}\\
    c_2(A,x)={\rm tr_E}V(x)+\frac{1}{6}\,k\,\mathcal{R}\,,\\
    c_4(A,x)=\frac{1}{360}\left(
    60\,{\rm tr_E}\triangle V(x)+60\,\mathcal{R}\,{\rm tr_E}V(x)+180\,
    {\rm tr_E}V^2(x)+\right.\nonumber\\ \left.\mbox{}+
    12\,k\,\triangle \mathcal{R}+5\,k\,\mathcal{R}^2-2\,k\,
    \mathcal{R}_{\mu\nu}\mathcal{R}^{\mu\nu}+2\,k\,
    \mathcal{R}_{\mu\nu\rho\sigma}\mathcal{R}^{\mu\nu\rho\sigma}
    +30\,{\rm tr_E}\Omega_{\mu\nu}\Omega^{\mu\nu}
    \right)\,,\label{coefi4}\\\nonumber\\
    c^b_0(A,x)=0\,,\\
    c^b_2(A,x)=\frac{1}{3}\left(k\,
    K+6\,k\,S\right)\,,\\\nonumber\\
    c^b_1(A,x)=\frac{1}{4}\,k\,\,,\\
    c^b_{3}(A,x)=\frac{1}{384}\left(
    96\,{\rm tr_E}V(x)+16\,k\,\mathcal{R}+8\,k\,
    \mathcal{R}^\mu_{m\mu m}+\right.\nonumber\\\left.\mbox{}+
    13\,k\,K^2+2\,k\,K_{\mu\nu}K^{\mu\nu}+96\,k\,S\,K+192\,k\,S^2
    \right)\,.\label{coefiult}
\end{eqnarray}
En estas expresiones ${\rm tr}_E$ representa la traza de los
operadores sobre la fibra, que es isomorfa a $\mathbb{C}^k$, y
$\Omega$ es la ``curvatura'' de la conexi{\'o}n $\omega$.

\bs

Por el contrario, los resultados obtenidos en el cap{\'\i}tulo
\ref{opesing} y verificados con ejemplos en el cap{\'\i}tulo \ref{apli}
indican que las funciones espectrales ${\rm Tr}\,(A-\lambda)^{-1}$
y ${\rm Tr}\,e^{-tA}$ del o\-pe\-ra\-do\-r (\ref{sch}) con un potencial $U_\nu(x)$ singular
admiten un de\-sa\-rro\-llo asint{\'o}tico en potencias de
$\lambda$ y $t$ cuyos exponentes dependen del par{\'a}metro $\nu$
involucrado en el t{\'e}rmino singular. Un resultado similar para
la traza de la resolvente ${\rm Tr}\,(A-\lambda)^{-1}$ del
operador singular (\ref{dir}) es ilustrado con algunos ejemplos en
el cap{\'\i}tulo \ref{dirope}.

\bs

En el Cap{\'\i}tulo \ref{opesing} estudiaremos, en particular, el operador de Sch\"odinger (\ref{sch}), definido sobre la variedad de base $\mathbb{R}^+$, cuyo un potencial $U_\nu(x)$ tiene un comportamiento singular en $x=0$ dado por,
\begin{equation}\label{mod}
    U_\nu(x)=\frac{\nu^2-1/4}{x^2}+V(x)\,,
\end{equation}
siendo $V(x)$ una funci{\'o}n anal{\'\i}tica e inferiormente
acotada. Mostraremos que el desarrollo asint{\'o}tico de la traza de su
heat-kernel contiene potencias de $t$ dependientes del
par{\'a}metro $\nu$ dadas por,
\begin{equation}\label{asin}
    \sum_{N=1}^\infty\sum_{n=1}^{\infty}
    b_{N,n}(A)\,\theta^N\,t^{\nu N +n/2-1/2}\,,
\end{equation}
donde $\theta$ es un par{\'a}metro que caracteriza la
condici{\'o}n de contorno en el origen. Daremos tambi{\'e}n una
t{\'e}cnica para determinar los coeficientes $b_{N,n}(A)$.

\bigskip

Algunas otras divergencias con respecto al resultado (\ref{resu})
ya han sido estudiadas bajo distintas hip{\'o}tesis. La presencia
de polos de multiplicidad mayor que uno en la funci{\'o}n
$\zeta_A(s)$ est{\'a} relacionada con desarrollos asint{\'o}ticos
de la traza de la resolvente y del heat-kernel que involucran
factores de la forma $\log{\lambda}$ y $\log{t}$ multiplicando a
las potencias de $\lambda$ y $t$, respectivamente. Se ha
demostrado \cite{Gr1,Gr2,GS,GS2,GS3} que el desarrollo
asint{\'o}tico para peque{\~n}os valores de $t$ de la traza del
heat-kernel ${\rm Tr}\,e^{-tA}$ de operadores diferenciales
definidos sobre funciones que satisfacen condiciones de contorno
espectrales\,\footnote{Condiciones de contorno espectrales son un
tipo de condiciones no locales que aparecen en el Teorema del
{\'\i}ndice de Atiyah-Patodi-Singer para variedades con borde
\cite{APS}.} presentan t{\'e}rminos logar{\'\i}tmicos en $t$
multiplicando a las potencias de $t$. Por lo tanto, la funci{\'o}n
$\zeta_A(s)$ correspondiente presenta polos de multiplicidad
ma\-yor. No obstante, la ubicaci{\'o}n de estas singularidades en
el plano complejo $s$ est{\'a} determinada, al igual que bajo las
hip{\'o}tesis del resultado (\ref{resu}), por el orden $d$ del
operador $A$ y la dimensi{\'o}n $m$ de la variedad de base $M$.

\bs

Existe, tambi{\'e}n, cierta controversia
\cite{DGK,Avramidi:2001ns,Seeley2,Dowker:2001pz} con respecto al
de\-sa\-rro\-llo a\-sin\-t{\'o}\-ti\-co del heat-kernel en
problemas con condiciones de contorno mixtas, esto es, definidas
por un operador de borde singular\,\footnote{Tambi{\'e}n llamadas
condiciones de contorno de Zaremba; {\it e.g.}, condiciones de
contorno del tipo Dirichlet en una subvariedad de $\partial M$ y
condiciones de contorno del tipo Neumann en su complemento.},
referida a las propiedades de los coeficientes e, incluso, a la
presencia de t{\'e}rminos logar{\'\i}tmicos en $t$. No obstante,
una vez m{\'a}s los exponentes de las potencias de $t$ y,
consecuentemente, la ubicaci{\'o}n de las singularidades de la
funci{\'o}n-$\zeta$ est{\'a}n determinadas por el orden del
operador diferencial y la dimensi{\'o}n de la variedad.

\bigskip

En 1980, C.\ Callias y C.H.\ Taubes \cite{Callias:1979fi}
conjeturaron que el desarrollo asint{\'o}tico de la traza del
heat-kernel de operadores diferenciales con coeficientes
singulares presentar{\'\i}a factores logar{\'\i}tmicos en $t$
multiplicando potencias de $t$ cuyos exponentes po\-dr\'\i\-an,
adem{\'a}s, depender de los par{\'a}metros que caracterizaran las
singularidades de los coeficientes. Sin embargo, no proveyeron en
esa ocasi{\'o}n argumentos o ejemplos que sostuvieran su
conjetura.

\bs

En 1983, C.\ Callias \cite{Callias1} estudi{\'o} el operador de
Schr\"odinger (\ref{sch}), con $x\in\mathbb{R^+}$, dado por el
potencial
\begin{equation}\label{cal}
    U_\nu(x)=\frac{\nu^2-1/4}{x^2}\,,
\end{equation}
con $\nu>1$ pero demostr{\'o} que la traza de su heat-kernel
admite un desarrollo asint{\'o}tico en potencias de $t$ cuyos
exponentes responden a la ecuaci{\'o}n (\ref{resu}).

\bs

Sin embargo, en el cap{\'\i}tulo \ref{apli} mostraremos que, para
o\-pe\-ra\-do\-res de la forma (\ref{sch}) con un potencial
(\ref{cal}) pero con $0\leq\nu<1$, los exponentes de las potencias
de $t$ dependen, en general, del par{\'a}metro $\nu$. La
diferencia se debe a que este operador admite una familia infinita
de extensiones autoadjuntas si $0\leq\nu< 1$ en tanto que es
esencialmente autoadjunto si $\nu\notin[0,1)$.

\bs

Si el operador es esencialmente autoadjunto, como es el caso
considerado en \cite{Callias1}, las condiciones de contorno en la
singularidad est{\'a}n un{\'\i}vocamente determinadas. Por el
contrario, el caso en el que $0\leq\nu <1$, es\-tu\-dia\-do en el
cap{\'\i}tulo \ref{apli} de esta Tesis, admite un conjunto
infinito de condiciones de contorno en $x=0$. Como veremos, la
presencia de esta va\-rie\-dad de condiciones de contorno posibles
es esencial para obtener potencias de $t$ con exponentes
dependientes de $\nu$.

\bs

La relevancia de las condiciones de contorno para el estudio de
operadores de la forma (\ref{sch}) con un potencial singular dado
por (\ref{mod}) se deduce de un argumento di\-men\-sio\-nal que
desarrollaremos con mayor detalle en el cap{\'\i}tulo \ref{proy}.
En efecto, como el orden de la singularidad $x^{-2}$ coincide con
el orden del operador ($d=2$), {\'e}ste es ``formalmente''
invariante de escala. Por consiguiente, el problema presenta esta
simetr{\'\i}a si el dominio del operador o,
e\-qui\-va\-len\-te\-men\-te, las condiciones de contorno que lo
definen son tambi{\'e}n invariantes de escala. Utilizando la
teor{\'\i}a de von Neumann para las extensiones autoadjuntas, que
expondremos en el cap{\'\i}tulo \ref{sae}, puede probarse que
existen dos condiciones de contorno particulares, esto es, dos
extensiones autoadjuntas, que presentan esta invariancia. Estas
dos condiciones de contorno no involucran, por lo
tanto, ning{\'u}n par{\'a}metro con dimensiones. Es plausible,
por ello, que los exponentes del desarrollo asint{\'o}tico del
heat-kernel respondan, para estas dos extensiones, al resultado
(\ref{resu}).

\bs

Para comprender esto observemos, en primer lugar, que los
au\-to\-va\-lo\-res $\lambda_n$ del o\-pe\-ra\-dor de
Schr\"odinger tienen dimensi{\'o}n correspondiente a la inversa
del cuadrado de una longitud, $L^{-2}$. Por lo tanto, el
par{\'a}metro $t$ en la traza ${\rm Tr\, e^{-tA}}$ tiene
dimensi{\'o}n $L^{2}$ y, consecuentemente, el coeficiente de la
potencia $t^{-s_n}$ en el desarrollo asint{\'o}tico del
heat-kernel tiene dimensiones $L^{2s_n}$. Si, con excepci{\'o}n
del t{\'e}rmino singular proporcional a $x^{-2}$, el potencial es
anal{\'\i}tico en la coordenada entonces el operador s{\'o}lo
involucra pa\-r{\'a}\-me\-tros cuyas dimensiones son potencias
enteras de $L$. Como, adem{\'a}s, las dos condiciones de contorno
invriantes de escala no involucran ning{\'u}n par{\'a}metro adicional
con dimensiones, se puede probar que las dimensiones de los
coeficientes de la potencia $t^{-s_n}$ son potencias enteras de
$L$. De modo que, para estas dos extensiones autoadjuntas
particulares, los exponentes $s_n$ y, en consecuentemente los
polos de la funci{\'o}n-$\zeta$, deben ser semienteros, como
indica la ecuaci{\'o}n (\ref{resu}).

\bs

Contrariamente, las dem{\'a}s extensiones autoadjuntas que admite
el operador di\-fe\-ren\-cial involucran un par{\'a}metro $\theta$
cuya dimensi{\'o}n, como se deduce de la condici{\'o}n de contorno
correspondiente (v{\'e}ase el cap{\'\i}tulo \ref{sae}), es una
potencia de $L$ dependiente del par{\'a}metro adimensional $\nu$
del t{\'e}rmino singular. Si los coeficientes del desarrollo
asint{\'o}tico del heat-kernel dependen de la extensi{\'o}n
autoadjunta est{\'a}n, entonces, relacionados con el par{\'a}metro
$\theta$ por lo que resulta plausible que sus dimensiones sean
potencias de $L$ que, al igual que el exponente de $t$, dependan
tambi{\'e}n de $\nu$.

\bs

En conclusi{\'o}n, existen dos extensiones autoadjuntas
invariantes de escala para las que los exponentes de $t$ en el desarrollo
asint{\'o}tico del heat-kernel obedecen a la e\-cua\-ci{\'o}n
(\ref{resu}). Las condiciones de contorno de las restantes
extensiones autoadjuntas est{\'a}n caracterizadas por un
par{\'a}metro $\theta$ cuya dimensi{\'o}n depende del coeficiente
del t{\'e}rmino singular del o\-pe\-ra\-dor diferencial. Para
estas extensiones autoadjuntas los coeficientes del desarrollo
asint{\'o}tico del heat-kernel involucran al par{\'a}metro
$\theta$ y, por consiguiente, los exponentes de las potencias de
$t$ tambi{\'e}n dependen del coeficiente del t{\'e}rmino singular
en el operador.

\bs

Esta dependencia, uno de los resultados centrales de esta Tesis,
ser{\'a} enunciada en el Teorema \ref{elthm11}.

\bs

La demostraci{\'o}n de este Teorema se basa en una
generalizaci{\'o}n de la f{\'o}rmula de Krein. Esta f{\'o}rmula
relaciona las resolventes de dos extensiones autoadjuntas de un
o\-pe\-ra\-dor regular. En el Teorema \ref{elthmenun} extendemos
este resultado al caso de un operador de Schr\"odinger
(\ref{sch}), con $x\in\mathbb{R}^+$, cuyo potencial presenta un
t{\'e}rmino singular proporcional a $x^{-2}$. De esta manera
podemos expresar la resolvente $(A^\theta-\lambda)^{-1}$ de una
extensi{\'o}n autoadjunta arbitraria $A^\theta$ del operador $A$
como una combinaci{\'o}n lineal de las resolventes de las dos
extensiones autoadjuntas caracterizadas por condiciones de contorno invariantes
de escala. Los coeficientes de esta combinaci{\'o}n lineal
presentan desarrollos asint{\'o}ticos en potencias de $\lambda$
cuyos exponentes dependen del coeficiente del t{\'e}rmino singular
del operador diferencial. Este es el origen de la dependencia con
la intensidad de la singularidad de los exponentes de las
potencias de $t$ en el desarrollo asint{\'o}tico del heat-kernel y
de la posici{\'o}n de los polos de la funci{\'o}n-$\zeta$ del
operador.

\section{Aplicaciones en Teor{\'\i}a Cu{\'a}ntica de Campos}

\subsection{Funciones espectrales en Teor{\'\i}a Cu{\'a}ntica de Campos}\label{fetcc}

Las funciones espectrales encuentran aplicaci{\'o}n en diversas
{\'a}reas de la f{\'\i}sica y la matem{\'a}tica. Completas
descripciones y referencias acerca de estas aplicaciones pueden
encontrarse en los trabajos de K.\ Kirsten \cite{Klaus1}, D.\
Vassilevich \cite{Vass1}, G.\ Esp{\'o}sito \cite{Esposito:1997mw}
y E.\ Elizalde {\it et al.\ }\cite{EORBZ}.

\bs

En Teor{\'\i}a Cu{\'a}ntica de Campos las amplitudes de
dispersi{\'o}n de part{\'\i}culas representadas por campos
cu{\'a}nticos $\phi(x)$ se calculan en t{\'e}rminos de las
funciones de Green definidas como el valor de expectaci{\'o}n de
vac{\'\i}o del producto ordenado temporalmente de esos campos. En
la formulaci{\'o}n basada en la integral funcional, esos valores
de expectaci{\'o}n est{\'a}n dados por,
\begin{equation}\label{valexp}
    <0|T\left\{\phi(x_1)\ldots\phi(x_n)\right\}|0>=
    \frac{1}{N}\int\mathcal{D}\phi\,\phi(x_1)\ldots\phi(x_n)\,e^{-\frac{1}{\hbar}\,S[\phi]}\,,
\end{equation}
donde $N$ es una constante de normalizaci{\'o}n y $S[\phi]$ es la
acci{\'o}n que rige la din{\'a}mica de los campos $\phi(x)$
evaluados en una variedad de base $M$. La integral funcional
(\ref{valexp}) puede calcularse perturbativamente, con respecto a
$\hbar$, utilizando la t{\'e}cnica de los diagramas de Feynmann.
Este c{\'a}lculo permite regularizar, perturbativamente, los
par{\'a}metros que intervienen en la acci{\'o}n $S[\phi]$ mediante
la introducci{\'o}n de contrat{\'e}rminos en el lagrangiano. Las
funciones espectrales proveen un mecanismo de regularizaci{\'o}n
al primer orden perturbativo, esto es, $O(\hbar)$, equivalente al
c{\'a}lculo de los diagramas de Feynmann con 1-loop.

\bs

Consideremos la funcional,
\begin{equation}\label{Z}
    Z[J]=\frac{1}{N}\int\mathcal{D}\phi\,e^{-\frac{1}{\hbar}\,S[\phi]+
    \frac{1}{\hbar}\,(J,\phi)}\,,
\end{equation}
donde $(\cdot,\cdot)$ es el producto interno en el espacio de
funciones sobre la variedad de base $M$. La funcional $Z[J]$ se
denomina funcional generatriz, pues sus derivadas funcionales
permite calcular los valores de expectaci{\'o}n (\ref{valexp}),
\begin{equation}
    <0|T\left\{\phi(x_1)\ldots\phi(x_n)\right\}|0>=\hbar^n
    \frac{\delta}{\delta J(x_1)}\ldots
    \frac{\delta}{\delta J(x_n)}\,Z[J]
\end{equation}
Por consiguiente, la informaci{\'o}n obtenida perturbativamente
utilizando los diagramas de Feynmann se encuentra contenida en la
funcional $Z[J]$. Las contribuciones de los diagramas conexos
est{\'a}n dadas por la funcional,
\begin{equation}
    W[J]=\hbar\log Z[J]\,.
\end{equation}
Por su parte, las contribuciones de los diagramas conexos e
irreducibles a una part{\'\i}cula est{\'a}n dados por la
transformada de Legendre $\Gamma[\phi_J]$ de $W[J]$,
\begin{equation}\label{tl}
    \Gamma[\phi_J]=(J,\phi_J)-W[J]\,,
\end{equation}
donde,
\begin{equation}
    \phi_J(x):= \frac{\delta}{\delta J(x)}W[J]\,.
\end{equation}
Esta ecuaci{\'o}n permite expresar a la fuente $J(x)$ en
t{\'e}rminos del campo $\phi_J(x)$, en virtud de lo cual podemos
eliminar en la ecuaci{\'o}n (\ref{tl}) la dependencia de la
funcional $\Gamma[\phi_J]$ con $J(x)$.

\bs

En consecuencia, el campo $\phi_J(x)$ satisface,
\begin{equation}\label{phij}
    \frac{\delta}{\delta\phi_J(x)}\Gamma[\phi_J]=J(x)\,.
\end{equation}
La funcional $\Gamma[\phi_J]$ se denomina acci{\'o}n efectiva pues
coincide, a orden dominante en $\hbar$, con la acci{\'o}n
cl{\'a}sica $S[\phi_J]$. Haciendo una traslaci{\'o}n en la
variable de integraci{\'o}n $\phi$ de (\ref{Z}), obtenemos la
siguiente expresi{\'o}n para la acci{\'o}n efectiva,
\begin{equation}
    \Gamma[\phi_J]=-\hbar\log\left[
    \frac{1}{N}\int\mathcal{D}\phi\, e^{-\frac{1}{\hbar}S[\phi_J+\phi]+
    \frac{1}{\hbar}(\phi,J)}
    \right]\,.
\end{equation}
Si desarrollamos la acci{\'o}n $S[\phi_J+\phi]$ alrededor del
campo $\phi_J(x)$ obtenemos,
\begin{equation}\label{ae}
    \Gamma[\phi_J]=-\hbar\log\left[\,
    \frac{1}{N}\,e^{-\frac{1}{\hbar}S[\phi_J]}\,
    \int\mathcal{D}\phi\,e^{-\frac{1}{2\hbar}(\phi,A\cdot\phi)+\ldots}\,
    \right]\,,
\end{equation}
siendo,
\begin{equation}\label{opa}
    A\cdot\phi(x):=\int_M\left.\frac{\delta^2\,S[\phi]}
    {\delta\phi(x)\delta\phi(x')}
    \right|_{\phi=\phi_J}
    \phi(x')\,.
\end{equation}
Como la acci{\'o}n es local el operador $A$ es un operador
diferencial. N{\'o}tese tambi{\'e}n que, al orden que estamos
considerando, los t{\'e}rminos lineales en los campos en
(\ref{ae}) se cancelan en virtud de la ecuaci{\'o}n (\ref{phij}).

\bs

Si representamos la variable de integraci{\'o}n $\phi$ en
(\ref{ae}) en t{\'e}rminos de las autofunciones del operador $A$ y
definimos la constante ``infinita'' $N$ convenientemente, la
integral en (\ref{ae}) satisface,
\begin{equation}
    \int\mathcal{D}\phi\,e^{-\frac{1}{2\hbar}(\phi,A\cdot\phi)}\sim
    {\rm Det}^{-1/2}(A)\,.
\end{equation}
Este resultado es v{\'a}lido para campo bos{\'o}nicos. En el caso
de campos fermi{\'o}nicos, la integral es proporcional al
determinante del operador $A$.

\bs

Obtenemos finalmente la aproximaci{\'o}n al orden de 1-loop de la
acci{\'o}n efectiva,
\begin{equation}\label{ae1}
    \Gamma[\phi_J]=S[\phi _J]+\frac{\hbar}{2}\log{\rm Det}(A)
    +O(\hbar^2)\,
\end{equation}
Como $A$ es un operador diferencial su determinante debe ser
correctamente definido. En efecto, si designamos por
$\{\lambda_n\}_{n\in\mathbb{N}}$ al espectro del operador, el
producto de sus autovalores $\prod _{n}\lambda_n$ resulta
divergente por lo que requiere una ``regularizaci{\'o}n'', esto
es, una definici{\'o}n consistente que conduzca a un resultado
finito. Esto equivale a las divergencias que se encuentran en el
c{\'a}lculo perturbativo a 1-loop. Las funciones espectrales
proveen algunas t{\'e}cnicas para regularizar el determinante de
un operador diferencial y, consecuentemente, regularizar la
acci{\'o}n efectiva.

\bs

Para introducir una de las regularizaciones del determinante de
operadores diferenciales com{\'u}nmente utilizadas consideremos la
identidad,
\begin{equation}
    \log{a}-\log{b}=-\int_0^\infty\frac{dt}{t}\,\left(e^{-ta}-e^{-tb}\right)\,,
\end{equation}
v{\'a}lida para $a,b\in\mathbb{R}^+$. Podemos entonces proponer,
\begin{equation}\label{logdet}
    \log{\rm Det}(A)\sim-\int_0^\infty\frac{dt}{t}\,
    {\rm Tr}\,e^{-tA}\,.
\end{equation}
No obstante, como ${\rm Tr}\,e^{-tA}\sim t^{-m/d}$ cuando
$t\rightarrow 0^+$ (v{\'e}ase la ecuaci{\'o}n (\ref{teohea})), la
integral en (\ref{logdet}) no es convergente y requiere una
regularizaci{\'o}n. Definimos entonces,
\begin{equation}
    \log{\rm Det}(A):=\left.-\mu^{2s}\int_0^\infty dt\,
    t^{s-1}{\rm Tr}\, e^{-tA}\right|_{s=0}\,,
\end{equation}
siendo $\mu$ un par{\'a}metro con dimensiones de energ{\'\i}a.

\bs

De acuerdo con la relaci{\'o}n entre la traza del heat-kernel y la
funci{\'o}n-$\zeta$ que probaremos en la secci{\'o}n
\ref{funcionesespectrales} (v{\'e}ase la ecuaci{\'o}n
(\ref{zethea})),
\begin{eqnarray}\label{deter}
    \log{\rm Det}(A)=
    \left.-\mu^{2s}\Gamma(s)\zeta(s)\right|_{s=0}=\nonumber\\=
    -\frac{\zeta(0)}{s}-
    2\log{\mu}\,\zeta(0)+\gamma_E\,\zeta(0)-\zeta'(0)\,.
\end{eqnarray}
Como la funci{\'o}n $\zeta(s)$ es regular en $s=0$
\cite{Seeley1,Seeley3,Seeley4}, las divergencias del determinante
del operador $A$ est{\'a}n representadas por el primer t{\'e}rmino
de la ecuaci{\'o}n (\ref{deter}), que es proporcional a
$\zeta(0)$. En la secci{\'o}n \ref{funcionesespectrales} se
demuestra que este valor de la funci{\'o}n-$\zeta$ est{\'a} dado
por uno de los coeficientes del desarrollo asint{\'o}tico de la
traza del heat-kernel (v{\'e}ase la ecuaci{\'o}n (\ref{zetcero})),
\begin{equation}
    \zeta(0)=c_m(A)\,,
\end{equation}
siendo $m$ la dimensi{\'o}n de la variedad de base $M$. Es
importante observar en este punto que, dado que el t{\'e}rmino
divergente en la acci{\'o}n efectiva resulta proporcional al
coeficiente $c_m(A,x)$, es esencial que este coeficiente dependa
localmente del campo de modo que la divergencia pueda removerse
mediante una redefinici{\'o}n de los par{\'a}metros del
lagrangiano (v{\'e}anse las expresiones
(\ref{coefi1}-\ref{coefiult}).)

\bs

A modo de ilustraci{\'o}n consideremos un campo escalar sin carga
$\phi(x)$ con un lagrangiano dado por,
\begin{equation}\label{lag}
    \mathcal{L}=\partial_\mu\phi\cdot\partial_\mu\phi+
    m^2\phi^2+\lambda\phi^4\,.
\end{equation}
Suponemos que el campo est{\'a} definido en una variedad $M$
eucl{\'\i}dea, plana y compacta. Omitiremos adem{\'a}s los efectos
del borde $\partial M$.

\bs

De acuerdo con las expresiones (\ref{opa}) y (\ref{lag}), el
operador $A$ resulta, en este caso,
\begin{equation}
    A=-\Delta+m^2+6\lambda\,\phi_J(x)\,.
\end{equation}
En consecuencia, si la variedad de base $M$ tiene dimensi{\'o}n
$m=4$ (v{\'e}ase la ecuaci{\'o}n (\ref{coefi4})),
\begin{equation}\label{zetacero}
    \zeta(0)=c_4(A)=\frac{3}{8\pi^2}\int_M
    \left(m^2\lambda\,\phi_J^2(x)
    +3\lambda^2\phi_J^4(x)
    \right)\,.
\end{equation}
Vemos que la dependencia de $\zeta(0)$ con el campo $\phi_J(x)$
permite remover los t{\'e}rminos divergentes de la acci{\'o}n
efectiva redefiniendo los par{\'a}metros $m$ y $\lambda$. De modo
que, teniendo en cuenta las ecuaciones (\ref{ae1}), (\ref{deter})
y (\ref{zetacero}), la acci{\'o}n efectiva resulta finita si
introducimos en el lagrangiano los contrat{\'e}rminos de modo que,
\begin{eqnarray}
    m^2\rightarrow m^2\left[
    1+\hbar\,\frac{3}{16\pi^2}\,\lambda
    \,\frac{1}{s}+O(\hbar^2)
    \right]\,,\\
    \lambda\rightarrow\lambda\left[
    1+\hbar\,\frac{9}{16\pi^2}\,\lambda
    \,\frac{1}{s}
    +O(\hbar^2)
    \right]\,.
\end{eqnarray}

\bs

Mencionamos tambi{\'e}n la regularizaci{\'o}n del tiempo propio
que consiste en modificar la integral (\ref{logdet}),
\begin{equation}\label{regu2}
    \log{\rm Det}(A)=-\int_{\Lambda^{-2}}^\infty\frac{dt}{t}\,
    {\rm Tr}\,e^{-tA}\,,
\end{equation}
que permite expresar los t{\'e}rminos divergentes de la acci{\'o}n
efectiva en t{\'e}rminos de la energ{\'\i}a de corte $\Lambda$. De
ese modo, las divergencias ultravioletas est{\'a}n determinadas
por los primeros t{\'e}rminos del desarrollo asint{\'o}tico del
heat-kernel.

\bs

Existe una regularizaci{\'o}n ``anal{\'\i}tica'' del determinante
basada en la funci{\'o}n $\zeta(s)={\rm Tr}\,A^{-s}$. Si el
operador $A$ act{\'u}a sobre un espacio de dimensi{\'o}n finita y
tiene un conjunto de autovalores dado por
$\{\lambda_n\}_{n=1,\ldots,N}$ entonces se verifica,
\begin{eqnarray}
    \log{\rm Det}\,(A/\mu^2)=
    \log\prod_{n=1}^{N}\frac{\lambda_n}{\mu^2}=
    \sum_{n=1}^{N}\log{\lambda_n}-2\log{\mu}\,N=\nonumber\\=
    -\left.\frac{d}{ds}\left(\sum_{n=1}^N\lambda_n^{-s}\right)
    -2\log{\mu}\sum_{n=1}^N \lambda_n^{-s}\ \right|_{s=0}\,.
\end{eqnarray}
Generalizando esta igualdad al caso de operadores en espacios de
dimensi{\'o}n infinita, definimos
\cite{Hawking:1976ja,Ray-Singer},
\begin{equation}
    \log{\rm Det}\,(A):= -\zeta'(0)-2\log{\mu}\,\zeta(0)\,.
\end{equation}
Esta definici{\'o}n conduce a una regularizaci{\'o}n finita de las
constantes de acoplamiento.

\bs

Las funciones espectrales poseen informaci{\'o}n acerca de otras
cantidades en Teor{\'\i}a Cu{\'a}ntica de Campos. Si el
lagrangiano s{\'o}lo posee t{\'e}rmino cuadr{\'a}ticos se puede
ver, completando cuadrados en la expresi{\'o}n (\ref{Z}), que la
funcional generatriz $Z_0[J]$ de los campos libres est{\'a} dada
por,
\begin{equation}
    Z_0[J]=\frac{1}{N}\,
    e^{-\frac{1}{4}\int_{M\times M}J(x)\,A^{-1}(x,x')\,J(x')}\,
    {\rm Det}^{-1/2}(A).
\end{equation}
Las funciones espectrales aportan informaci{\'o}n acerca de la
estructura de singularidades del propagador $A^{-1}(x,x')$. En
efecto, en la secci{\'o}n \ref{funcionesespectrales} veremos que
el n{\'u}cleo de la resolvente $(A-\lambda)^{-1}(x,x')$ es la
transformada de Laplace del n{\'u}cleo del heat-kernel
$e^{-tA}(x,x')$. En consecuencia, el propagador o funci{\'o}n de
Green $A^{-1}(x,x')$ est{\'a} dado por,
\begin{equation}
    A^{-1}(x,x')=\int_0^\infty dt\,e^{-tA}(x,x')\,.
\end{equation}
Por consiguiente, el desarrollo asint{\'o}tico del n{\'u}cleo del
heat-kernel para peque{\~n}os valores de $t$ determina las
singularidades que presenta el propagador en puntos coincidentes
$x=x'$.

\bs

Por otra parte, el c{\'a}lculo de la energ{\'\i}a de Casimir del
campo electromagn{\'e}tico cu{\'a}ntico tambi{\'e}n conduce a la
aparici{\'o}n de cantidades divergentes de la forma
$\hbar/2\cdot\sum_{n}\omega_n$ siendo
$\{\omega_n\}_{n\in\mathbb{N}}$ las frecuencias de los modos
electromagn{\'e}ticos. Las funciones espectrales son utilizadas,
en este contexto, para regularizar la energ{\'\i}a de Casimir. Una
des\-crip\-ci{\'o}n ilustrada con ejemplos de la
regularizaci{\'o}n de la energ{\'\i}a de las oscilaciones del
vac{\'\i}o puede consultarse en \cite{Marielosky}. En el trabajo
de M.\ Bordag {\it et al.\ }\cite{Bordag:2001qi} se encuentra un
estudio detallado del efecto Casimir y de sus aplicaciones. Los
resultado recientes pueden consultarse en el trabajo de K.A.\
Milton \cite{Milton:2004ya}.

\bs

Finalmente, mencionemos que las funciones espectrales permiten
tambi{\'e}n el c{\'a}lculo de anomal{\'\i}as en Teor{\'\i}a
Cu{\'a}ntica de Campos. Las anomal{\'\i}as representan una
variaci{\'o}n de la acci{\'o}n efectiva ante un grupo de
simetr{\'\i}a de la acci{\'o}n cl{\'a}sica. Teniendo en cuenta las
ecuaciones (\ref{ae1}) y (\ref{deter}) vemos que las
anomal{\'\i}as est{\'a}n determinadas por las propiedades de
transformaci{\'o}n de la funci{\'o}n $\zeta_A(s)$ ante el grupo de
simetr{\'\i}a\,\footnote{Estas identidades pueden demostrarse
formalmente en base a las t{\'e}cnicas de \cite{APS3,RS}.},
\begin{equation}\label{deltazeta}
    \delta\zeta(s)=\delta\left({\rm Tr}\,A^{-s}\right)=
    -s{\rm Tr}\,\left( \delta A\,A^{-s-1}\right)\,.
\end{equation}
Consideremos, por ejemplo, la anomal{\'\i}a conforme que est{\'a}
dada por la variaci{\'o}n de la acci{\'o}n efectiva frente a la
transformaci{\'o}n $g_{\mu\nu}(x)\rightarrow
e^{2\varphi(x)}g_{\mu\nu}(x)$ y puede escribirse en t{\'e}rminos
de la traza del tensor energ{\'\i}a impulso $T_{\mu\nu}(x)$,
\begin{eqnarray}\label{ano}
    \delta\Gamma[\phi]=\int_M\frac{\delta\Gamma[\phi]}
    {\delta g_{\mu\nu}(x)}\,
    \delta g^{\mu\nu}(x)\,d\mu(x)=-2
    \int_M\sqrt{g}\,T_{\mu\nu}(x)\,
    g^{\mu\nu}(x)\,\delta\varphi(x)\,d\mu(x)=\nonumber\\=
    \int_M\sqrt{g}\,T_{\mu}^\mu(x)\,\delta\varphi(x)\,d\mu(x)\,.
\end{eqnarray}
Si la teor{\'\i}a cl{\'a}sica posee simetr{\'\i}a conforme la
transformaci{\'o}n del operador $A$ est{\'a} dada por
$A\rightarrow e^{-2\varphi(x)}A$ de modo que la variaci{\'o}n
conforme de la funci{\'o}n $\zeta_A(s)$ es, de acuerdo con la
expresi{\'o}n (\ref{deltazeta}),
\begin{equation}
    \delta\zeta(s)=2s{\rm Tr}\,\left( \delta\varphi(x)A^{-s}\right)\,.
\end{equation}
Como la cantidad ${\rm Tr}\,\left( \delta\varphi(x)A^{-s}\right)$
es regular en $s=0$ \cite{Gilkey}, la transformaci{\'o}n conforme
de la funci{\'o}n $\zeta_A(s)$ en $s=0$ verifica,
\begin{equation}\label{zetaconf}
    \delta\zeta_A(0)=0\,,\qquad
    \delta\zeta'_A(0)=2\left.{\rm Tr}\,
    \left( \delta\varphi(x)A^{-s}\right)\right|_{s=0}\,.
\end{equation}
De acuerdo con las ecuaciones (\ref{ae1}) y (\ref{deter}), junto
con (\ref{zetaconf}), conclu{\'\i}mos que la anomal{\'\i}a
conforme no es divergente ni depende del par{\'a}metro de escala
$\mu$. Utilizando adem{\'a}s la ecuaci{\'o}n (\ref{ano}) y la
relaci{\'o}n entre ${\rm Tr}\,(\delta\varphi(x)A^{-s})$ y ${\rm
Tr}\,(\delta\varphi(x)e^{-tA})$, similar a la ecuaci{\'o}n
(\ref{zethea}), obtenemos la correcci{\'o}n cu{\'a}ntica a 1-loop
de la traza del tensor energ{\'\i}a impulso,
\begin{equation}
    T_\mu^\mu=-\hbar\,c_m(A,x)+O(\hbar^2)\,.
\end{equation}

\bs

La expresi{\'o}n (\ref{deltazeta}) puede aplicarse tambi{\'e}n al
c{\'a}lculo de la anomal{\'\i}a quiral\,\footnote{El m{\'e}todo de
K.\ Fujikawa para determinar la anomal{\'\i}a quiral \cite{fuji}
equivale a la regularizaci{\'o}n del determinante mediante la
ecuaci{\'o}n (\ref{regu2}).} y de esta manera se obtiene una
relaci{\'o}n entre las funciones espectrales y el Teorema del
{\'\i}ndice.

\subsection{Potenciales singulares}

Los fundamentos de la teor{\'\i}a de los {\'\i}ndices de
deficiencia de von Neumann para el estudio de las extensiones
autoadjuntas pueden consultarse en \cite{R-S,A-G}. La
contribuci{\'o}n de las singularidades aisladas a los {\'\i}ndices
de deficiencia de los operadores de Schr\"odinger es estudiada en
\cite{Bulla:kt}. En \cite{Bonneau:1999zq} se encuentra una
discusi{\'o}n pedag{\'o}gica de las extensiones autoadjuntas.

\bs

Los operadores de Schr\"odinger definidos por un potencial
singular $U_\nu(x)$ cuyo t{\'e}rmino dominante cerca de la
singularidad tiene la forma dada por la expresi{\'o}n (\ref{mod})
han sido estudiados como modelos de invariancia conforme en
mec{\'a}nica cu{\'a}ntica \cite{DA-F-F}. Con este objetivo, se
analiza en \cite{Camblong} este tipo de potenciales mediante una
regularizaci{\'o}n dimensional en tanto que en
\cite{Bawin:2003dm,Coon} se consideran regularizaciones del
potencial en las proxi\-mi\-da\-des de la singularidad. En
\cite{Tsutsui} se describen diversos aspectos de las extensiones
autoadjuntas debidas a la presencia de singularidades.

\bs

El modelo de Calogero es un sistema exactamente resoluble
\cite{Perelomov,Polychronakos:1992zk} que describe un conjunto de
part{\'\i}culas id{\'e}nticas en interacci{\'o}n mediante un
potencial de la forma (\ref{mod}). En sus trabajos, Calogero
\cite{Calogero} (v{\'e}ase tambi{\'e}n un tratamiento algebraico
en \cite{Brink:1992xr,Polychronakos:1996rj,Gurappa:1997jb}) impone
la anulaci{\'o}n de la funci{\'o}n de onda en los puntos en que
dos part{\'\i}culas coinciden. Sin embargo, condiciones de
contorno m{\'a}s generales, determinadas por las extensiones
autoadjuntas del hamiltoniano, han sido consideradas en este
modelo en presencia de potenciales confinantes \cite{Basu} y en
ausencia de ellos \cite{Basu1}.

\bs

Potenciales de la forma (\ref{mod}) adquieren especial inter{\'e}s
en teor{\'\i}as supersim{\'e}tricas pues proveen un mecanismo de
ruptura espont{\'a}nea de la supersimetr{\'\i}a en modelos de
mec{\'a}nica cu{\'a}ntica con superpotenciales singulares
\cite{FP2}. Este mecanismo se debe a que, en general, las
condiciones de contorno admisibles en la singularidad no respetan
la supersimetr{\'\i}a. La ruptura es\-pon\-t{\'a}\-ne\-a debida a
la presencia de singularidades \cite{Jevicki,RR,Cooper-K-S} ha
generado cierta controversia pues en
\cite{Pernice,Das,Gangopadhyaya:2002jr} se sostiene que una
regularizaci{\'o}n del potencial a\-pro\-pia\-da preserva la
supersimetr{\'\i}a.

\bs

Por otra parte, en virtud de la presencia de singularidades de la
forma (\ref{mod}) existe una descripci{\'o}n microsc{\'o}pica de
los agujeros negros en las proximidades del horizonte en
t{\'e}rminos de modelos de invariancia conforme
\cite{Claus,Gibbons,Solodukhin:1998tc}. {\it E.g.}, en
\cite{Birmingham} se considera un campo escalar en las
proximidades del horizonte de una m{\'e}trica de Schwarzschild
me\-dian\-te un m{\'e}todo algebraico y se estudia la relevancia
del {\'a}lgebra de Virasoro y la in\-va\-rian\-cia de escala en
relaci{\'o}n con las extensiones autoadjuntas. Las extensiones
autoadjuntas del operador de Klein-Gordon en distintos tipos de
agujeros negros son tambi{\'e}n consideradas en
\cite{Govindarajan,Moretti}.

\bs

La dispersi{\'o}n de part{\'\i}culas bos{\'o}nicas y
fermi{\'o}nicas por agujeros negros de\linebreak Schwarzschild ha
sido calculada en \cite{unruh} (v{\'e}anse tambi{\'e}n los
estudios detallados de \cite{Sanchez:1977vz,fhm}.) Sin embargo,
c{\'a}lculos recientes \cite{Kuchiev:2003ez,Kuchiev:2003fy}
basados en el comportamiento de las funciones de onda asociadas a
potenciales (\ref{mod}) muestran efectos cu{\'a}nticos que se
manifiestan en una reducci{\'o}n de la secci{\'o}n eficaz de
absorci{\'o}n y en la posibilidad de emisi{\'o}n de
part{\'\i}culas cl{\'a}sicamente confinadas al interior del
horizonte del agujero negro.

\bs

Adem{\'a}s de su aplicaci{\'o}n al c{\'a}lculo de la entrop{\'\i}a
de agujeros negros, las sin\-gu\-la\-ri\-da\-des de la forma
(\ref{mod}) tienen inter{\'e}s en el estudio de cuerdas
c{\'o}smicas \cite{Vilenkin:1984ib,Kim:2002ek} y en Teor{\'\i}a de
Cuerdas \cite{DJtH,DJ,Balasubramanian:2000rt}.

\bs

La construcci{\'o}n de una teor{\'\i}a cu{\'a}ntica de la gravedad
exige el desarrollo de la Teor{\'\i}a Cu{\'a}ntica de Campos en
espacios curvos \cite{B-O-S} en el contexto de la cual, como hemos
mencionado, las funciones espectrales determinan los
contrat{\'e}rminos del lagrangiano necesarios para renormalizar la
teor{\'\i}a al orden de 1-loop \cite{Wiesendanger:1993mw}.

\bs

Aunque se conoce el comportamiento de los desarrollos
asint{\'o}ticos de las funciones espectrales para el caso de
variedades de base suaves, s{\'o}lo han sido resueltos problemas
de Teor{\'\i}a Cu{\'a}ntica de Campos en espacios con
singularidades en algunos casos particulares (v{\'e}anse {\it
e.g.\ }\cite{Chang:1992fu,Fursaev:1993hm,Aurell:1993jb}.)

\bs

Operadores diferenciales con un potencial singular de la forma
(\ref{mod}) se obtienen a partir del estudio del laplaciano en
variedades con singularidades c{\'o}nicas; el par{\'a}metro $\nu$
que ca\-rac\-te\-ri\-za la intensidad de la singularidad est{\'a},
en este caso, relacionado con el {\'a}ngulo de deficiencia del
cono. El desarrollo asint{\'o}tico del heat-kernel del laplaciano
en variedades con singularidades c{\'o}nicas fue tratado,
quiz{\'a}s por primera vez, en \cite{carcone,somcone}. J.J.\
Cheeger \cite{checone} realiza un tratamiento detallado del
problema en el que se prueba que la contribuci{\'o}n de la
singularidad al desarrollo asint{\'o}tico de la traza del
heat-kernel puede calcularse en las proximidades de la
singularidad. En \cite{brusee} se calcula el desarrollo de la
traza del heat-kernel para ciertos operadores singulares que
incluyen al laplaciano en un cono para la extensi{\'o}n de
Friedrichs.

\bs

Existen numerosos trabajos en Teor{\'\i}a Cu{\'a}ntica de Campos
referidos a variedades con singularidades c{\'o}nicas. En
\cite{Cognola:1993qg} se estudia la acci{\'o}n efectiva al orden
de 1-loop de campos escalares con masa y autointeracci{\'o}n sobre
singularidades c{\'o}nicas. Se han calculado tambi{\'e}n
propiedades del heat-kernel para campos con spines mayores en
\cite{Larsen:1995ax} en conexi{\'o}n con las correcciones
cu{\'a}nticas a la entrop{\'\i}a de los agujeros negros.

\bs

En \cite{Fursaev:1996uz} se calculan las correcciones a los
primeros coeficientes del desarrollo a\-sin\-t{\'o}\-ti\-co del
heat-kernel del laplaciano para campos de spin 1/2, 3/2 y 2 y se
obtienen las divergencias ultravioletas cu{\'a}nticas para la
entrop{\'\i}a de agujeros negros. Es interesante notar que los
resultados no siempre se reducen al caso de una variedad suave
cuando se considera el l{\'\i}mite en el que el {\'a}ngulo de
deficiencia que define al cono tiende a creo. Se sugiere que esto
puede relacionarse con la formulaci{\'o}n de
\cite{Teitelboim:1995fr,Carlip:1993sa} en la que el {\'a}ngulo de
deficiencia es una variable cu{\'a}ntica conjugada al {\'a}rea del
horizonte.

\bs

El heat-kernel en casos especiales de variedades con
singularidades c{\'o}nicas tambi{\'e}n es tratado en detalle en
\cite{Dowker:1977zj,Fursaev:1993qk,Fursaev:1995ef,Bordag:1996fw,DeNardo:1996kp,Dowker:1998sc}.
Sin embargo, en ninguno de estos casos se consideran las
condiciones de contorno m{\'a}s generales que admite el operador
en la singularidad. De este modo, se calculan las contribuciones
de la singularidad a los coeficientes del desarrollo
asint{\'o}tico del heat-kernel en los casos en que los exponentes
de las potencias de $t$ son las usuales (v{\'e}ase la ecuaci{\'o}n
(\ref{teohea}).)

\bs

En \cite{Mooers} E.\ Mooers estudia las extensiones autoadjuntas
del operador laplaciano de\-fi\-ni\-do sobre las formas
diferenciales de una variedad con una singularidad c{\'o}nica y
encuentra que el desarrollo a\-sin\-t{\'o}\-ti\-co de la traza del
heat-kernel posee potencias que dependen del {\'a}ngulo de
deficiencia de la variedad. Este resultado coincide con nuestro
estudio del o\-pe\-ra\-dor (\ref{sch}) en el caso $V(x)=0$. Este
es, a nuestro entender, el {\'u}nico trabajo que describe un
desarrollo asint{\'o}tico con potencias dependientes de
par{\'a}metros externos. Por otra parte, la extensi{\'o}n de la
f{\'o}rmula de Krein que derivamos en la secci{\'o}n \ref{singu}
se basa en la formulaci{\'o}n planteada en \cite{Mooers}.

\section{Plan de la Tesis}

\subsubsection{Cap{\'\i}tulo \ref{sae}: Extensiones Autoadjuntas}

Como hemos mencionado, la posici{\'o}n inusual de los polos de la
funci{\'o}n-$\zeta$ de algunos operadores con coeficientes
singulares, dependiente de la intensidad de la singularidad, se
manifiesta en virtud de la variedad de condiciones de contorno que
hacen autoadjunto al operador. Los operadores sim{\'e}tricos no
son, en general, autoadjuntos; pero existe la posibilidad,
determinada por el valor de sus {\'\i}ndices de deficiencia, de
extender el dominio de un operador sim{\'e}trico de modo de
obtener un operador autoadjunto.

\bs

En la secci{\'o}n \ref{TvN} expondremos las ideas b{\'a}sicas
necesarias para la construcci{\'o}n de las extensiones
autoadjuntas de un operador cerrado y sim{\'e}trico. En primer
lugar, daremos las definiciones y teoremas b{\'a}sicos de la
teor{\'\i}a de los {\'\i}ndices de deficiencia de von Neumann.
Estos conceptos ser{\'a}n ilustrados mediante el ejemplo sencillo
del operador impulso $P=-i\partial_x$.

\bs

Existe una perspectiva que permite comprender algunas propiedades
topol{\'o}gicas del conjunto de extensiones autoadjuntas que
ser{\'a} presentada brevemente en la secci{\'o}n \ref{topo}. En
este contexto se definir{\'a}n las subvariedades de Cayley,
relacionadas con la to\-po\-lo\-g\'\i\-a no trivial del conjunto
de extensiones autoadjuntas, con la existencia de estados de borde
y con la ausencia de una cota inferior com{\'u}n a todas las
energ{\'\i}as de los estados fundamentales de las distintas
extensiones autoadjuntas.

\subsubsection{Cap{\'\i}tulo \ref{rupturaSUSY}: Ruptura Espont{\'a}nea de
SUSY en Mec{\'a}nica Cu{\'a}ntica}

En este cap{\'\i}tulo ilustraremos la importancia de
con\-si\-de\-rar el conjunto de todas las condiciones de contorno
que hacen autoadjunto a un operador a partir de un ejemplo en el
contexto de la mec{\'a}nica cu{\'a}ntica supersim{\'e}trica
unidimensional. La variedad de condiciones de contorno admisibles
en este caso provee un mecanismo de ruptura de la
supersimetr{\'\i}a. Esto es posible pues, aunque el operador
hamiltoniano conmuta ``formalmente'' con las supercargas, existen
condiciones de contorno que no preservan la supersimetr{\'\i}a.
Este es uno de los resultados originales de esta tesis.

\bs

En particular mostraremos que en el conjunto infinito de
extensiones autoadjuntas del hamiltoniano existen s{\'o}lamente
dos, definidas por condiciones de contorno invariantes de escala,
para las cuales puede realizarse el {\'a}lgebra de
supersimetr{\'\i}a N=2. En estos casos, el espectro del
hamiltoniano es doblemente degenerado. Sin embargo, para una de
estas extensiones la supersimetr{\'\i}a es manifiesta y existe un
estado fundamental con energ{\'\i}a nula, en tanto que para la
otra existe una ruptura espont{\'a}nea de la supersimetr{\'\i}a y
la energ{\'\i}a del estado fundamental es estrictamente positiva.

\bs

Para las restantes extensiones autoadjuntas los dominios de las
supercargas no coinciden, de modo que cada una de ellas, junto con
el hamiltoniano, constituye una realizaci{\'o}n del {\'a}lgebra de
supersimetr{\'\i}a N=1.

\subsubsection{Cap{\'\i}tulo \ref{fe}: Funciones Espectrales}

En la secci{\'o}n \ref{zc} introduciremos los espacios de Sobolev
y los operadores pseu\-do\-di\-fe\-ren\-cia\-les o de
Calder{\'o}n-Zygmund. En este contexto definiremos, en la
secci{\'o}n \ref{funcionesespectrales}, las funciones espectrales
que se obtienen de las trazas de los operadores $e^{-tA}$,
$(A-\lambda)^{-1}$ y $A^{-s}$, siendo $A$ un operador
diferencial\,\footnote{Recu{\'e}rdese que el operador $e^{-tA}$ se
define s{\'o}lamente para operadores positivos.}. Asimismo,
estableceremos las relaciones que las vinculan y las consecuentes
relaciones entre sus desarrollos asint{\'o}ticos.

\bs

Mostraremos que la traza de la resolvente ${\rm
Tr}\,(A-\lambda)^{-1}$ es la transformada de Laplace de la traza
del heat-kernel ${\rm Tr}\,e^{-tA}$ y que la funci{\'o}n
$\zeta_A(s):={\rm Tr}\, A^{-s}$ se obtiene a partir de la
transformada de Mellin de la traza del heat-kernel. Esto implica
ciertas relaciones entre sus comportamientos asint{\'o}ticos.
Veremos que, efectivamente, el comportamiento a peque\~nos valores
de $t$ de la traza del heat-kernel determina el comportamiento a
grandes valores de $|\lambda|$ de la traza de la resolvente y la
posici{\'o}n de los polos de la funci{\'o}n $\zeta_A(s)$ en el
plano complejo $s$.

\bs

En la secci{\'o}n \ref{deas} daremos una derivaci{\'o}n del
resultado (\ref{resu}) basada en la construcci{\'o}n del
desarrollo asint{\'o}tico del n{\'u}cleo de la resolvente
$(A-\lambda)^{-1}$ para grandes valores de $|\lambda|$. Esta
construcci{\'o}n, por su parte, se realiza a partir de una
aproximaci{\'o}n, para grandes valores de $|\lambda|$, del
s{\'\i}mbolo de la resolvente.

\subsubsection{Cap{\'\i}tulo \ref{opesing}: Operadores Singulares}

El cap{\'\i}tulo \ref{opesing} est{\'a} dedicado a la
descripci{\'o}n de un tipo de operadores de Schr\"odinger
singulares cuyas funciones-$\zeta$ tienen una estructura de polos
que se aparta del resultado (\ref{resu}). Estudiaremos, en
particular, operadores diferenciales $A$ de la forma:
\begin{equation}\label{delope}
    A=-\partial_x^2+\frac{\nu^2-1/4}{x^2}+V(x),
\end{equation}
donde $x$ pertenece a $\mathbb{R}^+$ o al intervalo $[0,1]$ y
$V(x)$ es una funci{\'o}n anal{\'\i}tica. El va\-lor del
par{\'a}metro $\nu\in(0,1)$, de manera que el o\-pe\-ra\-dor $A$
admite un conjunto infinito de extensiones autoadjuntas
caracterizado por un par{\'a}metro real $\theta$.

\bs

Debido a la presencia de un coeficiente singular en el operador,
los polos de la funci{\'o}n-$\zeta$ no est{\'a}n ubicados en
se\-mien\-te\-ros ne\-ga\-ti\-vos\,\footnote{T{\'e}ngase en cuenta
que, para un operador de Schr\"odinger unidimensional, $d=2$ y
$m=1$.}, como indicar{\'\i}a la ecuaci{\'o}n (\ref{resu}), sino
que sus posiciones dependen del par{\'a}metro $\nu$. En efecto, el
procedimiento descrito en la secci{\'o}n \ref{deas} para demostrar
(\ref{resu}) no puede ser aplicado a un operador con coeficientes
singulares. Por un lado, las cotas que permiten obtener un
desarrollo asint{\'o}tico para el n{\'u}cleo de la resolvente a
partir de una aproximaci{\'o}n de su s{\'\i}mbolo pierden validez
en presencia singularidades. Por otra parte, en el caso regular,
los coeficientes del desarrollo asint{\'o}tico de la traza de la
resolvente se expresan como integrales sobre la variedad de base y
sobre su borde de combinaciones lineales de potencias del
potencial y de sus derivadas\,\footnote{Estos coeficientes son, a
su vez, proporcionales a los coeficientes del desarrollo
asint{\'o}tico de la traza del heat-kernel que est{\'a}n dados por
las expresiones (\ref{666}).}; ciertamente, esto pierde sentido si
el potencial contiene un t{\'e}rmino no integrable de la forma
$x^{-2}$.

\bs

Para describir entonces la estructura de polos de la
funci{\'o}n-$\zeta$ del operador sin\-gu\-lar (\ref{delope})
tendremos en cuenta la existencia de dos extensiones autoadjuntas caracterizadas por condiciones de contorno
invariantes de escala. Los polos de las funciones-$\zeta$ de estas
dos extensiones autoadjuntas s\'\i\ est{\'a}n dados por la
ecuaci{\'o}n (\ref{resu}).

\bs

En la secci{\'o}n \ref{regu} presentaremos la f{\'o}rmula de Krein
que relaciona las resolventes de distintas extensiones
autoadjuntas de operadores regulares. A continuaci{\'o}n
cons\-trui\-re\-mos, en la secci{\'o}n \ref{singu}, una
f{\'o}rmula similar que relaciona las resolventes
co\-rres\-pon\-dien\-tes a distintas extensiones autoadjuntas del
operador con coeficientes singulares (\ref{delope}). A partir de
esta relaci{\'o}n expresaremos, en el Teorema \ref{elthmenun}, la
resolvente de una extensi{\'o}n autoadjunta arbitraria de $A$ como
combinaci{\'o}n lineal de las resolventes de las dos extensiones
caracterizadas por condiciones de contorno invarintes de escala. De esta manera, identificamos en los coeficientes
de esta combinaci{\'o}n el origen de los exponentes dependientes
de $\nu$ en el desarrollo en potencias de $\lambda$ de la
resolvente.

\bs

Finalmente, en la secci{\'o}n \ref{luego}, utilizamos el Teorema
\ref{elthmenun} para deducir el Teorema \ref{elthm11} que permite
calcular expl{\'\i}citamente el desarrollo asint{\'o}tico de la
resolvente del operador (\ref{delope}). De acuerdo con la
relaci{\'o}n entre el desarrollo asint{\'o}tico de la resolvente y
los polos de la funci{\'o}n-$\zeta$ demostrada en la secci{\'o}n
\ref{funcionesespectrales}, deducimos que la posici{\'o}n de los
polos de la funci{\'o}n-$\zeta$ del operador de Schr\"odinger
singular depende del par{\'a}metro $\nu$. En efecto, si $0<\nu<1$,
se prueba que la funci{\'o}n-$\zeta$ del operador (\ref{delope})
tiene una sucesi{\'o}n de polos en los puntos del plano complejo
dados por,
\begin{equation}\label{polos}
    s_{N,n}=-\nu\  N-\frac n 2\qquad N=1,2,\ldots\qquad n=0,1,\ldots
\end{equation}
cuyos residuos son proporcionales a $\theta^N$, siendo $\theta$ el
par{\'a}metro que caracteriza la extensi{\'o}n autoadjunta. La
funci{\'o}n-$\zeta$ posee, adem{\'a}s, una segunda serie de polos
en los puntos $1/2-n$ con $n=0,1,2,\ldots$, como indica el
resultado (\ref{resu}).

\bs

Finalizamos el cap{\'\i}tulo \ref{opesing} con algunas
consideraciones acerca del caso compacto para el que $x\in[0,1]$.
Aunque no se pudo obtener una forma expl{\'\i}cita de los polos de
la funci{\'o}n-$\zeta$ para este caso, las conclusiones de la
secci{\'o}n \ref{cccp} ser{\'a}n {\'u}tiles para verificar los
resultados del caso particular estudiado en la secci{\'o}n
\ref{Solari}.

\subsubsection{Cap{\'\i}tulo \ref{apli}: Ejemplos}

En este cap{\'\i}tulo resolveremos dos ejemplos particulares de
operadores diferenciales con coeficientes singulares para los que
el resultado (\ref{resu}) pierde validez. En ambos casos, el
estudio de las funciones espectrales se basa en la obtenci{\'o}n
de una resoluci{\'o}n espectral expl{\'\i}cita del operador. En
las Secciones \ref{Wipf} y \ref{Solari} estudiaremos operadores de
Schr\"odinger (\ref{sch}) con un t{\'e}rmino singular proporcional
a $x^{-2}$ sobre las variedades de base $\mathbb{R}^+$ y $[0,1]$,
respectivamente. Estos ejemplos ilustran los resultados obtenidos
en las Secciones \ref{singu} y \ref{luego}.

\begin{itemize}
\item

\noindent{{\bf Secci{\'o}n \ref{Wipf}: Un operador de
Schr\"odinger en una variedad de base no compacta}}

En la secci{\'o}n \ref{adjoint-H} estudiaremos el operador,
\begin{equation}\label{elo}
    A=-\partial_x^2+\frac{\nu^2-1/4}{x^2}+x^2
\end{equation}
definido sobre un subconjunto apropiado de
$\mathbf{L_2}(\mathbb{R}^+)$. El par{\'a}metro $\nu\in (0,1)$, de
modo que el operador diferencial (\ref{elo}) admite una familia de
extensiones autoadjuntas.

\bs

Como la teor{\'\i}a de von Neumann para las extensiones
autoadjuntas se aplica a o\-pe\-ra\-do\-res sim{\'e}tricos y
cerrados, definiremos primeramente un dominio de definici{\'o}n
sobre el cual (\ref{elo}) sea sim{\'e}trico. Luego calcularemos la
clausura del operador y posteriormente caracterizaremos el dominio
y la acci{\'o}n del operador adjunto. Resolviendo finalmente la
ecuaci{\'o}n de autovalores del operador adjunto determinaremos
los subespacios de deficiencia del operador que describen sus
extensiones autoadjuntas.

\bs

A partir de los resultados obtenidos en la secci{\'o}n
\ref{adjoint-H} y utilizando la teor{\'\i}a de von Neumann de los
{\'\i}ndices de deficiencia, calcularemos en la secci{\'o}n
\ref{SAE-H} las extensiones autoadjuntas del operador (\ref{elo}).
De esta manera, caracterizaremos las condiciones de contorno que
admite el problema y determinaremos el espectro de cada
extensi{\'o}n autoadjunta. En este punto, reconoceremos la
existencia de dos extensiones autoadjuntas particulares que, como
se ver{\'a}, corresponden a las condiciones de contorno
invariantes de escala a las que ya nos hemos referido. Por {\'u}ltimo,
estudiaremos el l{\'\i}mite regular $\nu\rightarrow 1/2$, esto es,
en ausencia del t{\'e}rmino singular en (\ref{elo});
verificaremos, en relaci{\'o}n con la secci{\'o}n \ref{esta}, la
existencia de estados de borde para las extensiones autoadjuntas
cercanas a la subvariedad de Cayley $\mathcal{C}_-$.

\bs

En la secci{\'o}n \ref{espoze} construiremos una
representaci{\'o}n integral de la funci{\'o}n $\zeta(s)$ del
operador (\ref{elo}) que permitir{\'a} obtener su estructura de
singularidades. De\-mos\-tra\-re\-mos que la funci{\'o}n-$\zeta$
presenta un polo simple en $s=1$ con residuo igual a $1/4$. Este
polo no obedece al resultado (\ref{resu}) debido a que la variedad
de base no es compacta. En efecto, en la secci{\'o}n
\ref{funcpart} del Ap{\'e}ndice mostraremos un argumento que
indica la posici{\'o}n del primer polo de la funci{\'o}n
$\zeta(s)$ de un operador de Schr\"odinger en una variedad de base
no compacta con un potencial homog{\'e}neo; este resultado predice
la existencia de un polo en $s=1$ si el operador est{\'a} definido
sobre $\mathbb{R}^+$ y el potencial es homog{\'e}neo de grado $2$
en el infinito. En consecuencia, no atribuiremos la existencia de
este polo que contradice el resultado (\ref{resu}) a la
singularidad del operador sino a la no compacidad de la variedad
de base.

\bs

La funci{\'o}n $\zeta(s)$ posee, adem{\'a}s, otras singularidades
ubicadas sobre el eje real que consisten en una sucesi{\'o}n
$s_{N,n}=-\nu\ N-2n$, con $N=1,2,\ldots$ y $n=0,1,\ldots$. Estos
polos simples confirman el resultado (\ref{polos}) calculado en la
secci{\'o}n \ref{luego} para un potencial arbitrario $V(x)$
considerando que, en el caso $V(x)=x^2$, los residuos de los polos
dados en (\ref{polos}) son no nulos s{\'o}lo si $n$ es un
m{\'u}ltiplo de $4$.

\bs

En la secci{\'o}n \ref{coasau} estudiaremos la relaci{\'o}n entre
la estructura de polos de la funci{\'o}n $\zeta(s)$ y el
comportamiento asint{\'o}tico de los autovalores mencionada al
finalizar la secci{\'o}n \ref{funcionesespectrales}.
Verificaremos, a partir de un estudio asint{\'o}tico de la
ecuaci{\'o}n de autovalores, la estructura de polos encontrada en
la secci{\'o}n \ref{espoze} mediante la representaci{\'o}n
integral de la funci{\'o}n $\zeta(s)$.

\bs

Finalizaremos el estudio del operador (\ref{elo}) considerando
algunos casos particulares; en la secci{\'o}n \ref{particular}
analizaremos las funciones $\zeta(s)$ de las extensiones
correspondientes a condiciones de contorno invariantes de escala y al
l{\'\i}mite regular $\nu\rightarrow 1/2$ de (\ref{elo}) que
corresponde al oscilador arm{\'o}nico en la semirrecta.

\item\noindent{{\bf Secci{\'o}n \ref{Solari}: Un operador de
Schr\"odinger en una variedad de base compacta}}

En la secci{\'o}n \ref{the-operator} estudiaremos las extensiones
autoadjuntas del operador singular,
\begin{equation}\label{elo2}
    A=-\partial_x^2+\frac{\nu^2-1/4}{x^2}
\end{equation}
definido sobre un subconjunto apropiado de $\mathbf{L_2}([0,1])$.
Nuevamente, el pa\-r{\'a}\-me\-tro $\nu\in(0,1)$ de modo que el
operador (\ref{elo2}) admite un conjunto infinito de extensiones
autoadjuntas.

\bigskip

En lugar de utilizar la teor{\'\i}a de von Neumann, como en la
secci{\'o}n \ref{Wipf}, para determinar las extensiones
autoadjuntas del operador (\ref{elo2}), describiremos las
condiciones de contorno en el origen en t{\'e}rminos de los mapeos
suryectivos definidos en el Teorema (\ref{k0}). Estos mapeos
definen una forma simpl{\'e}ctica respecto de la cual los
subespacios lagrangianos est{\'a}n identificados con las
extensiones autoadjuntas. En el extremo $x=1$ impondremos
condiciones de contorno de Dirichlet, de manera que las
extensiones autoadjuntas de (\ref{elo2}) resultan caracterizadas
por un par{\'a}metro real relacionado con el comportamiento de las
funciones en el origen; el espectro depende, consecuentemente, del
valor de este par{\'a}metro.

\bs

Al igual que en el ejemplo estudiado en la secci{\'o}n \ref{Wipf},
los espectros de las extensiones autoadjuntas del operador
(\ref{elo2}) no poseen una cota inferior uniforme, de modo que,
aunque el espectro de cada extensi{\'o}n est{\'a} acotado
inferiormente, todo n{\'u}mero real negativo corresponde al
autovalor del estado fundamental de alguna extensi{\'o}n
autoadjunta.

\bs

En la secci{\'o}n \ref{the-resolvent1} construiremos
expl{\'\i}citamente las resolventes de las extensiones
autoadjuntas del operador (\ref{elo2}). Para ello, reconoceremos
primeramente la existencia de dos extensiones autoadjuntas de\-fi\-ni\-das por condiciones de
contorno invariantes de escala, cuyas resolventes presentan el
desarrollo asint{\'o}tico usual. Posteriormente, expresaremos la
resolvente de una extensi{\'o}n autoadjunta arbitraria como
combinaci{\'o}n lineal de las resolventes de estas dos extensiones particulares. De esta manera, reconoceremos en los coeficientes
de la combinaci{\'o}n lineal el origen de las potencias con
exponentes dependientes de $\nu$ en el desarrollo asint{\'o}tico
de la resolvente.

\bs

A partir de las relaciones entre las funciones espectrales
presentada en la secci{\'o}n \ref{funcionesespectrales}, en la
secci{\'o}n \ref{spectral-functions} utilizaremos el desarrollo
asint{\'o}tico de la traza de la resolvente para demostrar que los
polos de la funci{\'o}n-$\zeta$ y los exponentes de las potencias
de $t$ en el desarrollo asint{\'o}tico del heat-kernel dependen
del par{\'a}metro $\nu$. En efecto, los resultados de la
secci{\'o}n \ref{spectral-functions} indican que la funci{\'o}n
$\zeta(s)$ posee polos en los puntos $s_k=-\nu\,k$ con
$k=1,2,\ldots$ Existe tambi{\'e}n una sucesi{\'o}n de polos en los
puntos $s_n=1/2-n$, con $n=0,1,\ldots$ que obedece al resultado
(\ref{resu}).

\bs

Como hemos mencionado, la presencia de polos de la
funci{\'o}n-$\zeta$ en posiciones dependientes de $\nu$ se
manifiesta en aquellas condiciones de contorno que no son
in\-va\-rian\-tes ante una transformaci{\'o}n de escala.
Finalizamos la secci{\'o}n \ref{spectral-functions} verificando
con un an{\'a}lisis dimensional la consistencia del resultado
obtenido para la dependencia de los re\-si\-duos de la
funci{\'o}n-$\zeta$ con el par{\'a}metro que caracteriza las
extensiones autoadjuntas.

\bs

Por {\'u}ltimo, como el comportamiento de las funciones en el
origen es esencialmente distinto cuando $\nu=0$, estudiaremos este
caso por separado en la secci{\'o}n \ref{nu=0}. A pesar de la
existencia de un conjunto infinito de extensiones autoadjuntas, la
construcci{\'o}n expl{\'\i}cita de la resolvente y de su
desarrollo asint{\'o}tico muestra que los polos de la
funci{\'o}n-$\zeta$ obedecen, en este caso, al resultado
(\ref{resu}).

\subsubsection{Cap{\'\i}tulo \ref{apli}: Operadores de Dirac}

En este cap{\'\i}tulo estudiaremos dos operadores de Dirac con
coeficientes singulares para los que el resultado (\ref{resu})
tambi{\'e}n pierde validez.

\bs

En la secci{\'o}n \ref{Seeley} estudiaremos un operador de Dirac
(\ref{dir}) sobre la variedad de base $[0,1]$ cuyo campo de gauge
posee un t{\'e}rmino singular proporcional a $x^{-1}$.

\bs

En la secci{\'o}n \ref{tubos} consideraremos el hamiltoniano de
una part{\'\i}cula cargada, sin masa y con spin en 2+1 dimensiones
en presencia de un campo magn{\'e}tico homog{\'e}neo y de un flujo
magn{\'e}tico de Aharonov-Bohm.

\item\noindent{{\bf Secci{\'o}n \ref{Seeley}: Un operador de
primer orden}}

En la secci{\'o}n \ref{the-operator1} estudiaremos las extensiones
autoadjuntas del operador de Dirac,
\begin{equation}\label{dir2}
  D=\left(\begin{array}{cc}
    0  & \displaystyle{-\partial_x+\frac{\alpha}{x}} \\
    \displaystyle{\partial_x+\frac{\alpha}{x}} & 0 \
  \end{array}\right)\, ,
\end{equation}
definido sobre un subconjunto de
$\mathbb{C}^2\otimes\mathbf{L_2}([0,1])$ sobre el cual $D$ sea
sim{\'e}trico. Estudiaremos el caso $|\alpha|<1/2$ para el que el
operador (\ref{dir2}) admite un conjunto infinito de extensiones
autoadjuntas.

\bs

Primeramente estudiaremos el comportamiento de las funciones del
dominio del o\-pe\-ra\-dor adjunto $D^{\dagger}$ en proximidades
de la singularidad $x=0$. Esto permitir{\'a} definir los mapeos
suryectivos referidos en el Teorema (\ref{k0}) que definen una
forma simpl{\'e}ctica respecto de la cual los subespacios
lagrangianos est{\'a}n identificados con las extensiones
autoadjuntas de $D$. En el extremo $x=1$ impondremos una
condici{\'o}n tipo Dirichlet para una de las componentes de las
funciones de $\mathbb{C}^2\otimes\mathbf{L_2}([0,1])$.

\bs

Posteriormente, determinaremos el espectro de cada una de las
extensiones autoadjuntas. Veremos que estos espectros s{\'o}lo son
sim{\'e}tricos respecto del origen para las extensiones
autoadjuntas caracterizadas por una condici{\'o}n
de contorno invariante de escala.

\bs

En la secci{\'o}n \ref{the-resolvent} calcularemos las resolventes
de las extensiones autoadjuntas del operador (\ref{dir2}). Para
ello, expresaremos la resolvente de una extensi{\'o}n arbitraria
como una combinaci{\'o}n lineal de las resolventes de las
extensiones caracterizadas por condiciones de contorno invariantes de escala, que admiten el desarrollo
asint{\'o}tico usual. Como se ver{\'a}, el comportamiento
asint{\'o}tico de los autovalores $\lambda_n$ de las extensiones
del operador (\ref{dir2}) satisface $\lambda_n\sim n$. Por
consiguiente, no existe la traza de las resolventes que son
operadores de Hilbert-Schmidt. En consecuencia, calcularemos las
trazas de los cuadrados de las resolventes.

\bs

Mostraremos entonces que la traza del cuadrado de la resolvente de
una extensi{\'o}n autoadjunta caracterizada por una condici{\'o}n de contorno que no es invariantes de escala admite un
desarrollo asint{\'o}tico en potencias de $\lambda$ cuyos
exponentes dependen del par{\'a}metro $\alpha$. El origen de estas
potencias reside en los coeficientes de la combinaci{\'o}n lineal
que expresa esta resolvente en t{\'e}rminos de las resolventes de
las extensiones correspondientes a condiciones de contorno invariantes de escala.

\bs

En la secci{\'o}n \ref{spectral-functions11} calculamos la
estructura de polos de las funciones espectrales $\zeta(s)$ y
$\eta(s)$ definidas en la secci{\'o}n \ref{funcionesespectrales}.
Comenzaremos estudiando los polos de una funci{\'o}n-$\zeta$
parcial, que se calcula teniendo en cuenta s{\'o}lo las
contribuciones de los autovalores positivos. Como veremos, la
funci{\'o}n $\zeta(s)$ de una extensi{\'o}n autoadjunta correpondientes a una condici{\'o}n de contorno que no es invariantes de escala presenta polos en posiciones dependientes del
par{\'a}metro $\alpha$ dadas por $s_k=-2|\alpha| k$ con
$k=1,2,\ldots$ Cabe se{\~n}alar que, a diferencia de los casos
considerados en las secciones anteriores, la funci{\'o}n
$\zeta(s)$ completa del operador de primer orden (\ref{dir2}) no
posee otras singularidades. Por consiguiente, las funciones
$\zeta(s)$ de las extensiones definidas por condiciones de contorno invariantes de escala son funciones
enteras.

\bs

Asimismo, las funciones $\eta(s)$ de estas extensiones
se anulan, en virtud de la simetr{\'\i}a de sus
espectros. Sin embargo, las funci{\'o}n $\eta(s)$ de una
extensi{\'o}n correspondiente a una condici{\'o}n de contorno que no es invariante de escala no se anula trivialmente y posee
polos en los puntos $s_k=-2|\alpha| (2k+1)$ con $k=0,1,\ldots$

\bs

Finalizaremos la secci{\'o}n \ref{spectral-functions11} con una
discusi{\'o}n acerca del comportamiento de las funciones
espectrales ante una transformaci{\'o}n de escala que nos
permitir{\'a} verificar la dependencia de los residuos de los
polos de la funci{\'o}n $\zeta(s)$ con la extensi{\'o}n
autoadjunta considerada. Se{\~n}alaremos adem{\'a}s que la
estructura de polos de la funci{\'o}n-$\zeta$ se reduce a la
indicada por el resultado (\ref{resu}) en el caso $\alpha=0$.

\item\noindent{{\bf Secci{\'o}n \ref{tubos}: El problema de
Aharonov-Bohm}}

En esta secci{\'o}n estudiaremos el hamiltoniano de Dirac de una
part{\'\i}cula con spin, de masa nula y carga $e$, en $1+2$
dimensiones, en presencia de un campo magn{\'e}tico homog{\'e}neo
y de un flujo magn{\'e}tico singular de Aharonov-Bohm. El
hamiltoniano de Dirac correspondiente es un operador del tipo
(\ref{dir}) en el que el campo de gauge $\mathcal{A}$ posee un
t{\'e}rmino singular en el origen $r=0$ de la forma $\Phi/2\pi r$,
siendo $\Phi$ el flujo magn{\'e}tico singular. Demostraremos, al
finalizar la secci{\'o}n \ref{tubos}, que el desarrollo asint{\'o}tico de la traza del heat-kernel\,\footnote{Como la variedad de base no es compacta, deberemos substraer cantidades divergentes. Para ello calculamos el desarrollo asint{\'o}tico de la traza del operador $e^{-tD^2}-e^{-t\underline{D}^2}$, siendo $\underline{D}$ el hamiltoniano en el caso $\Phi=0$, esto es, en ausencia del flujo singular.} $e^{-t D^2}$ correspondiente al cuadrado del hamiltoniano de Dirac $D$ presenta potencias de $t$ cuyos exponentes dependen del flujo singular $\Phi$.

\bs

En la secci{\'o}n \ref{tuboelo} calculamos las extensiones
autoadjuntas del hamiltoniano $D$. Como el campo
de gauge es invariante ante rotaciones, ser{\'a} suficiente
considerar la restricci{\'o}n $D_l$ del hamiltoniano a los
subespacios de momento angular caracterizados por el Casimir
$l+1/2$, con $l\in\mathbb{Z}$. Mostraremos que s{\'o}lamente la
restricci{\'o}n $D_0$ admite un conjunto infinito de extensiones
autoadjuntas en tanto que las restantes son esencialmente
autoadjuntas.

\bs

Los autovalores de los operadores $D_l$, con $l>0$ est{\'a}n dados
por $\lambda_n=\pm2\sqrt{n}$, por lo que las correspondientes
funciones $\zeta_l(s)$ son el producto de una funci{\'o}n entera
por la funci{\'o}n $\zeta_R(s/2)$ de Riemann, que posee un polo en
$s=2$.

\bs

Por su parte, los autovalores de los operadores $D_l$, con $l<0$
est{\'a}n dados por $\lambda_n=\pm2\sqrt{n+|l|+e\Phi/2\pi}$ por lo
que las correspondientes funciones $\zeta_l(s)$ son el producto de
una funci{\'o}n entera de $s$ por la funci{\'o}n
$\zeta_H(s/2,|l|+e\Phi/2\pi)$ de Hurwitz, que posee un polo simple
en $s=2$.

\bs

El espectro correspondiente a la restricci{\'o}n $D_0$ est{\'a}
dado por las soluciones de una ecuaci{\'o}n trascendente. A
diferencia del caso $l\neq 0$ el espectro no es sim{\'e}trico con
respecto al origen, excepto para las dos extensiones autoadjuntas
definidas por condiciones de contorno invariantes de escala.

\bs

En la secci{\'o}n \ref{zetabeta} calcularemos la estructura de polos
de la funci{\'o}n $\zeta^\beta(s)$ correspondiente a la extensi{\'o}n autoadjunta $D^\beta$ de la restricci{\'o}n $D_0$ del hamiltoniano de Dirac $D$ al subespacio caracterizado por $l=0$. Como
veremos, esta funci{\'o}n $\zeta^\beta(s)$ tiene un polo simple en
$s=1$. Aunque este polo no es previsto por la ecuaci{\'o}n
(\ref{resu}), no atribu{\'\i}mos esta discrepancia al t{\'e}rmino
singular en el operador sino a la no compacidad de la variedad de
base.

\bs

Los polos restantes de la funci{\'o}n $\zeta_\beta(s)$ est{\'a}n
ubicados en puntos dependientes del flujo magn{\'e}tico $\Phi$,
\begin{equation}\label{pop}
    s_{N,n}=-N\left(1-\frac{e\Phi}{\pi}\right)-2n\,,
    \qquad N=1,2,3,\ldots\qquad n=0,1,2,\ldots
\end{equation}
Como hemos mencionado, debido a la presencia de un campo de gauge
singular, la estructura de polos de la funci{\'o}n $\zeta(s)$
depende del flujo singular $\Phi$.

\bs

Finalizaremos la secci{\'o}n \ref{tubos} y con ella el cap{\'\i}tulo
\ref{dirope}, calculando el desarrollo asint{\'o}tico de la traza del heat-kernel $e^{-tD^2}-e^{-t\underline{D}^2}$, siendo $\underline{D}$ el hamiltoniano de Dirac correspondiente al caso $\Phi=0$. Mostraremos que este desarrollo presenta potencias de $t$ cuyos exponentes dependen del flujo singular $\Phi$ (v{\'e}ase la expresi{\'o}n (\ref{dhkfin}).)

\end{itemize}

\subsubsection{Cap{\'\i}tulos \ref{conc} y \ref{proy}: Conclusiones y
Problemas de inter{\'e}s}

En el cap{\'\i}tulo \ref{conc} resumiremos las principales
conclusiones de esta Tesis en tanto que algunos problemas que
merecen consideraci{\'o}n a partir de los resultados que hemos
obtenido ser{\'a}n discutidos en el cap{\'\i}tulo \ref{proy}.

\bs

En la secci{\'o}n \ref{functorial} describiremos una perspectiva
com{\'u}nmente adoptada en el estudio del desarrollo
asint{\'o}tico del heat-kernel. Se ha demostrado \cite{Gilkey} que
los coeficientes del desarrollo asint{\'o}tico en potencias de $t$
que admite el heat-kernel correspondiente a un operador regular
pueden escribirse como integrales sobre la variedad de base $M$, o
sobre su borde $\partial M$, de ciertas cantidades locales
(v{\'e}anse las ecuaciones (\ref{666}) y
(\ref{coefi1}-\ref{coefiult}).) Estas cantidades son combinaciones
lineales de los invariantes geom{\'e}tricos del problema y los
coeficientes de estas combinaciones lineales son, bajo las mismas
hip{\'o}tesis que las del resultado (\ref{resu}), constantes
universales independientes del problema en consideraci{\'o}n. Por
ello, ha sido motivo de intenso estudio la determinaci{\'o}n de
estos coeficientes universales.

\bs

Determinaremos luego, a partir de argumentos dimensionales, los
primeros t{\'e}rminos del desarrollo asint{\'o}tico del
heat-kernel del operador (\ref{delope}) y verificaremos algunos de
los resultados obtenidos en las secciones anteriores.
Consideraremos luego la posibilidad de extender el resultado
acerca de la universalidad de los coeficientes del desarrollo
asint{\'o}tico del heat-kernel al caso de este operador singular.

\bs

En la secci{\'o}n \ref{otrassing}, por su parte, analizaremos la
posibilidad de generalizar nuestros resultados a
o\-pe\-ra\-do\-res diferenciales con otro tipo de singularidad. En
ese sentido, estudiaremos un operador de Schr\"odinger en una
dimensi{\'o}n cuyo potencial posee un t{\'e}rmino singular
proporcional a $x^{-1}$. Como en este caso el orden de la
singularidad no coincide con el orden del operador diferencial, no
pueden aplicarse directamente las t{\'e}cnicas utilizadas en el
caso del operador (\ref{delope}). No obstante, el ejemplo que
trataremos admite una soluci{\'o}n expl{\'\i}cita y permite
observar que la estructura de polos de la correspondiente
funci{\'o}n-$\zeta$ es de una naturaleza distinta.

\bs

En efecto, aunque el operador admite una familia de extensiones
autoadjuntas, mos\-tra\-re\-mos que, a{\'u}n imponiendo
condiciones de contorno del tipo Dirichlet, {\it i.e.},
invariantes de escala, los polos de la funci{\'o}n $\zeta(s)$ no
obedecen el comportamiento (\ref{resu}). La funci{\'o}n $\zeta(s)$
posee un polo en $s=1/2$ con residuo $1/2\pi$ en coincidencia con
el caso regular.

\bs

Sin embargo, demostraremos tambi{\'e}n la existencia de un polo
doble en $s=-1/2$. Verificaremos posteriormente que la presencia
de este polo doble corresponde a un comportamiento asint{\'o}tico
de los autovalores dado por $\lambda_n\sim \log{n}/n$. Como hemos
mencionado, esto implica la presencia de t{\'e}rminos de la forma
$\log{t}$ en el desarrollo asint{\'o}tico del heat-kernel a
peque{\~n}os valores de $t$.

\bs

Los resultados originales contenidos en esta Tesis se encuentran en los trabajos:

\begin{itemize}

\item
H.\ Falomir y P.A.G.\ Pisani, ``Hamiltonian self-adjoint
extensions for (2+1)-dimensional Dirac particles,''\ J.\ Phys.\ A: Mathematical and General {\bf 34}, 1 (2001).

\item
H.\ Falomir, P.A.G.\ Pisani y A.\ Wipf, ``Pole structure of the
Hamiltonian $\zeta$-function for a singular potential,''\ J.\ Phys.\ A: Mathematical and General {\bf 35}, (2002) 5427.

\item
H.\ Falomir, M.A.\ Muschietti, P.A.G.\ Pisani y R.\ Seeley,
``Unusual poles of the $\zeta$-functions for some regular singular
differential operators,'' J.\ Phys.\ A:  Mathematical and General {\bf 36}, 9991 (2003).

\item
H.\ Falomir, M.A.\ Muschietti y P.A.G.\ Pisani, ``On the resolvent
and spectral functions of a second order differential operator
with a regular singularity,'' aceptado para su publicaci{\'o}n en
Journal of Math.\ Phys.; arXiv:math-ph/0404034.

\item
H.\ Falomir y P.A.G.\ Pisani, ``Self-adjoint extensions and SUSY breaking in Supersymmetric Quantum Mechanics,''; enviado para su publicaci{\'o}n al J.\ Phys.\ A (2004).

\end{itemize}


\part{Extensiones Autoadjuntas}\label{sae}

\vspace{5mm}\begin{flushright}
{\it In mathematics you don't understand things.\\
You just get used to them.\\
(John von Neumann.)}
\end{flushright}

\vspace{25mm}

\section{Introducci{\'o}n}

En la f{\'\i}sica cl{\'a}sica, las condiciones de contorno de un
problema sobre una variedad con borde est{\'a}n generalmente
determinadas por consideraciones fe\-no\-me\-no\-l{\'o}\-gi\-cas.
En ocasiones, estas condiciones de contorno son locales.
Encontramos, {\it e.g.}, condiciones tipo Dirichlet para el
potencial electrost{\'a}tico en una regi{\'o}n limitada por
conductores, condiciones tipo Neumann en el movimiento de una
cuerda finita con extremos libres o tipo Robin en la emisi{\'o}n
de calor de un cuerpo sumergido en un medio de temperatura
constante. No obstante, tambi{\'e}n se presenta la necesidad de
estudiar condiciones de contorno no locales, como las condiciones
de contorno peri{\'o}dicas en el caso de problemas sobre una
variedad con topolog{\'\i}a no trivial. Este tipo de condiciones
de contorno son relevantes en teor{\'\i}as de gauge,
 gravedad cu{\'a}ntica y teor{\'\i}a de cuerdas, en las que deben sumarse las
 contribuciones de distintas topolog{\'\i}as de la variedad de base.

\bigskip

En mec{\'a}nica cu{\'a}ntica las condiciones de contorno
apropiadas est{\'a}n relacionadas con la conservaci{\'o}n de la
probabilidad, garantizada por la unitariedad del o\-pe\-ra\-dor
evoluci{\'o}n temporal $\mathcal{U}(t)$. Las traslaciones
temporales infinitesimales est{\'a}n ge\-ne\-ra\-das por el
operador hamiltoniano $H$, de modo que el operador
$\mathcal{U}(t)=e^{i t H}$ est{\'a} \mbox{bien} definido y
representa un grupo unitario dependiente de un par{\'a}metro y
fuertemente continuo\,\footnote{La funci{\'o}n $U(t)$ con valores
en el espacio de operadores lineales sobre un espacio de Hilbert
$\mathcal{H} $ es un grupo de un par{\'a}metro fuertemente
continuo si $\mathcal{U}(t)\mathcal{U}(s)=\mathcal{U}(t+s)$ y
$\forall\, \phi\in\mathcal{H}$,\
$\mathcal{U}(t)\phi\rightarrow\mathcal{U}(0)\phi$ si $t\rightarrow
0$ (v{\'e}ase el Teorema de Stone \cite{Stone}.)} si el
hamiltoniano es autoadjunto.

\bs

En general, no es dif{\'\i}cil determinar un dominio de
definici{\'o}n sobre el cual el hamiltoniano sea sim{\'e}trico
pero puede ser necesario extender este dominio para que el
hamiltoniano sea tambi{\'e}n autoadjunto. Esta extensi{\'o}n
requiere un an{\'a}lisis cuidadoso que determine las condiciones
de contorno apropiadas del problema. Por otra parte, existen
operadores sim{\'e}tricos que admiten distintas extensiones de su
dominio que resultan en sendos o\-pe\-ra\-do\-res autoadjuntos.
Este conjunto de extensiones autoadjuntas determina una variedad
de condiciones de contorno admisibles que corresponden a sistemas
f{\'\i}sicos distintos y caracterizan el efecto de las propiedades
microsc{\'o}picas del borde.

\bigskip

El presente cap{\'\i}tulo est{\'a} dedicado al estudio de las
extensiones autoadjuntas de o\-pe\-ra\-do\-res sim{\'e}tricos. En
el transcurso de este estudio distinguiremos operadores
sim{\'e}tricos de o\-pe\-ra\-do\-res autoadjuntos y mostraremos,
en particular, que existen operadores sim{\'e}tricos que no son
autoadjuntos. En estos casos, extenderemos el dominio de
definici{\'o}n de modo de obtener operadores autoadjuntos. Veremos
el modo de determinar en qu{\'e} casos esta extensi{\'o}n es
posible y, si el operador admite m{\'a}s de una extensi{\'o}n
autoadjunta, estudiaremos c{\'o}mo caracterizarlas. Estos
planteamientos son resueltos por la teor{\'\i}a de von Neumann de
los {\'\i}ndices de deficiencia, que describiremos a
continuaci{\'o}n.

\bs

En la secci{\'o}n \ref{topo} daremos las propiedades
topol{\'o}gicas del conjunto de extensiones autoadjuntas que
admite un operador sim{\'e}trico y finalizaremos este
cap{\'\i}tulo considerando, en la secci{\'o}n \ref{rupturaSUSY},
un problema en mec{\'a}nica cu{\'a}ntica supersim{\'e}trica con un
superpotencial singular en el que la presencia de extensiones
autoadjuntas est{\'a} relacionada con la ruptura espont{\'a}nea de
la supersimetr{\'\i}a.

\section{Teor{\'\i}a de von Neumann}\label{TvN}

Dado un operador $A$ definido en un subespacio denso
$\mathcal{D}(A)$ de un espacio de Hilbert $\mathcal{H}$ dotado de
un producto interno $(\cdot,\cdot)$, decimos que $A$ es {\bf
sim{\'e}trico} si todo par de elementos
$\phi,\psi\in\mathcal{D}(A)$ satisface,
\begin{equation}\label{simetricos}
    (\psi,A\phi)=(A\psi,\phi)\,.
\end{equation}
En el caso de operadores diferenciales, la condici{\'o}n
(\ref{simetricos}) suele verificarse mediante una integraci{\'o}n
por partes; la anulaci{\'o}n de los t{\'e}rminos de borde que
surgen de esta integraci{\'o}n imponen restricciones sobre las
condiciones de contorno. No obstante, como hemos mencionado, las
condiciones de contorno apropiadas no s{\'o}lo deben asegurar que
el operador sea sim{\'e}trico en su dominio de definici{\'o}n sino
tambi{\'e}n autoadjunto.

\bigskip

Consideremos, por ejemplo, la propiedad (\ref{simetricos}) para el
caso del operador impulso de una part{\'\i}cula en un segmento,
\begin{equation}
    P=-i\partial_x\, .
\end{equation}
Si definimos su dominio $\mathcal{D}(P)$ como el conjunto denso
$C_0^{\infty}((0,1))$ de las funciones suaves del intervalo
$[0,1]\subset\mathbb{R}$ cuyo soporte no contiene a los extremos
entonces el operador $P$ resulta sim{\'e}trico con respecto al
producto interno usual en $\mathbf{L}_2{([0,1])}$:
\begin{eqnarray}\label{simetrico}
    (\psi,P\phi)=\int_0^1\psi^*(x)(-i\phi'(x))\,dx=\\
    -i\psi^*(x)\phi(x)|_{x=0}^{1}+\int_0^1(-i\psi'(x))^*\phi(x)\,dx
    =(P\psi,\phi)\, .\nonumber
\end{eqnarray}
En efecto, los t{\'e}rminos de borde se anulan en la ecuaci{\'o}n
anterior en virtud de la adecuada elecci{\'o}n del dominio de
definici{\'o}n $\mathcal{D}(P)=C_0^{\infty}((0,1))$.

\bigskip

La definici{\'o}n (\ref{simetricos}) tiene sentido si tanto $\psi$
como $\phi$ pertenecen al dominio de definici{\'o}n de $A$. No
obstante, podemos considerar el producto,
\begin{equation}\label{producto}
    (\psi,A\phi)\,,
\end{equation}
a{\'u}n para aquellos vectores $\psi$ que no pertenezcan al
dominio de definici{\'o}n de $A$, siempre que
$\phi\in\mathcal{D}(A)$. Nuestro objetivo es estudiar el
subconjunto $\mathcal{D}(A^{\dagger})$ de $\mathcal{H}$
constitu{\'\i}do por aquellos vectores $\psi$ para los cuales la
cantidad (\ref{producto}) es una funcional lineal y continua de
$\phi$. Esto es,
\begin{equation}\label{domadj}
    \mathcal{D}(A^{\dagger}):=\{\psi\in\mathcal{H}:\exists K>0,\ \forall
    \phi\in\mathcal{D}(A),\ |(\psi,A\phi)|\leq K \|\phi\|\}\,,
\end{equation}
donde $\|\cdot\|=\sqrt{(\cdot,\cdot)}$. La desigualdad en
(\ref{domadj}) indica que la cantidad $(\psi,A\phi)$ tiende a cero
si la norma de $\phi$ lo hace o que $(\psi,A\phi)$ es una
funcional lineal acotada sobre el conjunto
$\phi\in\mathcal{D}(A):\|\phi\|=1$, lo que garantiza su
continuidad en $\phi$.

\bigskip

Esta definici{\'o}n es relevante cuando tratamos con operadores no
acotados pues si el operador $A$ es acotado existe $\|A\|:={\rm
sup}_{\mathcal{D}(A)}\{\|A\phi\|/\|\phi\|\}$ y se verifica que
$\|A\phi\|\leq \|A\|\cdot\|\phi\|$ para todo
$\phi\in\mathcal{D}(A)$. Por consiguiente, en virtud de la
desigualdad de Cauchy-Schwartz, $|(\psi,A\phi)|\leq
\|\psi\|\cdot\|A\phi\|\leq \|\psi\|\cdot\|A\|\cdot\|\phi\|$ y, en
consecuencia, $\mathcal{D}(A^{\dagger})=\mathcal{H}$. En tanto
que, si el operador $A$ no es acotado, $\|A\phi\|$ no est{\'a}
acotada por una cantidad proporcional a $\|\phi\|$ y la
desigualdad en (\ref{domadj}) impone restricciones que no se
satisfacen para todo $\psi\in\mathcal{H}$.

\bigskip

No obstante, debemos notar que si el operador $A$ es sim{\'e}trico
entonces, utilizando nuevamente la desigualdad de Cauchy Schwartz,
vemos que $|(\psi,A\phi)|=|(A\psi,\phi)|\leq
\|A\psi\|\cdot\|\phi\|$ que implica,
\begin{equation}\label{relacion}
    \mathcal{D}(A)\subset\mathcal{D}(A^{\dagger})\,,
\end{equation}
que es tambi{\'e}n un subconjunto denso de $\mathcal{H}$. El {\bf
operador adjunto} $A^\dagger$ se define sobre el conjunto
(\ref{domadj}) que, por esta raz{\'o}n, hemos designado
$\mathcal{D}(A^\dagger)$.

\bs

Para definir la acci{\'o}n del operador adjunto $A^{\dagger}$
sobre los vectores de $\mathcal{D}(A^{\dagger})$, notemos que
$\forall\psi\in\mathcal{D}(A^{\dagger})$, existe, en virtud del
Lema de representaci{\'o}n de Riesz\,\footnote{El Lema de Riesz
\cite{Riesz} indica tambi{\'e}n que si la funcional $(\psi,
A\phi)$ est{\'a} acotada, entonces su norma, esto es, el
m{\'a}ximo de la funcional en el conjunto
$\phi\in\mathcal{D}(A):\|\phi\|=1$ coincide con
$\|\tilde{\psi}\|$.}, un vector $\tilde{\psi}$ en $\mathcal{H}$
que representa la funcional lineal y continua $(\psi,A\phi)$, esto
es,
\begin{equation}\label{der}
    (\psi,A\phi)=(\tilde{\psi},\phi)\,.
\end{equation}
Este vector es {\'u}nico puesto que $\mathcal{D}(A)$ es denso en
$\mathcal{H}$. Definimos entonces la acci{\'o}n de $A^\dagger$
sobre $\psi$ como,
\begin{equation}\label{der2}
A^{\dagger}\psi:=\tilde{\psi}\,.
\end{equation}

\bigskip

El operador $A$ es {\bf autoadjunto} si $A=A^\dagger$, lo que
requiere que $\mathcal{D}(A)=\mathcal{D}(A^\dagger)$. La
ecuaci{\'o}n (\ref{relacion}) muestra entonces que el operador
sim{\'e}trico $A$ no es autoadjunto toda vez que existan vectores
$\psi\in\mathcal{H}-\mathcal{D}(A)$ para los que la funcional
$(\psi, A\phi)$ sea continua para todo $\phi\in\mathcal{D}(A)$.

\bs

Podemos en este punto explicar el origen de la expresi{\'o}n {\bf
extensiones autoadjuntas}. Es inmediato ver, a partir de la
definici{\'o}n (\ref{domadj}), que si definimos una extensi{\'o}n
sim{\'e}trica de un operador entonces el dominio del operador
adjunto se reduce. En efecto, si dos operadores sim{\'e}tricos
satisfacen $A\subset\tilde{A}$, {\it i.e.},
$\mathcal{D}(A)\subset\mathcal{D}(\tilde{A})$ y
$A=\tilde{A}|_{\mathcal{D}(A)}$, entonces el conjunto
$\mathcal{D}(\tilde{A}^\dagger)$ de vectores $\psi\in\mathcal{H}$
para los que el producto $(\psi,\tilde{A}\,\cdot)$ es una
funcional lineal y continua en $\mathcal{D}(\tilde{A})$ est{\'a}
contenido en el conjunto de vectores para los que el mismo
producto es una funcional lineal y continua en $\mathcal{D}(A)$.
Por consiguiente,
\begin{equation}
    \mathcal{D}({A})\subset\mathcal{D}(\tilde{A})\subset
    \mathcal{D}(\tilde{A}^{\dagger})\subset\mathcal{D}(A^{\dagger})\,.
\end{equation}
De modo que si extendemos en forma sim{\'e}trica un operador,
reducimos el dominio del operador adjunto. Nuestro objetivo es
realizar esta extensi{\'o}n de modo que su dominio de
definici{\'o}n coincida con el dominio de su operador adjunto.
Obtendremos, entonces, una extensi{\'o}n que no s{\'o}lo es
sim{\'e}trica sino tambi{\'e}n autoadjunta. La teor{\'\i}a de los
{\'\i}ndices de deficiencia de von Neumann establece en qu{\'e}
casos es posible hacer coincidir los dominios de las extensiones
sim{\'e}tricas con los dominios de sus operadores adjuntos e
indica c{\'o}mo caracterizar las diversas extensiones posibles.

\bigskip

Consideremos nuevamente el operador sim{\'e}trico $P$ y
determinemos el dominio $\mathcal{D}(P^\dagger)$ de su adjunto,
que satisface,
\begin{equation}\label{sinoau}
    \mathcal{D}(P)=\mathcal{C}_0^\infty((0,1))
    \subset\mathcal{D}(P^\dagger)\subset\mathcal{H}=\mathbf{L_2}([0,1])\,.
\end{equation}
El dominio $\mathcal{D}(P^\dagger)$ es el conjunto de funciones
$\psi(x)\in\mathbf{L_2}([0,1])$ para las que la funcional
$(\psi,P\phi)$ es continua para todo
$\phi\in\mathcal{C}_0^\infty((0,1))$. Por el Lema de Riesz, esto
implica la existencia de una funci{\'o}n
$\tilde{\psi}(x)\in\mathbf{L_2}([0,1])$ que satisface,
\begin{equation}\label{sobolev}
    \int_0^{1}\psi^*(x)\cdot(-i\partial_x\phi(x))\,dx=
    \int_0^{1}\tilde{\psi}^*(x)\cdot\phi(x)\,dx\,.
\end{equation}
La funci{\'o}n $\psi$ pertenece entonces al espacio de Sobolev
$\mathbf{H_1}((0,1))$ (v{\'e}ase la definici{\'o}n (\ref{sobo}))
que coincide con el conjunto de funciones
$\psi\in\mathbf{L_2}([0,1])$ con derivada ge\-ne\-ra\-li\-za\-da
$\psi'\in\mathbf{L_2}([0,1])$ definida por\,\footnote{Las
funciones con derivada generalizada en $\mathbf{L_2}([0,1])$ son
aquellas que definen una distribuci{\'o}n regular cuya derivada
d{\'e}bil es otra distribuci{\'o}n regular.},
\begin{equation}\label{dergen}
    \int_0^{1}\psi'^*(x)\cdot\phi(x)\,dx=-
    \int_0^{1}\psi^*(x)\cdot\partial_x\phi(x)\,dx\,,\qquad
    \forall\phi\in\mathcal{C}_0^\infty((0,1))\,.
\end{equation}
Por consiguiente, $\mathcal{D}(P^{\dagger})=\mathbf{H_1}((0,1))$ y
la forma en que opera $P^{\dagger}$ est{\'a} dada por,
\begin{equation}\label{pedaga}
    P^{\dagger}\psi=\tilde{\psi}=-i\psi'\in\mathbf{L_2}([0,1])\,,
\end{equation}
donde $\psi'$ es la derivada generalizada o derivada (d{\'e}bil)
en el sentido de las distribuciones de la distribuci{\'o}n regular
$\psi$.

\bs

Vemos, entonces, que el operador $P=-i\partial_x$ definido sobre
$\mathcal{C}_0^{\infty}((0,1))$ es sim{\'e}trico pero no es
autoadjunto, puesto que
$\mathcal{D}(P^{\dagger})=\mathbf{H_1}((0,1))$ no coincide con
$\mathcal{D}(P)=\mathcal{C}_0^{\infty}((0,1))$ mas lo contiene.

\bs

Como hemos se\~nalado, debemos extender el dominio de
definici{\'o}n del operador si\-m{\'e}\-tri\-co $P$, reduciendo
as\'\i\ el dominio de su adjunto $P^\dagger$, de modo de lograr
$\mathcal{D}(P)=\mathcal{D}(P^\dagger)$. Como buscamos una
extensi{\'o}n sim{\'e}trica los t{\'e}rminos de borde que surgen
de la integraci{\'o}n por partes que conduce a la propiedad
(\ref{simetricos}) deben permanecer nulos. De todas maneras,
veremos que no existe una {\'u}nica extensi{\'o}n autoadjunta del
operador $P$ sino un conjunto infinito de ellas que tiene la
topolog{\'\i}a de $S^1$.

\bigskip

De acuerdo con la teor{\'\i}a de von Neumann, para caracterizar
las extensiones au\-to\-ad\-jun\-tas de un operador $A$ debe
resolverse la ecuaci{\'o}n de autovalores,
\begin{equation}\label{autodaga}
    A^\dagger\psi=\lambda\psi\,,
\end{equation}
con $\psi\in\mathcal{D}(A^\dagger)$ y $\mathbb{I}(\lambda)\neq 0$.
Esta expresi{\'o}n exige algunas observaciones. En primer lugar,
n{\'o}tese que si $A$ es un operador diferencial sim{\'e}trico
entonces, como indican las ecuaciones (\ref{der}) y (\ref{der2}),
$A^\dagger$ se obtiene ``formalmente'' reemplazando las derivadas
en $A$ por derivadas generalizadas sobre $\psi$ (v{\'e}ase, {\it
e.g.}, la ecuaci{\'o}n (\ref{pedaga}).) Por otra parte, en los
ejemplos que estudiaremos podr{\'a} verse que la propiedad
(\ref{autodaga}) implica que
$\psi\in\mathcal{C}^\infty\cap\mathbf{L_2}$, de manera que la
ecuaci{\'o}n (\ref{autodaga}) equivale a una ecuaci{\'o}n
diferencial sobre funciones infinitamente derivables cuyas
soluciones en $\mathcal{H}$ pertenecen a $\mathcal{D}(A^\dagger)$.

\bigskip

N{\'o}tese que un operador sim{\'e}trico no tiene autovalores
complejos puesto que si $\phi\neq 0$ es un autovector de $A$ con
autovalor $\lambda$ entonces,
\begin{equation}
    (\phi,A\phi)=(\phi,\lambda\phi)=\lambda(\phi,\phi)=
    (A\phi,\phi)=(\lambda\phi,\phi)=\lambda^*(\phi,\phi)\rightarrow
    \lambda\in\mathbb{R}\,.
\end{equation}
De modo que la existencia de soluciones de (\ref{autodaga}) con
$\lambda\notin\mathbb{R}$ indica que el conjunto
$\mathcal{D}(A^\dagger)-\mathcal{D}(A)$ no es vac{\'\i}o y, por
consiguiente, el operador $A$ es sim{\'e}trico pero no
autoadjunto. Como ${\rm dim}\,{\rm Ker}(A^\dagger-\lambda)$ es
constante en los semiplanos abiertos superior e inferior del plano
complejo-$\lambda$ \cite{teorX1}, ser{\'a} suficiente para
determinar la existencia y caracterizar las extensiones
autoadjuntas de $A$ considerar las soluciones de (\ref{autodaga})
con $\lambda=\pm i$. Por ello, damos las siguientes definiciones.
\begin{defn}
Los {\bf espacios de deficiencia} $\mathcal{K}_{\pm}$ son los
subespacios caracter{\'\i}sticos del operador $A^{\dagger}$ de
autovalor $\pm i$ respectivamente.
\end{defn}
Esto es,
\begin{equation}
    \mathcal{K}_{\pm}:= {\rm Ker}(A^{\dagger}\mp i)\,.
\end{equation}
\begin{defn}
Los {\bf {\'\i}ndices de deficiencia} $n_{\pm}$ son las
dimensiones de los subespacios de deficiencia.
\begin{equation}
    n_{\pm}:= {\rm dim}\,\mathcal{K}_{\pm}\,.
\end{equation}
\end{defn}
De acuerdo con la discusi{\'o}n anterior, si alguno de los
{\'\i}ndices de deficiencia es distinto de cero entonces el
operador no es autoadjunto. Presentamos ahora dos Teoremas. El
primero de ellos, Teorema \ref{auto}, establece la rec{\'\i}proca
de esta {\'u}ltima afirmaci{\'o}n, {\it i.e.}, si los {\'\i}ndices
de deficiencia son nulos, entonces el operador es autoadjunto. El
Teorema siguiente, Teorema \ref{222}, permite construir las
extensiones autoadjuntas si los {\'\i}ndices de deficiencia son
iguales y distintos de cero.
\begin{thm}\label{auto}
Si $A$ es un operador cerrado\,\footnote{Un operador es cerrado si
su gr{\'a}fica es un conjunto cerrado en $\mathcal{H}\oplus
\mathcal{H}$.}, sim{\'e}trico y densamente definido en un espacio
de Hilbert $\mathcal{H}$ entonces A es autoadjunto si y s{\'o}lo
si $n_{\pm}=0$.
\end{thm}

\noindent{\bf Demostraci{\'o}n:} N{\'o}tese primeramente que,
\begin{equation}\label{anterior}
{\rm Ker}(A^{\dagger} \pm i)={\rm Ran}^{\bot}(A \mp i).
\end{equation}
Esta propiedad es inmediata a partir de la siguiente igualdad,
\begin{equation}\label{ranker}
((A^{\dagger} \pm i)\psi,\phi)=(\psi,(A \mp i)\phi)\,,
\end{equation}
v{\'a}lida para todo par de vectores
$\psi\in\mathcal{D}(A^\dagger)$ y $\phi\in\mathcal{D}(A)$. La
ecuaci{\'o}n (\ref{ranker}) implica la ecuaci{\'o}n
(\ref{anterior}) en virtud de que $\mathcal{D}(A)$ es un
subespacio denso del espacio de Hilbert $\mathcal{H}$. Por lo
tanto, si $n_+=0$ entonces ${\rm Ran}(A + i)$, que para un
operador sim{\'e}trico y cerrado es un subespacio cerrado,
coincide con el espacio de Hilbert $\mathcal{H}$. En consecuencia
todo vector perteneciente a $\mathcal{H}$ es la imagen por $A + i$
de alg{\'u}n vector de $\mathcal{D}(A)$.

\bs

En particular, si $\phi \in \mathcal{D}(A^{\dagger})$ entonces
existe $\chi\in\mathcal{D}(A)$ tal que $(A^{\dagger}+ i)\phi=(A +
i)\chi$ o bien $A^{\dagger}(\phi-\chi)=- i(\phi-\chi)$ puesto que
$A$ es la restricci{\'o}n de $A^{\dagger}$ a $\mathcal{D}(A)$.
Pero si, adem{\'a}s, $n_{-}=0$ entonces no existen autovectores de
$A^\dagger$ con autovalor $-i$, por lo cual $\phi=\chi$. Por
consiguiente, $\phi \in \mathcal{D}(A)$ y el operador es
autoadjunto.\begin{flushright}$\Box$\end{flushright}

El Teorema \ref{auto} prueba que si los {\'\i}ndices de
deficiencia de un operador sim{\'e}trico y cerrado son nulos
entonces el o\-pe\-ra\-dor es autoadjunto. A continuaci{\'o}n,
mostraremos c{\'o}mo se debe proceder cuando los {\'\i}ndices de
deficiencia son no nulos. Veremos que si los {\'\i}ndices de
deficiencia no coinciden entonces no pueden construirse
extensiones autoadjuntas. Por el contrario, si los {\'\i}ndices de
deficiencia son no nulos e iguales, $n_+=n_-> 0$, entonces existe
un conjunto de extensiones autoadjuntas en correspondencia
biun{\'\i}voca con los elementos de $U(n_\pm)$.
\begin{thm}\label{222}
Sea $A$ un operador cerrado, sim{\'e}trico y densamente definido.
Las extensiones sim{\'e}tricas $\tilde{A}$ de $A$ est{\'a}n en
correspondencia biun{\'\i}voca con el conjunto de isometr{\'\i}as
lineales parciales de $\mathcal{K} _{+}$ en $\mathcal{K}
 _{-}$.

Si U es una de tales isometr{\'\i}as, con dominio
$\mathcal{D}(U)\subset \mathcal{K}_{+}$, entonces la extensi{\'o}n
sim{\'e}trica correspondiente $\tilde{A}$ tiene dominio\,\footnote{La
suma directa debe entenderse con respecto al producto interno
$(\cdot,\cdot)_A:= (\cdot,\cdot)+(A\,\cdot,A\,\cdot)$.},
\begin{equation}
    \mathcal{D}(\tilde{A})=\{\phi:
    \phi=\phi_{0}\oplus\phi_{+}\oplus U(\phi_{+});\ \phi_0\in\mathcal{D}(A)
    \wedge\phi_+\in\mathcal{D}(U)\}\,,
\end{equation}
y su acci{\'o}n queda definida por,
\begin{equation}
\tilde{A}(\phi_{0}+\phi_{+}+U(\phi_{+}))={A^\dagger}
(\phi_{0}+\phi_{+}+U(\phi_{+}))
=A(\phi_0)+i\phi_{+}-iU(\phi_{+})\,,
\end{equation}
donde $\phi_{0}\in \mathcal{D}(A)$ y $\phi_{+}\in
\mathcal{D}(U)\subset\mathcal{K}_{+}$. Adem{\'a}s, si los
{\'\i}ndices de deficiencia son finitos,
\begin{equation}
n_{\pm}(\tilde{A})=n_{\pm}(A)- {\rm dim}\, \mathcal{D}(U)\,.
\end{equation}
\end{thm}
V{\'e}ase la demostraci{\'o}n en \cite{R-S}.\fin
\begin{defn}
Un operador $A$ sim{\'e}trico y densamente definido es {\bf
esencialmente autoadjunto} si su clausura\,\footnote{El operador clausura $\overline{A}$ de un operador $A$ es aquel cuya gr{\'a}fica es la clausura de la gr{\'a}fica de $A$ con la norma inducida en $\mathcal{H}\otimes\mathcal{H}$. Puede probarse que $\overline{A}=(A^\dagger)^\dagger$.} $\overline{A}$ es autoadjunta.
\end{defn}
\begin{cor}\label{coro3}
Dado un operador sim{\'e}trico y densamente definido con {\'\i}ndices de
deficiencia finitos:
\begin{itemize}
\item Si sus {\'\i}ndices de deficiencia son distintos entonces no
admite extensiones autoadjuntas. \item Si sus {\'\i}ndices de
deficiencia son nulos entonces el operador es esencialmente
autoadjunto y admite una {\'u}nica extensi{\'o}n autoadjunta, que
est{\'a} dada por su clausura. \item Si sus {\'\i}ndices de
deficiencia son iguales y distintos de cero, $n_+=n_->0$, se
pueden establecer $n_\pm^2$ isometr{\'\i}as lineales con dominio
en todo el subespacio $\mathcal{K}_{+}$ cuyas im{\'a}genes son
$\mathcal{K}_{-}$. En ese caso, la extensi{\'o}n asociada a cada
isometr{\'\i}a tendr{\'a} {\'\i}ndices de deficiencia nulos y
ser{\'a}, por lo tanto, ella misma autoadjunta. El conjunto de
extensiones autoadjuntas est{\'a} entonces en correspondencia
biun{\'\i}voca con los elementos del grupo $U(n_{\pm})$.
\end{itemize}
\end{cor}

\bigskip

Finalizamos esta secci{\'o}n considerando nuevamente el operador
$P=-i\partial_x$ definido sobre
$\mathcal{D}(P)=\mathcal{C}_0^\infty((0,1))$ con el fin de
ilustrar los resultados del Corolario \ref{coro3}.

\bs

Ya hemos demostrado que se verifica,
\begin{equation}
    \mathcal{D}(P)=\mathcal{C}_0^\infty((0,1))
    \subset\mathcal{D}(P^\dagger)=\mathbf{H_1}((0,1))
    \subset\mathcal{H}=\mathbf{L_2}([0,1])\,,
\end{equation}
de modo que el operador $P$ no es esencialmente autoadjunto.

\bs

Para calcular los {\'\i}ndices de deficiencia del operador ${P}$
debemos resolver,
\begin{equation}\label{autov}
    {P}^{\dagger}\psi_{\pm}=-i\psi_{\pm}'=\pm i\psi_{\pm}\,,
\end{equation}
en $\mathcal{H}=\mathbf{L_2}([0,1])$. Recordemos que $\psi'_\pm$
representa, en principio, la derivada (d{\'e}bil) en el sentido de
las distribuciones. Sin embargo, como ya hemos adelantado, la
propiedad (\ref{autov}) implica que las autofunciones $\psi_\pm$
son infinitamente derivables. En efecto, la ecuaci{\'o}n
(\ref{autov}) indica que $\psi'_\pm\in\mathbf{L_2}([0,1])$; esto
significa que las funciones $\psi_\pm$ admiten una derivada
generalizada en $\mathbf{L_2}([0,1])$ o que las distribuciones que
definen admiten una derivada regular. Utilizando reiteradamente la
ecuaci{\'o}n (\ref{autov}) conclu{\'\i}mos que $\psi_\pm$ admiten
derivadas regulares de todo orden. Las funciones que admiten una
derivada generalizada en $\mathbf{L_2}([0,1])$ son funciones
absolutamente continuas y, consecuentemente, continuas. Por
consiguiente $\psi_\pm\in\mathcal{C}^{\infty}([0,1])$. En otros
t{\'e}rminos, como $\psi_\pm$ admiten derivadas generalizadas
regulares de todo orden, pertenecen a
$\bigcap_{d\in\mathbb{N}}\mathbf{H_d}((0,1))$; el Lema de Sobolev
indica, entonces, que $\psi$ admite infinitas derivadas en el
sentido usual\,\footnote{El Lema de Sobolev \cite{Gilkey2}
demuestra que si $\psi\in\mathbf{H_d}(M)$ entonces
$\psi\in\mathcal{C}^{k}(M)$ para todo $k<d-m/2$, siendo
$k\in\mathbb{N}$ y $m={\rm dim}\,M$.}. La ecuaci{\'o}n
(\ref{autov}) es, entonces, una ecuaci{\'o}n diferencial en el
sentido usual cuyas soluciones est{\'a}n generadas por,
\begin{eqnarray}
    \psi_{+}(x)=e^{1 - x}\,,\label{pma}\\
    \psi_{-}(x)=e^{ x}\,.\label{pme}
\end{eqnarray}
Los subespacios de deficiencia $\mathcal{K}_{\pm}$ son entonces
subespacios de una dimensi{\'o}n generados por las funciones
(\ref{pma}) y (\ref{pme}), respectivamente. Los {\'\i}ndices de
deficiencia resultan, en consecuencia,
\begin{equation}
    n_{\pm}=1\,.
\end{equation}
De acuerdo con el Corolario \ref{coro3}, el operador ${P}$ no es
esencialmente autoadjunto y admite extensiones autoadjuntas
identificadas con las isometr{\'\i}as de $\mathcal{K}_+\approx
\mathbb{C}$ en $\mathcal{K}_-\approx \mathbb{C}$. De modo que el
conjunto de extensiones autoadjuntas est{\'a} identificado con el
grupo $U(1)$ y sus elementos est{\'a}n ca\-rac\-te\-ri\-za\-dos
por un par{\'a}metro real $\gamma\in[0,2\pi)$. Para construir una
extensi{\'o}n autoadjunta particular consideramos una
isometr{\'\i}a $U_{\gamma}\in U(1)$ de $\mathcal{K}_+$ en
$\mathcal{K}_-$ que definimos de acuerdo con su acci{\'o}n sobre
el elemento de la base (\ref{pma}),
\begin{eqnarray}
    U_{\gamma}:\mathcal{K}_+\rightarrow\mathcal{K}_-\\
    U_{\gamma}(\psi_+):= e^{i\gamma}\psi_-\,.
\end{eqnarray}
De acuerdo con el Teorema \ref{222}, el dominio
$\mathcal{D}(P_{\gamma})$ de las extensi{\'o}n autoadjunta
$P_{\gamma}$ del operador impulso est{\'a} definido por,
\begin{equation}\label{dom}
    \mathcal{D}(P_{\gamma})=\{\psi:\psi=\phi+A\psi_+
    +e^{i\gamma}A\psi_-\,,\
    \phi\in\mathcal{D}(\overline{P}),\ A\in\mathbb{C}\}.
\end{equation}
Asimismo, resulta inmediato mostrar que,
\begin{equation}\label{dom-clausura}
    \mathcal{D}(\overline{P})=\{\phi\in\mathbf{H_1}((0,1)):\phi(0)=\phi(1)=0\}.
\end{equation}

Hemos se{\~n}alado que la condici{\'o}n de autoadjunto del
generador de las transformaciones unitarias determina las
condiciones de contorno apropiadas en mec{\'a}nica cu{\'a}ntica.
Estamos ahora en condiciones de establecer estas condiciones de
contorno para el operador impulso. Para ello calculamos los
valores de las funciones $\psi\in\mathcal{D}(P_{\gamma})$ en los
extremos del intervalo $[0,1]$ (v{\'e}anse las ecuaciones
(\ref{dom}), (\ref{pma}) y (\ref{pme}))\,\footnote{Puede
demostrarse con los m{\'e}todos de las Secciones \ref{rupturaSUSY}
y \ref{Wipf} que la funci{\'o}n $\phi\in\mathcal{D}(\overline{P})$
no contribuye al orden dominante.},
\begin{eqnarray}
    \psi(0)=A\,
    e+e^{i\gamma}A\,,\\
    \psi(1)=A
    +e^{i\gamma}A\, e\,.
\end{eqnarray}
A partir de estas expresiones es f{\'a}cil ver que,
\begin{equation}
    \psi(1)=e^{-i\gamma}\frac{1+e^{1+i\gamma}}{1+e^{1-i\gamma}}\,\psi(0)\,.
\end{equation}
Esta relaci{\'o}n indica que los valores de borde de la funciones
de onda difieren en una fase,
\begin{equation}\label{cond}
    \psi(1)=e^{i\alpha}\psi(0)\,,
\end{equation}
donde hemos definido,
\begin{equation}
    \alpha:= -\gamma+2\arctan{\left(\frac{\sin\gamma}{1+e\cos\gamma}\right)}\,.
\end{equation}

\bigskip

En conclusi{\'o}n, el operador $P$ definido sobre las funciones de
$\mathcal{C}_0^{\infty}((0,1))$ no es autoadjunto. Las funciones
de $\mathcal{C}_0^{\infty}((0,1))$ y todas sus derivadas se anulan
en los bordes y esto asegura que $P$ sea sim{\'e}trico pues se
cancelan las contribuciones de borde,
\begin{equation}\label{contrbor}
    -i(\psi^*(1)\phi(1)-\psi^*(0)\phi(0))\,,
\end{equation}
en la integraci{\'o}n por partes de la ecuaci{\'o}n
(\ref{simetrico}). Sin embargo, $\mathcal{C}_0^{\infty}((0,1))$ es
un conjunto muy restringido de funciones, por lo que el dominio de
$P^{\dagger}$ es demasiado amplio y el operador $P$ no resulta
autoadjunto.

\bs

Como el objetivo es encontrar extensiones sim{\'e}tricas de $P$,
el comportamiento de las funciones en los extremos debe ser
a{\'u}n tal que se anule la contribuci{\'o}n de los t{\'e}rminos
de borde (\ref{contrbor}). La condici{\'o}n de contorno apropiada
para el operador impulso, dada por la ecuaci{\'o}n (\ref{cond}),
es la misma que habr{\'\i}amos obtenido de haber impuesto el menor
n{\'u}mero posible de restricciones que exige la anulaci{\'o}n de
las contribuciones de borde $(\ref{contrbor})$. En efecto, la
anulaci{\'o}n de estos t{\'e}rminos en el caso particular en el
que $\psi=\phi$ implica que el valor de la funci{\'o}n en los
extremos difiere en una fase. Es inmediato ver, luego, que esa
fase debe ser la misma para todas las funciones del dominio del
operador sim{\'e}trico. En consecuencia, la condici{\'o}n
(\ref{cond}), a la que hemos arribado independientemente a partir
del Teorema \ref{222}, es la m{\'\i}nima necesaria para la
anulaci{\'o}n de (\ref{contrbor}).

\bs

Existe otro argumento en favor de la condici{\'o}n de contorno
(\ref{cond}). Como el operador $P$ es autoadjunto, el operador de
traslaci{\'o}n $e^{i a P}$ resulta unitario. Pero la
transformaci{\'o}n $e^{i a P}$ s{\'o}lo conserva la norma de las
funciones definidas sobre el intervalo $[0,1]$ si la funci{\'o}n
de onda $\psi(x)$ satisface $\|\psi(0)\|^2=\|\psi(1)\|^2$, que es
una consecuencia de la condici{\'o}n (\ref{cond}).

\bs

Finalizamos esta secci{\'o}n con dos {\'u}ltimas observaciones. En
primer lugar la condici{\'o}n de contorno no local (\ref{cond})
modifica la topolog{\'\i}a de la variedad de base $[0,1]$
identific{\'a}ndola con $S^1$. Si $\alpha=0$, la condici{\'o}n de
contorno describe el movimiento de una part{\'\i}cula en $S^1$ y
el espectro del operador impulso $k_n=2\pi n$ indica que la
{\'o}rbita de la part{\'\i}cula es un m{\'u}ltiplo de su longitud
de onda de de Broglie. Sin embargo, la existencia de una familia
de extensiones autoadjuntas del operador impulso, que implica la
necesidad de considerar condiciones de contorno m{\'a}s generales,
con $\alpha$ arbitrario, tiene consecuencias f{\'\i}sicas, {\it
i.e.}, afecta el resultado de las mediciones. En efecto, los
autovalores $k^{\alpha}_n$ del operador impulso definido por las
condiciones de contorno (\ref{cond}) est{\'a}n dados por
$k^\alpha_n=2\pi n+\alpha$, de modo que dependen de las
extensi{\'o}n autoadjunta. En este caso, el par{\'a}metro $\alpha$
representa el ``flujo magn{\'e}tico'' encerrado por $S^1$.

\bs

Es interesante, finalmente, observar que las condiciones de
contorno (\ref{cond}) para la funci{\'o}n de onda $\psi(x)$ de una
part{\'\i}cula confinada en el intervalo $[0,1]$ no incluyen la
condici{\'o}n tipo Dirichlet $\psi(0)=\psi(1)=0$. Esto significa que,
bajo esta condici{\'o}n, el operador impulso, si bien sim{\'e}trico, no es
autoadjunto. Como consecuencia, el operador impulso con
condiciones de contorno tipo Dirichlet no tiene autofunciones; de
hecho, si las hubiera no se verficar{\'\i}a el principio de incerteza.
En efecto, las autofunciones del operador impulso tienen
dispersi{\'o}n $\Delta p=0$ en tanto que $\Delta x\sim 1$; la
desigualdad de Heisenberg no se verifica pues es v{\'a}lida para
operadores autoadjuntos.

\section{ Topolog{\'\i}a del conjunto de extensiones autoadjuntas}\label{topo}

En esta secci{\'o}n presentaremos algunas propiedades
topol{\'o}gicas de la variedad
 $\mathcal{M}$ de las extensiones autoadjuntas de un operador diferencial
 de segundo orden a partir del estudio realizado por M.\ Asorey {\it et al.}
 \cite{Asorey:2004kk}.

\bs

Consideremos el operador $A=\Delta+V$ sobre secciones de un
fibrado vectorial $E$ de rango $k$ sobre una variedad de base $M$
de m{\'e}trica $g$ sobre la cual se define un potencial $V$ y un campo
de gauge $\mathcal{A}$. El laplaciano est{\'a} dado por
$\Delta=d_{\mathcal{A}}^{\dagger}d_{\mathcal{A}}$, siendo
$d_{\mathcal{A}}$ la derivada covariante exterior.

\bs

El espacio de estados f{\'\i}sicos $\mathcal{H}$ es isomorfo a
$\mathbf{L_2}(M)\otimes \mathbb{C}^k$. El producto interno de dos
estados $\psi^{(1)},\psi^{(2)}\in\mathcal{H}$ est{\'a} dado por,
\begin{equation}
    (\psi^{(1)},\psi^{(2)})=\int_M(\psi^{(1)},\psi^{(2)})_E\,\sqrt{g}\,dx\,,
\end{equation}
siendo $(\psi^{(1)},\psi^{(2)})_E$ el producto interno en la
fibra, que es isomorfa a $\mathbb{C}^k$.

\bs

El operador $A$ es sim{\'e}trico en $\mathcal{C}_0^{\infty}(M)$, el
subespacio de funciones de $\mathcal{H}$ cuyo soporte es disjunto
con el borde $\partial M$. El dominio $\mathcal{D}(A^{\dagger})$
del operador adjunto $A^{\dagger}$ es isomorfo al espacio de
Sobolev $\mathbf{H_2}(M)$.

\bs

Definimos ahora para toda funci{\'o}n $\psi$ del espacio
$\mathcal{H}$, sus valores de borde. Llamamos $\phi$ a la
restricci{\'o}n de $\psi$ al borde de $\partial M$ y $\dot{\phi}$
a la restricci{\'o}n de su derivada con respecto al versor normal
interior. Definimos tambi{\'e}n los valores de borde
$\phi_{\pm}:=\phi\pm i \dot{\phi}$ y la forma simpl{\'e}ctica
$\Sigma$ sobre el espacio $\mathbf{L_2}(\partial
M,E)\otimes\mathbf{L_2}(\partial M,E)$,
\begin{defn}
\begin{equation}\label{simple}
    \Sigma(\psi^{(1)},\psi^{(2)}):= \frac{i}{2}\int_{\partial M}
    \left\{(\phi_+^{(1)},\phi_+^{(2)})-
    (\phi_-^{(1)},\phi_-^{(2)})\right\}
\end{equation}
\end{defn}

Puede entonces probarse el siguiente Teorema,
\begin{thm}\label{valbor}
\begin{equation}
\psi^{(1)},\psi^{(2)}\in\mathcal{D}(A^{\dagger})\Rightarrow
(\psi^{(1)},A^{\dagger}\psi^{(2)})-
(A^{\dagger}\psi^{(1)},\psi^{(2)})=\Sigma(\psi^{(1)},\psi^{(2)})\,.
\end{equation}
\end{thm}

En consecuencia, el conjunto de extensiones autoadjuntas
$\mathcal{M}$ est{\'a} en correspondencia biun{\'\i}voca con el
conjunto de transformaciones unitarias de funciones del borde, que
anulan la forma simpl{\'e}ctica (\ref{simple}),
\begin{equation}
    \mathcal{M}\approx \mathcal{U}(\mathbf{L_2}(\partial M,E))\,.
\end{equation}
De este modo, a cada transformaci{\'o}n unitaria
 $U\in\mathcal{U}(\mathbf{L_2}(\partial M,E))$ le corresponde una extensi{\'o}n
 autoadjunta $H^U$; el dominio $\mathcal{D}(H^U)$ est{\'a} constitu{\'\i}do por las
 funciones $\psi\in\mathbf{L_2}(M,E)$ cuyos valores de borde satisfacen,
\begin{equation}
    \phi_-=U\cdot\phi_+\,.
\end{equation}

Esta caracterizaci{\'o}n se corresponde con la determinada por la
teor{\'\i}a de von Neumann. Esto es evidente para el caso
particular de aquellas variedades cuyos bordes est{\'a}n
constitu{\'\i}dos por $n$ puntos aislados. En ese caso, el espacio
de valores de borde es isomorfo a $\mathbb{C}^n$ y el grupo de
transformaciones unitarias es $\mathcal{U}=U(n)$. Por su parte, la
teor{\'\i}a de los {\'\i}ndices de deficiencia indica que cada uno
de estos bordes contribuye en una unidad al {\'\i}ndice de
deficiencia, de modo que de acuerdo con la teor{\'\i}a de von
Neumann el conjunto $\mathcal{M}$ de extensiones autoadjuntas
tambi{\'e}n se identifica con el grupo $U(n)$.

\bs

Puede verse, en forma inmediata, que las transformaciones
unitarias $U=\mp \mathbf{1}$ definen las extensiones autoadjuntas
caracterizadas por condiciones de contorno Dirichlet y Neumann,
respectivamente.

\bs

Por otra parte, para una extensi{\'o}n caracterizada por el
operador $U$ podemos
 definir los operadores autoadjuntos $B_{\pm}$,
\begin{eqnarray}
    B_-:= -i \frac{1-U}{1+U}\,,\label{da-}\\
    B_+:= i \frac{1+U}{1-U}\,.\label{da+}\
\end{eqnarray}
y caracterizar la condici{\'o}n de contorno por,
\begin{equation}\label{a-}
    \dot{\phi}=B_-\cdot \phi\,,
\end{equation}
o, equivalentemente, por,
\begin{equation}\label{a+}
    \phi=B_+\cdot\dot{\phi}\,.
\end{equation}
Sin embargo, no debe pensarse que el conjunto $\mathcal{M}$ de
extensiones autoadjuntas puede identificarse, mediante los
operadores $B_{+}$ o $B_-$, con el conjunto $\mathcal{L}(\partial
M,E)$ de operadores autoadjuntos de\-fi\-ni\-dos sobre el borde
$\partial M$. En efecto, el conjunto $\mathcal{L}$ tiene
topolog{\'\i}a trivial, en tanto que la topolog{\'\i}a del grupo
unitario $\mathcal{U}(\mathbf{L_2}(\partial M,E))$, que s\'\i\
est{\'a} identificado con $\mathcal{M}$, est{\'a} dada por,
\begin{equation}\label{top}
    \pi_n(\mathcal{U}(\mathbf{L_2}(\partial M,E)))=
    \left\{\begin{array}{cc}
        0&{\rm si\ n\ es\ par\,,}\\
        \mathbb{Z}&{\rm si\ n\ es\ impar\,.}
    \end{array}\right.
\end{equation}
Para comprender la diferencia entre los conjuntos
$\mathcal{L}(\partial M,E)$ y $\mathcal{M}$ definimos los
subconjuntos de $\mathcal{U}(\mathbf{L_2}(\partial M,E))$ cuyos
o\-pe\-ra\-do\-res contienen a $\mp 1$ en su espectro,
\begin{equation}
    \mathcal{C}_{\mp}:= \{U\in\mathcal{U}(\mathbf{L_2}(\partial M,E)):\mp
 1\in\sigma(U)\}.
\end{equation}
Estos subconjuntos se denominan {\bf subvariedades de Cayley}. De
acuerdo con las ecuaciones (\ref{da-}) y (\ref{da+}), si
$U\in\mathcal{C}_{\mp}$ entonces los operadores $B_{\mp}$ no
pueden definirse. En otros t{\'e}rminos, la caracterizaci{\'o}n de
las extensiones autoadjuntas o, equivalentemente, de las
condiciones de contorno mediante la ecuaci{\'o}n (\ref{a-})
((\ref{a+})) no es posible para aquellas extensiones autoadjuntas
pertenecientes al conjunto $\mathcal{C}_{-}$ ($\mathcal{C}_{+}$.)

\bs

Por su parte, el conjunto $\mathcal{U}(\mathbf{L_2}(\partial
M,E))-\mathcal{C}_-$ est{\'a} identificado con
$\mathcal{L}(\partial M,E)$ y define extensiones autoadjuntas que
pueden caracterizarse por una condici{\'o}n de contorno de la
forma (\ref{a-}). Lo mismo puede decirse de las extensiones
autoadjuntas pertenecientes a $\mathcal{U}(\mathbf{L_2}(\partial
M,E))-\mathcal{C}_+$ con respecto a la condici{\'o}n (\ref{a+}).

\bs

En conclusi{\'o}n, la topolog{\'\i}a de la variedad
$\mathcal{U}(\mathbf{L_2}(\partial M,E))$ no es trivial pero todos
los ciclos intersecan tanto a $\mathcal{C}_-$ como a
$\mathcal{C}_+$. Por consiguiente, las variedades,
\begin{equation}\label{Cay}
\mathcal{L}_\pm\approx\mathcal{U}(\mathbf{L_2}(\partial
M,E))-\mathcal{C}_{\pm}\,,
\end{equation}
tiene topolog{\'\i}a trivial.

\bs

En los ejemplos que veremos a lo largo de esta Tesis, las
extensiones autoadjuntas provienen del estudio de fibrados
lineales, {\it i.e.}, $E=\mathbb{C}$, sobre variedades
unidimensionales cuyo borde es un punto $P$ (a excepci{\'o}n del
ejemplo considerado en la secci{\'o}n \ref{tubos}.) En estos casos
la variedad $\mathcal{M}$ de extensiones autoadjuntas es isomorfa
al grupo de transformaciones unitarias
$\mathcal{U}(\mathbf{L_2}(P,\mathbb{C}))=U(1)$. N{\'o}tese que,
como indica la ecuaci{\'o}n (\ref{top}),
$\pi_1(U(1))\approx\mathbb{Z}$, todos los ciclos intersecan a las
variedades de Cayley $\mathcal{C}_{\mp}=\mp 1$ y las subvariedades
$\mathcal{L}_\pm$ (v{\'e}ase la ecuaci{\'o}n (\ref{Cay})) tienen,
en efecto, topolog{\'\i}a trivial.

\bs

Aunque no discutiremos los detalles, mencionamos adem{\'a}s que,
dado que el grupo fundamental de homotop{\'\i}a de
$\pi_1(\mathcal{M})$ es isomorfo a los enteros y que $\mathcal{M}$
es conexa, el grupo de cohomolog{\'\i}a
$H^1(\mathcal{M})\approx\mathbb{Z}$ y se puede entonces construir,
sobre el fibrado de determinantes, la 1-forma diferencial que lo
caracteriza.

\subsection{Estados de borde}\label{esta}

Consideraremos, ahora, una propiedad interesante de las
subvariedades de Cayley
 en relaci{\'o}n con la existencia de estados de borde \cite{Asorey:2004kk}.

\bs

En primer lugar, se\~nalemos que el operador
$\Delta=-d_\mathcal{A}^\dagger d_\mathcal{A}$ definido sobre
$\mathcal{C}_0^{\infty}(M,E)$ es positivo definido. No obstante
una extensi{\'o}n autoadjunta $\Delta^U$, caracterizada por un
operador unitario $U\notin \mathcal{C}_{-}$, puede no ser positiva
definida. Consideremos el producto interno de dos funciones
$\psi^{(1)},\psi^{(2)}\in \mathcal{D}(\Delta^U)$,
\begin{equation}
    (d_\mathcal{A}\psi^{(1)},d_\mathcal{A}\psi^{(2)})=(\dot{\phi}^{(1)},
    \phi^{(2)})
    +(\Delta^U\psi^{(1)},\psi^{(2)})=
    (B_-\cdot\phi^{(1)},\phi^{(2)})+(\Delta^U\psi^{(1)},\psi^{(2)}).
\end{equation}
En consecuencia, para todo vector $\psi\in\mathcal{D}(\Delta^U)$,
\begin{equation}\label{posi}
    (\Delta^U\psi,\psi)=(d_\mathcal{A}\psi,d_\mathcal{A}\psi)-
    (B_-\cdot\phi,\phi)\,.
\end{equation}
De modo que el operador $\Delta$ puede no resultar positivo
definido si el segundo t{\'e}rmino del miembro derecho de la
ecuaci{\'o}n anterior es suficientemente grande. El siguiente
Teorema relaciona esta posibilidad con la ``proximidad'' del
operador $U$ a la subvariedad de Cayley $\mathcal{C}_-$.
\begin{thm}\label{cayley}
    $U\in\mathcal{C}_-$ entonces $U_t:= e^{it}U$ define una
    extensi{\'o}n autoadjunta $\Delta^t$ que tiene un ``estado de
    borde'' de energ{\'\i}a ne\-ga\-ti\-va para $t$ suficientemente
    peque\~no. La energ{\'\i}a de este estado tiende a $-\infty$ cuando
    $t$ tiende a cero.
\end{thm}

\noindent{{\bf Demostraci{\'o}n:}} Daremos indicaciones para una
demostraci{\'o}n constructiva omitiendo los detalles t{\'e}cnicos.
El estado de borde $\psi_{\xi,t}$ est{\'a} dado por,
\begin{equation}
    \psi_{\xi,t}:= \xi(y)\exp{\left(-\frac{\tan{(x)}}{\tan{(t/2)}}\right)}\,,
\end{equation}
siendo $y$ la coordenada sobre el borde $\partial M$, $x$ la
coordenada normal al borde y $\xi(y)$ un estado de
$\mathbf{L_2}(\partial M,E)$ que satisface
\begin{equation}
    U\xi(y)=-\xi(y).
\end{equation}
El estado $\xi(y)$ existe puesto que $U\in\mathcal{C}_-$. El
estado de borde $\psi_{\xi,t}$, por su parte, pertenece a la
extensi{\'o}n autoadjunta caracterizada por el operador unitario
$e^{it} U$, como puede ve\-ri\-fi\-car\-se sin dificultad.
Asimismo, es f{\'a}cil ver que el estado $\psi_{\xi,t}$ se
concentra en el borde $\partial M$ a medida que $t$ tiende a cero.

\bs

Finalmente, puede calcularse el miembro derecho de la ecuaci{\'o}n
(\ref{posi}) para el estado de borde $\psi_{\xi,t}$ y probar que
tiende a $-\infty$ a medida que $t$ tiende a cero.\fin

En conlusi{\'o}n, podemos construir extensiones autoadjuntas
caracterizadas por un o\-pe\-ra\-dor $e^{it}U$, con
$U\in\mathcal{C}_-$ (e.g., condiciones Dirichlet\,\footnote{Si
hubier{\'a}mos realizado el mismo an{\'a}lisis a partir de
extensiones caracterizadas por operadores en $\mathcal{C}_+$
(e.g., condiciones Neumann) habr{\'\i}amos obtenido estados con
energ{\'\i}as arbitrariamente grandes que tienden, en el
l{\'\i}mite $t\rightarrow 0$ a modos ceros del operador
$\Delta^{t}$.}), cuyos estados fundamentales tienen energ{\'\i}as
que tienden a $-\infty$ cuando $t$ tiende a cero. Estas
extensiones contienen, adem{\'a}s, estados de borde cuyas normas
tienden a cero con $\sqrt{t}$.

\subsubsection{Ejemplo}

Consideremos el operador laplaciano en una dimensi{\'o}n
$-\partial_x^2$ con dominio de definici{\'o}n
$\mathcal{D}(-\partial_x^2)=\mathcal{C}_0^{\infty}((0,1))$. Como
el conjunto de funciones de borde $\mathbf{L_2}(\partial
M,\mathbb{C})$ es isomorfo a $\mathbb{C}^2$, el conjunto de
extensiones autoadjuntas est{\'a} en correspondencia
biu\-n{\'\i}\-vo\-ca con $U(2)$.

\bs

Construiremos un estado de borde a partir de una extensi{\'o}n en
$\mathcal{C}_-$. Sea entonces,
\begin{equation}
    U=\left(\begin{array}{cc}
    -1&0\\0&e^{i\beta}
    \end{array}\right)\,,
\end{equation}
que define condiciones de contorno Dirichlet en el origen y
condiciones de contorno Robin en $1$ caracterizadas por el
par{\'a}metro $\beta$. El autovector correspondiente $\xi$
est{\'a} dado por,
\begin{equation}
    \xi=\left(\begin{array}{c}
    \xi(0)\\\xi(1)
    \end{array}\right)
    =\left(\begin{array}{c}
    1\\0
    \end{array}\right)\,.
\end{equation}
De acuerdo con el Teorema \ref{cayley}, el estado de borde
est{\'a} entonces dado por,
\begin{equation}\label{estados}
    \psi_{\xi,t}=\left\{\begin{array}{cc}
    \exp{\left(-\epsilon\frac{\tan{(x/\epsilon)}}{\tan{(t/2)}}\right)}
    & {si}\ \ 0\leq x\leq \epsilon\frac{\pi}{2}
    \\0
    & {si}\ \ \epsilon\frac{\pi}{2}\leq x\leq 1\,,
    \end{array}\right.
\end{equation}
donde $0<\epsilon<1$.

\bs

Es inmediato verificar que $\psi_{\xi,t}$ pertenece a la
extensi{\'o}n caracterizada por $e^{it}U$ y que el valor de
expectaci{\'o}n del hamiltoniano con respecto a este estado
tiende\,\footnote{Si $\epsilon<1/\sqrt{2}$.} a $-\infty$ conforme
$t$ tiende a cero. La norma del estado (\ref{estados}), cuya
energ{\'\i}a es arbitrariamente negativa para valores
suficientemente peque\~nos de $t$, est{\'a} acotada por $t$, de
modo que este estado no existe en el l{\'\i}mite $t\rightarrow 0$.

\bs

N{\'o}tese que las extensiones autoadjuntas caracterizadas por
operadores unitarios $U$ en la intersecci{\'o}n
$\mathcal{C}_-\cap\mathcal{C}_+$ representa un cambio de
topolog{\'\i}a en la variedad de base $M$, {\it i.e.}, en este
caso, condiciones de contorno peri{\'o}dicas. Eventualmente, una
fase en los elementos de $U$ representa condiciones de contorno
pseudoperi{\'o}dicas, correspondientes a part{\'\i}culas con
estad{\'\i}stica fraccionaria.


\part{{Ruptura Espont{\'a}nea de SUSY en Mec{\'a}nica Cu{\'a}ntica.}}\label{rupturaSUSY}

\vspace{5mm}\begin{flushright}{\it I have done a terrible thing:
I have postulated a particle\\that cannot be detected.\\
(Wolfgang Pauli.)}
\end{flushright}

\vspace{25mm}

\section{Introducci{\'o}n}

La supersimetr{\'\i}a (SUSY) es considerada una extensi{\'o}n
natural de las teor{\'\i}as de gauge, contribuye a la
cancelaci{\'o}n de divergencias en la Teor{\'\i}a Cu{\'a}ntica de
Campos y es un ingrediente esencial de la Teor{\'\i}a de Cuerdas.
Sin embargo, la degeneraci{\'o}n de los niveles de energ{\'\i}a
asociada con esta simetr{\'\i}a implica la existencia de
compa{\~n}eros supersim{\'e}tricos de igual masa que no existen en
la naturaleza.

\bs

En consecuencia, esta simetr{\'\i}a s{\'o}lo puede realizarse bajo
un mecanismo de ruptura espont{\'a}nea (din{\'a}mica). Pero a
diferencia de lo que ocurre con las simetr{\'\i}as ordinarias, la
ruptura espont{\'a}nea de la SUSY es muy dif{\'\i}cil de
implementar.

\bs

Modelos de mec{\'a}nica cu{\'a}ntica supersim{\'e}trica (SUSYQM)
en una dimensi{\'o}n fueron estudiados primeramente por H.\
Nicolai \cite{Nicolai:xp} y E.\ Witten
\cite{Witten:nf,Witten:1982df,Witten:im}. En este contexto se
propuso la ruptura de la SUSY a partir de mecanismos no
perturbativos basados en potenciales asociados a soluciones de
instantones
\cite{Witten:nf,Witten:1982df,Salomonson:1981ug,Cooper-K-S,Freedman:1983as,Cooper-Freedman}.

\bs

M{\'a}s recientemente, A.\ Jevicki y J.P.\ Rodrigues
\cite{Jevicki} han sugerido que la SUSY tambi{\'e}n podr{\'\i}a
ser espont{\'a}neamente rota mediante superpotenciales singulares
por efecto de condiciones de contorno inusuales en la
singularidad. Para ello consideraron un operador diferencial de
segundo orden, estudiado previamente por L.\ Lathouwers
\cite{lath}, correspondiente a un hamiltoniano supersim{\'e}trico
derivado de un superpotencial con una sin\-gu\-la\-ri\-dad en el
origen $x=0$ proporcional a $x^{-1}$. Sin embargo, los autores no
han tenido en cuenta si las autofunciones analizadas corresponden
a un mismo hamiltoniano autoadjunto.

\bs

Posteriormente, A.\ Das y S.\ Pernice \cite{Das,Pernice} han
mostrado que una regularizaci{\'o}n del superpotencial conduce a
una SUSY expl{\'\i}cita, es decir, a un sistema con un estado
fundamental de energ{\'\i}a nula y estados excitados doblemente
degenerados.

\bs

En este cap{\'\i}tulo presentaremos uno de los resultados
originales de esta tesis referido a este superpotencial en SUSYQM
\cite{FP2}. En particular, estudiaremos la relevancia de las
extensiones autoadjuntas de las supercargas (generadores de la
SUSY) y del ha\-mil\-to\-nia\-no del sistema en relaci{\'o}n con
la ruptura espont{\'a}nea de la SUSY. Mostraremos que, en general,
las extensiones autoadjuntas poseen una SUSY din{\'a}micamente
rota por efecto de las condiciones de contorno y, por
consecuencia, sus espectros presentan un estado fundamental de
energ{\'\i}a no nula y estados excitados no degenerados.

\bs

El hamiltoniano y las supercargas de este sistema constituyen,
asimismo, ejemplos de inter{\'e}s en relaci{\'o}n con el objetivo
central de esta tesis, referido a las propiedades inusuales de las
funciones espectrales correspondientes a operadores singulares. En
la secci{\'o}n \ref{spectral-functions111} del Ap{\'e}ndice
calcularemos algunas de las funciones espectrales asociadas a este
problema.

\section{N=2 SUSYQM}

La mec{\'a}nica cu{\'a}ntica supersim{\'e}trica $N=2$ es la
realizaci{\'o}n en
 $0+1$-dimensiones del {\'a}lgebra de supersimetr{\'\i}a:
\begin{equation}\label{qqth0}\begin{array}{c}
    \displaystyle{\{Q,Q\}=\{Q^{\dagger},Q^\dagger\}=0\,,\qquad
    \left\{ Q, Q^{\dagger} \right\}=H\,,}\\ \\
    \displaystyle{
    \mbox{}[H,Q]=[H,Q^\dagger]=0\,.}
    \end{array}
\end{equation}
El generador de las traslaciones temporales es el hamiltoniano $H$
y los generadores $Q$ y $Q^\dagger$, adjuntos uno a otro, son las
supercargas que generan traslaciones en el superespacio y
constituyen el sector fermi{\'o}nico del {\'a}lgebra. Todos los
generadores act{\'u}an sobre un mismo espacio de Hilbert
$\mathcal{H}$. Si definimos las combinaciones lineales
autoadjuntas,
\begin{equation}
    Q_+=Q+Q^\dagger\qquad
    Q_-=i\left(Q-Q^\dagger\right)\,,
\end{equation}
el {\'a}lgebra (\ref{qqth0}) toma la forma,
\begin{equation}\label{qqth}\begin{array}{c}
    \displaystyle{\{Q_+,Q_-\}=0\,,\qquad
    \left\{ Q_+,Q_+ \right\}=\left\{ Q_-,Q_- \right\}=2H\,,}\\ \\
    \displaystyle{
    \mbox{}[H,Q_+]=[H,Q_-]=0\,.}
    \end{array}
\end{equation}

\bs

Una representaci{\'o}n de las relaciones (\ref{qqth}) est{\'a}
dada por,
\begin{eqnarray}\label{Q-Qtilde}
    Q_+=\left(\begin{array}{cc}
            0&D^\dagger\\
            D&0\
        \end{array}\right)\,,\quad
    Q_-=\left(\begin{array}{cc}
            0&-iD^\dagger\\
            iD&0\
        \end{array}\right)\,,
        \quad
    H=\left(\begin{array}{cc}
            D^\dagger\, D&0\\
            0&D\,D^\dagger\
        \end{array}\right)\,,
\end{eqnarray}
donde,
\begin{eqnarray}\label{A-Atilde}
    D=\frac{1}{\sqrt{2}}\left(-\partial_x+W(x)\right)\,,\quad
    D^\dagger=\frac{1}{\sqrt{2}}\left(\partial_x+W(x)\right)\,,
\end{eqnarray}
son operadores diferenciales definidos sobre un subespacio denso
de funciones de una variedad unidimensional en el cual la
composici{\'o}n est{\'a} bien definida. La funci{\'o}n $W(x)$ se
denomina superpotencial.

\bigskip

De las relaciones de anticonmutaci{\'o}n (\ref{qqth}) se puede ver
que el operador hamiltoniano $H$ es positivo definido y
autoadjunto. Adem{\'a}s, si el estado fundamental $\psi_0$ del
sistema es invariante ante las transformaciones de
supersimetr{\'\i}a, {\it id est},
\begin{equation}
    Q_+\psi_0=Q_-\psi_0=0,
\end{equation}
entonces $H\psi_0=0$, la energ{\'\i}a del estado fundamental es
cero. Rec{\'\i}procamente, si la energ{\'\i}a del estado
fundamental es distinta de cero, entonces existe una ruptura
espont{\'a}nea de la supersimetr{\'\i}a; el hamiltoniano conmuta
con las supercargas pero sus autoestados no son invariantes ante
una transformaci{\'o}n de supersimetr{\'\i}a.

\bs

El mecanismos de ruptura espont{\'a}nea de la supersimetr{\'\i}a
que estudiaremos consiste en considerar variedades con borde y
determinar la energ{\'\i}a del estado fundamental en t{\'e}rminos
de las condiciones de contorno. Con este objetivo, estudiaremos a
continuaci{\'o}n un superpotencial que posee una singularidad en
el borde de una variedad unidimensional. Las distintas condiciones
de contorno admisibles en la singularidad est{\'a}n determinadas
por las extensiones autoadjuntas del hamiltoniano. La
supersimetr{\'\i}a resultar{\'a} espont{\'a}neamente rota para
aquellas extensiones autoadjuntas cuyos estados fundamentales
tengan energ{\'\i}a distinta de cero.

\bs

Se debe tener presente que, aunque la representaci{\'o}n dada por
las expresiones (\ref{Q-Qtilde}) y (\ref{A-Atilde}) indican que
los operadores $Q^\dagger,D^\dagger$ son ``formalmente'' los
operadores adjuntos de $Q,D$, respectivamente, esta propiedad
est{\'a} condicionada por la adecuada elecci{\'o}n de los dominios
de definici{\'o}n de los operadores. Mostraremos que estos
dominios quedan determinados por el procedimiento que utilizaremos
para construir las extensiones autoadjuntas del hamiltoniano.

\section{Superpotencial singular}

Consideremos entonces el problema en SUSYQM definido por un
superpotencial $W(x)$, con una singularidad en el origen, dado
por,
\begin{equation}\label{W}
  W=\frac{\alpha}{x}-x\,,
\end{equation}
siendo $x\in\mathbb{R}^+$ y $\alpha\in\mathbb{R}$. Teniendo en
cuenta las expresiones (\ref{A-Atilde}) definimos los
o\-pe\-ra\-do\-res diferenciales,
\begin{eqnarray}
    D_1=\frac{1}{\sqrt{2}}\left(-\partial_x+\frac{\alpha}{x}-x\right)\,,
    \label{superA}\\
    D_2=\frac{1}{\sqrt{2}}\left(\partial_x+
    \frac{\alpha}{x}-x\right)\,,\label{superAt}
\end{eqnarray}
con dominio en el subespacio denso ${\cal
C}^{\infty}_{0}(\mathbb{R}^+)$ de funciones infinitamente
derivables y con soporte compacto disjunto del origen.

\bs

Es conveniente definir ahora el operador de supercarga $Q_+$ en el
subespacio ${\cal D}(Q_+)={\cal
C}^{\infty}_{0}(\mathbb{R}^+)\otimes \mathbb{C}^2$ cuya acci{\'o}n
sobre los vectores $\Phi$ de componentes $\phi_1,\phi_2$ est{\'a}
dada por,
\begin{equation}\label{Q_+-def}
    Q_+ \Phi=\left(\begin{array}{cc}
            0&D_2\\
            D_1&0\
        \end{array}\right)\left(\begin{array}{c}
          \phi_1 \\
          \phi_2 \\
        \end{array}\right)\,.
\end{equation}
Su cuadrado, que est{\'a} bien definido, permite definir,
\begin{equation}\label{susy-ham}
    H:=Q_+^2\,.
\end{equation}
El operador $Q_+$, al igual que $H$, es sim{\'e}trico debido a las
restrictivas condiciones de contorno de su dominio de
definici{\'o}n, sin embargo, no es autoadjunto (ni cerrado) sino
que admite una familia de extensiones autoadjuntas
$Q^{(\gamma)}_+$ caracterizada por un par{\'a}metro real $\gamma$
(la clausura del operador ser{\'a} determinada, posteriormente, en
la secci{\'o}n (\ref{closuresusy}).)

\bs

Ahora bien, en virtud de un teorema de von Neumann \cite{teor-vN},
el operador hamiltoniano definido por,
\begin{equation}\label{ham2}
    H^{(\gamma)}:= Q^{(\gamma)}_+\cdot Q^{(\gamma)}_+\,,
\end{equation}
es autoadjunto en el dominio de definici{\'o}n,
\begin{equation}\label{domsusy}
    \mathcal{D}(H^{(\gamma)})=\{
    \psi\in\mathcal{D}(Q^{(\gamma)}_+):Q^{(\gamma)}_+\,
    \psi\in\mathcal{D}(Q^{(\gamma)}_+)\}\,.
\end{equation}
En consecuencia, las extensiones autoadjuntas del operador $Q_+$
constituyen distintas re\-pre\-sen\-ta\-cio\-nes de la supercarga
caracterizadas por el par{\'a}metro $\gamma$ y determinan los
dominios sobre los cuales los operadores hamiltonianos
(\ref{ham2}) son autoadjuntos.

\bs

Por otra parte, se puede definir una segunda supercarga
linealmente independiente de $Q_+$ como,
\begin{equation}\label{Qmenos}
    Q_- = \left(%
\begin{array}{cc}
  0 & -i D_2 \\
  i D_1 & 0 \\
\end{array}%
\right)\, ,
\end{equation}
que tambi{\'e}n resulta sim{\'e}trica en
$\mathcal{C}_0^{\infty}(\mathbb{R}^+)\otimes \mathbb{C}^2$ y
verifica,
\begin{equation}\label{Qmenoscuadrado}
    Q_-^2 = Q_+^2=H\,,\qquad
    \left\{ Q_+ , Q_- \right\} = 0\,.
\end{equation}
Sin embargo, los generadores $Q_+,Q_-,H$ no son autoadjuntos en
$\mathcal{C}_0^{\infty}(\mathbb{R}^+)\otimes \mathbb{C}^2$.

\bs

Dado que $Q_-$ se obtiene de $Q_+$ mediante una transformaci{\'o}n
unitaria,
\begin{equation}\label{transf-unit}
    Q_- = e^{\displaystyle{i\pi\sigma_3/4}}\,  Q_+ \,
    e^{-\displaystyle{i\pi\sigma_3/4}}\,,\qquad
    \sigma_3 = \left(%
    \begin{array}{cc}
      1 & 0 \\
      0 & -1 \\
    \end{array}%
    \right)
\end{equation}
toda extensi{\'o}n autoadjunta $Q^{(\gamma)}_+$ de $Q_+$ determina
una extensi{\'o}n autoadjunta $Q^{(\gamma)}_-$ de $Q_-$ cuyo
dominio es la imagen de $\mathcal{D}(Q^{(\gamma)}_+)$ por la
transformaci{\'o}n unitaria $e^{{i\pi\sigma_3/4}}$. En
consecuencia, ambas extensiones presentan el mismo espectro.
M{\'a}s adelante estudiaremos la compatibilidad entre los dominios
de estas supercargas en relaci{\'o}n con la posibilidad de
realizar el {\'a}lgebra (\ref{qqth}).

\bs

El primer paso en la construcci{\'o}n de las extensiones
autoadjuntas de $Q_+$ consiste en determinar su adjunto
$Q_+^\dagger$. Este es el tema de la siguiente secci{\'o}n.

\section{El operador adjunto}

Para determinar las extensiones autoadjuntas de $Q_+$ debemos
estudiar sus
 subespacios de
 deficiencia,
\begin{equation}\label{defi-sub}
  {\cal K}_\pm
    := {\rm Ker}\left(Q_+^\dagger \mp i \right).
\end{equation}
Para ello, determinaremos el dominio y el espectro de
$Q_+^\dagger$.

\subsection{Dominio de $Q_+^\dagger$}

Un vector $\Psi\in\mathbf{L_2}(\mathbb{R}^+)\otimes \mathbb{C}^2$
pertenece al dominio de $Q_+^\dagger$,
\begin{equation}\label{dom-adjoint}
  \Psi=\left( \begin{array}{c}
  \psi_1   \\
  \psi_2
\end{array} \right) \in {\cal D}(Q_+^\dagger)\,,
\end{equation}
si $\left(\Psi, Q_+ \Phi\right)$ es una funcional lineal y
continua de $\Phi \in {\cal D}(Q_+)$. Esto implica, por el Teorema
de Riesz, la existencia de una funci{\'o}n $\widetilde{\Psi}$,
\begin{equation}
  \widetilde{\Psi}=\left(
    \begin{array}{c}
  \widetilde{\psi}_1   \\
  \widetilde{\psi}_2
    \end{array}\right)  \in
    \mathbf{L_2}(\mathbb{R^+})\,,
\end{equation}
tal que,
\begin{equation}\label{adjoint}
  \left(\Psi,Q_+ \Phi\right)=\left(\widetilde{\Psi},\Phi\right), \ \forall\,
 \Phi
  \in {\cal D}(Q_+)\,.
\end{equation}
La funci{\'o}n $\widetilde{\Psi}$ est{\'a} un{\'\i}vocamente
determinada en virtud de que ${\cal D}(Q_+)$ es un subespacio
denso. Consecuentemente, la acci{\'o}n de $Q_+^\dagger$, para cada
$\Psi \in {\cal D}(Q_+^\dagger)$, est{\'a} definida por,
\begin{equation}\label{qtilde}
Q_+^\dagger \Psi :=\widetilde{\Psi}\,.
\end{equation}
Se debe recordar que, como $Q_+$ es sim{\'e}trico, ${\cal D}(Q_+)
\subset {\cal D }(Q_+^\dagger)$.

\bigskip

Determinaremos ahora las propiedades de las funciones
pertenecientes a ${\cal D}(Q_+^\dagger)$, y el modo en que
$Q_+^\dagger$ act{\'u}a sobre ellas. La ecuaci{\'o}n
(\ref{adjoint}) implica,
\begin{eqnarray}\label{deriv-prim}
    -\psi_1'+\left(\frac{\alpha}{x}-x\right)\psi_1=\sqrt{2} \,
 \widetilde{\psi}_2\,, \\
    \psi_2'+\left(\frac{\alpha}{x}-x\right)\psi_2=\sqrt{2}\,
    \widetilde{\psi}_1\,,
\end{eqnarray}
donde las derivadas se consideran en el sentido generalizado. Esto
muestra que $\Psi'(x)$ es una distribuci{\'o}n regular, {\it
i.e.}, localmente integrable. Por lo tanto, $\Psi(x)$ es una
funci{\'o}n absolutamente continua para $x>0$, y el dominio de
$Q_+^\dagger$ resulta,
\begin{equation}\label{dom-Q_++}\begin{array}{c}
  {\cal D}(Q_+^\dagger)=\left\{
  \Psi \in AC(\mathbb{R^+}-\{0\}) \cap
  \mathbf{L_2}(\mathbb{R^+}); \right.
  \left. D_1\,\psi_1,\, D_2\,\psi_2 \in  \mathbf{L_2}(\mathbb{R^+})
  \right\}\,.
\end{array}
\end{equation}
Por consiguiente, podemos realizar una integraci{\'o}n por partes
en el miembro izquierdo de la ecuaci{\'o}n (\ref{adjoint}) y
concluir que la acci{\'o}n de $Q_+^\dagger$ sobre $\Psi \in  {\cal
D}(Q_+^\dagger)$ est{\'a} dada por,
\begin{equation}\label{Q_++-def}
    Q_+^\dagger \Psi=\left(\begin{array}{cc}
            0&D_2\\
            D_1&0\
        \end{array}\right)\left(\begin{array}{c}
          \psi_1 \\
          \psi_2 \
        \end{array}\right)\,,
\end{equation}
donde las derivadas deben interpretarse en el sentido
generalizado.

\subsection{Espectro de $Q_+^\dagger$} \label{spectrumqm}

Consideremos ahora el problema de autovalores de $Q_+^\dagger$,
\begin{equation}\label{Q_+-auto}
  Q_+^\dagger \Phi_\lambda = \lambda \Phi_\lambda,
\end{equation}
o equivalentemente,
\begin{equation}\label{eigen-problem}
  D_1\,\phi_1=\lambda\, \phi_2\,, \quad D_2\,\phi_2=\lambda\,
  \phi_1\,,
\end{equation}
con,
\begin{equation}\label{phi-eigen}
  \Phi_\lambda = \left(
    \begin{array}{c}
      \phi_1 \\
      \phi_2
    \end{array}
\right) \in {\cal D}(Q_+^\dagger)\,,
\end{equation}
y $\lambda \in \mathbb{C}$.

A partir de las ecuaciones (\ref{superA}), (\ref{superAt}) y
(\ref{eigen-problem}), se deduce que $\Phi_\lambda'(x)$ es
tambi{\'e}n una funci{\'o}n absolutamente continua. En
consecuencia, aplicaciones sucesivas de $Q_+^\dagger$ a ambos
lados de la ecuaci{\'o}n (\ref{Q_+-auto}), demuestran que
$\Phi_\lambda(x) \in {\cal C}^{\infty}(\mathbb{R}^+-\{0\})$, y la
ecuaci{\'o}n (\ref{eigen-problem}) resulta equivalente a un
sistema de ecuaciones diferenciales ordinarias.

\bigskip

Reemplazando $\phi_2$ en t{\'e}rminos de $\phi_1$ en la
ecuaci{\'o}n (\ref{eigen-problem}) obtenemos,
\begin{eqnarray}
    -\frac{1}{2}\,\phi_1''+\frac{1}{2}
            \left\{\frac{\alpha(\alpha-1)}{x^2}+x^2-1-2\alpha\right\}\phi_1=
            \lambda^2\, \phi_1\,,\label{sys1}\\
   \lambda\, \phi_2=\frac{1}{ \sqrt{2}}
    \left\{-\phi_1'+
    \left(\frac{\alpha}{x}-x\right)\phi_1\right\}\,. \label{sys2}
\end{eqnarray}
Haciendo la substituci{\'o}n,
\begin{equation}\label{phi1dex}
    \phi_1(x)=x^\alpha\, e^{{- {x^2/ 2}} }\, F(x^2)
\end{equation}
en la ecuaci{\'o}n (\ref{sys1}) obtenemos la ecuaci{\'o}n de
Kummer \cite{A-S} para
 $F(z)$,
\begin{equation}\label{kummer-eq}
    z\, F''(z)+(b-z)\,F'(z) -a\, F(z)=0,
\end{equation}
con,
\begin{equation}\label{a-b}
  a=- \frac{\lambda^2}{2}\,,\qquad b=\alpha+\frac{1}{2}\,.
\end{equation}

Para cualquier valor de los par{\'a}metros $a$ y $b$, la
ecuaci{\'o}n (\ref{kummer-eq}) tiene dos soluciones linealmente
independientes \cite{A-S} dadas por las funciones de Kummer,
\begin{equation}\label{LI-sol-1}
  \begin{array}{c}
    y_1(z)=U(a,b,z)= \\ \\
    \displaystyle{ \frac{\pi}{\sin \pi b}\left\{
    \frac{M(a,b,z)}{\Gamma(1+a-b)\Gamma(b)} -
    z^{1-b}\, \frac{M(1+a-b,2-b,z)}{\Gamma(a)\Gamma(2-b)}
    \right\} }\,,
  \end{array}
\end{equation}
y,
\begin{equation}\label{LI-sol-2}
  y_2(z)= e^z \, U(b-a,b,-z)\,.
\end{equation}
En la ecuaci{\'o}n (\ref{LI-sol-1}), $M(a,b,z)$ es la funci{\'o}n
hipergeom{\'e}trica confluente ${_1F_1(a;b;z)}$.

\bs

Para grandes valores de $|z|$ \cite{A-S},
\begin{equation}\label{U-asymp}
  U(a,b,z)=z^{-a}\left\{1+ {\cal O}(|z|^{-1})\right\}\,,
\end{equation}
de modo que solamente $y_1(x^2)$ conduce a una funci{\'o}n
$\phi_1(x) \in\mathbf{L_2}(1,\infty)$ al ser reemplazada en la
ecuaci{\'o}n (\ref{phi1dex}).

\bs

Por consiguiente, una de las componentes de $\Phi_\lambda$
est{\'a} dada por,
\begin{equation}\label{phi1-U}
  \phi_1(x)= x^\alpha\, e^{{- {x^2/ 2}} }\,
  U\left(-\frac{\lambda^2}{2}, \alpha+\frac{1}{2}, x^2\right)\,.
\end{equation}
Por su parte, reemplazando la ecuaci{\'o}n (\ref{phi1-U}) en la
ecuaci{\'o}n (\ref{sys2}), obtenemos para la otra componente de
$\Phi_\lambda$,
\begin{equation}\label{phi2-U}
    \phi_2(x)=-\frac{\lambda}{\sqrt{2}}\, x^{\alpha+1}\, e^{-{x^2}/{2}}
        \, U\left(1-\frac{\lambda^2}{2};\alpha+\frac{3}{2},x^2\right)\,,
\end{equation}
que tambi{\'e}n pertenece a $\mathbf{L_2}(1,\infty)$.

\bs

Sin embargo, debemos tambi{\'e}n considerar el comportamiento de
$\Phi_\lambda(x)$ cerca del origen; esto permitir{\'a} la
determinaci{\'o}n del espectro. De la ecuaci{\'o}n
(\ref{LI-sol-1}), y del desarrollo para peque\~nos argumentos de
las funciones de Kummer \cite{A-S}, se concluye que existen tres
casos que deben analizarse por separado, de acuerdo con el valor
del par{\'a}metro $\alpha$:

\begin{enumerate}

\item Si $\alpha\geq 1/2$, se puede ver que
$\Phi_\lambda(x)\in\mathbf{L_2}(0,1)$ si y s{\'o}lo si
$-\lambda^2/2 = -n$, con $n=0,1,2,\dots$

\bs

En este caso, teniendo en cuenta que $U(-n, b, z)$ se reduce al
polinomio de Laguerre de grado $n$ en $z$,
\begin{equation}\label{U-poly}
    U(-n, b, z) = (-1)^n \, n! \,
    L_n^{(b-1)}(z)\,,
\end{equation}
obtenemos $\phi_1(x)\sim x^\alpha$ y $\phi_2(x)\sim x^{\alpha+1}$
para $0<x\ll 1$. En consecuencia, si $\alpha\geq 1/2$,
$Q_+^\dagger$ tiene un espectro real no degenerado y sim{\'e}trico
respecto del origen, dado por los autovalores,
\begin{equation}\label{spectrum-1}
  \lambda_{0}=0,\qquad \lambda_{\pm,n}= \pm \sqrt{2 n},\qquad n=1,2,3,\dots
\end{equation}
que corresponden a las autofunciones,
\begin{equation}\label{eigenfunctions-1-0}
  \Phi_{0}=x^\alpha\, e^{{- {x^2/ 2}} }
  \left(\begin{array}{c}
    1 \\0
  \end{array}\right)\,,
\end{equation}
y,
\begin{equation}\label{eigenfunctions-1}
  \Phi_{\pm,n}=(-1)^n \, n! \,x^\alpha\, e^{{- {x^2/ 2}} }
  \left(\begin{array}{c}
    L_n^{(\alpha-\frac{1}{2})}(x^2)
  \\ \\
  \mp \displaystyle{\frac{x}{\sqrt{n}}} \,
   L_{n-1}^{(\alpha+\frac{1}{2})}(x^2)
  \end{array}\right)\,,
\end{equation}
respectivamente.

\item Por su parte, si $\alpha\leq -1/2$ se puede ver que
$\Phi_\lambda(x)\in \mathbf{L_2}(0,1)$ si y s{\'o}lo si
$-\lambda^2/2 =\alpha-\frac{1}{2}-n$, con $n=0,1,2,\dots$

\bs

En este caso, teniendo en cuenta la transformaci{\'o}n de Kummer
(v{\'e}ase \cite{A-S}, p{\'a}g.\ 505),
\begin{equation}\label{Kummer-transform}
  U(1-n-b,2-b,z)=z^{b-1}\,U(-n,b,z),
\end{equation}
y la ecuaci{\'o}n (\ref{U-poly}), obtenemos $\phi_1(x)\sim
x^{1-\alpha}$ y $\phi_2(x)\sim x^{-\alpha}$ para $0<x\ll 1$. Por
lo tanto, si $\alpha\leq -1/2$, $Q_+^\dagger$ tiene un espectro
real no degenerado y sim{\'e}trico respecto del origen, dado por
los autovalores,
\begin{equation}\label{spectrum-2}
 \lambda_{\pm,n}= \pm \sqrt{2 n+1-2\alpha},\qquad n=0,1,2,\dots,
\end{equation}
correspondientes a las autofunciones,
\begin{equation}\label{eigenfunctions-2}\begin{array}{c}
  \Phi_{\pm,n}= (-1)^n \, n!\, 
  x^{-\alpha}\, e^{{- {x^2/ 2}} }
  \left(\begin{array}{c}
   x\, L_n^{(\frac{1}{2}-\alpha)}(x^2)
  \\ \\
  \mp \sqrt{n+\frac{1}{2}-\alpha} \,
   L_{n}^{(-\alpha-\frac{1}{2})}(x^2)
  \end{array}\right).
\end{array}
\end{equation}
N{\'o}tese que no existen modos cero para este rango de valores
del par{\'a}metro $\alpha$.

\item Por {\'u}ltimo, si $-1/2 < \alpha < 1/2$ se puede ver, a
partir de (\ref{phi1-U}), (\ref{phi2-U}) y (\ref{LI-sol-1}), que
$\Phi_\lambda(x)\in\mathbf{L_2}(0,1)$, $\forall \lambda \in
\mathbb{C}$.

\bs

Esto implica que, si $|\alpha|<1/2$, todo n{\'u}mero complejo es
un autovalor no degenerado de $Q_+^\dagger$. En consecuencia, la
autofunci{\'o}n de $Q_+^\dagger$ co\-rres\-pon\-dien\-te a
$\lambda=i$ est{\'a} dada por,
\begin{equation}\label{Phi+}
  \Phi_{+}(x):=\Phi_{\lambda=i}(x)
  = x^\alpha\, e^{{- {x^2/ 2}} }
  \left(\begin{array}{c}
    U\left(\frac{1}{2},\alpha+\frac{1}{2},x^2\right)
  \\ \\
  -\frac{i}{\sqrt{2}}\, x\, U\left(\frac{3}{2},\alpha+\frac{3}{2},x^2\right)
  \end{array}\right)\,,
\end{equation}
en tanto que la autofunci{\'o}n correspondiente a $\lambda=-i$
est{\'a} dada por su complejo conjugado,
\begin{equation}\label{Phi-}
  \Phi_{-}(x):=\Phi_{\lambda=-i}(x)=\Phi_{+}(x)^*\,,
\end{equation}
dado que los coeficientes del operador diferencial de la
ecuaci{\'o}n (\ref{eigen-problem}) son reales.

\end{enumerate}

\bs

N{\'o}tese que, como hemos se{\~n}alado, la dimensi{\'o}n del
subespacio ${\rm Ker}\,(Q^\dagger_+-\lambda)$ es cons\-tan\-te en
cada uno de los semiplanos $\mathcal{I}(\lambda)\neq 0$. En la
secci{\'o}n siguiente determinaremos las extensiones autoadjuntas
de $Q_+$.

\section{Extensiones autoadjuntas de la supercarga}

Para construir las extensiones autoadjuntas de $Q_+$ debemos tener
en cuenta, de acuerdo con el Corolario \ref{coro3}, los
subespacios caracter{\'\i}sticos de $Q_+^\dagger$ correspondientes
al autovalor $\pm i$ calculados en la secci{\'o}n anterior.
Veremos que si $|\alpha|\geq 1/2$ los subespacios de deficiencia
son triviales y el operador $Q_+$ es esencialmente autoadjunto.
Por el contrario, si $|\alpha|< 1/2$ los subespacios de
deficiencia son undimensionales y el operador $Q_+$ admite un
conjunto infinito de extensiones autoadjuntas.

\subsubsection{Si $|\alpha|\geq 1/2$ el operador $Q_+$ es esencialmente
 autoadjunto}

Como hemos visto en la secci{\'o}n \ref{spectrumqm}, los
{\'\i}ndices de deficiencia de $Q_+$,
\begin{equation}\label{defi-indi}
  n_\pm = {\rm dim}\ {\rm Ker}\left(Q_+^\dagger \mp i \right),
\end{equation}
se anulan para $|\alpha| \geq 1/2$. Esto significa que $Q_+$ es
esencialmente autoadjunto y admite una {\'u}nica extensi{\'o}n
autoadjunta dada por su clausura $\overline{Q}_+$.

\bs

De acuerdo con las ecuaciones (\ref{susy-ham}) y (\ref{domsusy}),
el hamiltoniano admite entonces una {\'u}nica extensi{\'o}n
autoadjunta dada por,
\begin{equation}\label{SAE-H-ESA}
    H=\overline{Q}_+\cdot\overline{Q}_+\,,
\end{equation}
cuyo dominio de definici{\'o}n est{\'a} dado por,
\begin{equation}\label{domain-Q_+2}
  {\cal D} \left( {H} \right) =
  \left\{\psi \in {\cal D}\left(\overline{Q}_+\right):
  \overline{Q}_+ \psi \in {\cal D}\left(\overline{Q}_+\right) \right\}.
\end{equation}
N{\'o}tese que toda autofunci{\'o}n de $\overline{Q}_+$,
correspondiente a un autovalor $\lambda$, pertenece a ${\cal D}
\left( {H}\right)$. Por lo tanto, es tambi{\'e}n una
autofunci{\'o}n de ${H}$ con autovalor $E=\lambda^2$. Obtenemos
entonces la siguiente descripci{\'o}n del espectro del
hamiltoniano en el caso $|\alpha|\geq 1/2$:

\begin{itemize}

\item {Si $\alpha\geq 1/2$, las autofunciones de ${H}$ est{\'a}n
dadas por las ecuaciones (\ref{eigenfunctions-1-0}) y
(\ref{eigenfunctions-1}). N{\'o}tese que existe un {\'u}nico modo
cero, en tanto que los restantes autovalores de ${H}$,
\begin{equation}\label{eigen-H-1}
  E_n=2\,n, \quad n=1,2,3,\dots
\end{equation}
son positivos y tienen una degeneraci{\'o}n doble (v{\'e}ase la
ecuaci{\'o}n (\ref{spectrum-1}).)

\bs

Las combinaciones $\Phi_{+,n}\pm \Phi_{-,n}$ (v{\'e}ase la
ecuaci{\'o}n (\ref{eigenfunctions-1}) representan estados {\it
bo\-s{\'o}\-ni\-cos} y {\it fermi{\'o}nicos}, esto es, con la
componente inferior o superior nula, respectivamente. Para estos
valores del par{\'a}metro $\alpha$ el {\'\i}ndice de Witten es
$\Delta = 1$ y, por consiguiente, la supersimetr{\'\i}a es
expl{\'\i}cita.}

\item {Si $\alpha\leq -1/2$, las autofunciones de ${H}$ est{\'a}n
dadas por la ecuaci{\'o}n (\ref{eigenfunctions-2}). N{\'o}tese
que, en este caso, no existen modos cero. Los autovalores de
${H}$,
\begin{equation}\label{eigen-H-2}
  E_n=2\,n+1-2\,\alpha\geq 2, \quad n=0,1,2,\dots
\end{equation}
son todos estrictamente positivos y tienen una degeneraci{\'o}n
doble (ver ecuaci{\'o}n (\ref{spectrum-2})).

\bs

Las combinaciones $\Phi_{+,n}\pm \Phi_{-,n}$ (v{\'e}ase la
ecuaci{\'o}n (\ref{eigenfunctions-2}) representan estados
bo\-s{\'o}\-ni\-cos y fermi{\'o}nicos. Para estos valores de
$\alpha$, la supersimetr{\'\i}a est{\'a} espont{\'a}neamente rota
y, consecuentemente, el {\'\i}ndice de Witten es $\Delta = 0$.}

\end{itemize}

\subsubsection{Si $|\alpha|< 1/2$ el operador $Q_+$ admite extensiones
autoadjuntas no triviales}
 \label{non-ESA}

De acuerdo con las ecuaciones (\ref{Phi+}) y (\ref{Phi-}) de la
secci{\'o}n \ref{spectrumqm}, si $-1/2 < \alpha < 1/2$ los
{\'\i}ndices de deficiencia son $n_\pm = 1$. En consecuencia,
$Q_+$ admite una familia de extensiones autoadjuntas
${Q}^{(\gamma)}_+$ caracterizadas por un par{\'a}metro real
$\gamma$, que est{\'a} en correspondencia biun{\'\i}voca con el
grupo $U(1)$ de isometr{\'\i}as $\mathcal{U}(\gamma)$ de ${\cal
K}_+$ en ${\cal K}_-$,
\begin{equation}\label{isometries}
  {\cal U}(\gamma) \Phi_+(x):= e^{2 i \gamma} \Phi_-,
  \quad \gamma \in [0,\pi),
\end{equation}
siendo $\Phi_+$ y $\Phi_-$ las funciones dadas por las ecuaciones
(\ref{Phi+}) y (\ref{Phi-}), respectivamente.

\bs

N\' otese que la variedad $\cal M$ de extensiones autoadjuntas es
isomorfa a $U(1)$ y posee, en consecuencia, la topolog{\'\i}a
se\~nalada en la secci{\'o}n \ref{topo}: $\pi_0(\cal M)$ es
trivial y $\pi_1(\mathcal{M})=\mathbb{Z}$. Las subvariedades de
Cayley son $\mathcal{C}_-=\{ -1\}$, correspondiente a
$\gamma=\pi/2$, y $\mathcal{ C}_+=\{1\}$, correspondiente a
$\gamma=0$. Las variedades $\cal M -\mathcal{C}_{\mp}$ tienen, en
consecuencia, topolog{\'\i}a trivial.

\bs

El operador autoadjunto ${Q}^{(\gamma)}_+$ es la restricci{\'o}n
de $Q_+^\dagger$ al subespacio denso,
\begin{equation}\label{dom-Pgamma}
  {\cal D}({Q}^{(\gamma)}_+)\subset
  {\cal D}(Q_+^\dagger) =
  {\cal D}(\overline{Q}_+)\oplus{\cal K}_+
  \oplus{\cal K}_-\,,
\end{equation}
que consiste en las funciones de la forma,
\begin{equation}\label{func-dom-Pgamma}
  \Psi =\left( \begin{array}{c}
  \psi_1 \\
  \psi_2
\end{array}\right) =
\overline{\Psi}_0 + c \left( \Phi_+ + e^{2 i \gamma} \Phi_-
  \right)\,,
\end{equation}
con $\overline{\Psi}_0\in {\cal D}\left(\overline{Q}_+\right)$ y
$c \in \mathbb{C}$.

\bs

La acci{\'o}n del operador ${Q}^{(\gamma)}_+$ est{\'a} dada por,
\begin{equation}\label{accion-Pgamma}
  {Q}^{(\gamma)}_+ \Psi = Q_+^\dagger \overline{\Psi}_0 + i\, c \left(
  \Phi_+ - e^{2 i \gamma} \Phi_- \right)\,,
\end{equation}
con $Q_+^\dagger$ dado por la ecuaci{\'o}n (\ref{Q_++-def}).

\bigskip

La ecuaci{\'o}n (\ref{func-dom-Pgamma}) caracteriza completamente
el comportamiento en el origen de las funciones $\Psi \in {\cal
D}({Q_+}^{(\gamma)})$, de modo que permite determinar el espectro
de ${Q}^{(\gamma)}_+$. En efecto, en la secci{\'o}n
\ref{closuresusy} estudiaremos el dominio de la clausura
$\overline{Q}_+$ y mostraremos que,
\begin{equation}\label{phi0}
    \overline{\Psi}_{0}(x)\in\mathcal{D}(\overline{Q}_+)\rightarrow
    \overline{\Psi}_{0}(x)=
    \left(\begin{array}{c}
    {o}(x^\alpha)\\{o}(x^{-\alpha})
    \end{array}\right)\,,
\end{equation}
para $x\rightarrow 0^+$. Por su parte, se puede ver, a partir de
las ecuaciones (\ref{phi-eigen}), (\ref{phi1-U}), (\ref{phi2-U}) y
(\ref{LI-sol-1}), que el comportamiento en el origen de las
autofunciones $\Phi_\lambda$ de $Q_+^\dagger$ est{\'a} dado por,
\begin{equation}\label{eigen-asymp}
  \begin{array}{c}
    \phi_1(x) = \displaystyle{\frac{\Gamma\left(
    \frac12 - \alpha\right)}{\Gamma\left(
    \frac{1-\lambda^2}{2}-\alpha\right)}}\ x^\alpha + O(x^{1-\alpha}),\\ \\
    \phi_2(x) = \displaystyle{\frac{\sqrt{2}}{\lambda} \
    \frac{\Gamma\left(
    \frac12 + \alpha\right)}{\Gamma\left(
    -\frac{\lambda^2}{2}\right)}}\  x^{-\alpha} + O(x^{1+\alpha})\,.
  \end{array}
\end{equation}
Las ecuaciones (\ref{phi0}) y (\ref{eigen-asymp}) muestran que no
existen autofunciones de $Q_+^\dagger$ pertenecientes a ${\mathcal
{D}}\left(\overline{Q}_+\right)$. Por consiguiente, son las
contribuciones de $\Phi_\pm$ en la ecuaci{\'o}n
(\ref{func-dom-Pgamma}) las que determinan el espectro de
${Q_+}^{(\gamma)}$. En efecto, el l{\'\i}mite,
\begin{equation}\label{limit-ratio-phi}
  \lim_{x\rightarrow 0^+} \frac{x^{-\alpha}\, \phi_1(x)}{x^{\alpha}\,
  \phi_2(x)}=
    \displaystyle{ \frac{\lambda}{\sqrt{2}} \,
    \frac{\Gamma\left(-\frac{\lambda^2}{2}\right)}
    {\Gamma\left(
    \frac{1-\lambda^2}{2}-\alpha\right)}
    \frac{\Gamma\left(\frac12 - \alpha\right)}
    {\Gamma\left(\frac12 + \alpha\right)}   }\,,.
\end{equation}
obtenido a partir de la ecuaci{\'o}n (\ref{eigen-asymp}), debe
coincidir con,
\begin{equation}\label{limit-ratio-phi+}
  \lim_{x\rightarrow 0^+} \frac{x^{-\alpha}\,
   \psi_{1}(x)}
  {x^{\alpha}\, \psi_{2}(x)}=
    \displaystyle{- \sqrt{\frac{\pi}{2}} \,
    \frac{\cot(\gamma)}
    {\Gamma\left( 1-\alpha\right)}
    \frac{\Gamma\left(\frac12 - \alpha\right)}
    {\Gamma\left(\frac12 + \alpha\right)}   }\,,
\end{equation}
obtenido a partir de las ecuaciones (\ref{func-dom-Pgamma}) y
(\ref{phi0}).

\bs

Por lo tanto, los autovalores de ${Q}^{(\gamma)}_+$ son las
soluciones de la ecuaci{\'o}n trascendental,
\begin{equation}\label{trascendental}
    \displaystyle{ f(\lambda) :=
    \frac{\lambda \, \Gamma\left(-\frac{\lambda^2}{2}\right)}
    {\Gamma\left(
    \frac{1-\lambda^2}{2}-\alpha\right)}
    = -\frac{\sqrt{\pi} \,\cot(\gamma)}
    {\Gamma\left( 1-\alpha\right)} := \beta(\gamma)}\,.
\end{equation}
N{\'o}tese que $-\infty\leq \beta(\gamma)< \infty$ para $0 \leq
\gamma <\pi$.
\begin{figure}\label{spectrumsusy}
\center
    \epsffile{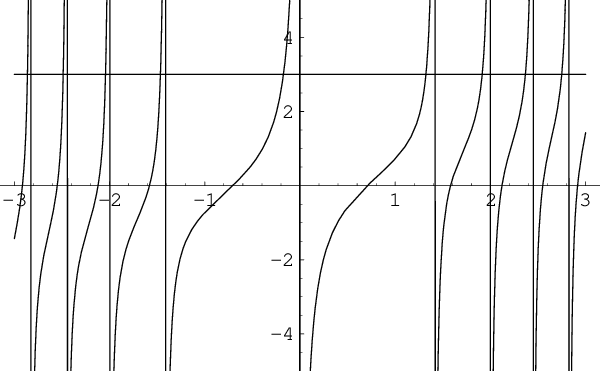} \caption{{\small Las
    intersecciones de la funci{\'o}n $f(\lambda)$, representada en la
    figura para $\alpha=1/4$, con una recta horizontal determina
    el espectro de la extensi{\'o}n autoadjunta correspondiente;
    consideramos en la figura el caso $\beta(\gamma)= 3$.}}
    \label{espectrosusy}
\end{figure}
La funci{\'o}n $f(\lambda)$ es impar en la variable $\lambda$ y ha
sido representada en la Figura 1 para un valor de $\alpha=1/4$.
Los autovalores de ${Q_+}^{(\gamma)}$ est{\'a}n determinados por
las {\it abscis\ae} de las intersecciones de la gr{\'a}fica de
$f(\lambda)$ con la recta horizontal correspondiente a la
constante $\beta(\gamma)$. En la Figura 1 se ha representado
adem{\'a}s la recta horizontal $\beta(\gamma)=3$. Las
autofunciones correspondientes se obtienen de las ecuaciones
(\ref{phi-eigen}), (\ref{phi1-U}) y (\ref{phi2-U}). Cabe
se{\~n}alar que, de acuerdo con las expresiones
(\ref{eigen-asymp}), estas autofunciones son singulares en el
origen.

\bs

Los autovalores de $Q^{(\gamma)}_+$ son, en general, no
degenerados. En efecto, el espectro es sim{\'e}trico con respecto
al origen s{\'o}lo para las extensiones autoadjuntas
correspondientes a $\beta=-\infty$ ($\gamma=0$) y $\beta=0$
($\gamma=\pi/2$) cuyos autovalores est{\'a}n dados por,
\begin{equation}\label{beta-infty}
  \lambda_{\pm,n}=\pm\sqrt{2n},\quad n=1,2,3,\dots
\end{equation}
si $\beta=-\infty$ y por,
\begin{equation}\label{beta-0}
  \lambda_{\pm,n}=\pm\sqrt{2n+1-2\alpha},\quad n=0,1,2,\dots
\end{equation}
si $\beta=0$.

\bs

En general los autovalores correspondientes al espectro de
cualquier extensi{\'o}n autoadjunta est{\'a}n contenidos entre
as{\'\i}ntotas contiguas de $\Gamma\left(-\lambda^2/2 \right)$,
\begin{equation}\label{eigen-cotas}
  \sqrt{2 n}<\left| \lambda_{\pm,n} \right|<\sqrt{2(n+1)}\,.
\end{equation}

\bs

Por su parte, las extensiones autoadjuntas $H^{(\gamma)}$ del
hamiltoniano est{\'a}n dadas por,
\begin{equation}\label{SAE-HQ}
  H^{(\gamma)} ={Q}^{(\gamma)}_+\cdot{Q}^{(\gamma)}_+\,,
\end{equation}
sobre los dominios definidos por la ecuaci{\'o}n (\ref{domsusy}).
Estos dominios contienen, en par\-ti\-cu\-lar, a las autofunciones
de ${Q_+}^{(\gamma)}$ que resultan, por consiguiente,
autofunciones de $H^{(\gamma)}$.

\bigskip

Describimos el espectro del hamiltoniano considerando
separadamente distintas extensiones autoadjuntas:

\begin{itemize}

\item{La extensi{\'o}n caracterizada por $\gamma=0$
($\beta=-\infty$) es la {\'u}nica extensi{\'o}n que posee un modo
cero que es, por otra parte, no degenerado. El estado fundamental
est{\'a} dado por la ecuaci{\'o}n (\ref{eigenfunctions-1-0}). La
ecuaci{\'o}n (\ref{beta-infty}) muestra que los autovalores no
nulos de $H^{(0)}$ son doblemente degenerados,
\begin{equation}\label{eigen-gamma-0}
   E_{n} = \left( \lambda_{\pm,n} \right)^2 = 2n, \quad
  n=1,2,3,\dots
\end{equation}
Las combinaciones $\Phi_{+,n}\pm \Phi_{-,n}$ (v{\'e}ase la
ecuaci{\'o}n (\ref{eigenfunctions-1}) representan autoestados
bos{\'o}nicos y fermi{\'o}nicos del hamiltoniano, respectivamente.

\bs

Por lo tanto, las condiciones que satisfacen las funciones del
dominio de ${Q}^{(0)}_+$ dan lugar a una extensi{\'o}n autoadjunta
supersim{\'e}trica del hamiltoniano $H$. El {\'\i}ndice de Witten
es, en este caso, $\Delta=1$. }

\item{La extensi{\'o}n caracterizada por $\gamma=\pi/2$
($\beta=0$) no posee modos cero y tiene un espectro doblemente
degenerado. En efecto, de la ecuaci{\'o}n (\ref{beta-0}) se deduce
que las energ{\'\i}as de $H^{(\pi/2)}$ son,
\begin{equation}\label{eigen-gamma-pi}
  E_{n} = \left( \lambda_{\pm,n} \right)^2 =
  2n+1-2\alpha, \quad n=0,1,2,\dots
\end{equation}
que son todas estrictamente positivas. Las combinaciones
$\Phi_{+,n}\pm \Phi_{-,n}$ (v{\'e}ase la ecuaci{\'o}n
(\ref{eigenfunctions-2}) representan autoestados bos{\'o}nicos y
fermi{\'o}nicos del hamiltoniano, respectivamente.

\bs

Las condiciones impuestas sobre las funciones del domino de
${Q}^{(\pi/2)}_+$ rompen es\-pon\-t{\'a}\-ne\-a\-men\-te la
supersimetr{\'\i}a pero preservando la degeneraci{\'o}n del
espectro. En consecuencia, el {\'\i}ndice de Witten es $\Delta=0$.
}

\item{Los valores de $\gamma\neq 0,\pi/2$ definen extensiones
autoadjuntas del hamiltoniano que no poseen modos cero y que
tienen un espectro no degenerado. Los autovalores de
$H^{(\gamma)}$ est{\'a}n dados por los cuadrados de las soluciones
de la ecuaci{\'o}n (\ref{trascendental}).

\bs

Estas extensiones autoadjuntas, a diferencia de $H^{(0)}$ y
$H^{(\pi/2)}$, no poseen autofunciones correspondientes a estados
bos{\'o}nicos ni fermi{\'o}nicos (v{\'e}anse las ecuaciones
(\ref{phi1-U}) y (\ref{phi2-U}).) La condici{\'o}n de contorno que
define el dominio de ${Q}^{(\gamma)}_+$ no solo rompe la
supersimetr{\'\i}a, sino tambi{\'e}n la degeneraci{\'o}n del
espectro. El {\'\i}ndice de Witten es, por consiguiente,
$\Delta=0$. }

\end{itemize}

{\small

\subsection{L{\'\i}mite regular}

Es interesante considerar el caso $\alpha=0$, en el que el
superpotencial, dado por la ecuaci{\'o}n (\ref{W}), es regular en
el origen.

\bs

En este caso, las funciones en ${\cal D}({Q}^{(\gamma)}_+)$
tienden a valores finitos para $x\rightarrow 0^+$, como se deduce
de la ecuaci{\'o}n (\ref{eigen-asymp}),
\begin{equation}\label{BC-alpha=0}
  \begin{array}{c}
    \overline{\Phi}_0 (x)= o(x^0), \\ \\
    \Phi_+(x) + e^{2 i \gamma} \Phi_-(x) = 2^{3/2}\, e^{i \gamma}
    \left( \begin{array}{c}
      \sqrt{\frac{\pi}{2}}\, \cos\gamma \\
      -\sin \gamma
    \end{array} \right)+ O(x)\,.
  \end{array}
\end{equation}
Por lo tanto, el dominio de ${Q_+}^{(\gamma)}$ puede ser
caracterizado por una condici{\'o}n de contorno local de la forma,
\begin{equation}\label{BBCC-alpha=0}
  \Phi \in {\cal D}( {Q_+}^{(\gamma)} ) \Rightarrow
  \left( \begin{array}{cc}
    \sin \gamma & \sqrt{\frac{\pi}{2}}\,\cos\gamma
  \end{array} \right) \cdot \left( \begin{array}{c}
    \phi_1(0) \\
    \phi_2(0)
  \end{array} \right)=0.
\end{equation}
Los valores particulares $\gamma=0$ y $\gamma=\pi/2$ conducen a
las condiciones de contorno $\phi_2(0)=0$ y $\phi_1(0)=0$,
respectivamente.

\bs

Como hemos mencionado en la secci{\'o}n \ref{non-ESA}, para
$\gamma=0$ la supersimetr{\'\i}a es expl{\'\i}cita. El modo cero
de $H^{(0)}$ est{\'a} dado por,
\begin{equation}\label{zero-mode-alpha0}
  \Phi_0 = \left( \begin{array}{c}
    e^{-x^2/2} \\
    0
  \end{array} \right)\,.
\end{equation}
Por su parte, los autovalores del hamiltoniano est{\'a}n dador por
$E_{n}=2n$, $n=1,2,\dots$ (v{\'e}ase la ecuaci{\'o}n
(\ref{eigen-gamma-0})), son doblemente degenerados y las
correspondientes autofunciones toman la forma (v{\'e}anse
ecuaciones (\ref{phi1-U}) y (\ref{phi2-U})),
\begin{equation}\label{eigen-alpha0-gamma0}
  \Phi_{\pm,n}(x)=\frac{e^{-x^2/2}}{2^{2n}} \left( \begin{array}{c}
    H_{2n}(x) \\
    \pm 2 \sqrt{n} \, H_{2n-1}(x)
  \end{array} \right)\,,
\end{equation}
donde $H_n(x)$ son los polinomios Hermite. N{\'o}tese que la
componente inferior y la derivada de la componente superior de los
autovectores se anulan en el origen.

\bs

Para $\gamma=\pi/2$ la supersimetr{\'\i}a est{\'a}
espont{\'a}neamente rota pues no existen modos cero. Los
autovalores de $H^{(\pi/2)}$ est{\'a}n dados por $E_{n}=2n+1$,
$n=0,1,\dots$ (v{\'e}ase la ecuaci{\'o}n (\ref{eigen-gamma-pi})),
son doblemente degenerados y las correspondientes autofunciones
toman la forma (v{\'e}anse las ecuaciones (\ref{phi1-U}) y
(\ref{phi2-U})),
\begin{equation}\label{eigen-alpha0-gammapi}
  \Phi_{\pm,n}(x)=\frac{e^{-x^2/2}}{2^{2n+1}} \left( \begin{array}{c}
    H_{2n+1}(x) \\
    \mp \sqrt{4n+2} \, H_{2n}(x)
  \end{array} \right)\,.
\end{equation}
En este caso, la componente superior y la derivada de la
componente inferior del autovector se anula en el origen.

\bs

Para valores del par{\'a}metro $\gamma\neq 0,\pi/2$, la
supersimetr{\'\i}a tambi{\'e}n est{\'a} espont{\'a}neamente rota
pues no existen modos cero pero el espectro es no degenerado.
Vemos entonces que, a excepci{\'o}n de $\gamma=0$, las condiciones
de contorno en el origen que definen el operador autoadjunto
${Q}^{(\gamma)}_+$ y el operador autoadjunto $H^{(\gamma)}$ rompen
la supersimetr{\'\i}a. }

\section{Realizaci{\'o}n del {\'a}lgebra de N=2 SUSYQM}

Como hemos se{\~n}alado, las extensiones autoadjuntas
$Q_-^{(\gamma)}$ de la supercarga $Q_-$ se obtienen a partir de la
transformaci{\'o}n unitaria,
\begin{equation}\label{tranf-unit-SAE}
    Q_-^{(\gamma)} = e^{\displaystyle{i\pi\sigma_3/4}}\,
    Q_+^{(\gamma)} \,
    e^{-\displaystyle{i\pi\sigma_3/4}}\,.
\end{equation}
Los correspondientes dominios de definici{\'o}n est{\'a}n dados
por,
\begin{equation}\label{dominio-SAE-Qmenos}
    \mathcal{D}(Q_-^{(\gamma)}) = \left\{
    \psi : e^{-{i\pi\sigma_3/4}}\psi \in
    \mathcal{D}(Q_+^{(\gamma)})\right\}
    = e^{\displaystyle{i\pi\sigma_3/4}} \left(
    \mathcal{D}(Q_+^{(\gamma)}) \right)\, .
\end{equation}
En consecuencia, $Q_-^{(\gamma)}$  constituye una
representaci{\'o}n equivalente de la supercarga que, desde luego,
presenta el mismo espectro que $Q_+^{(\gamma)}$. Asimismo, su
cuadrado, definido sobre el conjunto,
\begin{equation}\label{dominio-Qmenos-cuadrado}
    \begin{array}{c}
      \mathcal{D}\left((Q_-^{(\gamma)})^2 \right)
    =\left\{
    \psi\in \mathcal{D}(Q_-^{(\gamma)})
     : Q_-^{(\gamma)} \psi \in
    \mathcal{D}(Q_-^{(\gamma)})\right\} = \\ \\
      =\left\{\psi :
    e^{{-i\pi\sigma_3/4}}\psi\in \mathcal{D}(Q_+^{(\gamma)})
     \, \wedge\,  e^{{- i\pi\sigma_3/4}}Q_-^{(\gamma)} \psi
     = Q_+^{(\gamma)}  e^{{-i\pi\sigma_3/4}}\psi
     \in
    \mathcal{D}(Q_+^{(\gamma)})\right\} =\\ \\
    = e^{\displaystyle{i\pi\sigma_3/4}} \left(
    \mathcal{D}(H^{(\gamma)}) \right)\, ,
    \end{array}
\end{equation}
constituye una representaci{\'o}n equivalente de la  extensi{\'o}n
autoadjunta del hamiltoniano $H^{(\gamma)}$.

\bs

Ahora bien, las dos representaciones equivalentes del hamiltoniano
$H^{(\gamma)}$ dadas por los cuadrados de las supercargas
$Q^{(\gamma)}_\pm$ coinciden {\'u}nicamente si el dominio
$\mathcal{D}(Q^{(\gamma)}_+)$ es invariante ante la
transformaci{\'o}n unitaria $e^{{i\pi\sigma_3/4}}$ y, en virtud de
la relaci{\'o}n (\ref{limit-ratio-phi+}), esto s{\'o}lo es
v{\'a}lido para las extensiones autoadjuntas correspondientes a
$\gamma = 0, \pi/2$.

\bs

En consecuencia, s{\'o}lo para los valores $\gamma=0,\pi/2$ tienen
sentido las composiciones,
\begin{equation}\label{composicionQmasQmenos}
    Q_+^{(\gamma)}  Q_-^{(\gamma)}\qquad {\rm y} \qquad
    Q_-^{(\gamma)} Q_+^{(\gamma)}\,,
\end{equation}
en el dominio $\mathcal{D}(H^{(\gamma)})$, donde se satisface
adem{\'a}s el {\'a}lgebra de la SUSY con $N=2$,
\begin{equation}\label{algebra-SUSY-0}
    \left\{ Q_+^{(\gamma)} , Q_-^{(\gamma)} \right\} = 0\,,\qquad
      H^{(\gamma)} =  \left(Q_+^{(\gamma)}\right)^2 =
      \left(Q_-^{(\gamma)}\right)^2\,.
\end{equation}

Para los restantes valores de $\gamma$, el dominio
$\mathcal{D}(Q_+^{(\gamma)})$ no es invariante frente a
$e^{{i\pi\sigma_3/4}}$ y no existe un dominio denso del espacio de
Hilbert donde puedan definirse las composiciones
(\ref{composicionQmasQmenos}). Por lo tanto, para estos casos
existe una {\'u}nica supercarga y el {\'a}lgebra de la SUSY se
reduce a,
\begin{equation}\label{algebra-SUSY-gamma}
      H^{(\gamma)} =  \left(Q_+^{(\gamma)}\right)^2 \, .
\end{equation}

\bs

Cabe se{\~n}alar que la doble degeneraci{\'o}n de los autovalores
no nulos de $H^{(\gamma)}$ para $\gamma = 0, \pi/2$ es
consecuencia de la existencia de dos supercargas. En efecto, si,
\begin{equation}
    Q^{(\gamma)}_+\phi_\lambda=\lambda\phi_\lambda\,,
\end{equation}
para $\phi_\lambda\in \mathcal{D}(Q^{(\gamma)}_+)$ y $\lambda\neq
0$, entonces las relaciones (\ref{algebra-SUSY-0}) implican que,
\begin{equation}\label{QmasQmenospsi}
    Q_+^{(\gamma)}\left(Q_-^{(\gamma)} \phi_\lambda\right)
    =- Q_-^{(\gamma)}\left(Q_+^{(\gamma)} \phi_\lambda\right)
    =- \lambda \left(Q_-^{(\gamma)} \phi_\lambda\right)\, ,
\end{equation}
por lo que $Q_-^{(\gamma)} \phi_\lambda$ es un
autovector de autovalor $-\lambda$ de $Q_+^{(\gamma)}$ y, en consecuencia, $Q_-^{(\gamma)} \phi_\lambda \perp \phi_\lambda$, con,
\begin{equation}\label{Qmenos-phi-norma}
    \parallel Q_-^{(\gamma)} \phi_\lambda \parallel^2
    =\left(\phi_\lambda,\left(Q_-^{(\gamma)}\right)^2
       \phi_\lambda\right) = \lambda^2 \,
       \parallel \phi_\lambda \parallel^2 \, \neq 0\,.
\end{equation}

En conclusi{\'o}n, las condiciones de contorno en $x=0$ rompen, en
general, la SUSY N=2, eliminando un generador. Esto implica que el
espectro de la supercarga no sea sim{\'e}trico y el del
hamiltoniano resulte no degenerado.

\bs

Las {\'u}nicas excepciones son a las extensiones autoadjuntas
correspondientes a $\gamma = 0, \pi/2$, para las que se obtiene
SUSY con $N=2$ (dos generadores de la SUSY). Las dos supercargas
presentan espectros sim{\'e}tricos y los estados excitados del
hamiltoniano son doblemente degenerados.

\bs

Finalmente, para $\gamma = 0$ el estado fundamental tiene
energ{\'\i}a nula y la SUSY es ma\-ni\-fies\-ta, en tanto que para
$\gamma= \pi/2$ el estado fundamental tiene energ{\'\i}a
estrictamente positiva y la SUSY est{\'a} espont{\'a}neamente
rota.

\section{Clausura del operador} \label{closuresusy}

En esta secci{\'o}n estudiaremos la clausura del operador $Q_+$
definido sobre ${\cal D}(Q_+)={\cal
C}^{\infty}_{0}(\mathbb{R}^+)\otimes \mathbb{C}^2$ y demostraremos
que, cerca del origen, las funciones $\overline{\Phi}_0(x)\in
{\cal D}\left(\overline{Q}_+\right)$ verifican el comportamiento
(\ref{phi0}), para todo $|\alpha|<1/2$. Esto permiti{\'o} omitir
las contribuciones de las funciones pertenecientes a ${\cal
D}\left(\overline{Q}_+\right)$ en el l{\'\i}mite $x\rightarrow
0^+$ del miembro derecho de la ecuaci{\'o}n
(\ref{func-dom-Pgamma}) y arribar a la ecuaci{\'o}n
(\ref{limit-ratio-phi+}).

\bs

Para determinar la clausura de la gr{\'a}fica de $Q_+$ debemos
considerar las sucesiones de Cauchy,
\begin{equation}\label{cauchy}
  \left\{ \Phi_n = \left( \begin{array}{c}
    \phi_{1,n} \\
    \phi_{2,n}
  \end{array} \right) \right\}_{n \in \mathbb{N}} \subset
  {\cal D}\left(Q_+\right) := {\cal C}_0^\infty(\mathbb{R}^+)\,,
\end{equation}
tales que $\left\{ Q_+\Phi_n \right\}_{n \in \mathbb{N}}$ son
tambi{\'e}n sucesiones de Cauchy.

\bigskip

En este caso, $\left\{ \phi_{1,n} \right\}_{n \in\mathbb{N}}$,
$\left\{ \phi_{2,n} \right\}_{n \in \mathbb{N}}$, $\left\{ D_1
\phi_{1,n} \right\}_{n \in \mathbb{N}}$ y $\left\{D_2 \phi_{2,n}
\right\}_{n \in \mathbb{N}}$ son sucesiones de Cauchy en ${\mathbf
L_2}([0,1])$, siendo $D_1$ y $D_2$ los operadores diferenciales
dados por las ecuaciones (\ref{superA}) y (\ref{superAt})
respectivamente.

\bs

Por otra parte, como $x$ est{\'a} acotado en $[0,1]$ y la suma de
dos sucesiones de Cauchy es tambi{\'e}n una sucesi{\'o}n de
Cauchy, se deduce que,
\begin{equation}
\left\{ \phi_{1,n}'(x)- \frac{\alpha}{x}\, \phi_{1,n}(x)
\right\}_{n \in \mathbb{N}}\,,
\end{equation}
y,
\begin{equation}
\left\{ \phi_{2,n}'(x)+ \frac{\alpha}{x}\, \phi_{2,n}(x)
\right\}_{n \in \mathbb{N}}\,,
\end{equation}
son sucesiones de Cauchy en ${ \mathbf L_2}(0,1)$.

\bigskip

Asimismo, como $x^{\pm \alpha}\in {\mathbf L_2}([0,1])$ para todo
$-1/2 < \alpha
 <1/2$ se verifica,
\begin{equation}\label{cauchy2}
  \left\{ x^{-\alpha} \left( \phi_{1,n}'(x)-\frac{\alpha}{x}
  \phi_{1,n}(x) \right)\right\}_{n \in \mathbb{N}} = \left\{\left(
  x^{-\alpha} \phi_{1,n}(x) \right)'\right\}_{n \in \mathbb{N}}
\end{equation}
y
\begin{equation}\label{cauchy3}
  \left\{ x^{\alpha}  \left( \phi_{2,n}'(x)+\frac{\alpha}{x}
  \phi_{2,n}(x) \right)\right\}_{n \in \mathbb{N}} = \left\{ \left(
  x^{\alpha} \phi_{2,n}(x) \right)'\right\}_{n \in \mathbb{N}}
\end{equation}
son sucesiones de Cauchy en ${\mathbf{L_1}}([0,1])$.

\bigskip

Ahora bien, teniendo en cuenta que estas funciones se anulan
id{\'e}nticamente en una vecindad del origen, se puede probar que
$\left\{x^{-\alpha}\, \phi_{1,n}(x) \right\}_{n \in \mathbb{N}}$ y
$\left\{x^{\alpha}\, \phi_{2,n}(x) \right\}_{n \in \mathbb{N}}$
convergen uniformemente en $[0,1]$. En efecto, $\forall\, x \in
[0,1] $ se verifica,
\begin{equation}\label{conv-unif}\begin{array}{c}
  \Big|\, x^{\pm\alpha} \left[ \phi_{1,n}(x)-\phi_{1,m}(x) \right] \Big|
  = \\ \\{\displaystyle
  = \left| \int_0^x \left(y^{\pm\alpha} \left[ \phi_{1,n}(y)-
  \phi_{1,m}(y) \right]\right)' \, dy\right| \leq }\\ \\{\displaystyle
  \leq \left\| \left(x^{\pm\alpha} \phi_{1,n}(x)\right)' -
    \left(x^{\pm\alpha} \phi_{1,m}(x)\right)' \right\|_{{\mathbf{L_1}}(0,1)}
    \rightarrow_{n,m\rightarrow \infty} 0\,.}
\end{array}
\end{equation}
Por lo tanto, existen dos funciones continuas,
$x^{-\alpha}\overline{\phi}_{1}(x)$ y $x^\alpha
\overline{\phi}_{2}(x)$, que son sus l{\'\i}mites uniformes en
$[0,1]$,
\begin{equation}\label{limits}
  \begin{array}{c}
    x^{-\alpha} \, \overline{\phi}_{1}(x) =\displaystyle{\lim_{n\rightarrow
    \infty}} \
    x^{-\alpha} \, \phi_{1,n}(x)\,,
     \\ \\
    x^{\alpha} \, \overline{\phi}_{2}(x) = \displaystyle{\lim_{n\rightarrow
    \infty}}\
    x^{\alpha} \, \phi_{2,n}(x)\,.
  \end{array}
\end{equation}
Por consiguiente,
\begin{equation}\label{lim0}
  \begin{array}{c}
    \displaystyle{\lim_{x \rightarrow 0}}\  x^{-\alpha} \,
    \overline{\phi}_{1}(x) =0, \\ \\
    \displaystyle{\lim_{x \rightarrow 0}}\  x^{\alpha} \,
    \overline{\phi}_{2}(x) =0\,.
  \end{array}
\end{equation}

\bs

Por otra parte, el l{\'\i}mite de la sucesi{\'o}n $\left\{\Phi_n
\right\}_{n \in \mathbb{N}}$ en ${\mathbf{L_2}}[(0,1)]$ est{\'a}
dado por,
\begin{equation}\label{lim-L2}
  \lim_{n \rightarrow \infty} \Phi_n =
  \overline{\Phi}_0=\left( \begin{array}{c}
  \overline{\phi}_{1} \\
  \overline{\phi}_{2}
\end{array} \right) .
\end{equation}
En efecto, teniendo en cuenta que, para todo $\varepsilon>0$,
\begin{equation}\label{inif-conv-lim}
  \Big|\, x^{-\alpha}\left[ \phi_{1,n}(x) - \overline{\phi}_{1}(x)
  \right] \Big|<
  \varepsilon, \quad \forall\, x\in [0,1],
\end{equation}
si $n$ es suficientemente grande, se deduce,
\begin{equation}\label{lim-L2-1}\begin{array}{c}
  \left\| \phi_{1,n} - \overline{\phi}_{1} \right\|_{{\mathbf{L_2}}(0,1)}^2 =
  \\ \\{\displaystyle
  = \int_0^1 x^{2\alpha} \Big|\, x^{-\alpha} \left(  \phi_{1,n}(x) -
  \overline{\phi}_{1}(x) \Big)  \right|^2 <  }\\ \\ < \varepsilon^2 \,
  \left\|x^\alpha\right\|_{{\mathbf{L_2}}(0,1)}^2\,.
\end{array}
\end{equation}
Se procede an{\'a}logamente para la componente inferior. La
ecuaci{\'o}n (\ref{phi0}) se deduce de las ecuaciones
(\ref{lim-L2}) y (\ref{lim0}).

\bigskip

Como la gr{\'a}fica de $Q^\dagger_+$ es cerrada \cite{R-S},
contiene a la gr{\'a}fica de $\overline{Q}_+$. Verificaremos, para
finalizar, que efectivamente la funci{\'o}n $\overline{\Phi}_0$
pertenece a ${\cal D}\left({Q_+^\dagger}\right)$.

\bs

Sea $\rho_1(x)$ el l{\'\i}mite en ${\mathbf{L_1}}(0,1)$ de la
sucesi{\'o}n de Cauchy dada por la ecuaci{\'o}n (\ref{cauchy2}),
\begin{equation}\label{rho1}
  \rho_1(x)=\lim_{n \rightarrow \infty} {\left(x^{-\alpha}
 \phi_{1,n}(x)\right)'} \,.
\end{equation}
Entonces, dado $\varepsilon>0$, $\forall\, x \in [0,1]$, se
verifica,
\begin{equation}\label{rho2}\begin{array}{c}
  {\displaystyle \Big| \, x^{-\alpha} \phi_{1,n}(x) - \int_0^x \rho_1(y) \,
  dy \Big| = }\\ \\{\displaystyle
  =\left| \int_0^x \left[ \left( y^{-\alpha} \phi_{1,n}(y) \right)'
  -\rho_1(y) \right] dy \right|\leq }\\ \\{\displaystyle
  \leq \left\| \left( y^{-\alpha} \phi_{1,n}(y) \right)'
  -\rho_1(y) \right\|_{{\mathbf{L_1}}(0,1)} < \varepsilon\,,}
\end{array}
\end{equation}
para $n$ suficientemente grande.

\bs

Como el l{\'\i}mite uniforme es {\'u}nico, se deduce de las
ecuaciones (\ref{limits}) y (\ref{rho2}),
\begin{equation}\label{rho-phi}
  \overline{\phi}_{1}(x) = x^\alpha \int_0^x \rho_1(y)\, dy\,,
\end{equation}
con $\rho_1 \in {{\mathbf{L_1}}(0,1)}$. Por consiguiente,
$\overline{\phi}_{1}(x)$ es absolutamente continua para $x>0$. La
misma conclusi{\'o}n se obtiene an{\'a}logamente para la
componente inferior $\overline{\phi}_{1}(x)$ de
$\overline{\Phi}_0$.


\part{Funciones Espectrales}\label{fe}

\vspace{5mm}\begin{flushright}{\it The divergent series are the
invention of the devil,\\and it is a shame to base on them any
demonstration whatsoever.
\\By using them, one may draw any conclusion he pleases\\and that is why these
series have produced\\so many fallacies and so many paradoxes.\\
(Niels H. Abel.)}
\end{flushright}

\vspace{25mm}

\section{Operadores pseudodiferenciales}\label{zc}

Los trabajos precursores en la teor{\'\i}a de los operadores
pseudodiferenciales u o\-pe\-ra\-do\-res de Calder{\'o}n-Zygmund
se deben a Mikhlin \cite{mik} y Calder{\'o}n y Zygmund \cite{cz}.
Re\-fe\-ren\-cias acerca del desarrollo de esta teor{\'\i}a pueden
encontrarse en los trabajos de P.B.\ Gilkey \cite{Gilkey} y S.G.\
Krantz \cite{Krantz}.

\bs

En este cap{\'\i}tulo haremos una breve presentaci{\'o}n de los operadores
pseu\-do\-di\-fe\-ren\-cia\-les como generalizaci{\'o}n de los
operadores diferenciales. El orden de un o\-pe\-ra\-dor
diferencial es, desde luego, un entero positivo, en tanto que el
orden de los operadores pseudodiferenciales puede tomar cualquier
valor real. En este sentido, veremos que, en una dimensi{\'o}n, los
operadores pseudodiferenciales de orden menor que $-1$ representan
operadores integrales con n{\'u}cleos continuos. De modo que al
considerar el conjunto ${\bf \Psi^d}$ de operadores
pseu\-do\-di\-fe\-ren\-cia\-les de orden $d\in\mathbb{R}$ podremos
tratar en un mismo formalismo a los operadores diferenciales y a
sus inversos.

\bs

Sin embargo, as\'\i\ como los operadores diferenciales de orden
$d\in\mathbb{N}$ act{\'u}an sobre funciones que admiten $d$
derivadas, para tratar con operadores de orden $d\in\mathbb{R}$
deberemos definir, primeramente, un conjunto de funciones que
admitan ``derivada'' de orden $d\in\mathbb{R}$. Este conjunto se
denomina espacio de Sobolev $\mathbf{H_d}$. Tanto la
definici{\'o}n de los espacios de Sobolev $\mathbf{H_d}$ como la
de los operadores pseudodiferenciales ${\bf \Psi^d}$ se expresan
en t{\'e}rminos de la transformaci{\'o}n de Fourier $\mathcal{F}$.

\bs

Sean $d\in\mathbb{R}$ y $f(x)$ una funci{\'o}n que pertenece al
espacio de Schwartz $\mathcal{S}\subset\mathbf{L_2}(\mathbb{R})$,
esto es, tal que admite derivadas continuas de todo orden que
decrecen m{\'a}s r{\'a}pido que cualquier potencia de $x$ cuando
$|x|\rightarrow\infty$. Definimos a continuaci{\'o}n la
transformada de Fourier $\mathcal{F}\{f\}(p)$, la norma
$\|\cdot\|_d$ en $S$ y los espacios de Sobolev
$\mathbf{H_d}(\mathbb{R})$.
\begin{defn}\label{fourier} Dado que $\mathcal{S}\subset \mathbf{L_1}
(\mathbb{R})$, definimos la {\bf transformada de Fourier}
$\mathcal{F}\{f\}(p)$ de la funci{\'o}n $f(x)$,
\begin{equation}
    \mathcal{F}\{f\}(p):=\int_{\mathbb{R}}e^{-ipx}f(x)\,\frac{dx}
    {\sqrt{2\pi}}\,.
\end{equation}
\end{defn}
\begin{defn}
\begin{equation}\label{norma}
    \|f\|_d^2:= \int_{\mathbb{R}}(1+|p|^2)^d\
    |\mathcal{F}\{f\}(p)|^2\,dp\,.
\end{equation}
\end{defn}
\begin{defn}\label{sobo}
    El {\bf espacio de Sobolev} $\mathbf{H_d}(\mathbb{R})$ es la
    clausura de $S$
    con respecto a la norma $\|\cdot\|_d$.
\end{defn}
Enunciamos, sin demostraci{\'o}n, dos propiedades de los espacios
de Sobolev. , si $f(x)\in\mathbf{H_d}(\mathbb{R})$ entonces
$f(x)\in\mathcal{C}^k(\mathbb{R})$, para todo $k<d-1/2$ (Lema de
Sobolev \cite{Gilkey2}.) Esta propiedad indica que una funci{\'o}n
admite derivadas continuas de cierto orden si pertenece a un
espacio de Sobolev con {\'\i}ndice suficientemente grande.

\bs

Por otra parte, si $d$ es un entero positivo,
$\mathbf{H_d}(\mathbb{R})$ coincide con el conjunto de funciones
de $\mathbf{L_2}(\mathbb{R})$ que admiten derivadas generalizadas
en $\mathbf{L_2}(\mathbb{R})$ de orden $d$ (v{\'e}ase la
ecuaci{\'o}n (\ref{dergen}).)

\bigskip

Consideremos, ahora, un operador diferencial regular $A$ de orden
$d\in\mathbb{N}$ que act{\'u}a sobre elementos de
$\mathbf{H_d}(\mathbb{R})\otimes \mathbb{C}^k$, {\it i.e.}, sobre
funciones con dominio en $\mathbb{R}$ que toman valores en un
espacio vectorial de dimensi{\'o}n $k$ y que admiten $d$ derivadas
en $\mathbf{L_2}(\mathbb{R})$. El operador $A$ puede expresarse de
la siguiente manera,
\begin{equation}
    A=\sum_{n=0}^{d}A_n(x) (-i\partial_x)^n\,.
\end{equation}
Los coeficientes $A_n(x)$ son funciones infinitamente derivables
sobre $\mathbb{R}$ que toman valores en el espacio de matrices
$\mathbb{C}^{k\times k}$.

\bs

Es conveniente, antes de introducir el concepto de operador
pseudodiferencial, definir el {\bf s{\'\i}mbolo}
$\sigma\{A\}(x,p)$ del operador diferencial $A$:
\begin{equation}\label{simbolo}
    \sigma\{A\}(x,p):= \sum_{n=0}^{d}A_n(x) p^n\,.
\end{equation}
De acuerdo con esta definici{\'o}n, la acci{\'o}n de $A$ sobre una
funci{\'o}n $f(x)$ del espacio de Schwartz $S$ est{\'a}
representada por la acci{\'o}n de su s{\'\i}mbolo
$\sigma\{A\}(x,p)$ sobre su transformada de Fourier
$\mathcal{F}\{f\}(p)$,
\begin{equation}
    Af(x)=\int_{\mathbb{R}}e^{ipx}\sigma\{A\}(x,p)\cdot
    \mathcal{F}\{f\}(p)\,\frac{dp}{\sqrt{2\pi}}\,.
\end{equation}
El s{\'\i}mbolo $\sigma\{A\}(x,p)$ de un operador diferencial $A$
de orden $d$ es un polinomio de grado $d$ en la variable $p$. Para
definir operadores pseudodiferenciales simplemente extendemos la
definici{\'o}n de s{\'\i}mbolo.
\begin{defn}
    Sean $d\in\mathbb{R}$ y
    $S^d\subset\mathcal{C}^\infty(\mathbb{R}^2)$ un subconjunto de
    funciones $\xi(x,p)$ para las que, dado un par de enteros
    positivos $m,n$, existe una constante $C_{m,n}$ tal que,
\begin{equation}
    |\partial_x^{m}\partial_p^{n}\xi(x,p)|\leq
    C_{m,n}(1+|p|)^{d-n}.
\end{equation}
    Sea, tambi{\'e}n, $S^{-\infty}:=\bigcap_d S^d$.
\end{defn}
N{\'o}tese que el s{\'\i}mbolo $\sigma\{A\}(x,p)$ de un operador
diferencial $A$, que es un polinomio en $p$ de grado
$d\in\mathbb{N}$, pertenece al conjunto $S^d$. Asimismo, toda
funci{\'o}n $\xi(x,p)$ que sea un polinomio de grado $d$ en $p$
con coeficientes dependientes de $x$ tiene asociado un
o\-pe\-ra\-dor diferencial de orden $d$ cuyo s{\'\i}mbolo est{\'a}
dado por $\xi(x,p)$. Definiremos el conjunto $\Psi$ de operadores
pseu\-do\-di\-fe\-ren\-cia\-les asociando un operador a toda
funci{\'o}n $\xi(x,p)\in S^d$.
\begin{defn}\label{pseudo}
    Dados $\xi(x,p)\in S^d$, $d\in\mathbb{R}$ y $f(x)\in
    \mathcal{S}$, definimos el {\bf operador pseudodiferencial}
    $P$ como aquel que asigna a la funci{\'o}n $f(x)$ la imagen,
\begin{equation}\label{pse}
    Pf(x)=\int_{\mathbb{R}}e^{ipx}\xi(x,p)\cdot\mathcal{F}\{f\}(p)\,
    \frac{dp}{\sqrt{2\pi}}\,.
\end{equation}
El conjunto de operadores as{\'\i} definido ser{\'a} denotado por
${\bf \Psi^d}$. Llamaremos, a su vez, ${\bf \Psi^{-\infty}}$ al
conjunto de operadores definido an{\'a}logamente por funciones
$\xi(x,p)\in S^{-\infty}$.
\end{defn}
Aunque la definici{\'o}n \ref{pseudo} asigna a cada funci{\'o}n
$\xi(x,p)\in S^d$ un operador pseu\-do\-di\-fe\-ren\-cial $P$ que
act{\'u}a sobre las funciones de $\mathcal{S}$, es posible
extender el dominio de definici{\'o}n de $P$ a los espacios de
Sobolev.

\bs

Para extender la acci{\'o}n del operador $P$ a las funciones de
$\mathbf{H_n}(\mathbb{R})$, con $n\in\mathbb{R}$, enunciamos sin
demostraci{\'o}n la siguiente propiedad \cite{Gilkey}. Si $f\in
\mathcal{S}$, entonces, se verifica $\|Pf(x)\|_{n-d}<C\,\|f\|_n$
(v{\'e}ase la ecuaci{\'o}n (\ref{norma})), donde $C$ es una
constante in\-de\-pen\-dien\-te de $f$. Por consiguiente, si una
sucesi{\'o}n de funciones en $\mathcal{S}$ converge a una
funci{\'o}n en la norma $\|\cdot\|_n$, entonces sus im{\'a}genes
por $P$ convergen en la norma $\|\cdot\|_{n-d}$. De esta manera,
la acci{\'o}n de $P$ puede extenderse a las funciones de
$\mathbf{H_n}(\mathbb{R})$.

\bs

Por otra parte, se verifica \cite{Gilkey},
\begin{equation}\label{deri}
    P\in\Psi^{d}\Rightarrow P:\mathbf{H_n}(\mathbb{R})\rightarrow
    \mathbf{H_{n-d}}(\mathbb{R})\,.
\end{equation}
Esta propiedad es bien conocida en el caso de operadores
diferenciales cuando $n>d$. Los operadores diferenciales de orden
$d$ act{\'u}an sobre funciones que admiten $n$ derivadas
ge\-ne\-ra\-li\-za\-das y sus im{\'a}genes son funciones que
admiten $n-d$ derivadas. Esta propiedad se extiende al caso de
operadores pseudodiferenciales definidos sobre espacios de
Sobolev.

\bs

Cabe destacar que, de acuerdo con (\ref{deri}), los operadores
pseudodiferenciales de orden negativo hacen m{\'a}s ``suaves'' a
las funciones. Veremos, por ejemplo, que ${\bf \Psi^{-\infty}}$
contiene a los o\-pe\-ra\-do\-res integrales de n{\'u}cleo
infinitamente derivable y puede probarse sin dificultad que la
imagen de una funci{\'o}n por uno de tales operadores admite a su
vez infinitas derivadas.

\bs

En el presente cap{\'\i}tulo deseamos estudiar algunas propiedades del
o\-pe\-ra\-dor integral resolvente $(A-\lambda)^{-1}$
correspondiente a un o\-pe\-ra\-dor diferencial $A$. La teor{\'\i}a de
los o\-pe\-ra\-do\-res pseudodiferenciales nos permite obtener una
aproximaci{\'o}n del s{\'\i}mbolo del o\-pe\-ra\-dor $(A-\lambda)^{-1}$ en
t{\'e}rminos del s{\'\i}mbolo del operador $A$. Para ello necesitamos una
manera de obtener el s{\'\i}mbolo de la composici{\'o}n de dos operadores
pseudodiferenciales. Una vez m{\'a}s, comenzaremos por estudiar el
s{\'\i}mbolo de la composici{\'o}n de dos operadores diferenciales.

\bs

Si $P$ y $Q$ son dos operadores diferenciales de orden $d$ y $e$,
respectivamente,
\begin{eqnarray}
    P=\sum_{n=0}^{d}P_n(x) (-i\partial_x)^n\,,\\
    Q=\sum_{n=0}^{e}Q_n(x) (-i\partial_x)^n\,,\
\end{eqnarray}
sus s{\'\i}mbolos est{\'a}n dados por,
\begin{eqnarray}
    \sigma\{P\}(x,p)=\sum_{n=0}^{d}P_n(x) p^n\,,\\
    \sigma\{Q\}(x,p)=\sum_{n=0}^{e}Q_n(x) p^n\,.\
\end{eqnarray}
De acuerdo con la regla de Leibnitz,
\begin{equation}
    (-i\partial_x)^n(f(x)g(x))=\sum_{m=0}^n
    \left(\begin{array}{c}
    n\\m\end{array}\right)
    (-i\partial_x)^m f(x)
    (-i\partial_x)^{n-m} g(x)\,.
\end{equation}
Por lo tanto, la composici{\'o}n $PQ$ es un operador diferencial
de orden $e+d$ cuyo s{\'\i}mbolo $\sigma\{PQ\}$ est{\'a} dado por,
\begin{eqnarray}\label{sim}
    \sigma\{PQ\}=\sigma\left\{\sum_{n=0}^{d}\sum_{m=0}^{e}P_n(x)
 (-i\partial_x)^n
    \left(Q_m(x) (-i\partial_x)^m\right)\right\}=\nonumber\\
    =\sum_{n=0}^{d}\sum_{m=0}^{e}P_n(x) \sum_{k=0}^n
    \left(\begin{array}{c}
    n\\k\end{array}\right)
    (-i\partial_x)^k
    \left(Q_m(x)\right) p^{m+n-k}=\nonumber\\
    =\sum_{n=0}^{d} \sum_{k=0}^n
    \left(\begin{array}{c}
    n\\k\end{array}\right)
    P_n(x)\frac{(n-k)!}{n!}\partial_p^k p^{n}
    (-i\partial_x)^k\left(\sigma\{Q\}\right)=\nonumber\\
    =\sum_{k=0}^{d} \sum_{n=k}^d
    \frac{1}{k!}
    P_n(x)\partial_p^k p^{n}
    (-i\partial_x)^k\left(\sigma\{Q\}\right)=\nonumber\\
    =\sum_{k=0}^{d}
    \frac{1}{k!}\,
    \partial_p^k \left(\sigma\{P\}\right)\cdot
    (-i\partial_x)^k\left(\sigma\{Q\}\right)\,.
\end{eqnarray}

Cabe preguntarse si el s{\'\i}mbolo de la composici{\'o}n de dos
operadores pseudodiferenciales est{\'a} dado por una ecuaci{\'o}n
similar a (\ref{sim}).

\bs

Puede probarse \cite{Gilkey} que los s{\'\i}mbolos de los
operadores pseudodiferenciales satisfacen la siguiente
relaci{\'o}n,
\begin{equation}\label{sim2}
    \sigma\{PQ\}\sim\sum_{k=0}^{\infty}
    \frac{1}{k!}\,
    \partial_p^k \left(\sigma\{P\}\right)\cdot
    (-i\partial_x)^k\left(\sigma\{Q\}\right)\,,
\end{equation}
en la que $\sim$ significa que para todo $N$, no importa cu{\'a}n
grande, existe un $K(N)$ suficientemente grande tal que la
diferencia,
\begin{equation}
    \sigma\{PQ\}-\sum_{k=0}^{K(N)}
    \frac{1}{k!}\,
    \partial_p^k \left(\sigma\{P\}\right)\cdot
    (-i\partial_x)^k\left(\sigma\{Q\}\right)\in S^{-N}\,.
\end{equation}
En otros t{\'e}rminos, la relaci{\'o}n (\ref{sim2}) significa que
la diferencia entre el operador $PQ$ y el operador
pseudodiferencial cuyo s{\'\i}mbolo est{\'a} dado por un
n{\'u}mero finito de t{\'e}rminos de la suma del miembro derecho
de (\ref{sim2}) pertenece a la clase $\Psi^{-N}$, para $N$
arbitrariamente grande, si se consideran un n{\'u}mero suficiente
de t{\'e}rminos de la suma. La ecuaci{\'o}n (\ref{sim2}) no
implica la convergencia de la serie sino que provee un desarrollo
asint{\'o}tico del s{\'\i}mbolo de la composici{\'o}n de dos
operadores pseudodiferenciales; esto ser{\'a} suficiente para
nuestro prop{\'o}sito de estudiar el desarrollo asint{\'o}tico del
n{\'u}cleo del operador integral $(A-\lambda)^{-1}$ para grandes
valores de $|\lambda|$.

\subsection{Operadores integrales}

Antes de definir las funciones espectrales de un operador
diferencial $A$, entre las que se cuenta la resolvente
$(A-\lambda)^{-1}$, presentaremos a los operadores integrales
desde la perspectiva de los operadores pseudodiferenciales.

\bs

La acci{\'o}n de un operador integral $P$ sobre una funci{\'o}n
$f(x)$ de su dominio est{\'a} dada por,
\begin{equation}\label{nucleo}
    P f(x)=\int_{\mathbb{R}}K(x,x')f(x')\,dx'\,.
\end{equation}
La funci{\'o}n $K(x,y)$ es el {\bf n{\'u}cleo} del operador $P$.
Como los operadores integrales pertenecen al conjunto de
operadores pseudodiferenciales, cabe preguntarse qu{\'e}
relaci{\'o}n existe entre el n{\'u}cleo $K(x,y)$ del operador $P$
y su s{\'\i}mbolo $\sigma\{P\}(x,p)$.

\bs

Esta relaci{\'o}n se obtiene f{\'a}cilmente a partir de las
siguientes expresiones,
\begin{eqnarray}
    P f(x)=\int_{\mathbb{R}}\,
    e^{-ipx}\,\sigma\{P\}(x,p)\cdot\mathcal{F}\{f\}(p)
    \ \frac{dp}{\sqrt{2\pi}}=\nonumber\\
    =\int_{\mathbb{R}}\,
    e^{-ipx}\,\sigma\{P\}(x,p)\,\int_{\mathbb{R}}
    \,e^{ipx'}f(x')\ \frac{dx'}{\sqrt{2\pi}}
    \ \frac{dp}{\sqrt{2\pi}}=\nonumber\\
    =\int_{\mathbb{R}}\int_{\mathbb{R}}
    \,e^{-ip(x-x')}\,\sigma\{P\}(x,p)\ \frac{dp}{2\pi}
    \ f(x')\ dx'\,.\label{simbolos}
\end{eqnarray}
Comparando las ecuaciones (\ref{simbolos}) y (\ref{nucleo})
deducimos que la relaci{\'o}n entre el n{\'u}cleo y el
s{\'\i}mbolo del operador $P$ est{\'a} dada por,
\begin{equation}\label{nuc}
    K(x,x')=\int_{\mathbb{R}}
    \,e^{-ip(x-x')}\,\sigma\{P\}(x,p)\ \frac{dp}{2\pi}\,.
\end{equation}
De acuerdo con esta expresi{\'o}n, el valor del n{\'u}cleo
$K(x,x')$ en la diagonal, esto es, para $x=x'$, se escribe,
\begin{equation}\label{nuc-sim}
    K(x,x)=\int_{\mathbb{R}}
    \,\sigma\{P\}(x,p)\ \frac{dp}{2\pi}\,.
\end{equation}
Debe observarse que estas ecuaciones son v{\'a}lidas en tanto
exista la posibilidad de intercambiar el orden de integraci{\'o}n
en la ecuaci{\'o}n (\ref{simbolos}). De acuerdo con el Teorema de
Fubini, esto es posible si el integrando converge absolutamente,
esto es, si $d<-1$.

\bs

En consecuencia, como hemos mencionado, los operadores pseu\-do\-di\-fe\-ren\-cia\-les de orden $d<-1$ en una dimensi{\'o}n son operadores integrales. La relaci{\'o}n entre el n{\'u}cleo y el s{\'\i}mbolo del operador est{\'a} dada
por la ecuaci{\'o}n (\ref{nuc}) en la cual la integral es convergente.

\section{Funciones espectrales}\label{funcionesespectrales}

Consideremos, sobre una variedad de base $M$ de dimensi{\'o}n $m$
con borde, un o\-pe\-ra\-dor diferencial autoadjunto $A$ de orden
$d$ que admite un conjunto de autovalores reales
$\{\lambda_n\}_{n\in\mathbb{N}}$ correspondientes a autovectores
$\{\phi_n\}_{n\in\mathbb{N}}$ que forman una base ortonormal y
completa del espacio de Hilbert $\mathcal{H}$. El dominio
$\mathcal{D}(A)$ del operador est{\'a} caracterizado por
condiciones de contorno que aseguran que $A$ sea autoadjunto.

\bs

Definiremos tres funciones espectrales de nuestro inter{\'e}s
asociadas al o\-pe\-ra\-dor $A$. Todas ellas corresponden a la
traza de sendos operadores integrales: $(A-\lambda)^{-1}$,
$e^{-tA}$, $A^{-s}$; siendo
$\lambda\in\mathbb{C}-\{\lambda_n\}_{n\in\mathbb{N}}$,
$t\in\mathbb{R}^+$ y $s\in\mathbb{C}$ con $\mathcal{R}(s)$
suficientemente grande. Estos operadores est{\'a}n caracterizados
por sus respectivos n{\'u}cleos: $G(x,x',\lambda)$, $K(x,x',t)$,
$\zeta_A(x,x',s)$. El operador $(A-\lambda)^{-1}$ es la {\bf
resolvente}, $K(x,x',t)$ es el {\bf heat-kernel} y $\zeta_A(s):={\rm Tr}\,A^{-s}$ es
la {\bf funci{\'o}n-$\zeta$} del operador diferencial $A$.

\bs

Enunciamos a continuaci{\'o}n algunas propiedades de estos
operadores integrales:

\begin{itemize}
\item Dada una funci{\'o}n $f(x)\in\mathcal{H}$, el operador
$(A-\lambda)^{-1}$ permite resolver la ecuaci{\'o}n diferencial:
\begin{equation}\label{prores}
    (A-\lambda)\phi(x)=f(x)\,,
\end{equation}
en la que $\phi(x)$ satisface las condiciones de contorno sobre el
borde $\partial M$ que definen el dominio del operador $A$. La
soluci{\'o}n de este problema est{\'a} dada por,
\begin{equation}
    \phi(x)=(A-\lambda)^{-1}f(x)\,,
\end{equation}
siendo $(A-\lambda)^{-1}$ un operador integral cuyo n{\'u}cleo es,
\begin{equation}\label{fg}
    G(x,x',\lambda)=\sum_{n\in\mathbb{N}}\frac{\phi_n(x)\phi^*_n(x')}
    {\lambda_n-\lambda}\,.
\end{equation}
En consecuencia, la traza de la resolvente est{\'a} dada por,
\begin{equation}\label{trre}
    {\rm Tr}(A-\lambda)^{-1}=\int_M G(x,x,\lambda)=\sum_{n\in\mathbb{N}}
    \frac{1}{\lambda_n-\lambda}\,.
\end{equation}
La convergencia de la serie en la ecuaci{\'o}n (\ref{trre})
est{\'a} condicionada por el comportamiento asint{\'o}tico de los
autovalores de $A$ que, como veremos en esta misma secci{\'o}n, se
rige por la estructura de singularidades de la funci{\'o}n
$\zeta_A(s)$.

\bs

En el caso de operadores regulares sobre variedades compactas, las
singularidades de la funci{\'o}n $\zeta_A(s)$ est{\'a}n dadas por
la ecuaci{\'o}n (\ref{resu}) que, como mostraremos, implican un
comportamiento asint{\'o}tico de los autovalores de la forma
$\lambda_n\sim n^{d/m}$. En consecuencia, la serie de la
ecuaci{\'o}n (\ref{trre}) converge, para operadores regulares, si
el orden $d$ del operador diferencial es mayor que la
dimensi{\'o}n $m$ de la variedad de base\,\footnote{ En la
secci{\'o}n \ref{funcpart} del Ap{\'e}ndice, mostraremos que si la
variedad de base no es compacta, el comportamiento asint{\'o}tico
de los autovalores se rige por el comportamiento de los
coeficientes del operador diferencial en el infinito.}.

\bs

Si el operador diferencial $A$ est{\'a} definido sobre funciones
del intervalo $[0,1]\subset\mathbb{R}$, existe otra manera de
expresar la traza de la resolvente que ser{\'a}, en lo sucesivo,
de mayor utilidad. El n{\'u}cleo $G(x,x',\lambda)$ del operador
$(A-\lambda)^{-1}$, denominado tambi{\'e}n {\bf funci{\'o}n de
Green} del operador $A-\lambda$, satisface, en virtud de la
ecuaci{\'o}n (\ref{fg}),
\begin{equation}\label{reso}
    (A-\lambda)G(x,x',\lambda)=\delta(x-x')\,.
\end{equation}
Esta ecuaci{\'o}n define, junto con las condiciones de contorno
apropiadas en $x=0$ y $x=1$, la funci{\'o}n $G(x,x',\lambda)$.

\bs

La soluci{\'o}n de la ecuaci{\'o}n (\ref{reso}) est{\'a} dada por,
\begin{equation}\label{solres}
    G(x,x',\lambda)=-\frac{\theta(x'-x)L(x,\lambda)R(x',\lambda)+
    \theta(x-x')L(x',\lambda)R(x,\lambda)}{W[L,R](\lambda)}.
\end{equation}
Las funciones $L(x,\lambda),R(x,\lambda)$ pertenecen al n{\'u}cleo
del operador diferencial $A-\lambda$ pero no al dominio de $A$,
pues no satisfacen las condiciones de contorno que
ca\-rac\-te\-ri\-zan a $\mathcal{D}(A)$. En efecto, $L(x,\lambda)$
satisface la condici{\'o}n de contorno apropiada en $x=0$ pero no
en $x=1$ en tanto que $R(x,\lambda)$ satisface la condici{\'o}n de
contorno en $x=1$ pero no lo hace en $x=0$. La funci{\'o}n
$\theta(\cdot)$ es la funci{\'o}n Heaviside y
$W[L(x,\lambda),R(x,\lambda)]$ es el wronskiano,
\begin{equation}
    W[L,R](\lambda)=L(x,\lambda)\partial_xR(x,\lambda)-
    \partial_xL(x,\lambda),R(x,\lambda)\,,
\end{equation}
que es no nulo si $\lambda$ no pertenece al espectro de $A$ y,
para los operadores dados por las expresiones (\ref{sch}) y
(\ref{dir}), no depende de $x$.

\bs

De acuerdo con la ecuaci{\'o}n (\ref{solres}), la traza de la
resolvente resulta,
\begin{equation}
    {\rm Tr}(A-\lambda)^{-1}=-\frac{1}{W[L,R(\lambda)]}\int_0^1
    L(x,\lambda)R(x,\lambda)\,dx\,.
\end{equation}

\item Si $A$ es un operador diferencial positivo definido, podemos
resolver el sistema:
\begin{equation}\label{ci}
    \begin{array}{c}
    (\partial_t+A)\phi(x,t)=0\,,\\ \\
    \phi(x,t=0)=f(x)\,,
    \end{array}
\end{equation}
donde $x\in M$, $t\in\mathbb{R}^+$ y $\phi(x,t)$ satisface las
condiciones de contorno apropiadas sobre $\partial M$. El problema
planteado por las expresiones (\ref{ci}) puede invertirse de la
siguiente manera,
\begin{equation}
    \phi(x,t)=e^{-t A}f(x)\,,
\end{equation}
donde $e^{-t A}$ es un operador integral cuyo n{\'u}cleo est{\'a}
dado por,
\begin{equation}
    K(x,x',t)=\sum_{n\in\mathbb{N}}e^{-t\lambda_n}\,\phi_n(x)\phi^*_n(x')\,,
\end{equation}
y su traza por,
\begin{equation}\label{trhe}
    {\rm Tr}\,e^{-tA}=\int_M K(x,x,t)\,dx=\sum_{n\in\mathbb{N}}e^{-t\lambda_n}.
\end{equation}

\item En \cite{Seeley1}, R.T.\ Seeley defini{\'o} las potencias
$A^{-s}$ de un operador diferencial $A$ y demostr{\'o} que si
$\mathcal{R}(s)$ es suficientemente grande, $A^{-s}$ tiene un
n{\'u}cleo continuo $\zeta_A(x,x',s)$. Adem{\'a}s, si $x\neq x'$
el n{\'u}cleo $\zeta_A(x,x',s)$ es una funci{\'o}n entera de $s$,
en tanto que $\zeta_A(x,x,s)$ admite una extensi{\'o}n meromorfa
al plano complejo $s$ con polos simples dados por la ecuaci{\'o}n
(\ref{resu}).

\bs

La traza del operador $A^{-s}$ est{\'a} dada por,
\begin{equation}\label{113}
    \zeta_A(s):= {\rm Tr}\,A^{-s}=\int_M \zeta_A(x,x,s)\,dx=
    \sum_{n\in\mathbb{N}}\lambda_n^{-s}\,.
\end{equation}

\item En \cite{APS} se define, en conexi{\'o}n con el Teorema del
{\'\i}ndice para variedades con borde, la funci{\'o}n $\eta(s)$ de
un operador $A$ cuyo espectro est{\'a} dado por los autovalores
$\{\lambda_n\}_{n\in\mathbb{N}}$. Si $\mathcal{R}(s)$ es
suficientemente grande, la funci{\'o}n $\eta(s)$ est{\'a} dada
por,
\begin{equation}\label{eta}
    \eta(s)=\sum_{\lambda_n>0}\lambda_n^{-s}-\sum_{\lambda_n<0}|
    \lambda_n|^{-s}\,.
\end{equation}

\end{itemize}

\subsection{Relaci{\'o}n entre la distintas funciones espectrales}\label{relfe}

En la secci{\'o}n anterior hemos definido las trazas ${\rm
Tr}\,(A-\lambda)^{-1}$, ${\rm Tr}\,e^{-tA}$ y ${\rm Tr}\,A^{-s}$,
que son funciones de los par{\'a}metros $\lambda,t,s$ y est{\'a}n
determinadas por el espectro del operador diferencial $A$. Cabe,
entonces, esperar que existan relaciones entre estas funciones
espectrales; veremos, en particular, que los desarrollos
asint{\'o}ticos de ${\rm Tr}\,(A-\lambda)^{-1}$ y ${\rm
Tr}\,e^{-tA}$ para grandes valores de $|\lambda|$ y peque{\~n}os
valores de $t$, respectivamente, est{\'a}n determinados por las
singularidades de la funci{\'o}n $\zeta_A(s)$.

\bs

Se{\~n}alamos, en primer lugar, que las funciones espectrales
correspondientes a un o\-pe\-ra\-dor diferencial positivo definido
$A$ satisfacen las siguientes relaciones:
\begin{eqnarray}
    {\rm Tr}\,(A-\lambda)^{-1}=\int_0^{\infty}e^{t\lambda}\,
    {\rm Tr}\,e^{-tA}\,dt\,,\label{reshea}\\
    {\rm Tr}\,A^{-s}=\frac{1}{\Gamma(s)}\int_0^{\infty}t^{s-1}\,
    {\rm Tr}\,e^{-tA}\,dt\,,\label{zethea}
\end{eqnarray}
si $\mathcal{R}(\lambda)$ es menor que todos los autovalores de
$A$ y $\mathcal{R}(s)>m/d$, respectivamente. De acuerdo con estas
expresiones la transformada de Laplace de la traza del heat-kernel
es la traza de la resolvente y su transformada de Mellin es la
funci{\'o}n $\zeta_A(s)$. Estas relaciones pueden demostrarse
estableciendo relaciones similares entre cada uno de los
t{\'e}rminos de la series (\ref{trhe}), (\ref{trre} y (\ref{113}).

\bs

Mediante una integraci{\'o}n por partes en la ecuaci{\'o}n
(\ref{zethea}) y utilizando el desarrollo asint{\'o}tico
(\ref{teohea}) se puede probar,
\begin{equation}\label{zetcero}
    \zeta(0)=c_m(A)\,.
\end{equation}

\bs

Cabe se{\~n}alar que para un operador diferencial autoadjunto $A$
que no sea positivo definido, para el cual no existe un operador
$e^{-tA}$ asociado, se pueden, sin embargo, definir los operadores
integrales $A^{-s}$ y $(A-\lambda)^{-1}$, cuyas trazas est{\'a}n
relacionadas por,
\begin{equation}\label{zetres}
    {\rm Tr}\,A^{-s}=- \frac{1}{2\,\pi\,i} \oint_{\mathcal{C}}
  {\lambda^{-s}} \, {\rm Tr}\left(A-\lambda\right)^{-1}
  \, d\lambda\, ,
\end{equation}
donde $\mathcal{C}$ es una curva que encierra a los autovalores de
$A$ en sentido antihorario. La relaci{\'o}n (\ref{zetres}) puede
demostrarse en forma inmediata a partir de la ecuaci{\'o}n
(\ref{trre}), que indica que la traza de la resolvente tiene polos
simples en los autovalores de $A$ con re\-si\-duos igual a $1$.

\bigskip

Supongamos ahora que la traza del heat-kernel admite un desarrollo
asint{\'o}tico para valores peque{\~n}os de $t$ de la forma,
\begin{equation}\label{desaheat}
    {\rm Tr}\,e^{-tA}\sim \sum_{n=0}^{\infty}c_n(A)\cdot t^{j_n}\,.
\end{equation}
Reemplazando este desarrollo asint{\'o}tico en la expresi{\'o}n
(\ref{reshea}) obtenemos el siguiente desarrollo asint{\'o}tico
para la traza de la resolvente,
\begin{equation}\label{desareso}
    {\rm Tr}\,(A-\lambda)^{-1}\sim \sum_{n=0}^{\infty}\Gamma(j_n+1)\,
    c_n(A)\cdot\lambda^{-j_n-1}\,.
\end{equation}

El desarrollo asint{\'o}tico de la traza del heat-kernel describe
tambi{\'e}n las singularidades de la funci{\'o}n $\zeta_A(s)$. En
efecto, las ecuaciones (\ref{desaheat}) y (\ref{zethea}) indican
que los polos de la funci{\'o}n $\zeta_A(s)$ est{\'a}n ubicados en
los puntos $s_n$,
\begin{equation}\label{polres}
    s_n=-j_n\,,
\end{equation}
y los residuos correpondientes est{\'a}n dados por,
\begin{equation}\label{desazeta}
    {\rm Res}\{\zeta_A(s)\}|_{s=s_n}=\frac{c_n(A)}{\Gamma(-j_n)}\,.
\end{equation}

En consecuencia, el desarrollo asint{\'o}tico de la traza del
heat-kernel para peque{\~n}os va\-lo\-res de $t$, el desarrollo
asint{\'o}tico de la traza de la resolvente para grandes valores
de $|\lambda|$ y las singularidades de la funci{\'o}n $\zeta_A(s)$
est{\'a}n relacionados en virtud de las ecuaciones (\ref{reshea})
y (\ref{zethea}). En particular, si la traza del heat-kernel
admite un desarrollo en potencias de $t$ (v{\'e}ase la
ecuaci{\'o}n (\ref{desaheat})) entonces la traza de la resolvente
admite un desarrollo en potencias de $\lambda$ cuyos exponentes y
coeficientes se pueden expresar en t{\'e}rminos de los exponentes
y coeficientes del primero. La funci{\'o}n $\zeta_A(s)$, por su
parte, presenta polos simples en puntos del eje real que est{\'a}n
determinados por los exponentes de $t$ y sus residuos est{\'a}n, a
su vez, determinados por los coeficientes de las potencias de $t$.

\bs

Por consiguiente, de acuerdo con el resultado (\ref{resu}),
v{\'a}lido para operadores diferenciales $A$ con coeficientes
regulares, definidos por condiciones de contorno locales sobre el
borde de una variedad compacta, las potencias $j_n$ de $t$ en los
desarrollos (\ref{desaheat}) y (\ref{desareso}) son,
\begin{equation}\label{resu2}
    j_n=\frac{n-m}{d}\,,
\end{equation}
siendo $d$ el orden del operador $A$ y $m$ la dimensi{\'o}n de la
variedad de base $M$ (v{\'e}ase, {\it e.g.}, la ecuaci{\'o}n
(\ref{teohea}).) El resultado (\ref{resu}) indica que los
exponentes de las potencias de $\lambda$ y de $t$ en los
desarrollos asint{\'o}ticos de las trazas de la resolvente y del
heat-kernel co\-rres\-pon\-dien\-tes a operadores regulares
est{\'a}n determinados por el orden del operador diferencial y la
dimensi{\'o}n de la variedad.

\bs

Antes de finalizar esta secci{\'o}n haremos un par de
observaciones con respecto a la relaci{\'o}n entre las distintas
funciones espectrales. Las ecuaciones (\ref{reshea}) y
(\ref{zethea}) permiten relacionar las singularidades de la
funci{\'o}n $\zeta_A(s)$ con los desarrollos asint{\'o}ticos de
las trazas de la resolvente y del heat-kernel a{\'u}n cuando estos
desarrollos contengan logaritmos de $|\lambda|$ y de $t$,
respectivamente. En ese caso, puede verse que la funci{\'o}n
$\zeta_A(s)$ presenta polos de multiplicidad mayor.

\bs

Es interesante notar que los desarrollos asint{\'o}ticos de las
funciones espectrales o las singularidades de la funci{\'o}n
$\zeta_A(s)$ guardan relaci{\'o}n, a su vez, con el desarrollo
asint{\'o}tico de los autovalores $\lambda_n$ del operador diferencial
$A$ para grandes valores de $n$. Para ilustrar esta relaci{\'o}n
mostraremos que si los autovalores de un operador diferencial
suave con condiciones de contorno locales sobre el borde de una
variedad compacta satisfacen el siguiente comportamiento
asint{\'o}tico:
\begin{equation}\label{newnew}
    \lambda_n\sim \beta\,n^\alpha+\ldots
\end{equation}
entonces los coeficientes $\alpha$ y $\beta$ est{\'a}n dados por,
\begin{equation}\label{Ike}
    \lambda_n\sim \left[\frac{m\Gamma(m/d)}{d\,c_0(A)}\right]^{d/m}\,n^{d/m}\,,
\end{equation}
donde $c_0(A)$ es el primer coeficiente del desarrollo asint{\'o}tico
(\ref{desaheat}). En efecto, reemplazando el desarrollo asint{\'o}tico
(\ref{newnew}) en la ecuaci{\'o}n (\ref{113}) obtenemos,
\begin{eqnarray}
    \zeta_A(s)=\sum_{n=0}^{\infty}(\beta\,n^\alpha+\ldots)^{-s}=
    \beta^{-s}\sum_{n=0}^{\infty}n^{-\alpha s}(1+\ldots)^{-s}=\nonumber\\
    =\beta^{-s}\sum_{n=0}^{\infty}n^{-\alpha s}+\ldots=\beta^{-s}
    \zeta_R(\alpha s)+\ldots
\end{eqnarray}
donde $\zeta_R(z)$ es la funci{\'o}n-$\zeta$ de Riemann, que tiene
un {\'u}nico polo simple en $z=1$ con residuo $1$. Por lo tanto,
la funci{\'o}n $\zeta_A(s)$ tiene un polo en $s=\alpha^{-1}$ con
residuo $\alpha^{-1}\beta^{-1/\alpha}$. Esto determina, en
conjunto con las ecuaciones (\ref{polres}) y (\ref{resu2}), los
valores $\alpha=d/m$ y $\beta=(m\Gamma(m/d)/a_0d)^{d/m}$, de
acuerdo con lo expresado en (\ref{Ike}).

\bigskip

En esta Tesis estudiaremos algunos operadores diferenciales con
coeficientes singulares cuyas funciones-$\zeta$ presentan polos
que no obedecen al resultado (\ref{resu}). Mostraremos que las
trazas de la resolvente y del heat-kernel correspondientes a estos
operadores presentan desarrollos asint{\'o}ticos cuyas potencias
no est{\'a}n determinadas, consecuentemente, por el orden del
operador diferencial y la dimensi{\'o}n de la variedad de base. En
efecto, veremos que los exponentes de estas potencias dependen, en
general, del coeficiente que caracteriza la intensidad de la
singularidad del operador diferencial.

\section{Desarrollos asint{\'o}ticos}\label{deas}

El desarrollo asint{\'o}tico de la traza de la resolvente
$(A-\lambda)^{-1}$ para grandes valores de $|\lambda|$ ha sido
estudiado para operadores diferenciales con coeficientes regulares
definidos sobre variedades suaves de dimensi{\'o}n arbitraria
\cite{Seeley1,Seeley3,Seeley4}.

\bs

Sin embargo, ser{\'a} suficiente para nuestros prop{\'o}sitos
considerar un operador diferencial $A$ de orden $d$ que act{\'u}a
sobre elementos de $\mathbf{H_d}(\mathbb{R}^+)\otimes
\mathbb{C}^k$, {\it i.e.}, funciones de $\mathbb{R}^+$ con valores
en un espacio vectorial de dimensi{\'o}n $k$ que admiten $d$
derivadas,
\begin{equation}
    A=\sum_{n=0}^{d}A_n(x) (-i\partial_x)^n\,.
\end{equation}
Los coeficientes $A_n(x)$ son funciones en
$\mathcal{C}^{\infty}(\mathbb{R})$ con valores en el espacio de
matrices $\mathbb{C}^{k\times k}$.

\bs

Consideremos tambi{\'e}n $kd/2$ operadores de borde $B_j$,
\begin{equation}\label{opbo}
    B_j=\sum_{k=1}^d B_{jk}(-i\partial_x)^{d-k}\,,
\end{equation}
con $j=1,\ldots,kd/2$ y $B_{jk}$ un conjunto de matrices $1\times
k$.

\bs

\begin{defn}\label{deuno}
    El operador diferencial $A$ es un operador {\bf el{\'\i}ptico} si,
    \begin{equation}
    \mathit{det}(A_d(x))\neq 0\,.
    \end{equation}
\end{defn}
\begin{defn}\label{dedos}
    El operador diferencial $A$ satisface la {\bf condici{\'o}n de Agmon}
    sobre el rayo $R_{\theta}:=\{\mathit{arg}(\lambda)=\theta\}$ si,
    \begin{equation}
    \mathit{det}(A_d(x)p^d-\lambda)\neq 0\,,
    \end{equation}
    $\forall \lambda\in R_{\theta},p\neq 0$.
\end{defn}
\begin{defn}\label{decuatro}
El sistema $\{A,B_1,\ldots,B_{kd/2}\}$ satisface la condici{\'o}n
de Agmon sobre el rayo
$R_{\theta}:=\{\mathit{arg}(\lambda)=\theta\}$ si $A$ satisface la
condici{\'o}n de Agmon sobre el rayo
$R_{\theta}:=\{\mathit{arg}(\lambda)=\theta\}$ y si para todo
$\lambda\in R_{\theta}$ y todo elemento $(g_1,\ldots,g_{kd/2})$ de
$\mathbb{C}^{kd/2}$ existe una {\'u}nica funci{\'o}n $f(x)$ con
valores en $\mathbb{C}^{k}$ que satisface,
\begin{eqnarray}
    [A_{d}(0)(-i\partial_x)^d-\lambda]\,f(x)=0\,,\\
    \lim_{x\rightarrow\infty}f(x)=0\,,\\
    \left.B_j\,f(x)\right|_{x=0}=g_j\,.
\end{eqnarray}
\end{defn}

\bs

Consideremos entonces un sistema el{\'\i}ptico de operadores
$\{A,B_1,\ldots,B_{kd/2}\}$ que sa\-tis\-fa\-ce la condici{\'o}n
de Agmon sobre el rayo $R_{\theta}$. Los operadores de borde $B_j$
ca\-rac\-te\-ri\-zan las condiciones de contorno locales que
satisfacen las funciones pertenecientes al dominio
$\mathcal{D}(A)$ de $A$,
\begin{equation}\label{condcont}
    \mathcal{D}(A)=\{\phi(x)\in\mathbf{H_d}(\mathbb{R}^+)\otimes
    \mathbb{C}^k:
    B_j\phi(x)|_{x=0}=0\}\,.
\end{equation}
Obtendremos a continuaci{\'o}n un desarrollo asint{\'o}tico para
el s{\'\i}mbolo de la resolvente $\sigma\{(A-\lambda)^{-1}\}$, de
acuerdo con las condiciones de contorno definidas en
(\ref{condcont}), de la forma,
\begin{equation}\label{ces}
    \sigma\{(A-\lambda)^{-1}\}\sim\sum_{n=0}^{\infty}c_{-d-n}(x,p,\lambda)\,
\end{equation}
en la que los coeficientes $c_{-d-n}(x,p)$ son homog{\'e}neos de
grado $-d-n$ en las variables $(p,\lambda^{1/d})$,
\begin{equation}\label{homo}
    c_{-d-n}(x,t p,t^{d} \lambda)=t^{-d-n}c_{-d-n}(x,p,\lambda)\,.
\end{equation}
Para ello, calculamos los s{\'\i}mbolos de cada miembro de la
expresi{\'o}n,
\begin{equation}\label{nada0}
    (A-\lambda)\cdot(A-\lambda)^{-1}=\mathbf{1}\,.
\end{equation}
La regla de composici{\'o}n (\ref{sim2}) permite escribir,
\begin{eqnarray}\label{nada}
    \sigma\{(A-\lambda)\cdot(A-\lambda)^{-1}\}\sim\nonumber\\
    \sim\sum_{k=0}^{\infty}
    \frac{1}{k!}\,
    \partial_p^k \left(\sigma\{(A-\lambda)\}\right)\cdot
    (-i\partial_x)^k\left(\sigma\{(A-\lambda)^{-1}\}\right)\sim\nonumber\\
    \sim1.
\end{eqnarray}
Si reemplazamos ahora el s{\'\i}mbolo de la resolvente
$\sigma\{(A-\lambda)^{-1}\}$ en la ecuaci{\'o}n (\ref{nada}) por
el Ansatz (\ref{ces}) obtenemos,
\begin{eqnarray}\label{detces}
    \sum_{k=0}^{\infty}\frac{1}{k!}\,
    \partial_p^{k}\left(\sum_{h=0}^{d}A_h(x) p^h-\lambda\right)\cdot
(-i\partial_x)^k\left(\sum_{n=0}^{\infty}c_{-d-n}(x,p,\lambda)\right)=
\nonumber\\
    =\sum_{k=0}^{d}\frac{1}{k!}\,
    \partial_p^{k}\left(\sum_{h=0}^{d}A_h(x) p^h-\lambda\right)\cdot
   \left(\sum_{n=0}^{\infty}(-i\partial_x)^kc_{-d-n}(x,p,\lambda)\right)=
   \nonumber\\
    =\left(A_d(x) p^d-\lambda+\dots+A_0(x)\right)\cdot
    \left(c_{-d}(x,p,\lambda)+c_{-d-1}(x,p,\lambda)+\ldots\right)-\nonumber\\
    -i\left(A_d(x)d p^{d-1}+\dots+A_1(x)\right)\cdot
\left(\partial_xc_{-d}(x,p,\lambda)+\partial_xc_{-d-1}(x,p,\lambda)+
\ldots\right)+\nonumber
\\
    +\ldots+\nonumber\\
    +A_d(x)\cdot
(-i)^{d}\left(\partial_x^dc_{-d}(x,p,\lambda)+\partial_x^dc_{-d-1}
(x,p,\lambda)+\ldots
 \right)\sim
    \nonumber\\
    \sim 1.\nonumber\\
\end{eqnarray}
A partir de esta expresi{\'o}n y de la propiedad de homogeneidad
(\ref{homo}) obtenemos una relaci{\'o}n de recurrencia que permite
determinar los coeficientes $c_{-n-2}(x,p,\lambda)$ en
t{\'e}rminos de los coeficientes del operador diferencial,
\begin{eqnarray}
    c_{-d}(x,p,\lambda)=\left(A_d(x)p^d-\lambda\right)^{-1},\\
    c_{-d-n}(x,p,\lambda)=-\left(A_d(x)p^d-\lambda\right)^{-1}\times\nonumber\\
    \times
    \sum_{l=0}^{n-1}\sum_{h=d-n+l}^{d}
    \left(\begin{array}{c}
    h\\d-n+l
    \end{array}\right)\,
    A_h(x)p^{d-n+l}\cdot
    (-i\partial_x)^lc_{-d-l}(x,p,\lambda)\,.\label{coefi}
\end{eqnarray}
Se puede verificar por inducci{\'o}n que los coeficientes
$c_{-d-n}(x,p,\lambda)$, dados por (\ref{coefi}), son
homog{\'e}neos de grado $-d-n$ en las variables
$(p,\lambda^{1/d})$, tal como lo expresa la relaci{\'o}n
(\ref{homo}).

\bs

Asimismo, puede tambi{\'e}n verificarse que se satisface,
\begin{equation}
    \sigma\{A-\lambda\}\left(\sum_{n=0}^{J}c_{-d-n}(x,p,\lambda)\right)
    -1\in S^{-J-1}.
\end{equation}

\bs

Se puede demostrar, adem{\'a}s, que el operador pseudodiferencial
asociado al s{\'\i}mbolo (\ref{ces}) satisface asint{\'o}ticamente
la ecuaci{\'o}n (\ref{nada0}). Sin embargo, no hemos garantizado
que la imagen de este operador satisfaga la condici{\'o}n de
contorno (\ref{condcont}). Para esto, es necesario agregar nuevos
t{\'e}rminos a la expresi{\'o}n (\ref{ces}) de modo que el
s\'imbolo de la resolvente satisfaga la condici{\'o}n de contorno
(\ref{condcont}) sin modificar la relaci{\'o}n (\ref{nada0}).

\bs

Proponemos entonces para el s{\'\i}mbolo de la resolvente una
aproximaci{\'o}n asint{\'o}tica de la forma,
\begin{equation}\label{desa}
    \sigma\{(A-\lambda)^{-1}\}\sim
    \sum_{n=0}^{\infty}c_{-d-n}(x,p,\lambda)-
    \sum_{n=0}^{\infty}d_{-d-n}(x,p,\lambda)\,,
\end{equation}
en la que los coeficientes $d_{-d-n}(x,p,\lambda)$ quedan
definidos por las condiciones,
\begin{eqnarray}
    (A-\lambda)\cdot\sum_{n=0}^{\infty}d_{-d-n}(x,p,\lambda)=0,\label{ecu}\\
    \lim_{x\rightarrow\infty}d_{-d-n}(x,p,\lambda)=0,\label{infi}\\
    \left.B_j\,d_{-d-n}(x,p,\lambda)\right|_{x=0}=
    \left.\sigma\{B_j\}\,c_{-d-n}(x,p,\lambda)\right|_{x=0}\,.\label{cero}
\end{eqnarray}
La ecuaci{\'o}n (\ref{ecu}) representa un sistema de ecuaciones
diferenciales ordinarias para las cantidades
$d_{-d-n}(x,p,\lambda)$ como funciones de $x$ con condiciones de
contorno en $x\rightarrow\infty$ dadas por (\ref{infi}) y en el
origen por (\ref{cero}). La resoluci{\'o}n de estas ecuaciones
diferenciales, de manera recurrente, permite probar que los
coeficientes $d_{-d-n}(x,p,\lambda)$ son homog{\'e}neos de grado
$-d-n-1$ en las variables $(p,\lambda^{1/d})$.

\bs

A partir del desarrollo (\ref{desa}) y de la homogeneidad de los
coeficientes $c_{-d-n}(x,p,\lambda)$ y $d_{-d-n}(x,p,\lambda)$ en
las va\-ria\-bles $(p,\lambda^{1/d})$ podemos deducir el
desarrollo asint{\'o}tico del n{\'u}cleo de la resolvente en la
diagonal $G(x,x,\lambda)$ . Reemplazamos, para ello, el desarrollo
(\ref{desa}) en la ecuaci{\'o}n (\ref{nuc-sim}).

\bs

La integral del miembro derecho de (\ref{nuc-sim}) para el
t{\'e}rmino correspondiente a \linebreak$c_{-d-n}(x,p,\lambda)$
resulta,
\begin{eqnarray}\label{cg}
    \int_{\mathbb{R}}\,c_{-d-n}(x,p,\lambda)\ \frac{dp}{2\pi}=\nonumber\\
    =\int_{\mathbb{R}}\,c_{-d-n}(x,\lambda^{-1/d}\,p,1)\lambda^{-1-n/d}\
    \frac{dp}{2\pi}=\nonumber\\
    =\int_{\mathbb{R}}\,c_{-d-n}(x,p,1)\lambda^{-1-n/d+1/d}\ \frac{dp}
    {2\pi}=\nonumber\\
    =: \gamma_n(A,x)\cdot \lambda^{\frac{1-n}{d}-1}\,.
\end{eqnarray}
Para arribar a la segunda ecuaci{\'o}n hemos utilizado la
propiedad de homogeneidad de los coeficientes
$c_{-d-n}(x,p,\lambda)$, en tanto que la ecuaci{\'o}n siguiente se
obtuvo mediante una redefinici{\'o}n de la va\-ria\-ble de
integraci{\'o}n.

\bs

Si calculamos la integal dada por la ecuaci{\'o}n (\ref{nuc-sim})
correspondiente al t{\'e}rmino de borde dado por
$d_{-d-n}(x,p,\lambda)$ obtenemos,
\begin{equation}\label{dd}
    \int_{\mathbb{R}}\,d_{-d-n}(x,p,\lambda)\ \frac{dp}{2\pi}
    := -\delta_n(A,x)\cdot \lambda^{-\frac{n}{d}-1}\,.
\end{equation}

Finalmente, conclu{\'\i}mos que el n{\'u}cleo de la resolvente en
la diagonal $G(x,x,\lambda)$ admite un desarrollo asint{\'o}tico
para grandes valores de $\lambda$ dado por,
\begin{equation}\label{teor}
    G(x,x,\lambda)\sim \sum_{n=0}^{\infty}\gamma_n(A,x)\cdot
    \lambda^{\frac{1-n}{d}-1}+
    \sum_{n=0}^{\infty}\delta_n(A,x)\cdot \lambda^{-\frac{n}{d}-1}\,.
\end{equation}
Los coeficientes locales $\gamma_n(A,x)$ y $\delta_n(A,x)$
est{\'a}n dados por las integrales de
\linebreak$c_{-d-n}(x,p,1)$ y $d_{-d-n}(x,p,1)$,
respectivamente (v{\'e}anse las ecuaciones (\ref{cg}) y
(\ref{dd}).) N{\'o}tese que las potencias de $\lambda$ en el
desarrollo asint{\'o}tico de \mbox{$G(x,x,\lambda)$} s{\'o}lo
dependen del orden del operador diferencial.

\bigskip

Como estudiaremos la validez de este resultado en referencia a
operadores diferenciales sobre variedades de base
unidimensionales, nos hemos limitado a describir el comportamiento
asin{\'o}tico de la resolvente de un operador $A$ definido sobre
funciones de $\mathbb{R}^+$. Sin embargo, como hemos mencionado,
el mismo an{\'a}lisis puede repetirse para operadores
diferenciales regulares sobre variedades de base de dimensi{\'o}n
mayor\,\footnote{En el Ap{\'e}ndice \ref{hke} estudiamos el
desarrollo asint{\'o}tico del heat-kernel para un operador de
segundo orden en una variedad de dimensi{\'o}n arbitraria.}.

\bs

Si el operador diferencial regular est{\'a} definido sobre una
variedad de base de dimensi{\'o}n $m$ el n{\'u}cleo de la
resolvente en la diagonal admite un desarrollo asint{\'o}tico en
potencias de $\lambda$ de la forma,
\begin{equation}\label{desaheadesareso}
    \lambda^{\frac{m-n}{d}-1}\,.
\end{equation}

\bs

Es importante observar que el resultado (\ref{teor}) representa un
desarrollo asint{\'o}tico local puesto que corresponde a una
cantidad que depende de la coordenada $x\in\mathbb{R}^+$. Sin
embargo, como el operador diferencial $A$ est{\'a} definido sobre
una variedad de base no compacta, la expresi{\'o}n (\ref{teor}) no
implica  que la traza de la resolvente, si existiera, admita un
desarrollo asint{\'o}tico dado por la integral t{\'e}rmino a
t{\'e}rmino de su miembro derecho.

\bs

De todas maneras, en el caso de un operador diferencial regular
$A$ definido sobre una variedad de base $M$ compacta, el
desarrollo asint{\'o}tico (\ref{teor}) permite demostrar, mediante
una parametrizaci{\'o}n local de la variedad, que la traza de la
resolvente admite un desarollo asint{\'o}tico dado por,
\begin{eqnarray}\label{teordesa}
    {\rm Tr}(A-\lambda)^{-1}=\int_{M}G(x,x,\lambda)\,dx\sim\nonumber\\
    \sim \sum_{n=0}^{\infty}\gamma_n(A)\cdot \lambda^{\frac{1-n}{d}-1}+
    \sum_{n=0}^{\infty}\delta_n(A)\cdot \lambda^{-\frac{n}{d}-1}\,,
\end{eqnarray}
donde los coeficientes $\gamma_n(A)$ y $\delta_n(A)$ son las
integrales de los coeficientes locales $\gamma_n(A,x)$ y
$\delta_n(A,x)$ sobre $M$ y $\partial M$, respectivamente.

\bs

El desarrollo asint{\'o}tico (\ref{teordesa}), junto con las
relaciones (\ref{desaheat}), (\ref{desareso}) y (\ref{desazeta}),
permiten probar las validez de los resultados (\ref{resu}) y
(\ref{teohea}).

\bs

Insistimos en que las potencias del desarrollo asint{\'o}tico de
la traza de la resolvente dependen s{\'o}lamente del orden del
o\-pe\-ra\-dor diferencial y de la dimensi{\'o}n de la variedad de
base. La variaci{\'o}n de cualquier otro par{\'a}metro del
problema relacionado con los coeficientes del operador diferencial
o con las condiciones de contorno, se manifiesta en los
coeficientes $\gamma_n,\delta_n$ sin afectar las potencias del
desarrollo asint{\'o}tico.


\part{Operadores Singulares}\label{opesing}

\vspace{5mm}\begin{flushright}{\it If only I had the theorems!\\
Then I should find the proofs easily enough. \\
(G.F.\ Bernhard Riemann.)}
\end{flushright}

\vspace{25mm}

\section{Introducci{\'o}n}

En el cap{\'\i}tulo anterior hemos presentado algunas de las
propiedades que son conocidas acerca de los desarrollos
asint{\'o}ticos de las funciones espectrales correspondientes a
operadores diferenciales con coeficientes regulares. Para ello
hemos utilizado resultados de la teor{\'\i}a de los operadores
pseudodiferenciales.

\bs

No obstante, la derivaci{\'o}n del desarrollo asint{\'o}tico dado por
(\ref{teordesa}) que hemos presentado no es v{\'a}lida si el operador
diferencial posee coeficientes singulares.

\bigskip

En el presente cap{\'\i}tulo consideraremos operadores
diferenciales que poseen un coeficiente con un tipo especial de
singularidad, por lo que sus funciones espectrales presentan
desarrollos asint{\'o}ticos distintos del resultado
(\ref{teordesa}). En efecto, veremos que la traza de la resolvente
de estos operadores admite un desarrollo asint{\'o}tico en
potencias de $\lambda$ cuyos exponentes dependen de las
caracter{\'\i}sticas de la singularidad.

\bigskip

El primer paso en la obtenci{\'o}n de estos desarrollos
asint{\'o}ticos consiste en reconocer la existencia de una familia
de extensiones autoadjuntas. Este punto es esencial pues, como se
ver{\'a}, la validez del resultado (\ref{teordesa}), a{\'u}n en
presencia de una singularidad, depende de las condiciones de
contorno del problema. En efecto, para algunas condiciones de
contorno el resultado (\ref{teordesa}) es v{\'a}lido, en tanto que
existe un conjunto infinito de condiciones de contorno para las
que los exponentes de las potencias del desarrollo asint{\'o}tico
de la traza de la resolvente no est{\'a}n determinados por el
orden del operador y la dimensi{\'o}n de la variedad de base.

\bigskip

En general, la diversidad de condiciones de contorno de inter{\'e}s
f{\'\i}sico que admite un operador di\-fe\-ren\-cial sim{\'e}trico est{\'a}
caracterizada por sus extensiones autoadjuntas. En virtud de la
relevancia de las distintas extensiones autoadjuntas para estudiar
el desarrollo asint{\'o}tico de la traza de la resolvente, hemos
constru{\'\i}do una relaci{\'o}n entre las resolventes correspondientes a
distintas extensiones autoadjuntas. Esta relaci{\'o}n es conocida para
el caso de operadores regulares: entre 1944 y 1946 M.G.\ Krein
\cite{Krein1,Krein2} demostr{\'o} una manera de obtener la resolvente
de cualquier extensi{\'o}n autoadjunta si se conoce la resolvente de
una extensi{\'o}n autoadjunta particular. En la secci{\'o}n \ref{singu}
extenderemos este resultado al caso de los operadores
diferenciales con coeficientes singulares que hemos considerado.
Posteriormente, en la secci{\'o}n \ref{luego}, utilizaremos esta
extensi{\'o}n de la f{\'o}rmula de Krein para encontrar aquellas
condiciones de contorno para las que el desarrollo asint{\'o}tico de
la traza de la resolvente posee potencias cuyos exponentes
dependen de las caracter{\'\i}sticas de la singularidad.

\section{F{\'o}rmula de Krein para operadores regulares}\label{regu}

Describiremos en esta secci{\'o}n la f{\'o}rmula de Krein
\cite{Krein1,Krein2} (v{\'e}ase tambi{\'e}n \cite{A-G}), que
relaciona las resolventes de las distintas extensiones
autoadjuntas de un operador regular. Aunque esta relaci{\'o}n no
ha sido necesaria para estudiar el desarrollo asint{\'o}tico de la
resolvente, ser{\'a} de gran utilidad construir una relaci{\'o}n
equivalente para el caso de operadores con coeficientes
singulares.

\bs

Para caracterizar las extensiones autoadjuntas del operador
ser{\'a} {\'u}til el siguiente Teorema \cite{Gorba}, que
enunciamos sin demostraci{\'o}n.

\begin{thm}\label{k0}
Sea $A$ un operador sim{\'e}trico definido sobre un dominio denso
de un espacio de Hilbert $\mathcal{H}$, para el cual los
{\'\i}ndices de deficiencia sean iguales, $n_+=n_-=n<\infty$.
Entonces:
\begin{itemize}
\item Existen dos mapeos suryectivos
$K_1,K_2:\mathcal{D}(A^\dagger)\rightarrow \mathbb{C}^n$ que
satisfacen, para todo par de funciones
$\phi,\psi\in\mathcal{D}(A^\dagger)$,
\begin{equation}
    (\phi,A^\dagger\psi)-(A^\dagger\phi,\psi)=(K_1\phi,K_2\psi)-
    (K_2\phi,K_1\psi)\,,
\end{equation}
donde $(\cdot,\cdot)$ representa el producto usual en
$\mathcal{H}$ y en $\mathbb{C}^n$,
respectivamente\,\footnote{V{\'e}ase por ejemplo el Teorema
\ref{valbor} en el cual $K_1\phi$ y $K_2\phi$ est{\'a}n dados por
los valores de borde.}.

\item Las extensiones autoadjuntas $A^{(M,N)}$ de $A$ est{\'a}n
caracterizadas por las matrices $M,N\in\mathbb{C}^{n\times n}$
para las que $M\cdot N^\dagger$ es herm{\'\i}tica y
$(M|N)\in\mathbb{C}^{n\times 2n}$ tiene rango $n$. De acuerdo con
esta definici{\'o}n, el dominio de $A^{(M,N)}$ est{\'a} determinado por,
\begin{equation}\label{cc1}
    \mathcal{D}(A^{(M,N)})=\{\phi\in\mathcal{D}(A^\dagger):
    M K_1\phi=N K_2\phi\}\,.
\end{equation}
\end{itemize}
\end{thm}
N{\'o}tese que el conjunto de matrices $(M,N)$ que definen las
extensiones autoadjuntas de $A$ est{\'a} caracterizado por
$n^2={\rm dim}\,U(n)$ par{\'a}metros, en acuerdo con la
teor{\'\i}a de von Neumann de los {\'\i}ndices de deficiencia.

\bs

Para enunciar la f{\'o}rmula de Krein necesitamos a{\'u}n un par
de definiciones. Como se puede probar que las restricciones de
$K_1,K_2$ al subespacio ${\rm Ker}(A^\dagger-\lambda)$ son
invertibles, definimos entonces,
\begin{defn}\label{d1}
\begin{eqnarray}
    K_1^{-1}(\lambda):= \left(K_1|_{{\rm Ker}(A^\dagger-\lambda)}
    \right)^{-1}\,,\\
    K(\lambda):= -K_2\cdot K_1^{-1}(\lambda)\,.
\end{eqnarray}
\end{defn}

La f{\'o}rmula de Krein permite escribir la resolvente de una
extensi{\'o}n autoadjunta $A^{(M,N)}$ en t{\'e}rminos de la
resolvente de la extensi{\'o}n autoadjunta caracterizada por
$M=\mathbf{1}$ y $N=\mathbf{0}$.
\begin{thm}
\begin{equation}\label{krein0}
    \left(A^{(M,N)}-\lambda\right)^{-1}=
    \left(A^{(\mathbf{1},\mathbf{0})}-\lambda\right)^{-1}+
    K_1^{-1}(\lambda)\cdot\frac{N}{
    (M+N\,K(\lambda))}
    \cdot\left(K_1^{-1}(\lambda^*)\right)^\dagger\,.
\end{equation}
\end{thm}

A modo de ilustraci{\'o}n, aplicaremos la f{\'o}rmula de Krein a
un operador diferencial regular de la forma,
\begin{equation}\label{kreinope}
    A=-\partial_x^2+U(x)\,,
\end{equation}
definido sobre un subespacio de $\mathbf{L_2}(\mathbb{R}^+)$ para
el cual los {\'\i}ndices de deficiencia de $A$ sean $n_\pm=1$. En
primer lugar, definimos los operadores $K_1,K_2$ referidos en el
primer enunciado del Teorema \ref{k0},
\begin{eqnarray}
    K_1\phi(x):=\phi(0)\,,\\
    K_2\phi(x):=\phi'(0)\,.
\end{eqnarray}
De acuerdo con el segundo enunciado del Teorema \ref{k0}, el
dominio $\mathcal{D}(A^\theta)$ de las extensiones autoadjuntas
est{\'a} caracterizado por un par{\'a}metro real $\theta$,
\begin{equation}\label{cc}
    \mathcal{D}(A^\theta)=\{\phi\in\mathcal{D}(A^\dagger):
    \phi'(0)-\theta\,\phi(0)=0\}\,.
\end{equation}
La condici{\'o}n de contorno de Robin de la ecuaci{\'o}n
(\ref{cc}) se obtiene observando, en primer lugar, que las
matrices $M,N$ del Teorema \ref{k0} son n{\'u}meros reales y
definiendo, luego, en la expresi{\'o}n (\ref{cc1}) el
par{\'a}metro $\theta:= M^{-1}\,N$. La extensi{\'o}n $M=0$, que
corresponde a condiciones de contorno tipo Dirichlet, est'a
caracterizada por $\theta=\infty$.

\bs

La f{\'o}rmula de Krein puede escribirse, en este caso, de la
siguiente manera,
\begin{equation}\label{krein}
    \left(A^{\theta}-\lambda\right)^{-1}-
    \left(A^{\infty}-\lambda\right)^{-1}=
    \frac{\left(A^{0}-\lambda\right)^{-1}-
    \left(A^{\infty}-\lambda\right)^{-1}}{1+\theta\, K(\lambda)}\,.
\end{equation}
Determinemos ahora el factor $K(\lambda)$ de acuerdo con la
definici{\'o}n \ref{d1}.

\bs

Como los {\'\i}ndices de deficiencia del operador $A$ son
$n_\pm=1$, el subespacio de deficiencia ${\rm
Ker}(A^\dagger-\lambda)$ est{\'a} generado por una funci{\'o}n
normalizada que denotamos por $\phi_\lambda$. Esto es,
\begin{equation}
    (A^\dagger-\lambda)\phi_\lambda=0\,.
\end{equation}
En consecuencia, de acuerdo con la definici{\'o}n \ref{d1},
\begin{eqnarray}
    K_1^{-1}(\lambda):\mathbb{C}\rightarrow {\rm Ker}(A^\dagger-\lambda)\,,\\
    K_1^{-1}(\lambda)\cdot c=c\cdot \phi_\lambda(x)/\phi_\lambda(0)\,;
\end{eqnarray}
Por lo tanto, el operador $K(\lambda)$ est{\'a} dado por,
\begin{equation}
    K(\lambda)\cdot c=-c\cdot \phi_\lambda'(0)/\phi_\lambda(0)\,,
\end{equation}
esto es,
\begin{equation}\label{fak}
    K(\lambda)=-\frac{\phi_{\lambda}'(0)}{\phi_\lambda(0)}\,.
\end{equation}

En la secci{\'o}n siguiente determinaremos una f{\'o}rmula
an{\'a}loga a la de Krein que relaciona las resolventes de las
distintas extensiones autoadjuntas del operador (\ref{kreinope})
en el caso en el que el potencial $U(x)$ posee una singularidad en
el origen. Obtendremos una ecuaci{\'o}n similar a (\ref{krein})
(v{\'e}ase la ecuaci{\'o}n (\ref{elthm})) en la que el factor
$K(\lambda)$ no co\-rres\-pon\-de al valor dado por (\ref{fak})
pero est{\'a} tambi{\'e}n relacionado con el comportamiento de las
funciones de $\mathcal{D}(A^\dagger)$ en el origen.

\bs

En los ejemplos que trataremos en el cap{\'\i}tulo \ref{apli},
veremos que en el l{\'\i}mite en el que el t{\'e}rmino singular
del operador tiende a creo, el facto $K(\lambda)$ que obtendremos
en la secci{\'o}n siguiente se reduce al dado por la ecuaci{\'o}n
(\ref{fak}).

\section{F{\'o}rmula de Krein para operadores singulares}\label{singu}

En esta secci{\'o}n comenzamos nuestro estudio de los operadores
diferenciales con coeficientes singulares. Algunos de los
resultados que aqu\'\i\ presentamos fueron obtenidos en base a la
t{\'e}cnica utilizada por E.A.\ \mbox{Mooers} \cite{Mooers} en su
estudio del heat-kernel del laplaciano sobre variedades con
singularidades c{\'o}nicas.

\bs

Consideraremos, en particular, un operador di\-fe\-ren\-cial $A$
en una dimensi{\'o}n dado por (\ref{sch}) cuyo potencial
$U_\nu(x)$ tiene un comportamiento singular en el origen dado por
la expresi{\'o}n (\ref{mod}).

\bs

En primer lugar, caracterizaremos las extensiones autoadjuntas del
operador di\-fe\-ren\-cial $A$. Posteriormente, determinaremos una
relaci{\'o}n entre las resolventes $(A-\lambda)^{-1}$
co\-rres\-pon\-dien\-tes a distintas extensiones autoadjuntas.

\bs

Veremos que si el operador se define sobre la variedad de base
compacta $[0,1]\subset\mathbb{R}$ y se imponen condiciones de
contorno locales, las extensiones autoadjuntas de $A$ resultan
caracterizadas por dos par{\'a}metros reales; uno de ellos
des\-cri\-be el comportamiento de las funciones en $x=0$, el otro
en $x=1$. Sin embargo, debemos se{\~n}alar que los c{\'a}lculos de
esta secci{\'o}n pueden repetirse, an{\'a}logamente, si el
operador se define sobre la variedad de base no compacta
$\mathbb{R}^+$. En este caso, las extensiones autoadjuntas
est{\'a}n caracterizadas por un {\'u}nico par{\'a}metro real que
describe la condici{\'o}n de contorno en $x=0$.

\bs

Consecuentemente, utilizaremos los resultados de las Secciones
\ref{sec1} y \ref{sec2} cuando estudiemos el operador $A$ sobre la
variedad de base no compacta $\mathbb{R}^+$.

\subsection{Extensiones autoadjuntas}\label{sec1}

Consideremos el operador,
\begin{equation}\label{1}
    A=-\partial_x^2+\frac{\nu^2-1/4}{x^2}+V(x)\,,
\end{equation}
definido sobre el subespacio denso
$\mathcal{C}^\infty_0((0,1))\subset\mathbf{L}_2([0,1])$.

\bs

Supondremos que $\nu\in(0,1)$ y que $V(x)$ es una funci{\'o}n
anal{\'\i}tica en el intervalo $[0,1]\subset\mathbb{R}$. El
siguiente Teorema describe el comportamiento de las funciones de
$\mathcal{D}(A^\dagger)$ en las proximidades de la singularidad.

{\thm Si $\psi\in\mathcal{D}(A^{\dagger})$ entonces existe una
constante $\theta_{\psi}\in\mathbb{R}$ tal que el comportamiento
de $\psi$ en las proximidades de $x=0$ est{\'a} dado por,
\begin{equation}
    \psi(x)=C[\psi]\,\left(
    x^{-\nu+1/2}+
    \theta_{\psi}\,x^{\nu+1/2}\right)+O(x^{3/2})\,,\qquad
    C[\psi]\in\mathbb{C}\,.
\end{equation}}\label{comenelori}

{\noindent\bf Demostraci{\'o}n:} Por el Lema de representaci{\'o}n
de Riesz \cite{Riesz},
\begin{equation}
\psi\in\mathcal{D}(A^{\dagger})\rightarrow
\exists\,\tilde{\psi}\in\mathbf{L}_2([0,1])\,:(\psi,A\phi)=(\tilde{\psi},\phi)\quad
\forall \phi\in\mathcal{D}(A)\,.
\end{equation}
Adem{\'a}s,
\begin{equation}
A^{\dagger}\psi:=\tilde{\psi}\,.
\end{equation}
Esta expresi{\'o}n puede escribirse en t{\'e}rminos de $\chi:=
x^{-\nu-1/2}\psi$ de la siguiente manera,
\begin{equation}
    \partial_x(x^{2\nu+1}\partial_x\chi)=-x^{\nu+1/2}(\tilde{\psi}-V(x)\psi)
    \in\mathbf{L}_1([0,1])\,.
\end{equation}
En consecuencia, existe una constante $C_1\in\mathbb{R}$ tal que,
\begin{equation}
    \partial_x\chi=C_1 x^{-1-2\nu}-x^{-1-2\nu}\int_0^x
    y^{\nu+1/2}\left(-\partial_y^2+\frac{\nu^2-1/4}{y^2}\right)\psi\,dy\,.
\end{equation}
Por medio de la desigualdad de Cauchy-Schwartz se puede demostrar
que,
\begin{equation}
    \left|\,x^{-1-2\nu}\int_0^x
    y^{\nu+1/2}\left(-\partial_y^2+\frac{\nu^2-1/4}{y^2}\right)\psi\,dy\,\right|\leq
    C_2\,\|\left[-\partial_y^2+\frac{\nu^2-1/4}{y^2}\right]\psi\|_{(0,x)}\,x^{-\nu}\,,
\end{equation}
para alguna constante $C_2\in\mathbb{R}$. Por consiguiente,
\begin{equation}
    \left|\,\int^x z^{-1-2\nu}\int_0^z
    y^{\nu+1/2}\left(-\partial_x^2+\frac{\nu^2-1/4}{x^2}\right)\psi\,dy\,dz\,\right|\leq
    C_3+C_4\,x^{1-\nu}\,,
\end{equation}
donde $C_3,C_4\in\mathbb{R}$. Se verifica entonces que, para
algunas constantes $C_5,C_6\in\mathbb{R}$, el comportamiento de
$\psi$ en el origen est{\'a} dado por,
\begin{equation}
    \psi=C_5\,x^{-\nu+1/2}+C_6\,x^{\nu+1/2}+O(x^{3/2})\,.
\end{equation}
De esta {\'u}ltima expresi{\'o}n obtenemos el resultado del
Teorema.\begin{flushright}$\Box$\end{flushright}

{\cor \label{Corooo}
\begin{equation}\label{corolario}
    \phi,\psi\in\mathcal{D}(A^{\dagger})\rightarrow
    (\phi,A^{\dagger}\psi)-(A^{\dagger}\phi,\psi)=C[\phi^*]C[\psi]\left(
    \theta_{\phi}-\theta_{\psi}\right)+\left(\partial_x\phi^*\ \psi-
    \phi^*\ \partial_x\psi
    \right)|_{x=1}\,.
\end{equation}}

\medskip

{\noindent\bf Demostraci{\'o}n:} La expresi{\'o}n
(\ref{corolario}) se obtiene directamente mediante una
integraci{\'o}n por partes y utilizando el Teorema
\ref{comenelori}. N{\'o}tese que este Corolario verifica el primer
e\-nun\-cia\-do del Teorema
\ref{k0}.\begin{flushright}$\Box$\end{flushright}

Como consecuencia del Corolario \ref{Corooo}, el operador $A$
admite una familia de extensiones autoadjuntas $\mathcal{M}$
isomorfa a $U(2)$. Como estamos interesados en condiciones de
contorno locales, el conjunto de extensiones autoadjuntas se
reduce a una subvariedad isomorfa a $U(1)\otimes U(1)$.

\bs

Cada extensi{\'o}n $A^{\theta}_{\beta}\in\mathcal{M}$,
caracterizada por los par{\'a}metros reales $\theta,\beta$,
est{\'a} definida sobre el conjunto,
\begin{equation}\label{saesing}
    \mathcal{D}(A^{\theta}_{\beta})=
    \left\{\phi\in\mathbf{L}_2([0,1]):\theta_{\phi}=
    \theta\,,\ \left(\partial_x\phi-\beta\,\phi\right)|_{x=1}=0\right\}\,.
\end{equation}
La cantidad $\theta_\phi$ est{\'a} definida de acuerdo con el
Teorema \ref{comenelori}. Por lo tanto, los par{\'a}metros
$\theta,\beta$ determinan las condiciones de contorno.

\bs

Existe adem{\'a}s una extensi{\'o}n autoadjunta, que denotaremos
por $A^{\infty}_{\beta}$, cuyo dominio es el conjunto,
\begin{eqnarray}\label{beta}
    \mathcal{D}(A^{\infty}_{\beta})=
    \left\{\phi\in\mathbf{L}_2([0,1]):\right.\nonumber\\\left.
    \phi(x)=C[\phi]\,x^{\nu+1/2}+O(x^{3/2})\,,\ {\rm con\ }C[\phi]\in
    \mathbb{C}\,,\
    \left(\partial_x\phi-\beta\,\phi\right)|_{x=1}=0 \right\}\,.
\end{eqnarray}

N{\'o}tese que, en el l{\'\i}mite regular $\nu\rightarrow 1/2$, el
par{\'a}metro $\theta$ que caracteriza el comportamiento en el
origen de las funciones del dominio de la extensi{\'o}n
autoadjunta $A^\theta_\beta$ coincide con el par{\'a}metro
$\theta$ que caracteriza las condiciones de contorno Robin dadas
por (\ref{cc}) para un operador regular en el origen.

\bs

An{\'a}logamente, se puede probar que el operador $A$ (v{\'e}ase
la ecuaci{\'o}n (\ref{1})) definido sobre
$\mathcal{C}_0^\infty(\mathbb{R}^+)$ satisface el Teorema
\ref{comenelori} y admite una familia de extensiones autoadjuntas
$A^\theta$, caracterizadas por un par{\'a}metro real $\theta$,
cuyos dominios de definici{\'o}n est{\'a}n dados por,
\begin{equation}\label{saesingnocomp}
    \mathcal{D}(A^{\theta})=
    \left\{\phi\in\mathbf{L}_2(\mathbb{R}^+):\theta_{\phi}=
    \theta\right\}\,.
\end{equation}
Existe adem{\'a}s una otra extensi{\'o}n autoadjunta, que
denotaremos por $A^{\infty}$, cuyo dominio es el conjunto,
\begin{eqnarray}\label{betanocomp}
    \mathcal{D}(A^{\infty})=
    \left\{\phi\in\mathbf{L}_2(\mathbb{R}^+):
    \phi(x)=C[\phi]\,x^{\nu+1/2}+O(x^{3/2})\,,\ {\rm con\ }
    C[\phi]\in\mathbb{C}\right\}\,.
\end{eqnarray}

\subsection{Relaci{\'o}n entre las distintas resolventes.}\label{sec2}

De acuerdo con la secci{\'o}n anterior, el operador diferencial
$A$ dado por la ecuaci{\'o}n (\ref{1}) definido sobre
$\mathcal{C}_0^\infty((0,1))$ admite una familia de extensiones
autoadjuntas caracterizadas por dos par{\'a}metros $\theta,\beta$.
El par{\'a}metro $\theta$ describe el comportamiento de las
funciones en las proximidades del punto $x=0$ en tanto que $\beta$
define condiciones de contorno tipo Robin en $x=1$.

\bs

El objetivo de esta secci{\'o}n es establecer una relaci{\'o}n
entre las resolventes de las extensiones autoadjuntas
$A^{\theta}_{\beta}$ correspondientes a distintos valores de
$\theta$ y a un mismo valor de $\beta$. Por consiguiente,
omitiremos, en adelante, el sub{\'\i}ndice $\beta$.

\bs

En este sentido, obtendremos una f{\'o}rmula similar a la de Krein
presentada en la secci{\'o}n \ref{regu} (v{\'e}ase la ecuaci{\'o}n
(\ref{krein}).) En la secci{\'o}n siguiente mostraremos que el
desarrollo asint{\'o}tico de la traza de la resolvente del
operador con coeficientes singulares $A$ no res\-pon\-de, en
general, al comportamiento indicado por (\ref{teordesa}) sino que
presenta potencias de $\lambda$ dependientes de $\nu$.

\bs

No obstante, la presencia de estas potencias depende fuertemente
de la variedad de condiciones de contorno admisibles en la
singularidad. En efecto, la resolvente de la extensi{\'o}n de
Friedrichs ($\theta=\infty$), que es una de las condiciones de
contorno invariantes de escala, s\'\i\ verifica el comportamiento
(\ref{teordesa}). Por este motivo, ser{\'a} {\'u}til obtener una
expresi{\'o}n para la resolvente de una extensi{\'o}n autoadjunta
general en t{\'e}rminos de la resolvente correspondiente a
$\theta=\infty$. De esta manera, podremos identificar el origen de
la dependencia de los exponentes de $\lambda$ con el par{\'a}metro
$\nu$ en el desarrollo asint{\'o}tico de la resolvente.

\bs

Comenzamos por establecer, sin demostraci{\'o}n, la existencia y
unicidad de la resolvente de una extensi{\'o}n autoadjunta.

\begin{thm}\label{uni}
Para toda funci{\'o}n $f(x)\in\mathbf{L_2}([0,1])$ y $\lambda$ no
perteneciente al espectro de $A^\theta$ existe una {\'u}nica
$\phi^{\theta}(x,\lambda)\in\mathcal{D}(A^{\theta})$ tal que,
\begin{equation}\label{pro}
    (A^{\theta}-\lambda)\phi^{\theta}(x,\lambda)
    =f(x)\,.
\end{equation}
Adem{\'a}s,
\begin{equation}\label{inter}
    \phi^{\theta}(x,\lambda)=
    \int_0^{1}G_{\theta}(x,x',\lambda)f(x')\,dx'\,,
\end{equation}
siendo $G_{\theta}(x,x',\lambda)$ el n{\'u}cleo de la resolvente
$(A^{\theta}-\lambda)^{-1}$.
\end{thm}

\bigskip

El n{\'u}cleo $G_{\theta}(x,x',\lambda)$ de la resolvente admite
una expresi{\'o}n dada por la ecuaci{\'o}n (\ref{solres}). El
comportamiento en el origen de la funci{\'o}n $L(x,\lambda)$ que
figura en esta expresi{\'o}n est{\'a} dado por,
\begin{equation}
    L(x,\lambda)=x^{\nu+1/2}+O(x^{3/2})\,,
\end{equation}
para el caso $\theta=\infty$ (v{\'e}ase la ecuaci{\'o}n
(\ref{beta})), y por,
\begin{equation}
    L(x,\lambda)=x^{-\nu+1/2}+O(x^{3/2})\,,
\end{equation}
para el caso $\theta=0$ (v{\'e}ase la ecuaci{\'o}n
(\ref{saesing}).) Tiene sentido entonces, establecer la
si\-guien\-te definici{\'o}n:
\begin{defn}
\begin{eqnarray}
    G_{\infty}(x',\lambda):=\lim_{x\rightarrow
    0}x^{-\nu-1/2}\,G_{\infty}(x,x',\lambda)\,,\label{Rxp}\\
    G_{0}(x',\lambda):=\lim_{x\rightarrow
    0}x^{\nu-1/2}\,G_{0}(x,x',\lambda)\,.
\end{eqnarray}
\end{defn}

Los n{\'u}cleos $G_{\infty}(x,\lambda),G_{0}(x,\lambda)$ permiten
obtener el comportamiento en el origen, para los casos
$\theta=\infty,0$, respectivamente, de las soluciones del problema
(\ref{pro}) en t{\'e}rminos de la inhomogenidad $f(x)$. En efecto,
no es dif{\'\i}cil probar que,
\begin{eqnarray}
    \phi^{\infty}(x,\lambda)=
    \int_0^{1}G_{\infty}(x,x',\lambda)f(x')\,dx'=
    \phi^{\infty}(\lambda)\,x^{\nu+1/2}+O(x^{3/2})\,,\label{fii}\\
    \phi^{0}(x,\lambda)=
    \int_0^{1}G_{0}(x,x',\lambda)f(x')\,dx'=
    \phi^{0}(\lambda)\,x^{-\nu+1/2}+O(x^{3/2})\,,\label{fi0}
\end{eqnarray}
donde,
\begin{eqnarray}
    \phi^{\infty}(\lambda):=
    \int_0^{1}G_{\infty}(x',\lambda)f(x')\,dx'\,,\label{fiienelori}\\
    \phi^{0}(\lambda):=
    \int_0^{1}G_{0}(x',\lambda)f(x')\,dx'\,.\label{fi0enelori}
\end{eqnarray}

Los lemas \ref{lem1} y \ref{lem2}, que presentamos a
continuaci{\'o}n, establecen una primera relaci{\'o}n entre dos
soluciones al problema (\ref{pro}) con la misma inhomogeneidad
$f(x)$ pero pertenecientes al dominio de las extensiones
$\theta=\infty$ y $\theta=0$.

\begin{lem}\label{lem1}
Sea $\varphi_0(x)\in\mathcal{D}(A^{0})$ tal que
$\varphi_0(x)=x^{-\nu+1/2}+O(x^{3/2})$ cuando $x\rightarrow 0^+$.
Entonces las funciones (\ref{fii}) y (\ref{fi0}) satisfacen,
\begin{equation}\label{teo}
    \phi^{\infty}(x,\lambda)=\phi^{0}(x,\lambda)
    -\phi^{0}(\lambda)
    \left[\varphi_0(x)
    -\int_0^{1}G_{\infty}(x,x',\lambda)
    (A^0-\lambda)\varphi_0(x')\,dx'\right]\,.
\end{equation}
\end{lem}

\bigskip

\noindent{\bf Demostraci{\'o}n:} Por un lado,
\begin{equation}
    (A^\dagger-\lambda)\,\phi^\infty(x,\lambda)=f(x)\,.
\end{equation}
Asimismo,
\begin{eqnarray}
    (A^\dagger-\lambda)\,\left\{
    \phi^{0}(x,\lambda)
    -\phi^{0}(\lambda)
    \left[\varphi_0(x)
    -\int_0^{1}G_{\infty}(x,x',\lambda)
    (A^0-\lambda)\varphi_0(x')\,dx'\right]
    \right\}=\nonumber\\=
    f(x)-\phi^{0}(\lambda)\left[(A^0-\lambda)\varphi_0(x)
    -(A^0-\lambda)\varphi_0(x)\right]=f(x)\,.\nonumber\\
\end{eqnarray}
Por otra parte, para $x\rightarrow 0^+$,
\begin{equation}
    \phi^\infty(x,\lambda)=\phi^{\infty}(\lambda)\,x^{\nu+1/2}+O(x^{3/2})\,,
\end{equation}
en tanto que,
\begin{eqnarray}
    \left\{
    \phi^{0}(x,\lambda)
    -\phi^{0}(\lambda)
    \left[\varphi_0(x)
    -\int_0^{1}G_{\infty}(x,x',\lambda)
    (A^0-\lambda)\varphi_0(x')\,dx'\right]
    \right\}=\nonumber\\=
    \phi^{0}(\lambda)\,x^{-\nu+1/2}-\mbox{}
    \phi^{0}(\lambda)
    \left[x^{-\nu+1/2}
    -x^{\nu+1/2}\int_0^{1}G_{\infty}(x',\lambda)
    (A^0-\lambda)\varphi_0(x')\,dx'\right]+\nonumber\\\mbox{}
    +O(x^{3/2})=
    \phi^{0}(\lambda)
    \left[\int_0^{1}G_{\infty}(x',\lambda)
    (A^0-\lambda)\varphi_0(x')\,dx'\right]\cdot x^{\nu+1/2}
    +O(x^{3/2})
    \,.
    \nonumber\\
\end{eqnarray}
En consecuencia, ambos miembros de la expresi{\'o}n (\ref{teo})
pertenecen a $\mathcal{D}(A^\infty)$ y satisfacen la ecuaci{\'o}n
(\ref{pro}) para $\theta=\infty$. La igualdad (\ref{teo}) queda
entonces demostrada en virtud de la unicidad establecida en el
Teorema \ref{uni}.\begin{flushright}$\Box$\end{flushright}
\begin{lem}\label{lem2}
\begin{equation}
    \varphi_0(x)-\int_0^{1}G_{\infty}(x,x',\lambda)(A^0-\lambda)
    \varphi_0(x')\,dx'=
    2\nu\, G_{\infty}(x,\lambda)\,.
\end{equation}
\end{lem}

\noindent{\bf Demostraci{\'o}n:} Como los n{\'u}cleos de las
resolventes $G_{\infty}(x,x',\lambda),G_{0}(x,x',\lambda)$ son
sim{\'e}tricos (v{\'e}ase la ecuaci{\'o}n (\ref{solres})) se
verifica,
\begin{equation}
    A^\dagger\,\left[G_{0}(x,x',\lambda)-G_{\infty}(x,x',\lambda)\right]=0\,,
\end{equation}
tanto si el operador $A^\dagger$ act{\'u}a sobre la variable $x$
como sobre $x'$. En consecuencia,
\begin{eqnarray}
    \varphi_0(x)-\int_0^1G_\infty(x,x',\lambda)(A^0-\lambda)\varphi_0(x')\,
    dx'=\nonumber\\=
    \int_0^1\left[G_{0}(x,x',\lambda)-G_{\infty}(x,x',\lambda)\right]
    (A^0-\lambda)\varphi_0(x')\,dx'=\nonumber\\=
    \lim_{x\rightarrow 0^+}\left\{
    \left[G_{0}(x,x',\lambda)-G_{\infty}(x,x',\lambda)\right]\cdot
    \varphi_0'(x)-\right.\nonumber\\
    \left.\mbox{}-
    \partial_{x'}\left[G_{0}(x,x',\lambda)-G_{\infty}(x,x',\lambda)\right]\cdot
    \varphi_0'(x)
    \right\}=\nonumber\\=
    \left[x^{-\nu+1/2}\,G_{0}(x,\lambda)-x^{\nu+1/2}\,G_{\infty}(x,\lambda)\right]
    \cdot(-\nu+1/2)\,x^{-\nu-1/2}-\nonumber\\
    \mbox{}-
    \left[(-\nu+1/2)\,x^{-\nu-1/2}\,G_{0}(x,\lambda)-
    (\nu+1/2)\,x^{\nu-1/2}\,G_{\infty}(x,\lambda)\right]\cdot
    x^{-\nu+1/2}=\nonumber\\=2\nu\,G_{\infty}(x,\lambda)\,.
\end{eqnarray}

\begin{flushright}$\Box$\end{flushright}
Los lemas \ref{lem1} y \ref{lem2} conducen al siguiente resultado.
\begin{lem}\label{lemcon}
\begin{equation}\label{0vsinf}
    \phi^{0}(x,\lambda)=\phi^{\infty}(x,\lambda)+2\nu\, G_{\infty}
    (x,\lambda)\phi^{0}(\lambda)\,.
\end{equation}
\end{lem}

Analizando el comportamiento en las proximidades del origen de los
t{\'e}rminos de la ecuaci{\'o}n (\ref{0vsinf}) se puede obtener el
comportamiento de $G_{\infty}(x,\lambda)$ cuando $x\rightarrow
0^+$,
\begin{equation}\label{Renelori}
    G_{\infty}(x,\lambda)=\frac{1}{2\nu}
    \left(x^{-\nu+1/2}-K(\lambda)^{-1}x^{\nu+1/2}\right)+O(x^{3/2})\,,
\end{equation}
donde
\begin{equation}\label{Renelori2}
    K(\lambda):=\frac{\phi^{0}(\lambda)}{\phi^{\infty}(\lambda)}\,.
\end{equation}

\bs

Las ecuaciones (\ref{Renelori}) y (\ref{Renelori2}) son de gran
utilidad para nuestro objetivo. El factor $K(\lambda)$ definido en
(\ref{Renelori2}) relaciona informaci{\'o}n relativa a las dos
extensiones autoadjuntas correspondientes a $\theta=\infty$ y
$\theta=0$ (comp{\'a}rese con el factor $K(\lambda)$ definido en
(\ref{fak}) para el caso regular.) La ecuaci{\'o}n
(\ref{Renelori}) permite calcular el factor $K(\lambda)$ a partir
del comportamiento en el origen del n{\'u}cleo de la resolvente de
la extensi{\'o}n $\theta=\infty$.

\bs

De este modo, a partir del Lema \ref{lemcon} podemos expresar
cantidades correspondientes a la extensi{\'o}n $\theta=0$ en
t{\'e}rminos de cantidades obtenidas para la extensi{\'o}n
$\theta=\infty$,
\begin{equation}\label{comp}
    \phi^{0}(x,\lambda)=\phi^{\infty}(x,\lambda)+2\nu K(\lambda)\,
    G_{\infty}(x,\lambda)\phi^{\infty}(\lambda)\,.
\end{equation}
Como esta ecuaci{\'o}n es v{\'a}lida para cualquier inhomogeneidad
$f(x)$, utilizando las ecuaciones (\ref{fii}), (\ref{fi0}) y
(\ref{fiienelori}), obtenemos uno de los resultados m{\'a}s
importante de esta secci{\'o}n:

\begin{thm}\label{uno}
\begin{equation}\label{R0vsRinf}
    G_{0}(x,x',\lambda)=G_{\infty}(x,x',\lambda)+2\nu K(\lambda)\,
    G_{\infty}(x,\lambda)G_{\infty}(x',\lambda)\,.
\end{equation}
\end{thm}
El Teorema \ref{uno} expresa el n{\'u}cleo de la resolvente
correspondiente a la extensi{\'o}n autoadjunta $\theta=0$ en
t{\'e}rminos del n{\'u}cleo de la resolvente de la extensi{\'o}n
$\theta=\infty$.

\bs

Resta entonces escribir una relaci{\'o}n similar que permita
calcular el n{\'u}cleo de la resolvente para una extensi{\'o}n
autoadjunta general en funci{\'o}n del n{\'u}cleo de la resolvente
de la extensi{\'o}n $\theta=\infty$. Una extensi{\'o}n conveniente
de la ecuaci{\'o}n (\ref{comp}) conduce al siguiente Lema.
\begin{lem}\label{rem}
\begin{equation}\label{fithe}
    \phi^{\theta}(x,\lambda)=\phi^{\infty}(x,\lambda)+2\nu
 \left(K(\lambda)^{-1}+\theta\right)^{-1}G_{\infty}(x,\lambda)
 \phi^{\infty}(\lambda)\,.
\end{equation}
\end{lem}

\noindent{\bf Demostraci{\'o}n:} La ecuaci{\'o}n (\ref{0vsinf})
permite probar que la diferencia entre ambos miembros de la
expresi{\'o}n (\ref{fithe}) pertenece a ${\rm
Ker}(A^\dagger-\lambda)$.

Por otra parte, ambos miembros de la expresi{\'o}n (\ref{fithe})
pertenecen a $\mathcal{D}(A^\theta)$ puesto que el comportamiento
para $x\rightarrow 0^+$ del segundo miembro est{\'a} dado por
(v{\'e}anse las ecuaciones (\ref{fii}), (\ref{Renelori}) y
(\ref{Renelori2})),
\begin{equation}
    \frac{\phi^0(\lambda)\phi^\infty(\lambda)}{\phi^\infty(\lambda)+
    \theta\,\phi^0(\lambda)}\,
    \left(x^{-\nu+1/2}+\theta\,x^{\nu+1/2}\right)+O(x^{3/2})\,.
\end{equation}
Una vez m{\'a}s, la unicidad establecida en el Teorema \ref{uni}
nos conduce a la igualdad
(\ref{fithe}).\begin{flushright}$\Box$\end{flushright}

El Lema \ref{rem}, junto con las ecuaciones (\ref{inter}),
(\ref{fii}) y (\ref{fiienelori}), permiten demostrar el siguiente
Teorema que expresa el n{\'u}cleo de la resolvente de una
extensi{\'o}n autoadjunta general en t{\'e}rminos del n\'ucleo de
la extensi{\'o}n $\theta=\infty$.

{\thm
\begin{equation}\label{RthevsRinf}
    G_{\theta}(x,x',\lambda)=G_{\infty}(x,x',\lambda)+2\nu
    \left(K(\lambda)^{-1}+\theta\right)^{-1}G_{\infty}(x,\lambda)
    G_{\infty}(x',\lambda)\,.
\end{equation}}

Ser{\'a} conveniente resumir el resultado de las ecuaciones
(\ref{R0vsRinf}) y (\ref{RthevsRinf}) en la siguiente
expresi{\'o}n,
\begin{equation}\label{mira}
 G_{\theta}(x,x',\lambda)-G_{\infty}(x,x',\lambda)=
 \frac{G_{0}(x,x',\lambda)-G_{\infty}(x,x',
\lambda)}
    {1+\theta\, K(\lambda)}\,,
\end{equation}
que puede escribirse en t{\'e}rminos de los correspondientes
operadores,
\begin{equation}\label{kreinsing}
    \left(A^{\theta}-\lambda\right)^{-1}-
    \left(A^{\infty}-\lambda\right)^{-1}=
    \frac{\left(A^{0}-\lambda\right)^{-1}-
    \left(A^{\infty}-\lambda\right)^{-1}}{1+\theta\, K(\lambda)}\,.
\end{equation}
Esta expresi{\'o}n coincide formalmente con la f{\'o}rmula de
Krein (\ref{krein}) v{\'a}lida para o\-pe\-ra\-do\-res regulares,
pero el factor $K(\lambda)$ en ambas expresiones es distinto. Para
el caso regular est{\'a} dado por la ecuaci{\'o}n (\ref{fak}) en
tanto que para el caso singular est{\'a} dado por la ecuaci{\'o}n
(\ref{Renelori2}).

\bs

Establecemos finalmente el siguiente Teorema que pemitir{\'a}
estudiar el desarrollo asint{\'o}tico de la traza de la resolvente
de una extensi{\'o}n autoadjunta general.
\begin{thm}\label{elthmenun}
\begin{eqnarray}\label{elthm}
\begin{picture}(300,40)(25,10)
    \put(-5,0){\line(0,1){45}}
    \put(370,0){\line(0,1){45}}
    \put(-5,0){\line(1,0){375}}
    \put(-5,45){\line(1,0){375}}
    \put(15,18){$\displaystyle{{\rm Tr}\left\{(A^{\theta}-\lambda)^{-1}-
    (A^{\infty}-\lambda)^{-1}\right\}
    =\frac{{\rm Tr}\left\{(A^{0}-\lambda)^{-1}
    -(A^{\infty}-\lambda)^{-1}\right\}}
    {1+\theta\, K(\lambda)}}\,.$}
\end{picture}\nonumber\\\nonumber \\
\end{eqnarray}
\end{thm}
Esta ecuaci{\'o}n permitir{\'a} demostrar que el desarrollo
asint{\'o}ntico para grandes valores de $|\lambda|$ de la traza
${\rm Tr}\left\{(A^{\theta}-\lambda)^{-1}\right\}$ presenta
potencias de $\lambda$ cuyos exponentes dependen del par{\'a}metro
$\nu$. En efecto, mostraremos, en primer lugar, que las trazas de
los operadores pseudodiferenciales $(A^{0}-\lambda)^{-1}$ y
$(A^{\infty}-\lambda)^{-1}$, correspondientes a condiciones de
contorno sobre la singularidad invariantes de escala, admiten un
desarrollo asint{\'o}tico en potencias semienteras negativas de
$\lambda$. Veremos luego que, por el contrario, el desarrollo
asint{\'o}tico del factor $K(\lambda)$ presenta potencias de
$\lambda$ dependientes del par{\'a}metro $\nu$.

\bs

Se{\~n}alemos finalmente que el Teorema \ref{elthmenun} puede
demostrarse an{\'a}logamente para el operador dado por la
ecuaci{\'o}n (\ref{1}) pero definido sobre la variedad de base no
compacta $\mathbb{R}^+$. En ese caso las extensiones est{\'a}n
caracterizadas por un {\'u}nico par{\'a}metro $\theta$.

\section{ Desarrollo asint{\'o}tico de la resolvente}\label{luego}

En las Secciones \ref{nocom} y \ref{cccp} estudiaremos el
desarrollo asint{\'o}tico para grandes va\-lo\-res de $|\lambda|$
de las cantidades involucradas en la expresi{\'o}n (\ref{elthm})
para una variedad de base no compacta y compacta, respectivamente.

\subsection{Caso no compacto}\label{nocom}

Aplicaremos ahora los resultados de la secci{\'o}n anterior para
obtener un desarrollo asint{\'o}tico de la traza de la resolvente
del operador,
\begin{equation}\label{opea}
    A=-\partial^2_x+\frac{\nu^2-1/4}{x^2}+V(x)\,,
\end{equation}
donde $x\in\mathbb{R}^+$ y el potencial $V(x)$ es anal{\'\i}tico
en $x$ e inferiormente acotado. Como la variedad de base es, en
este caso, no compacta, es conveniente estudiar la resolvente para
valores de $\lambda$ en el semieje real negativo del plano
complejo. Consideraremos entonces las soluciones $\psi$ de la
ecuaci{\'o}n,
\begin{equation}\label{asiecu}
    (A+z)\psi=0\,,
\end{equation}
mediante un desarrollo de $\psi$ para grandes valores de
$z\in\mathbb{R}^+$. De acuerdo con el Teorema \ref{elthmenun}
ser{\'a} suficiente con encontrar una soluci{\'o}n que satisfaga
las condiciones de contorno correspondientes a $\theta=\infty$.

\bs

En virtud de la similitud ante transformaciones de escala de los
dos primeros t{\'e}rminos del operador (\ref{opea}) es conveniente
definir una nueva variable $y:=\sqrt{z}\,x\in\mathbb{R}^+$. De
este modo, la soluci{\'o}n de la ecuaci{\'o}n (\ref{asiecu}) puede
escribirse,
\begin{equation}
    \psi=\psi(\sqrt{z}x,z)\,,
\end{equation}
siendo $\psi(\sqrt{z}x,z)$ una soluci{\'o}n de,
\begin{equation}\label{asiecucamvar}
    \left(-\partial^2_y+\frac{\nu^2-1/4}{y^2}+1+\frac 1 z
    \,V(y/\sqrt{z})\right)\psi(y,z)=0\,.
\end{equation}

Nuestra t{\'e}cnica consiste en proponer un desarrollo
asint{\'o}tico para grandes valores de $z$ para la soluci{\'o}n de
(\ref{asiecucamvar}),
\begin{equation}
    \psi(y,z)=\phi(y)+\sum_{n=0}^{\infty}\phi_n(y)z^{-1-n/2}\,,
\end{equation}
consistente con el desarrollo en serie de potencias para el
potencial,
\begin{equation}\label{pottay}
    V(x)=\sum_{n=0}^{\infty}V_nx^{n}\,,
\end{equation}
siendo $V_n:= V^{(n)}(0)/n!$.

Si reemplazamos estos desarrollos en la ecuaci{\'o}n
(\ref{asiecucamvar}) obtenemos el siguiente sistema de ecuaciones
diferenciales,
\begin{eqnarray}
    \left(-\partial^2_y+\frac{\kappa}{y^2}+1\right)\phi(y)=0\,,\label{green}\\
    \left(-\partial^2_y+\frac{\kappa}{y^2}+1\right)\phi_0(y)=-V_0\phi(y)\,,\\
    \left(-\partial^2_y+\frac{\kappa}{y^2}+1\right)\phi_1(y)=-V_1y\phi(y)\,,\\
     \left(-\partial^2_y+\frac{\kappa}{y^2}+1\right)\phi_2(y)=
     -V_2y^2\phi(y)-V_0\phi_0
    (y)\,,\\
    \ldots\nonumber\\
    \left(-\partial^2_y+\frac{\kappa}{y^2}+1\right)\phi_n(y)=-V_n y^n\phi(y)-
    \sum_{h+k+2=n}\!\!\!\!V_h y^h\phi_k(y)\,.\label{resto}
\end{eqnarray}
La soluci{\'o}n $\phi(y)$ de la primera de estas ecuaciones es una
combinaci{\'o}n lineal de las funciones de Bessel
$\sqrt{y}I_{\nu}(y)$ y $\sqrt{y}K_{\nu}(y)$, de modo que la
soluci{\'o}n de la ecuaci{\'o}n (\ref{asiecucamvar}) de cuadrado
integrable en $y\rightarrow \infty$ est{\'a} dada por,
\begin{equation}\label{tam2}
    R(y,z)=\sqrt{y}K_\nu(y)+\sum_{n=0}^{\infty}\phi_n(y)z^{-1-n/2}.
\end{equation}

Por su parte, las soluciones de las ecuaciones de la forma
(\ref{resto}) pueden escribirse como,
\begin{equation}\label{phin}
    \phi_n(y)=\int_0^{\infty}G_{\infty}^{V=0}(y,y',1)
    \left[-V_n y'^n\phi(y')-
    \sum_{h+k+2=n}\!\!\!\!V_h y'^h\phi_k(y')\right]\,dy'\,,
\end{equation}
siendo,
\begin{eqnarray}\label{gredir}
    G_{\infty}^{V= 0}(y,y',1)= \sqrt{yy'}\left[
    \theta(y'-y)I_{\nu}(y)K_{\nu}(y')+
    \theta(y-y')I_{\nu}(y')K_{\nu}(y)\right]\,,\nonumber\\
\end{eqnarray}
la funci{\'o}n de Green del operador de la ecuaci{\'o}n
(\ref{green}) (v{\'e}ase la ecuaci{\'o}n (\ref{solres}).)
Reemplazando (\ref{gredir}) en (\ref{phin}) obtenemos una
f{\'o}rmula de recurrencia para $\phi_n(y)$,
\begin{eqnarray}\label{fin}
    \phi_n(y)=\nonumber\\
    -y^{1/2}K_\nu(y)\int_{0}^{y}\left[V_ny'^{(n+1/2)}K_{\nu}(y')+
    \!\!\!\!\sum_{l+m=n-2}\!\!\!\!
    V_{l}y'^l\phi_{m}(y')\right]\sqrt{y'}I_\nu(y')\,dy' -\nonumber\\
    -y^{1/2}I_\nu(y)\int_{y}^{\infty}\left[V_ny'^{(n+1/2)}K_{\nu}(y')+
    \!\!\!\!\sum_{l+m=n-2}\!\!\!\!
    V_{l}y'^l\phi_{m}(y')\right]\sqrt{y'}K_\nu(y')\,dy'\,.\nonumber\\
\end{eqnarray}
A partir de esta ecuaci{\'o}n podemos obtener el comportamiento de
$\phi_n(y)$ para $y\rightarrow 0^+$,
\begin{eqnarray}
    \phi_n(y\sim 0)= -\frac{y^{\nu+1/2}}{2^{\nu}\Gamma(\nu+1)}\times
    \nonumber\\
    \times\int_{0}^{\infty}\left[V_ny'^{(n+1/2)}K_{\nu}(y')+\!\!\!\!
    \sum_{l+m=n-2}\!\!\!\!
    V_{l}y'^l\phi_{m}(y')\right]\sqrt{y'}K_\nu(y')\,dy'+\ldots
\end{eqnarray}

Podemos calcular entonces el comportamiento en $y\rightarrow 0^+$
de la soluci{\'o}n $R(y,z)$ dada por la ecuaci{\'o}n (\ref{tam2}),
\begin{equation}\label{venelori}
    R(y,z)=\frac{\Gamma(\nu)}{2^{1-\nu}}\ y^{-\nu+1/2}+
    \frac{\Gamma(-\nu)}{2^{1+\nu}}\
    H(z)\cdot y^{\nu+1/2}
    +\ldots
\end{equation}
siendo,
\begin{eqnarray}\label{hache}
    H(z):= 1+\frac{2\sin(\pi \nu)}{\pi}
    \times\nonumber\\
    \times\sum_{n=0}^{\infty}
    z^{-1-n/2}
    \int_{0}^{\infty}\left[V_ny^{n+1/2}K_{\nu}(y)+\!\!\!\!\sum_{l+m=n-2}\!\!\!\!
    V_{l}y^l\phi_{m}(y)\right]\sqrt{y}K_\nu(y)\,dy\,.\nonumber\\
\end{eqnarray}
Es importante observar que el desarrollo de la funci{\'o}n $H(z)$
para grandes valores de $z$ s{\'o}lo presenta potencias
semienteras de $z$. Mostraremos a continuaci{\'o}n que el factor
$K(z)$ en la ecuaci{\'o}n (\ref{elthm}) puede expresarse en
t{\'e}rminos de la funci{\'o}n $H(z)$.

\bs

Para ello, primeramente advertimos que el n{\'u}cleo de la
resolvente $G_{\infty}(x,x',z)$, para $x<x'$, est{\'a} dado por
(v{\'e}ase la ecuaci{\'o}n (\ref{solres})),
\begin{equation}
    G_{\infty}(x,x',z)=\left.-\frac{z^{-1/2}}{W[L,R](z)}L(y,z)R(y',z)
    \right|_{y=\sqrt{z}x,y'=\sqrt{z}x'}\,,
\end{equation}
siendo $L(y,z)$ una soluci{\'o}n de la ecuaci{\'o}n
(\ref{asiecucamvar}) cuyo comportamiento en el origen es
proporcional a $y^{\nu+1/2}$. $W[L,R](z)$ es el wronskiano de
$L(y,z)$ y $R(y,z)$. En consecuencia, en virtud de la
definici{\'o}n (\ref{Rxp}),
\begin{equation}\label{307}
    G_{\infty}(x',z)=\left.-\frac{z^{-1/2}y^{-\nu-1/2}}{W[L,R](z)}L(y,z)R(y',z)
    \right|_{y=0,y'=\sqrt{z}x'}\,.
\end{equation}
Reemplazando (\ref{venelori}) en esta {\'u}ltima ecuaci{\'o}n
obtenemos el comportamiento de $G_{\infty}(x,z)$ para
$x\rightarrow 0^+$,
\begin{eqnarray}\label{re2}
    G_{\infty}(x\sim 0,z)=-\frac{\left.z^{-1/2}y^{-\nu-1/2}L(y,z)\right|_{y=0}}
    {W[L,R](z)}\times\nonumber\\\times\,
    \left[\frac{\Gamma(\nu)}{2^{1-\nu}}(\sqrt{z}x)^{-\nu+1/2}+
    \frac{\Gamma(-\nu)}{2^{1+\nu}}\
    H(z)\cdot (\sqrt{z}x)^{\nu+1/2}\right]
    +\ldots
\end{eqnarray}
Comparando las ecuaciones (\ref{Renelori}) y (\ref{re2}) obtenemos
la relaci{\'o}n entre el factor $K(z)$ y la funci{\'o}n $H(z)$,
\begin{equation}\label{kyh}
    K(z)=4^{\nu}\frac{\Gamma(1+\nu)}{\Gamma(1-\nu)}\;z^{-\nu}H(z)^{-1}\,.
\end{equation}
Como $H(z)$ admite un desarrollo asint{\'o}tico en potencias
semienteras de $z$, vemos que $K(z)$ admite un desarrollo
asint{\'o}tico que presenta potencias de $z$ dependientes del
pa\-r{\'a}\-me\-tro $\nu$. Estas potencias est{\'a}n presentes
tambi{\'e}n en el desarrollo asint{\'o}tico de la traza de la
resolvente en virtud de la ecuaci{\'o}n (\ref{elthm}), que toma la
forma,
\begin{eqnarray}\label{elthm1}
\begin{picture}(300,40)(25,20)
    \put(-5,0){\line(0,1){60}}
    \put(360,0){\line(0,1){60}}
    \put(-5,0){\line(1,0){365}}
    \put(-5,60){\line(1,0){365}}
    \put(15,30){$\displaystyle{{\rm Tr}\left\{(A^{\theta}+z)^{-1}-
    (A^{\infty}+z)^{-1}\right\}
    =\frac{{\rm Tr}\left\{(A^{0}+z)^{-1}-(A^{\infty}+z)^{-1}\right\}}
    { 1+4^{\nu}\displaystyle{ \frac{\Gamma(1+\nu)}{\Gamma(1-\nu)} }\;
    \theta\;z^{-\nu}H(z)^{-1} } } \,.$}
\end{picture}\nonumber\\\nonumber\\\nonumber \\
\end{eqnarray}
La traza del miembro derecho de la ecuaci{\'o}n (\ref{elthm1}),
que relaciona las resolventes de las extensiones correspondientes
a $\theta=0$ y $\theta=\infty$, puede obtenerse a partir de la
ecuaci{\'o}n (\ref{R0vsRinf}),
\begin{equation}\label{r0yinf}
    {\rm Tr}\left\{(A^{0}+z)^{-1}-(A^{\infty}+z)^{-1}\right\}=2\nu
    K(z)\int_0^\infty
    G_{\infty}^2(x,z)\,dx\,.
\end{equation}
Comparando las ecuaciones (\ref{Renelori}) y (\ref{re2})
obtenemos,
\begin{equation}
    -\frac{\left.z^{-1/2}y^{-\nu-1/2}L(y,z)\right|_{y=0}}
    {W[L,R](z)}=\frac{1}{2^{\nu}\nu\Gamma(\nu)}\sqrt{z}^{\nu-1/2}\,.
\end{equation}
De modo que la ecuaci{\'o}n (\ref{307}) resulta,
\begin{equation}
    G_{\infty}(x',z)=\frac{1}{2^{\nu}\nu\Gamma(\nu)}\sqrt{z}^{\nu-1/2}
    R(\sqrt{z}x',z) \,.
\end{equation}
Reemplazando esta ecuaci{\'o}n, junto con (\ref{kyh}), en la
ecuaci{\'o}n (\ref{r0yinf}) obtenemos,
\begin{eqnarray}\label{tam}
    {\rm Tr}\left\{(A^{0}+z)^{-1}-(A^{\infty}+z)^{-1}\right\}=
    \frac{2H(z)^{-1}\,z^{-1/2}}{\Gamma(\nu)\Gamma(1-\nu)}
    \int_0^\infty
    R(\sqrt{z}x,z)^2\,dx\,.\nonumber\\
\end{eqnarray}
De esta {\'u}ltima ecuaci{\'o}n puede verse que ${\rm
Tr}\left\{(A^{0}+z)^{-1}-(A^{\infty}+z)^{-1}\right\}$ admite un
desarrollo asint{\'o}tico en potencias semienteras de $z$.

\bs

Resumimos estos c{\'a}lculos en uno de los resultados centrales de
esta Tesis:

\bs

\begin{thm}\label{elthm11}
El desarrollo asint{\'o}tico de la traza de la diferencia entre
las resolventes $(A^{\theta}-\lambda)^{-1}$ y
$(A^{\infty}-\lambda)^{-1}$ tiene la forma,
\begin{eqnarray}\label{resuinf}
    {\rm Tr}\left\{(A^{\theta}-\lambda)^{-1}-(A^{\infty}-
    \lambda)^{-1}\right\}\sim
    \sum_{n=2}^\infty \alpha_n(\nu,V)\,\lambda^{-\frac n 2}+\nonumber\\
    +\sum_{N,n=1}^\infty \beta_{N,n}(\nu,V)\,\theta^N\,
    \lambda^{-\nu N-\frac n 2-\frac 1 2}\,.
\end{eqnarray}
Los coeficientes $\alpha_n(\nu,V),\beta_n(\nu,V)$ dependen del
coeficiente $\nu$ de la singularidad y est{\'a}n determinados por
los coeficientes $V_n$ que provienen del potencial suave $V(x)$
mediante las ecuaciones (\ref{elthm1}), (\ref{tam}), (\ref{tam2}),
(\ref{fin}) y (\ref{hache}) con $z=e^{i\pi}\lambda$.
\end{thm}\fin
La ecuaci{\'o}n (\ref{resuinf}), junto con las ecuaciones
(\ref{desareso}) y (\ref{polres}), implica que las singularidades
de la diferencia $\zeta_{A}^\theta(s)-\zeta_{A}^\infty(s)$ de las
funciones-$\zeta$ correspondientes a las extensiones autoadjuntas
$A^\theta$ y $A^\infty$ del operador $A$ consisten en polos
simples en los puntos $s_n$ y $s_{N,n}$ cuyas posiciones y
residuos est{\'a}n dadas por,
\begin{eqnarray}
    s_{n}=\frac 1 2-n\qquad n=1,2,3,\ldots\label{po1}\\\nonumber\\
    {\rm Res}\{\zeta_A^\theta(s)-\zeta_A^\infty(s)\}|_{s=s_n}=
    \frac{(-1)^n}{\pi}\,\alpha_{2n+1}(\nu,V)\,;\label{resu22}
    \\\nonumber\\\nonumber\\
    s_{N,n}=-\nu N-\frac n 2+\frac 1 2\qquad N,n=1,2,\ldots \label{po2}
    \\\nonumber\\
    {\rm Res}\{\zeta_A^\theta(s)-\zeta_A^\infty(s)\}|_{s=s_{N,n}}=
    \frac{\cos{[\pi(\nu N+ n/2)]}}{\pi}\,\theta^N\,\beta_{N,n}(\nu,V)
    \,.\label{resu3}
\end{eqnarray}

\subsubsection{Ejemplo: $V(x)=x^2$}

Consideremos, a continuaci{\'o}n, la estructura de polos de la
funci{\'o}n $\zeta_A^\theta(s)-\zeta_A^\infty(s)$ dada por las
ecuaciones (\ref{po1}) y (\ref{po2}) para el caso $V(x)=x^2$. El
resultado que obtendremos ser{\'a} ve\-ri\-fi\-ca\-do luego, en la
secci{\'o}n \ref{Wipf}, utilizando otras t{\'e}cnicas.

\bigskip

N{\'o}tese, en primer lugar, que si $V(x)=x^2$ entonces s{\'o}lo
uno de los coeficientes $V_n$ definidos en (\ref{pottay}) es
distinto de cero,
\begin{equation}
    V_n=\delta_{n,2}\,.
\end{equation}
En consecuencia, las {\'u}nicas funciones $\phi_n(y)$, definidas
en (\ref{fin}), que no se anulan trivialmente corresponden a los
valores $n=2+4k$ con $k=0,1,\ldots$

\bigskip

De acuerdo con las ecuaciones (\ref{hache}) y (\ref{tam2}),
\begin{eqnarray}
    H(z)^{-1}\sim 1+\sum_{k=1}^{\infty}C_k(\nu)\,z^{-2k}\,,\label{hasim}\\
    R(y,z)\sim \sqrt{y}K_\nu(y)+\sum_{k=1}^{\infty}C'_k(\nu,y)\,z^{-2k}\,.
\end{eqnarray}
No determinaremos la forma de los coeficientes
$C_k(\nu),C'_k(\nu,y)$. Reemplazando estas ecuaciones en
(\ref{tam}) obtenemos,
\begin{eqnarray}\label{fina}
    {\rm Tr}\left\{(A^{0}+z)^{-1}-(A^{\infty}+z)^{-1}\right\}\sim
    \nu\,z^{-1}+\sum_{k=0}^{\infty}C''_k(\nu)\,z^{-3-2k}\,,
\end{eqnarray}
donde el coeficiente $C''_k(\nu)$ puede expresarse en t{\'e}rminos
de $C_k(\nu),C'_k(\nu,y)$.

\bigskip

Substituyendo, finalmente, las ecuaciones (\ref{fina}) y
(\ref{hasim}) en (\ref{elthm1}) obtenemos las potencias de $z$ en
el desarrollo asint{\'o}tico de la traza de las diferencias entre
las resolventes $(A^{\theta}+z)^{-1}$ y $(A^{\infty}+z)^{-1}$,
\begin{eqnarray}\label{desarro}
    {\rm Tr}\left\{(A^{\theta}+z)^{-1}-(A^{\infty}+z)^{-1}\right\}\sim\nonumber\\
    \sim \left[\nu\,z^{-1}+
    \sum_{k=0}^{\infty}C''_k(\nu)\,z^{-3-2k}\right]\times\nonumber\\
    \times
    \sum_{N=0}^{\infty}
    (-1)^N 4^{N\nu}\left[\frac{\Gamma(1+\nu)}
    {\Gamma(1-\nu)}\right]^N \theta^N\,z^{-N\nu}
    \left[1+\sum_{k=1}^{\infty}C_k(\nu)\,z^{-2k}\right]^N\,.
\end{eqnarray}
Comparando las ecuaciones (\ref{resuinf}) y (\ref{desarro}) se
observa que,
\begin{equation}
    \alpha_{2n+1}(\nu,V)=0\,.
\end{equation}
Concluimos entonces que los polos de la funci{\'o}n
$\zeta_A^{\theta}(s)-\zeta_A^{\infty}(s)$ indicados en la
expresi{\'o}n (\ref{po1}) no est{\'a}n presentes para el potencial
$V(x)=x^2$.

\bigskip

Sin embargo, el desarrollo asint{\'o}tico (\ref{desarro}) presenta
tambi{\'e}n potencias de la forma,
\begin{equation}
    z^{-N\nu-1-2n}\,,
\end{equation}
con $n=0,1,\ldots$ En consecuencia, en virtud de la ecuaci{\'o}n
(\ref{po2}), la funci{\'o}n
$\zeta_A^{\theta}(s)-\zeta_A^{\infty}(s)$ presenta polos simples
en los puntos $s_{N,n}$ del plano complejo dados por,
\begin{equation}\label{toconf}
    s_{N,n}=-N\nu-2n\,,
\end{equation}
con $N=1,2,\ldots$ y $n=0,1,\ldots$ En particular, de acuerdo con
el t{\'e}rmino dominante en (\ref{desarro}) y con la expresi{\'o}n
(\ref{resu3}), existe un polo simple en el punto,
\begin{equation}
    s_{1,0}=-\nu\,,
\end{equation}
cuyo residuo est{\'a} dado por,
\begin{equation}\label{alfin}
    \frac{4^{\nu}}{\Gamma^2(-\nu)}\,
    \theta\,.
\end{equation}
Este resultado ser{\'a} verificado luego en la secci{\'o}n
\ref{Wipf}.

\subsection{Caso compacto}\label{cccp}

Finalizamos este cap{\'\i}tulo aplicando los resultados de la
secci{\'o}n \ref{singu} para estudiar el desarrollo asint{\'o}tico
de la traza de la resolvente del operador,
\begin{equation}\label{ope}
    A=-\partial^2_x+\frac{\nu^2-1/4}{x^2}+V(x)\,,
\end{equation}
donde $x\in[0,1]\in\mathbb{R}$ y el potencial es anal{\'\i}tico en
$x$.

\bs

Los c{\'a}lculos que presentaremos a continuaci{\'o}n son
similares a los desarrollados para el caso del operador
diferencial (\ref{opea}) definido sobre la variedad de base no
compacta $\mathbb{R}^+$ en la secci{\'o}n \ref{nocom}. Sin
embargo, no obtendremos expl{\'\i}citamente el desarrollo
asint{\'o}tico de la resolvente, tal como est{\'a} expresado en la
ecuaci{\'o}n (\ref{resuinf}), sino que calcularemos una
expresi{\'o}n similar a (\ref{elthm1}) que verificaremos luego, en
la secci{\'o}n \ref{Solari}, para el caso particular $V(x)=0$.

\bigskip

De acuerdo con la ecuaci{\'o}n (\ref{saesing}), las extensiones
autoadjuntas del operador co\-rres\-pon\-dien\-tes a condiciones
de contorno locales est{\'a}n caracterizadas por dos
par{\'a}metros reales $\theta,\beta$ que describen el
comportamiento de las funciones $\phi(x)$ pertenecientes al
dominio $\mathcal{D}(A)$ en los extremos de $[0,1]$.

\bs

Nuestro prop{\'o}sito es obtener el factor $K(\lambda)$ a partir
del comportamiento en el origen de la funci{\'o}n
$G_{\infty}(x,\lambda)$ (v{\'e}ase la ecuaci{\'o}n
(\ref{Renelori}).) Por su parte, esta funci{\'o}n est{\'a}
de\-fi\-ni\-da, por medio de la ecuaci{\'o}n (\ref{Rxp}), en
t{\'e}rminos de la resolvente correspondiente a la extensi{\'o}n
$\theta=\infty$.

\bs

Si utilizamos la expresi{\'o}n (\ref{solres}) para el n{\'u}cleo
de la resolvente $G_{\infty}(x',\lambda)$,
\begin{eqnarray}
    G_{\infty}(x',\lambda):=\lim_{x\rightarrow0}x^{-\nu-1/2}
    G_{\infty}(x,x',\lambda)=\nonumber\\
    =-\frac{\lim_{x\rightarrow0}x^{-\nu-1/2}L(x,\lambda)}
    {W[L,R](\lambda)}R(x',\lambda)\,.
\end{eqnarray}
En consecuencia, el factor $K(\lambda)$ est{\'a} determinado por
el comportamiento de $R(x',\lambda)$ en el origen. En efecto,
\begin{equation}\label{ayb}
    R(x\sim 0,\lambda)\sim x^{-\nu+1/2}-K(\lambda)^{-1}(\lambda)
    x^{\nu+1/2}+\ldots
\end{equation}

El resto de esta secci{\'o}n est{\'a} dedicado a calcular el
factor $K(\lambda)$ a partir del comportamiento en el origen de la
autofunci{\'o}n $R(x,\lambda)$ del operador (\ref{ope}) que
satisface la condici{\'o}n de contorno determinada por $\beta$ en
el extremo $x=1$ del intervalo $[0,1]\subset\mathbb{R}$.

\bs

Consideremos entonces la soluci{\'o}n $R(x,\lambda=\mu^2)$ al
si\-guien\-te problema,
\begin{eqnarray}
    \left(-\partial^2_x +\frac{\nu^2-1/4}{x^2}+V(x)-\mu^2\right)
    R(x,\mu^2)=0\,,\label{ecdi}\\\nonumber\\
    \left.\left[\partial_x R(x,\mu^2)-\beta\, R(x,\mu^2)\right]
    \right|_{x=1}=0\,.\label{ccecdi}
\end{eqnarray}
Derivaremos una soluci{\'o}n asint{\'o}tica de la ecuaci{\'o}n
(\ref{ecdi}), para grandes valores de $\mu$.

\bs

Al igual que en la secci{\'o}n anterior, es conveniente definir
una transformaci{\'o}n de escala $y:= \mu\,x$, de modo que la
ecuaci{\'o}n (\ref{ecdi}) tome la forma,
\begin{equation}
    \left(-\partial^2_y +\frac{\nu^2-1/4}{y^2}-1\right)R(y/\mu,\mu^2)=
    -V(y/\mu)R(y/\mu,\mu^2)\,.\label{ecdi2}
\end{equation}

Calcularemos la ecuaci{\'o}n integral asociada a la igualdad
(\ref{ecdi2}). Para ello, debemos encontrar la funci{\'o}n de
Green $G_0(y,y')$ del operador diferencial del miembro izquierdo
de (\ref{ecdi2}). Definimos entonces las funciones,
\begin{eqnarray}
    l(y):= \sqrt{y}J_\nu(y)\,,\\
    r(y):= \sqrt{y}\left(\gamma \,J_\nu(y)-\frac{\pi}{2\gamma}\,
    N_{\nu}(y)\right)\,,
\end{eqnarray}
donde,
\begin{equation}
    \gamma:= \frac{\pi}{2}\frac{(1/2-\beta)N_\nu(\mu)+\mu N_\nu'(\mu)}
    {(1/2-\beta)J_\nu(\mu)+\mu J_\nu'(\mu)}\,.
\end{equation}
Las funciones $l(y),r(y)$ pertenecen al n{\'u}cleo del operador
diferencial del miembro izquierdo de (\ref{ecdi2}). Por otra
parte, $r(\mu,x)$ satisface la condici{\'o}n de contorno
(\ref{ccecdi}).

\bs

De acuerdo con la ecuaci{\'o}n (\ref{solres}), la funci{\'o}n de
Green $G^{(V=0)}(y,y')$ que buscamos est{\'a} dada por,
\begin{equation}\label{solres00}
    G^{(V=0)}(y,y')=-\frac{\theta(y'-y)l(y)r(y')+
    \theta(y-y')l(y')r(y)}{W[l,r]}\,,
\end{equation}
donde $W[l,r]$ es el wronskiano,
\begin{equation}
    W[l,r]=l(y)r'(y)-l'(y)r(y)=-1\,.
\end{equation}

En consecuencia, la soluci{\'o}n $R(y/\mu,\mu^2)$ de la
ecuaci{\'o}n (\ref{ecdi2}) est{\'a} dada por,
\begin{eqnarray}
    R(x,\mu^2)=-\frac {1} {\mu^2} \int_0^\mu G^{(V=0)}(\mu\,x,y')
    V(y'/\mu)\,dy'=\nonumber\\
    =r(\mu x)+\frac 1 \mu \int_0^{1}K(x,x')R(x',\mu^2)\,dx'\,;\label{solu}
\end{eqnarray}
siendo,
\begin{eqnarray}
    K(x,x'):= \theta(x'-x)\cdot\{\mu x,\mu x'\}\cdot V(x')\,,\\
    \{x,x'\}:= -l(x)r(x')+l(x')r(x)\,.
\end{eqnarray}
La funci{\'o}n $R(x,\mu^2)$ cuyo comportamiento en el origen
determina el factor $K(\lambda)$ satisface, entonces, una
ecuaci{\'o}n integral de Volterra cuyo n{\'u}cleo $K(x,x')$ es
continuo en $(0,1)\times(0,1)$ e integrable en $[0,1]\times[0,1]$.

\bs

Si designamos por $*$ la convoluci{\'o}n de funciones, la
ecuaci{\'o}n (\ref{solu}) puede escribirse,
\begin{equation}
    R-\frac{1}{\mu}K*R=r\,,
\end{equation}
cuya soluci{\'o}n est{\'a} dada por,
\begin{equation}\label{serie}
    R=\sum_{n=0}^{\infty}K_n*r\cdot\mu^{-n}\,,
\end{equation}
siendo,
\begin{eqnarray}
    K_0:=1\,,\\
    K_1:= K\,,\\
    K_{n+1}:= K*K_n\,.
\end{eqnarray}
La serie (\ref{serie}) converge uniformemente para $\mu$
suficientemente grande (v\'ease, {\it e.g.}, \cite{C-H}) por lo
que provee un desarrollo asint{\'o}tico para el comportamiento en
el origen de $R(x,\mu^2)$ y, en consecuencia, para $K(\lambda)$.

\bigskip

Examinando el comportamiento en el origen de los distintos
t{\'e}rminos de la serie en (\ref{serie}) podemos obtener los
coeficientes correspondientes a las potencias $x^{-\nu+1/2}$ y
$x^{\nu+1/2}$ cuyo cociente determina, de acuerdo con la
ecuaci{\'o}n (\ref{ayb}), el factor $K(\lambda)$. De esta
ma\-ne\-ra obtenemos el desarrollo asint{\'o}tico,
\begin{eqnarray}\label{cha}
    K(\mu^2)\sim 4^{\nu}\frac{\Gamma(1+\nu)}{\Gamma(1-\nu)}\,
    e^{i\epsilon\pi\nu}
    \mu^{-2\nu}
    \left\{1+\frac{\delta\,\sum_{n=1}^{\infty}\mu^{-2n}\,S_r^n(\mu)}
    {1+\sum_{n=1}^{\infty}\mu^{-2n}\left(S_l^n(\mu)-\delta S_r^n(\mu)
    \right)}\right\}\,,\nonumber\\
\end{eqnarray}
siendo,
\begin{equation}
    \delta:= \left(\gamma-\frac \pi 2 \cot{\pi\nu}\right)^{-1}\sim
    -\frac{2}{\pi}\sin{(\pi\nu)}e^{-i\epsilon \pi\nu}\,,
\end{equation}
y,
\begin{eqnarray}
    S_r^n(\mu):=
    \int_0^{\mu}dx_1\int_{x_1}^{\mu}dx_2\ldots\int_{x_{n-1}}^\mu
    dx_n\,r(x_1)\,r(x_2)
    \ldots r(x_n)\times\nonumber\\\times\,V(x_1/\mu)\ldots
    V(x_n/\mu)
    \cdot\{x_1,x_2\}\ldots \{x_{n-1},x_n\}\,\\\nonumber\\
    S_l^n(\mu):=
    \int_0^{\mu}dx_1\int_{x_1}^{\mu}dx_2\ldots\int_{x_{n-1}}^\mu
    dx_n\,l(x_1)\,r(x_2)
    \ldots r(x_n)\times\nonumber\\\times\, V(x_1/\mu)\ldots
    V(x_n/\mu)
    \cdot\{x_1,x_2\}\ldots \{x_{n-1},x_n\}\,.
\end{eqnarray}
El factor $\epsilon$ es $\pm 1$ si $\lambda$ est{\'a} en el
semiplano complejo superior o inferior, respectivamente.

\bigskip

La expresi{\'o}n (\ref{cha}) muestra que el factor $K(\lambda)$
admite un desarrollo asint{\'o}tico en potencias de $\lambda$
cuyos exponentes dependen del par{\'a}metro $\nu$. En la
secci{\'o}n \ref{Solari} con\-si\-de\-ra\-re\-mos el caso
particular $V(x):= 0$ y obtendremos, con otras t{\'e}cnicas, el
factor $K(\lambda)$.

\bs

De acuerdo con (\ref{cha}), si $V(x):= 0$ entonces
$S_r^n(\mu)=S_l^n(\mu)=0$ y resulta,
\begin{equation}\label{chau}
    K(\lambda)\sim 4^{\nu}\frac{\Gamma(1+\nu)}{\Gamma(1-\nu)}
    \,e^{i\epsilon\pi\nu}
    \lambda^{-\nu}\,.
\end{equation}
Esta expresi{\'o}n ser{\'a} verificada con los c{\'a}lculos
desarrollados en la secci{\'o}n \ref{Solari}.

\bigskip

A partir de las ecuaciones (\ref{elthm}) y (\ref{cha}) se puede
ver que el desarrollo asint{\'o}tico de la traza de la resolvente
${\rm Tr}(A-\lambda)^{-1}$ presenta potencias enteras de
$\lambda^{-\nu}$. En consecuencia, la posici{\'o}n de los polos de
la funci{\'o}n-$\zeta$ del operador dado por la expresi{\'o}n
(\ref{ope}) depende del par{\'a}metro $\nu$. Esto difiere del
resultado (\ref{resu}) en el que se establece que la
funci{\'o}n-$\zeta$ correspondiente a un operador diferencial
regular de segundo orden $A$ definido sobre funciones del
intervalo $[0,1]$ tiene polos en los semienteros. El motivo de
esta discrepancia consiste en la presencia del t{\'e}rmino
singular $(\nu^2-1/4)/x^2$ en el operador (\ref{ope}).


\part{Ejemplos: Operadores de Schr\"odinger}\label{apli}

\vspace{5mm}\begin{flushright}{\it The art of doing mathematics
consists in finding\\that special case which contains\\all the
germs
of generality.\\
(David Hilbert.)}
\end{flushright}

\vspace{25mm}

En este cap{\'\i}tulo resolveremos dos ejemplos particulares de
operadores de Schr\"odinger con coeficientes singulares.
Primeramente, reproduciremos en la secci{\'o}n \ref{Wipf} el
trabajo desarrollado en \cite{FPW}, en el que se estudia el
operador de Schr\"odinger dado por la ecuaci{\'o}n (\ref{1}) para
el caso $V(x)=x^2$ definido sobre la variedad de base
unidimensional no compacta $\mathbb{R}^+$. Los resultados de esta
secci{\'o}n deben compararse con los de la secci{\'o}n
\ref{nocom}.

\bs

En la secci{\'o}n \ref{Solari} consideraremos el mismo operador de
Schr\"odinger correspondiente al caso $V(x)=0$ pero definido sobre
una variedad de base unidimensional compacta
$[0,1]\subset\mathbb{R}$. Este problema ha sido resuelto en
\cite{FPM} y confirma los resultados de la secci{\'o}n \ref{cccp}.

\bs

En ambos ejemplos encontraremos la resoluci{\'o}n espectral del
operador, obteniendo una ecuaci{\'o}n trascendente para los
autovalores y una forma expl{\'\i}cita para las autofunciones.
Esta resoluci{\'o}n permitir{\'a} calcular el desarrollo
asint{\'o}tico de la resolvente o, e\-qui\-va\-len\-te\-men\-te,
las singularidades de la funci{\'o}n-$\zeta$.

\bs

Mostraremos que los polos de las funciones-$\zeta$ no est{\'a}n
determinados por el orden del operador y la dimensi{\'o}n de la
variedad de base, como en el caso de los operadores con
coeficientes regulares, sino que pueden depender de los
par{\'a}metros que caracterizan la singularidad. Se{\~n}alamos,
asimismo, que este resultado est{\'a} relacionado con la presencia
de un conjunto infinito de extensiones autoadjuntas.

\section{{Un operador de Schr\"odinger en una variedad de base no compacta}}
\label{Wipf}

\subsection{El operador y su adjunto}\label{adjoint-H}

En esta secci{\'o}n consideraremos el operador,
\begin{equation}\label{ham666}
    A=-\partial_x^2+\frac{\nu^2-1/4}{x^2}+x^2\,,
\end{equation}
siendo $x\in\mathbb{R}^+$, como un caso particular de
(\ref{opea}), con el fin de verificar la expresi{\'o}n
(\ref{elthm1}).

\bigskip

En primer lugar definimos $A$ sobre el conjunto denso ${ \mathcal
D}(A):={\mathcal C}_0^\infty(\mathbb{R}^+)$ de funciones con
derivadas continuas de todo orden cuyo soporte es compacto y no
contiene al origen. Es f{\'a}cil ver que  en este dominio de
definici{\'o}n el operador $A$ es sim{\'e}trico. Sin embargo, $A$
no es autoadjunto.

\bs

Para construir sus extensiones autoadjuntas debemos calcular el
operador adjunto $A^{\dagger}$ y determinar los subespacios de
deficiencia $\mathcal{K_{\pm}}$. Pero, dado que $A$ no es cerrado
en ${\mathcal C}_0^\infty(\mathbb{R}^+)$, calcularemos
primeramente su clausura $\overline{A}$, esto es, extenderemos el
dominio de $A$ a partir del estudio de la clausura de su
gr{\'a}fica. Determinaremos, luego, el comportamiento en el origen
de las funciones que pertenecen al dominio de $\overline{A}$.

\subsubsection{Clausura del operador} \label{closure}

Mostraremos ahora que si $\phi\in{\mathcal D}(\overline{A})$
entonces,
\begin{equation}\label{clau}
    \phi(x)=o(x^{\nu+1/2}) \quad {\rm y}\quad
    \phi'(x)=o(x^{\nu-1/2})\,,
\end{equation}
para $\rightarrow 0^+$ y $\nu<1$.

\bs

Para determinar la clausura de la gr{\'a}fica del operador $A$,
consideramos aquellas sucesiones de Cauchy $\{\varphi_n\}_{n\in
\mathbb{N}}$ contenidas en ${\mathcal D}(A)={\mathcal C}_0^\infty
(\mathbb{R^+})$ tales que $\{A\varphi_n\}_{n\in \mathbb{N}}$ sean
tambi{\'e}n sucesiones de Cauchy. N{\'o}tese que, puesto que los
coeficientes de $A$ son reales, podemos considerar solamente
funciones reales.

\bs

Sea $\varphi:=\varphi_n - \varphi_m$, con $n,m\in\mathbb{N}$.
Entonces $\varphi \rightarrow 0$ y $A\varphi\rightarrow 0$ para
$n,m \rightarrow \infty$. Consideremos ahora el producto interno,
\begin{equation}\label{phi-Hphi}\begin{array}{c}
  \displaystyle{\left( \varphi,A\varphi \right) =
  \int_0^\infty \varphi \left( - \varphi'' + \frac{\nu^2-1/4}{x^2}\, \varphi +
  x^2 \varphi \right)\, dx = } \\ \\
  \displaystyle{=  \int_0^\infty \left( \varphi'^2 + \frac{\nu^2-1/4}{x^2}\,
  \varphi^2 +
  x^2 \varphi^2 \right)\, dx  \leq ||\varphi|| \cdot ||A\varphi||
  \rightarrow 0}\,,
\end{array}
\end{equation}
para $n,m \rightarrow \infty$. Por lo tanto, para $\nu > 1/2$,
\begin{equation}\label{C-seq}
  \{\varphi'_n(x)\}_{n\in \mathbb{N}}\,,\quad
  \displaystyle{\left\{\frac{\varphi_n(x)}{x}\right\}_{n\in
    \mathbb{N}}}\quad {\rm y}\quad  \{x\, \varphi_n(x)\}_{n\in \mathbb{N}}\,,
\end{equation}
son tambi{\'e}n sucesiones de Cauchy.

\bs

\begin{lem}\label{llee}
Sea $\{\varphi_n\}_{n\in \mathbb{N}}$ una sucesi{\'o}n de Cauchy
en ${\mathcal D}(A)={\mathcal C}_0^\infty(\mathbb{R^+})$ tal que,
para $\nu>1/2$, $1\leq a<2$ y
 $\nu\neq a$,
\begin{equation}
  \{A\varphi_n\}_{n\in \mathbb{N}}\,,\quad
  \displaystyle{\left\{\frac{\varphi_n(x)}{x^a}\right\}_{n\in
  \mathbb{N}}}\quad {\rm y} \quad
 \displaystyle{\left\{\frac{\varphi_n'(x)}{x^{a-1}}\right\}_{n\in
  \mathbb{N}}}\,,
\end{equation}
sean a su vez sucesiones de Cauchy. Entonces,
\begin{equation}
\displaystyle{\left\{\frac{\varphi_n(x)}{x^{1+{a}/{2}}}\right\}_{n\in
    \mathbb{N}}}\quad{\rm y} \quad
\displaystyle{\left\{\frac{\varphi_n'(x)}{x^{{a}/{2}}}\right\}_{n\in
    \mathbb{N}}}\,,
\end{equation}
son tambi{\'e}n sucesiones de Cauchy.
\end{lem}

\noindent {\bf Demostraci{\'o}n:} Sea $\varphi:=\varphi_n
-\varphi_m$. N{\'o}tese
 que, para $1\leq a<2$,
\begin{equation}\label{l0}\begin{array}{c}
  \displaystyle{\int_0^\infty \left( x^{1-a/2}\, \varphi(x) \right)^2\,
  dx \leq
  \int_0^1 \left(  \varphi(x) \right)^2\, dx
  +  \int_1^\infty  \left( x \, \varphi(x) \right)^2\,
  dx\leq}\\ \\ \displaystyle{
     \leq || \varphi(x) ||^2 + || x\,\varphi(x) ||^2\,.}
\end{array}
\end{equation}
Entonces, de  (\ref{C-seq}), vemos que $\displaystyle{ \left\{
x^{1-a/2}\, \varphi_n(x) \right\}_{n\in \mathbb{N}} }$ es
tambi{\'e}n una
 sucesi{\'o}n de Cauchy.

\bs

Un c{\'a}lculo sencillo muestra que,
\begin{equation}\label{l1}\begin{array}{c}
  \displaystyle{ \left(\frac{\varphi(x)}{x^a}, A \varphi(x) \right) =
  \int_0^\infty
  \left\{ \left(\frac{\varphi'(x)}{x^{a/2}}\right)^2 + \right.} \\ \\
  \displaystyle{\left. + \left[ \nu^2-\frac14 - \frac{a(a+1)}{2} \right]
  \left(\frac{\varphi(x)}{x^{1+a/2}}\right)^2 +
  \left( x^{1-a/2}\, \varphi(x) \right)^2\right\}\, dx}\,,
\end{array}
\end{equation}
\begin{equation}\label{l2}\begin{array}{c}
  \displaystyle{ \left(\frac{\varphi'(x)}{x^{a-1}},
  A \varphi(x) \right) = \int_0^\infty
  \left\{-\left(\frac{a-1}{2} \right)
  \left(\frac{\varphi'(x)}{x^{a/2}}\right)^2 + \right.} \\ \\
  \displaystyle{\left. \mbox{}+\left(\nu^2-\frac14\right) \left( \frac{a+1}{2}
  \right)
  \left(\frac{\varphi(x)}{x^{1+a/2}}\right)^2 + \left( \frac{a-3}{2} \right)
  \left( x^{1-a/2}\, \varphi(x) \right)^2\right\}\, dx}\,.
\end{array}
\end{equation}
Teniendo en cuenta que la suma de dos sucesiones de Cauchy es
tambi{\'e}n una sucesi{\'o}n de Cauchy, se verifica, para
cualquier par de n{\'u}meros reales $C$ y $C'$,
\begin{equation}\label{l3}
   \left(C\,\frac{\varphi(x)}{x^a}+ C'\,
    \frac{\varphi'(x)}{x^{a-1}}\, ,   A \varphi(x) \right) \rightarrow 0\,,
\end{equation}
si $n,m\rightarrow \infty$.

\bs

Reemplazando las ecuaciones (\ref{l1}) y (\ref{l2}) en (\ref{l3}),
los coeficientes de
$\displaystyle{\left(\frac{\varphi'(x)}{x^{a/2}}\right)^2}$ y
$\displaystyle{\left(\frac{\varphi(x)}{x^{1+a/2}}\right)^2}$
resultan,
\begin{equation}\label{l4}
  \begin{array}{c}
    \displaystyle{C -C'\left(\frac{a-1}{2} \right)} \quad {\rm y} \quad
     \displaystyle{
     C \left(\nu^2-\frac14- \frac{a(a+1)}{2} \right)+C'\,
     \left(\nu^2-\frac14\right) \left( \frac{a+1}{2}
     \right)}
  \end{array}\,,
\end{equation}
respectivamente. Puede verse que, si $\nu\neq a$, estos
coeficientes s{\'o}lo se anulan simult{\'a}neamente para $C=C'=0$.
Eligiendo entonces $C,C'$ adecuadamente se prueba el enunciado del
lema.\begin{flushright}$\Box$\end{flushright}

Teniendo en cuenta (\ref{C-seq}), el Lema \ref{llee}, para $a=1$,
impica que,
\begin{equation}\label{l5}
  \displaystyle{\left\{\frac{\varphi_n(x)}{x^{3/{2}}}\right\}_{n\in
    \mathbb{N}}} \quad {\rm y} \quad
    \displaystyle{\left\{\frac{\varphi_n'(x)}{x^{{1}/{2}}}\right\}_{n\in
    \mathbb{N}}}\,,
\end{equation}
son sucesiones de Cauchy. En consecuencia, si $\nu$ es un
n{\'u}mero irracional, aplicando entonces iterativamente el Lema
\ref{llee} puede probarse que, para cualquier entero positivo $k$,
\begin{equation}\label{l6} \displaystyle{\left\{\frac{\varphi_n(x)}{x^{2\left[
1- \left(   1/2 \right)^k \right]}}\right\}_{n\in \mathbb{N}}}\,,
\quad {\rm y} \quad
\displaystyle{\left\{\frac{\varphi_n'(x)}{x^{2\left[ 1- \left(
  1/2 \right)^k \right]-1}}\right\}_{n\in
    \mathbb{N}}}\,,
\end{equation}
son sucesiones de Cauchy.

\bs

Como, en general, para cualquier $\varepsilon >0$ existen enteros
$k_1$ y $k_2$ tales que $(1/2)^{k_1} \leq \varepsilon \leq
(1/2)^{k_2}$, teniendo en cuenta que,
\begin{equation}\label{l7}
  \begin{array}{c}
    \displaystyle{  \frac{1}{x^{2-\varepsilon}} \leq
    \frac{1}{x^{2\left[ 1- \left(
  1/2 \right)^{k_1} \right]}}\,, \quad {\rm para} \ 0<x\leq 1}\,,\\ \\
    \displaystyle{\frac{1}{x^{2-\varepsilon}} \leq
    \frac{1}{x^{2\left[ 1- \left(
  1/2 \right)^{k_2} \right]}}\,,\quad {\rm para}\ x\geq 1}\,,
  \end{array}
\end{equation}
se concluye inmediatamente que,
\begin{equation}\label{conclu1}
    \displaystyle{\left\{\frac{\varphi_n(x)}{x^{2-\varepsilon}}\right\}_{n\in
    \mathbb{N}}}\,,
\end{equation}
es una sucesi{\'o}n de Cauchy. La misma conclusi{\'o}n se obtiene
an{\'a}logamente para la sucesi{\'o}n,
\begin{equation}\label{conclu2}
    \displaystyle{\left\{\frac{\varphi_n'(x)}{x^{1-\varepsilon}}\right\}_{n\in
    \mathbb{N}}}\,.
\end{equation}

\bs

Si $\nu$ es un n{\'u}mero racional, entonces, a partir de
(\ref{C-seq}) y (\ref{l5}), se ve que existe un irracional
$a\in(1,3/2)$ para el cual el Lema \ref{llee} pueda aplicarse
iterativamente para demostrar que las sucesiones (\ref{conclu1}) y
(\ref{conclu2}) son sucesiones de Cauchy.

\bigskip

Determinaremos ahora el comportamiento de las funciones en
$\mathcal{D}(\overline{A})$ cerca del origen. Para cualquier
$\varepsilon>0$, podemos escribir,
\begin{equation}\label{1pri}\begin{array}{c}
  \displaystyle{ x^{-\nu-1/2} \, \varphi(x) =
  \int_0^x \left( y^{-\nu-1/2}  \, \varphi(y)
  \right)'\, dy = } \\ \\
  \displaystyle{=\int_0^x y^{{-\nu-1/2} +1-\varepsilon}\left\{ {-\nu-1/2}
  \,   \frac{\varphi(y)}{y^{2-\varepsilon}} +
  \frac{\varphi'(y)}{y^{1-\varepsilon}} \right\} \, dy \,.}
\end{array}
\end{equation}
Por lo tanto, para $x\leq 1$, ${\nu} < 1$ y $\varepsilon$
suficientemente peque\~no, se verifica,
\begin{equation}\label{2pri}\begin{array}{c}
  \displaystyle{\left| x^{-\nu-1/2}  \, \varphi(x) \right|\leq
  \left(
  \int_0^1 y^{2({-\nu-1/2} +1-\varepsilon)}dy \right)^{1/2}\times}\\ \\
    \displaystyle{\times\,
  |{\nu+1/2} |\,\left|\left|
  \frac{\varphi(y)}{y^{2-\varepsilon}} \right|\right|\cdot
  \left|\left|
  \frac{\varphi'(y)}{y^{1-\varepsilon}} \right|\right|
  \rightarrow_{n,m \rightarrow \infty} 0\,.}
\end{array}
\end{equation}
Por consiguiente, la sucesi{\'o}n $\{x^{-\nu-1/2}
\,\varphi_n(x)\}_{n\in\mathbb{N}}$, con ${\nu}<1$, es
uniformemente convergente en $[0,1]$ y su l{\'\i}mite es una
funci{\'o}n continua que se anula en el origen,
\begin{equation}\label{3}
  \displaystyle{x^{-\nu-1/2}  \,\phi(x):=
    \lim_{n\rightarrow\infty}\left( x^{-\nu-1/2}
  \,\varphi_n(x) \right)
  \,,}
\end{equation}
\begin{equation}\label{4}
  \displaystyle{\lim_{x\rightarrow 0^+} \left(x^{-\nu-1/2}
  \,\phi(x)\right)=0\,.}
\end{equation}
En particular,para ${\nu}= -1/2$ tenemos el l{\'\i}mite uniforme
\begin{equation}\label{4-5}
 \lim_{n\rightarrow\infty}\varphi_n(x) = \phi(x),
\end{equation}
que coincide con el l{\'\i}mite de esta sucesi{\'o}n en
$\mathbf{L_2}(\mathbb{R^+})$.

\bigskip

Por otra parte, tambi{\'e}n podemos escribir,
\begin{equation}\label{6}\begin{array}{c}
  \displaystyle{\int_0^x y^{-\nu+1/2}\,A\varphi(y)\,dy =
  -x^{-\nu+1/2}\,\varphi'(x) + } \\ \\
  \displaystyle{+ \int_0^x y^{{-\nu+1/2} -\varepsilon} \left\{
  ({-\nu+1/2} )\,
  \frac{\varphi'(y)}{y^{1-\varepsilon}}+ \kappa\,
  \frac{\varphi(y)}{y^{2-\varepsilon}} \right\}\, dy + } \\ \\
  \displaystyle{+ \int_0^x y^{{-\alpha} +2}\, y\, \varphi(y)\, dy}\,.
\end{array}
\end{equation}
Por lo tanto, para $ x \leq 1$,  ${\nu} < 1$ y $\varepsilon$
suficientemente peque\~no, se verifica,
\begin{equation}\label{7}\begin{array}{c}
  \displaystyle{\left| x^{-\nu+1/2}\,\varphi'(x) \right| \leq
  \left( \int_0^1 y^{{-2\nu+1}}\, dy \right)^{1/2} \left|\left|
  A\varphi(y) \right|\right| + } \\ \\
  \displaystyle{\left( \int_0^1 y^{{-2\nu} +1-2\varepsilon}\,
  dy \right)^{1/2}
  \left\{ |{\nu} +1/2| \left|\left| \frac{\varphi'(y)}
  {y^{1-\varepsilon}}
  \right|\right|+
  \kappa \left|\left| \frac{\varphi(y)}{x^{2-\varepsilon}} \right|\right|
  \right\}+ }\\ \\
  \displaystyle{+ \left( \int_0^1 y^{{-2\nu} +1}\, dy \right)^{1/2}
  \left|\left|
  y\,\varphi(y) \right|\right| \rightarrow_{n,m\rightarrow \infty} 0\,.}
\end{array}
\end{equation}
Consecuentemente, la sucesi{\'o}n
$\{x^{{-\nu}+1/2}\,\varphi_n'(x)\}_{n\in\mathbb{N}}$, con ${\nu}<
1$, es uniformemente convergente en $[0,1]$ y su l{\'\i}mite es
una funci{\'o}n continua que se anula en el origen,
\begin{equation}\label{8}
    \displaystyle{   x^{{-\nu} +1/2} \,\chi(x):=
    \lim_{n\rightarrow\infty}\left(   x^{{-\nu} +1/2}\,\varphi_n'(x)
\right)\,,}
\end{equation}
\begin{equation}\label{9}   \displaystyle{\lim_{x\rightarrow 0^+} \left(
x^{{-\nu} +1/2}   \,\chi(x)\right)=0\,.}
\end{equation}
En particular, para ${\nu} =1/2$, obtenemos el l{\'\i}mite
uniforme,
\begin{equation}\label{10}
  \lim_{n\rightarrow\infty}\varphi'_n(x) = \chi(x)\,,
\end{equation}
que coincide con el l{\'\i}mite de esta sucesi{\'o}n en
$\mathbf{L_2}(\mathbb{R^+})$ (ver (\ref{C-seq})).

\bs

Probemos ahora que $\chi(x)=\phi'(x)$. En efecto, para $x\leq 1$,
tenemos,
\begin{equation}\label{11}\begin{array}{c}
  \displaystyle{ \left| \phi(x)-\int_0^x \chi(y)\,dy \right|\leq }
  \\ \\
  \displaystyle{\leq \left| \phi(x)-\varphi_n(x) \right| +
  \left| \int_0^x \left(\chi(y)-\varphi'_n(y)\right)\,dy
  \right|\leq } \\ \\
  \displaystyle{\leq \left| \phi(x)-\varphi_n(x) \right| + \left|\left|
   \chi-\varphi'_n \right|\right| \rightarrow_{n \rightarrow \infty}
   0}\,.
\end{array}
\end{equation}
Entonces, $\phi(x)$ es una funci{\'o}n diferenciable cuya primera
derivada es $\chi(x)$.

\bs

Finalmente, las ecuaciones (\ref{4}) y (\ref{9}) implican que,
dado $\varepsilon_1>0$ y $\nu<1$,
\begin{equation}\label{5}
  \left| \phi(x) \right| < \varepsilon_1 \, x^{{\nu} }\quad {\rm
  and}\quad \left| \phi'(x) \right| < \varepsilon_1 \, x^{{\nu+1/2} }
\end{equation}
si $x<\delta$, para alg{\'u}n $\delta>0$ suficientemente
peque\~no. Esto prueba las expresiones (\ref{clau}).

\subsubsection{El operador adjunto}

Calcularemos ahora el dominio, la forma y las autofunciones del
operador adjunto $A^\dagger$ con el fin de determinar los
subespacios de deficiencia $\mathcal{K}_{\pm}$.

\bs

El operador $A^{\dagger}$ est{\'a} definido en el subespacio de
funciones $\psi (x)\in\mathbf{L_2}(\mathbb{R}^+)$ para las cuales
$(\psi, A \varphi)$ es una funcional continua de
$\varphi\in{\mathcal D}(A)$. Esto implica la existencia de una
funci{\'o}n $\tilde{\psi}(x)\in \mathbf{L_2}(\mathbb{R}^+)$ tal
que $(\psi, A \varphi)=(\tilde{\psi},\varphi), \forall\,
\varphi\in {\mathcal D}(A)$.

\bs

La funci{\'o}n $\tilde{\psi}$ est{\'a} un{\'\i}vocamente definida
pues ${\mathcal D}(A)$ es un subespacio denso de
$\mathbf{L_2}(\mathbb{R}^+)$. Definimos entonces $A^\dagger \psi
:= \tilde{\psi}$.

\bs

Por otra parte, si $\psi\in {\mathcal D}(A^\dagger)$ se satisface,
$\forall\,\varphi\in{\mathcal D}(A)$,
\begin{equation}\label{distrib}
  \begin{array}{c}
    \displaystyle{(\psi,A\varphi) = \int_0^{\infty}\psi(x)^*
    \left(-\varphi''(x)+\left[\frac{\nu^2-1/4}{x^2}+x^2\right]\,
    \varphi(x)\right)\, dx = }\\ \\
    \displaystyle{
    = \left(-\psi''+\left[\frac{\nu^2-1/4}{x^2}+x^2\right]\psi,\varphi \right) =
    (\tilde{\psi},\varphi)\,,}
  \end{array}
\end{equation}
donde las derivadas de $\psi$ corresponden a la derivada
generalizada en $\mathbf{L_2}(\mathbb{R}^+)$ o derivada en el
sentido de distribuciones.

\bs

La ecuaci{\'o}n (\ref{distrib}) implica que $\psi''(x)$ es
localmente integrable. Por consiguiente, su primitiva $\psi'(x)$
es absolutamente continua para $x>0$. En consecuencia, el dominio
de $A^\dagger$ es el espacio de funciones $\psi(x)$ de cuadrado
integrable con una derivada absolutamente continua y que
satisfacen \cite{Prop.2},
\begin{equation}\label{Hmas}
  A^\dagger\psi(x)= -\psi''(x)+\left(\frac{\nu^2-1/4}{x^2}+x^2\right)\,
   \psi(x)\in
\mathbf{L_2}(\mathbb{R}^+)\,,
\end{equation}
N{\'o}tese que esto no impone condici{\'o}n de contorno alguna
sobre $\psi(x)$ en $x=0$.

\bigskip

Estudiaremos ahora las autofunciones de $A^{\dagger}$ con el
prop{\'o}sito de determinar los sub\-es\-pa\-cios de deficiencia
${\mathcal K}_{\pm}= {\rm Ker}(A^\dagger \mp i )$ de $A$.

\subsubsection{Subespacios de deficiencia} \label{deficiency-sub}

Para calcular los subespacios de deficiencia debemos resolver el
problema de au\-to\-va\-lo\-res,
\begin{equation}\label{eigequ}
    A^\dagger\phi_{\lambda}=-\phi_\lambda''(x)+
    \left(\frac{\nu^2-1/4}{x^2}+x^2\right)\phi_\lambda(x)=\lambda
 \phi_{\lambda}\,,
\end{equation}
para $\phi_{\lambda}\in {\mathcal D}(A^\dagger)$ y
$\lambda\in\mathbb{C}$, con su parte imaginaria $\Im(\lambda)\neq
0$.

\bs

Mediante el siguiente Ansatz (sugerido por el comportamiento de la
soluciones de (\ref{eigequ}) para $x \rightarrow 0^+$ y
$x\rightarrow \infty$),
\begin{equation}
    \phi=x^{\nu+1/2}e^{-\frac{x^2}{2}}F(x^2)\,,
\end{equation}
la ecuaci{\'o}n (\ref{eigequ}) toma la forma de la ecuaci{\'o}n de
Kummer para $F(z)$,
\begin{equation}\label{Kum}
    zF''(z)+(b-z)F'(z)-aF(z)=0\,,
\end{equation}
donde $a={(2\nu+2-\lambda)}/{4}$ y $b=\nu+1$.

\bs

Como $\nu$ es real, la {\'u}nica soluci{\'o}n de la ecuaci{\'o}n
(\ref{Kum}) que conduce a una soluci{\'o}n de (\ref{eigequ}) de
cuadrado integrable en el infinito est{\'a} dada por la
funci{\'o}n de Kummer $U(a;b;z)$ (v{\'e}ase la ecuaci{\'o}n
(\ref{LI-sol-1}).) Por lo tanto, las autofunciones de $A^\dagger$
correspondientes al autovalor $\lambda$ son proporcionales a,
\begin{equation}\label{eigfun}
    \phi_{\lambda}(x)=x^{\nu +1/2}\, e^{-\frac{x^2}{2}}\,
    U\left(\frac{2\nu+2-\lambda}{4};
    \nu+1;x^2\right)\,.
\end{equation}

Debemos ahora estudiar el comportamiento de $\phi_{\lambda}$ cerca
del origen \cite{A-S}. Para ello consideraremos por separado los
siguientes casos, de acuerdo con el valor del par{\'a}metro $\nu$:
\begin{itemize}
\item Si $\nu\geq 1$, $\phi_{\lambda}\in {\mathbf L_2}({\mathbb
R}^+) \Leftrightarrow  a ={(2\nu +2-\lambda)}/{4}=-n$, con $n\in
\mathbb{N}$. En consecuencia, si $\lambda\notin \mathbb{R}$
entonces $\phi_{\lambda}\notin L_2(\mathbb{R}^+)$ y los
subespacios de deficiencia son triviales.

\bs

Esto implica que si $\nu\geq 1$ el operador $A$ es esencialmente
autoadjunto, esto es, admite una {\'u}nica extensi{\'o}n
autoadjunta que coincide con su clausura $\overline{A}$. Su
espectro est{\'a} dado por la condici{\'o}n $-a\in\mathbb{N}$,
{\it i.e.},
\begin{equation}\label{spec-ESA}
    \lambda_n=4n +2\nu+2\,,
\end{equation}
con $n=0,1,2,\ldots$ Las autofunciones correspondientes son,
\begin{equation}\label{eigenfunc}
    \phi_n=(-1)^n \, n! \,x^{\nu+1/2}e^{-\frac{x^2}{2}}
    L_n^{(\nu)}\left(x^2\right)\,.
\end{equation}

\item Por su parte, si $0\leq\nu<1$ se puede ver que
$\phi_{\lambda}\in L_2(\mathbb{R}^+), \forall
\lambda\in\mathbb{C}$ \cite{A-S}. Los subes\-pa\-cios de
deficiencia ${\mathcal K}_\pm$ son, entonces, unidimensionales
($n_\pm=1$) y el operador $A$ admite una familia de extensiones
autoadjuntas caracterizada por un par{\'a}metro
real\,\footnote{Esto responde al criterio de Weyl \cite{R-S}
seg{\'u}n el cual, para un potencial $V(x)$ continuo, el operador
$A=-\partial_x^2+V(x)$ es esencialmente autoadjunto si y s{\'o}lo
si est{\'a} en el caso de punto l{\'\i}mite tanto en el origen
como en infinito.

\medskip

Si $V(x)\geq M>0$, para $x$ suficientemente grande, entonces $A$
est{\'a} en el caso punto l{\'\i}mite en infinito. Por lo tanto,
en nuestro ejemplo, $A$ es esencialmente autoadjunto si y s{\'o}lo
si est{\'a} en el caso punto l{\'\i}mite en el origen.

\medskip

En particular, para $V(x)$ positivo, si $V(x)\geq 3/4\ x^{-2}$
para $x$ suficientemente cerca del origen entonces $A$ est{\'a} en
el caso de punto l{\'\i}mite en el origen. Por el contrario, si
$V(x)\leq (3/4-\eps)\ x^{-2}$, para alg{\'u}n $\eps>0$, entonces
$H$ est{\'a} en el caso c{\'\i}rculo l{\'\i}mite en el origen.

\medskip

Esto confirma nuestro resultado con respecto a las extensiones
autoadjuntas del operador $A$ en funci{\'o}n de los valores del
par{\'a}metro $\nu$.}.
\end{itemize}

\subsection{Extensiones autoadjuntas\label{SAE-H} }

Puesto que, para $0\leq\nu<1$, los {\'\i}ndices de deficiencia
satisfacen $n_{\pm}=1$, existe una familia de extensiones
autoadjuntas de $A$ en correspondencia biun{\'\i}voca con el
conjunto de isometr{\'\i}as de ${\mathcal K}_+$ en $\mathcal{K}_-$
que est{\'a}, por lo tanto, caracterizada por un par{\'a}metro
real.

\bs

En efecto, los subespacios de deficiencia ${\mathcal K}_+$ y
${\mathcal K}_-$ est{\'a}n generados por las funciones $\phi_+:=
\phi_{\lambda=i}$ y $\phi_-:= \phi_{\lambda=-i}=\phi_{+}^*$,
respectivamente. Por lo tanto, cada isometr{\'\i}a ${\mathcal
U}_{\gamma}:{\mathcal K}_+\rightarrow {\mathcal K}_-$ puede
identificarse con el par{\'a}metro $\gamma\in [0,\pi)$ definido
por,
\begin{equation}
     {\mathcal
    U}_{\gamma}     \phi_+ = e^{-2i\gamma}\phi_-\,.
\end{equation}
El operador autoadjunto correspondiente $A_{\gamma}$ est{\'a}
definido en el subespacio denso,
\begin{equation}
     {\mathcal D}(A_{\gamma})\subset     {\mathcal D}(A^{\dagger})=
     {\mathcal D}(\overline{A})\oplus     {\mathcal K}_+\oplus {\mathcal K}_- \,,
\end{equation}
constituido por las funciones $\phi\in{\mathcal D}(A^\dagger)$ que
tiene la forma,
\begin{equation}\label{phi-suma}
     \phi=\phi_0+C\, \left( \phi_+ +
    e^{-2i\gamma}\phi_+^* \right)\,,
\end{equation}
con $\phi_0\in {\mathcal D}(\overline{A})$ y $C$ una constante en
$\mathbb{C}$. Como $A_{\gamma}$ es una restricci{\'o}n de
$A^{\dagger}$ se verifica,
\begin{equation}\label{H-gamma}
  A_{\gamma}\,\phi = A^{\dagger}\, \phi =
  \overline{A}\,\phi_0 + i \,C\, \left( \phi_+ - e^{-2i\gamma}\phi_+^*
  \right).
\end{equation}

\bs

Mostraremos ahora que la condici{\'o}n (\ref{phi-suma}) determina
el comportamiento de $\phi\in {\mathcal D}(A_{\gamma})$ cerca del
origen. Esto permitir{\'a} determinar luego el espectro de cada
extensi{\'o}n autoadjunta. En adelante consideraremos el caso
$1/2\leq \nu<1$

\bs

De acuerdo con la ecuaci{\'o}n (\ref{phi-suma}), la derivada
logar{\'\i}tmica de $\phi$ est{\'a} dada por,
\begin{equation}\label{vNc}
\frac{\phi'}{\phi}=\frac{e^{i\gamma}\phi_0'+ 2A\,
\Re\left(e^{i\gamma}\phi_+'\right)}     {e^{i\gamma}\phi_0+2A\,
    \Re\left(e^{i\gamma}\phi_+\right)}\,.
\end{equation}
El l{\'\i}mite para $x\to 0^+$ de la ecuaci{\'o}n (\ref{vNc})
determina la condici{\'o}n de contorno que define el dominio de la
extensi{\'o}n autoadjunta caracterizada por el par{\'a}metro
$\gamma$. En virtud de las expresiones (\ref{clau}), el
t{\'e}rmino dominante para $x\sim 0$ est{\'a} dado por la
funci{\'o}n $\phi_+$. Por consiguiente, teniendo en cuenta el
comportamiento en el origen de la autofunci{\'o}n (\ref{eigfun})
para $\lambda=i$ \cite{A-S}, obtenemos de la ecuaci{\'o}n
(\ref{vNc}),
\begin{equation}\label{bc}
\begin{array}{c}\displaystyle{
    \left.\frac{\phi'(x)}{\phi(x)}\right|_{x\sim 0}=\frac{1/2-\nu}{x}+
    2\nu\frac{\Gamma
    (-\nu)}    {\Gamma (\nu)}     }
    \displaystyle{    \frac{\cos{(\gamma-\gamma_1)}}
    {\cos{(\gamma-\gamma_2)}}\cdot x^{2\nu-1}+
    o(x^{2\nu-1})}\,,
\end{array}
\end{equation}
donde $\gamma_1=\arg\left\{\Gamma[(-2\nu+2-i)/4]\right\}$ y
$\gamma_2=\arg\left\{\Gamma[(2\nu+2-i)/4]\right\}$.

\bs

La condici{\'o}n de contorno (\ref{bc}) caracteriza la
extensi{\'o}n autoadjunta correspondiente a un dado valor del
par{\'a}metro $\gamma$. Como es usual, esta condici{\'o}n de
contorno permite determinar el espectro del operador $A_{\gamma}$.
Como distintos valores de $\gamma$ corresponden a distintas
condiciones de contorno, o equivalentemente, a distintas
extensiones autoadjuntas, es de esperar que exista un espectro
distinto asociado a cada valor del par{\'a}metro $\gamma$.

\subsubsection{El espectro} \label{spectrum-SAE}

Para determinar el espectro de una extensi{\'o}n autoadjunta
$A_\gamma$ debemos estudiar las soluciones $\phi_{\lambda}$ de la
ecuaci{\'o}n (\ref{eigequ}), dadas por (\ref{eigfun}) con
$\lambda\in\mathbb{R}$, que satisfacen la condici{\'o}n de
contorno (\ref{bc}). El comportamiento en el origen de las
funciones (\ref{eigfun}) est{\'a} dado por \cite{A-S},
\begin{equation}\label{eigatori}
\begin{array}{c}
    \displaystyle{\left.\frac{\phi_{\lambda}'(x)}{\phi_{\lambda}(x)}
    \right|_{x\sim0}=
    \frac{1/2-\nu}{x}+2\nu   \frac{\Gamma (-\nu)}   {\Gamma
    (\nu)}}
   \displaystyle{ \frac{\Gamma
    \left[\frac{2\nu+2-\lambda}{4}\right]}   {\Gamma
    \left[\frac{-2\nu+2-\lambda}{4}\right]}\cdot    x^{2\nu-1}+
    o(x^{2\nu-1})\,. } \end{array}
 \end{equation}
Comparando las ecuaciones (\ref{bc}) y (\ref{eigatori}) obtenemos,
\begin{equation}\label{spe666}
    \frac{\Gamma \left(\frac{\nu}{2}+\frac{1}{2}-\frac{\lambda}{4}\right)}
    {\Gamma\left(-\frac{\nu}{2}+\frac{1}{2}-\frac{\lambda}{4}\right)}=
    \frac{\Gamma(\nu)}{\Gamma(-\nu)}\,\theta\,,
\end{equation}
donde hemos definido,
\begin{equation}\label{spe2}
    \theta:=\frac{\Gamma(-\nu)}{\Gamma(\nu)}\,\frac{\cos{(\gamma-\gamma_1)}}
    {\cos{(\gamma-\gamma_2)}}
    \,.
\end{equation}
Teniendo en cuenta que el comportamiento de las autofunciones
(\ref{eigfun}) en el origen est{\'a} dado por \cite{A-S},
\begin{equation}\label{eigfunen0}
    \phi_{\lambda}(x\sim 0)=
    \frac{\Gamma(\nu)}{\Gamma(\frac{\nu+1}{2}-\frac{\lambda}{4})}
    \,x^{-\nu +1/2}+
    \frac{\Gamma(-\nu)}{\Gamma(\frac{-\nu+1}{2}-\frac{\lambda}{4})}
    \,x^{\nu +1/2}+\ldots
    \,,
\end{equation}
puede verse que el par{\'a}metro $\theta$ que se define en
(\ref{spe2}) coincide con el par{\'a}metro $\theta$ definido en la
ecuaci{\'o}n (\ref{saesing}).

\bs

El par{\'a}metro $\theta\in \mathbb{R}\cup \{\infty\}$ determina,
mediante la ecuaci{\'o}n (\ref{spe666}), un espectro discreto para
cada extensi{\'o}n autoadjunta, que designaremos a partir de ahora
por $A^{\theta}$. En la Figura 2 se muestran ambos miembros de la
ecuaci{\'o}n (\ref{spe666}) como funci{\'o}n de $\lambda$, para
$\nu=3/5$ y $\theta=1$. Las absciss\ae\ de las intersecciones de
estas funciones representan el espectro de la extensi{\'o}n
autoadjunta correspondiente.
\begin{figure}\label{figspe222}
\center
    \epsffile{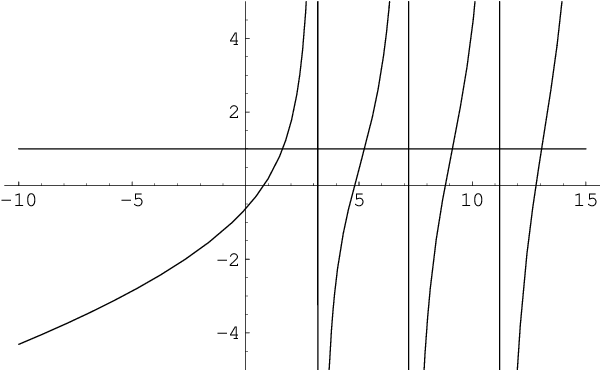} \caption{{\small $F(\lambda):=
    \frac{\Gamma(-\nu)}{\Gamma{(\nu)}}\frac{\Gamma
    \left(\frac{\nu}{2}+\frac{1}{2}-\frac{\lambda}{4}\right)}
    {\Gamma\left(-\frac{\nu}{2}+\frac{1}{2}-\frac{\lambda}{4}\right)}$
    como funci{\'o}n de $\lambda$, para $\nu=3/5$. Las soluciones de
    $F(\lambda)=\theta$ determinan el espectro de la extensi{\'o}n
    autoadjunta $A^{\theta}$.}}
\end{figure}

\bigskip

De acuerdo con la definici{\'o}n (\ref{simbolo}), el s{\'\i}mbolo
del operador $A^{\theta}$ est{\'a} dado por,
\begin{equation}
    \sigma\{A^\theta\}(x,p)=p^2+\frac{\nu^2-1/4}{x^2}+x^2\,.
\end{equation}
Si el t{\'e}rmino dominante del s{\'\i}mbolo de un operador
autoadjunto, en nuestro caso $p^2$, es positivo definido entonces
el espectro del operador est{\'a} acotado inferiormente
\cite{Gilkey1}. En efecto, el espectro de cada uno de los
operadores $A^{\theta}$ tiene, como puede apreciarse de la Figura
2, una cota inferior. Sin embargo, aquellas extensiones
autoadjuntas $A^\theta$ para las cuales,
\begin{equation}
\theta<-\frac{\Gamma(1-\nu)\Gamma(\frac{1+\nu}{2})}{\Gamma(1+\nu)
\Gamma(\frac{1-\nu}{2})}\,,
\end{equation}
poseen un autovalor negativo, a{\'u}n cuando el potencial
$\frac{\nu^2-1/4}{x^2}+x^2$ es estrictamente po\-si\-ti\-vo.
Incluso, no existe un l{\'\i}mite inferior uniforme, {\it i.e.},
com{\'u}n a todas las extensiones autoadjuntas; por el contrario,
cualquier n{\'u}mero negativo pertenece al espectro de alguna
extensi{\'o}n autoadjunta.

\bigskip

Para cualquier valor de $\nu$ existen dos extensiones particulares
cuyo espectro puede obtenerse expl{\'\i}citamente a partir de la
ecuaci{\'o}n (\ref{spe666}):
\begin{itemize}
\item {Si $\theta=0$  el espectro est{\'a} dado por
\begin{equation}\label{beta=0}
    \lambda_n=4(n+1/2-\nu/2),
\end{equation}
con $n=0,1,2,\ldots$}
  \item {Si $\theta=  \infty$  el espectro est{\'a} dado por
\begin{equation}\label{beta=-infty}
    \lambda_n=4(n+1/2+\nu/2),
\end{equation}
con $n=0,1,2,\ldots$}
\end{itemize}

\bs

N{\'o}tese que, en general, para cualquier valor de $\theta$ los
autovalores crecen linealmente con $n$,
\begin{equation}\label{linear-n}
  4(n-1/2+\nu/2)\leq \lambda_{n}\leq 4(n+1/2+\nu/2).
\end{equation}


 {\small
\subsubsection*{L{\'\i}mite regular}

Es interesante considerar el caso particularmente simple del
oscilador arm{\'o}nico en la se\-mi\-rrec\-ta, en el que el
potencial no es singular. En primer lugar, para $\nu=1/2$, la
condici{\'o}n de contorno (ecuaci{\'o}n (\ref{bc})) toma la forma,
\begin{equation}
    \frac{\phi'(x)}{\phi(x)}=\theta + O(x)\,,
\end{equation}
o equivalentemente,
\begin{equation}\label{robin}
    \lim_{x\rightarrow  0^+}\left\{\phi'(x)-\theta\,\phi(x)\right\}=0\,,
\end{equation}
que corresponde a las condiciones de contorno Robin en el origen.
Las condiciones de contorno Dirichlet y Neumann corresponden a
$\theta=\infty$ y $\theta=0$, respectivamente.

\bigskip

Consideremos ahora, a partir de la condici{\'o}n de contorno
(\ref{robin}), las autofunciones y au\-to\-va\-lo\-res de las
extensiones autoadjuntas de $A^{\theta}$ correspondientes a
distintos valores de $\theta$.

\begin{itemize}
\item Si $\theta=\infty$ las condiciones de contorno son del tipo
Dirichlet y los autovalores (v{\'e}ase la ecuaci{\'o}n
(\ref{beta=-infty})) est{\'a}n dados por,
\begin{equation}
    \lambda_n=4n+3\,,
\end{equation}
con $n=0,1,2,\ldots$ Como el operador (\ref{ham666}) corresponde,
para $\nu=1/2$, al hamiltoniano de un oscilador arm{\'o}nico de
masa $m=1/2$ y frecuencia $\omega=2$, los autovalores pueden
escribirse $\lambda_n=\omega[(2n+1)+1/2]$, que representa la parte
del espectro asociada a los autovectores impares del oscilador
arm{\'o}nico en $\mathbb{R}$. En efecto, las autofunciones
(\ref{eigfun}) est{\'a}n dadas por \cite{A-S},
\begin{equation}
    \phi_n=2^{-2n-1}e^{-\frac{x^2}{2}}H_{2n+1}(x)\,.
\end{equation}

\item Si $\theta=0$ las condiciones de contorno son del tipo
Neumann y los autovalores (v{\'e}ase (\ref{beta=0})) est{\'a}n
dados por,
\begin{equation}
    \lambda_n=4n+1\,,
\end{equation}
donde $n=0,1,2,\ldots$. Estos autovalores pueden escribirse como
$\lambda_n=\omega(2n+1/2)$, que corresponde a los autovalores del
sector de autofunciones pares del oscilador arm{\'o}nico en
$\mathbb{R}$. Las autofunciones (\ref{eigfun}) est{\'a}n dadas
por,
\begin{equation}
    \phi_n=2^{-2n}e^{-\frac{x^2}{2}}H_{2n}(x)\,.
\end{equation}

\item Si $\theta\neq 0,\infty$ las condiciones de contorno son del
tipo Robin y los autovalores est{\'a}n determinados por la
ecuaci{\'o}n trascendental,
\begin{equation}
    -2\,\frac{\Gamma \left(\frac{3-\lambda}{4}\right)}
    {\Gamma\left(\frac{1-\lambda}{4}\right)}=
    \theta\,.
\end{equation}
Las correspondientes autofunciones (\ref{eigfun}) est{\'a}n dadas
por,
\begin{equation}\label{autobor}
    \phi_{\lambda}=x \,e^{-\frac{x^2}{2}}U\left(\frac{3-\lambda}{4};
    \frac{3}{2};x^2\right)\,.
\end{equation}
Cabe observar que, para condiciones de contorno Robin y
$\theta<-2\,\Gamma(\frac{3}{4})/\Gamma(\frac{1}{4})$, la
energ{\'\i}a del estado fundamental es negativa y, por lo tanto,
menor que el m{\'\i}nimo del potencial.
\end{itemize}

\bigskip

De acuerdo con lo presentado en la secci{\'o}n (\ref{topo}), la
subvariedad de Cayley $\mathcal{C}_-$ est{\'a} identificada con la
condici{\'o}n de contorno Dirichlet. Si consideramos una familia
de extensiones autoadjuntas $A^\theta$ con $\theta\rightarrow
-\infty$, obtenemos la condici{\'o}n de contorno y el espectro
corr\-es\-pon\-dien\-tes a $\mathcal{C}_-$. De acuerdo con el
Teorema (\ref{topo}), el autovalor del estado fundamental tiende a
$-\infty$ junto con $\theta$ y la autofunci{\'o}n se concentra en
el borde de la variedad. La Figura 3 representa los estados
fundamentales correspondientes a cuatro extensiones $A^{\theta}$ y
muestra que estos estados tienden a un ``estado de borde'' a
medida que la extensi{\'o}n correspondiente se acerca a
$\mathcal{C}_-$, o equivalentemente, que el autovalor del estado
fundamental tiende a $-\infty$.

\begin{figure}\label{orde1}
\center
    \epsffile{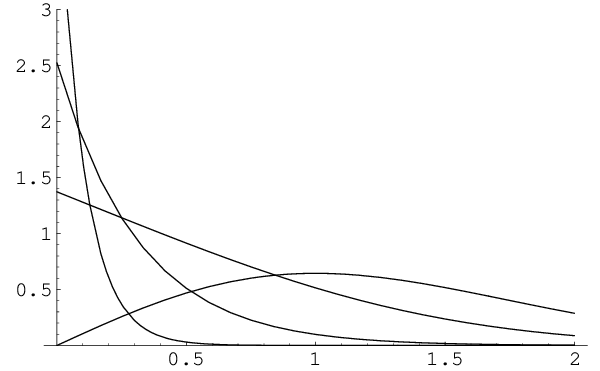} \caption{{\small Autofunciones de los
    estados fundamentales, de acuerdo con (\ref{autobor}) de
    cuatro distintas extensiones autoadjuntas del operador $A$,
    dado por (\ref{ham666}) para el caso $\nu=1/2$. Puede
    observarse que a medida que el autovalor tiende a $-\infty$, o
    que la condici{\'o}n de contorno tiende a la subvariedad de
    Cayley, $\mathcal{C}_-$ las autofunciones se concentran en el
    borde de la variedad. Los autovalores co\-rres\-pon\-dien\-tes
    a los estados graficados son $\lambda_0=3,0,-10,-100$. }}
\end{figure}}


\subsection{Estructura de polos de la funci{\'o}n-$\zeta$}\label{espoze}

\subsubsection{Representaci{\'o}n integral de la funci{\'o}n-$\zeta$.}
\label{integ-rep}

El espectro de cada extensi{\'o}n autoadjunta $A^\theta$ del
operador dado por la expresi{\'o}n (\ref{ham666}), est{\'a}
determinado por la ecuaci{\'o}n (\ref{spe666}), para cualquier
$\theta\in(-\infty,\infty]$. En esta secci{\'o}n, estudiaremos las
estructura de polos de la funci{\'o}n $\zeta^\theta_A(s)$ asociada
al operador $A^\theta$,
\begin{equation}\label{zeta-def}
  \zeta^{\theta}_A(s):= {\rm Tr}\left\{\left(A^{\theta}\right)^{-s}\right\} =
  \sum_{n} \lambda_{n}^{-s}\,.
\end{equation}
N{\'o}tese que, como la ecuaci{\'o}n (\ref{linear-n}) indica que
los autovalores crecen linealmente con $n$, la funci{\'o}n
$\zeta^\theta_A(s)$ es anal{\'\i}tica en el semiplano $\Re(s)>1$.

\bigskip

Es conveniente definir la funci{\'o}n entera,
\begin{equation}
    f(\lambda)=\frac{1}
    {\Gamma\left(-\frac{\nu}{2}+\frac{1}{2}-\frac{\lambda}{4}\right)}-
    \frac{\Gamma(\nu)}{\Gamma(-\nu)}\,
    \frac
    {\theta}{\Gamma \left(\frac{\nu}{2}+\frac{1}{2}-\frac{\lambda}{4}\right)}\,,
\end{equation}
con $\frac{1}{2}\leq \nu<1$. Los autovalores del operador
autoadjunto $A^{\theta}$ corresponden a los ceros de $f(\lambda)$
que son, por consiguiente, reales. Estos ceros son tambi{\'e}n
positivos a excepci{\'o}n, eventualmente, del primero de ellos.

\bigskip

Asimismo, los ceros de $f(\lambda)$ son simples. Supongamos que
esto no sea cierto, {\it i.e.}, que existe $\lambda\in \mathbb{R}$
tal que $f(\lambda)=f'(\lambda)=0$. Teniendo en cuenta que,
\begin{equation}\begin{array}{c}
  \displaystyle{f'(\lambda)=\frac{\psi\left(-\frac{\nu}{2}+\frac{1}{2}-
  \frac{\lambda}{4}\right)}
        {4\Gamma\left(-\frac{\nu}{2}+\frac{1}{2}-\frac{\lambda}{4}\right)}-
        \theta\,\frac{\Gamma(\nu)}{\Gamma(-\nu)}\,
        \frac{\psi\left(\frac{\nu}{2}+\frac{1}{2}-\frac{\lambda}{4}\right)}
        {4\Gamma\left(\frac{\nu}{2}+\frac{1}{2}-\frac{\lambda}{4}\right)}=
        }\\ \\{\displaystyle
  =\frac{1}{4}
            \left\{\frac{\left[\psi\left(-\frac{\nu}{2}+\frac{1}{2}-
            \frac{\lambda}{4}\right)-
            \psi\left(\frac{\nu}{2}+\frac{1}{2}-\frac{\lambda}{4}\right)\right]}
            {\Gamma\left(-\frac{\nu}{2}+\frac{1}{2}-\frac{\lambda}{4}\right)}+
            \psi\left(\frac{\nu}{2}+\frac{1}{2}-\lambda/4\right)f(\lambda)
            \right\}\,,}
\end{array}
\end{equation}
siendo $\psi(z)$ la funci{\'o}n poligamma, vemos, por
consiguiente, que,
\begin{equation}
    \psi\left(-\frac{\nu}{2}+\frac{1}{2}-\lambda/4\right)=
            \psi\left(\frac{\nu}{2}+\frac{1}{2}-\lambda/4\right)\,.
\end{equation}
Pero esto no se cumple para ning{\'u}n $\lambda\in \mathbb{R}$, si
$\frac12 \leq \nu <1$. N{\'o}tese que todos los residuos de la
funci{\'o}n $f'(\lambda)/f(\lambda)$ valen 1, de modo que su
estructura de singularidades coincide con la de la traza de la
resolvente ${\rm Tr}\,(A^\theta-\lambda)^{-1}$.

\bigskip

En consecuencia, la funci{\'o}n $\zeta_A^\theta$ admite una
representaci{\'o}n integral de la forma,
\begin{equation}\label{z-rep}
  \zeta_A^\theta(s)=\frac{1}{2\pi
  i}\oint_{\mathcal{C}}\lambda^{-s}\frac{f'(\lambda)}{f(\lambda)}+
  \theta(-\lambda_{0}) \lambda_{0}^{-s}\,,
\end{equation}
donde $\mathcal{C}$ es una curva que encierra los ceros positivos
de $f(\lambda)$ en sentido antihorario y $\theta(\cdot)$ es la
funci{\'o}n de Heaviside.

\bigskip

Consideremos ahora el comportamiento asint{\'o}tico dominante del
cociente,
\begin{equation}\label{int666}
\begin{array}{c}\displaystyle{
  \frac{f'(\lambda)}{f(\lambda)} } \displaystyle{ =
              \frac{\left[\psi\left(-\frac{\nu}{2}+\frac{1}{2}-
              \frac{\lambda}{4}\right)-
            \psi\left(\frac{\nu}{2}+\frac{1}{2}-\frac{\lambda}{4}\right)\right]}
            {4\left(1-\theta\,\frac{\Gamma(\nu)}{\Gamma(-\nu)}
            \,\frac{\Gamma\left(-\frac{\nu}{2}+\frac{1}{2}-\frac{\lambda}{4}
            \right)}
            {\Gamma\left(\frac{\nu}{2}+\frac{1}{2}-\frac{\lambda}{4}\right)}
            \right)}+
            \frac{1}{4}\,
            \psi\left(\frac{\nu}{2}+\frac{1}{2}-\lambda/4\right)
             }\,.
\end{array}
\end{equation}
Para $|\arg(-\lambda)|<\pi$ y $|\lambda|\rightarrow \infty$,
\begin{eqnarray}
    \psi\left(\frac{\nu}{2}+\frac{1}{2}-\lambda/4\right)=\log{(-\lambda)}+O(1),
    \label{fun1}\\ \nonumber \\
    \psi\left(-\frac{\nu}{2}+\frac{1}{2}-\lambda/4\right)-
    \psi\left(\frac{\nu}{2}+\frac{1}{2}-\lambda/4\right)=O(\lambda^{-1}),
    \label{fun2}\\ \nonumber \\ \label{fun3666}
    \frac{\Gamma\left(-\frac{\nu}{2}+\frac{1}{2}-\frac{\lambda}{4}\right)}
            {\Gamma\left(\frac{\nu}{2}+\frac{1}{2}-\frac{\lambda}{4}\right)}=
            O(\lambda^{-\nu})\,.
\end{eqnarray}
En consecuencia, si $\Re(s)>1$, el camino de integraci{\'o}n en la
expresi{\'o}n (\ref{z-rep}) puede hacerse coincidir con el eje
imaginario,
\begin{equation}\label{z-rep-imag}
  \zeta_A^\theta(s)=-\frac{1}{2\pi
  i}\int_{-i\infty+0}^{i\infty+0}\lambda^{-s}
  \frac{f'(\lambda)}{f(\lambda)}\, d\lambda+
  h(s)\,,
\end{equation}
donde $h(s)$ es una funci{\'o}n entera proveniente de la
contribuci{\'o}n de un eventual autovalor negativo.

\subsubsection{Polos de la funci{\'o}n-$\zeta$} \label{pole-structure}

La integral en (\ref{z-rep-imag}), que es una funci{\'o}n
anal{\'\i}tica en el semiplano $\Re(s)>1$ que admite una
extensi{\'o}n meromorfa a todo el plano complejo $s$, puede
expresarse como,
\begin{equation}\begin{array}{c}
        \displaystyle{
        \zeta_A^\theta (s)=
        -\frac{1}{2\pi i}\int_{i}^{i\infty}
        \frac{f'(\lambda)}{f(\lambda)}\,
        \lambda^{-s}\,d\lambda\, }
        \displaystyle{
        -\frac{1}{2\pi i}
        \int_{-i\infty}^{-i} \frac{f'(\lambda)}{f(\lambda)}\,
        \lambda^{-s}\,d\lambda + h_1(s)=} \\ \\
        \displaystyle{
        =  -\frac{e^{-i s \pi/2}}{2\pi}  \int_{1}^{\infty}
        \frac{f'(i\mu)}{f(i\mu)}\,
        \mu^{-s}\,d\mu  }
        \displaystyle{
        -\frac{e^{i s \pi/2}}{2\pi}
        \int_{1}^{\infty} \frac{f'(-i\mu)}{f(-i\mu)}\,
        \mu^{-s}\,d\mu + h_1(s)}\,,\label{zetint666}\
\end{array}
\end{equation}
donde $h_1(s)$ es una funci{\'o}n entera.

\bigskip

El desarrollo asint{\'o}tico de $f'(\lambda)/f(\lambda)$
(v{\'e}ase el Ap{\'e}ndice (\ref{asymptotic})) est{\'a} dado por,
\begin{eqnarray}\label{te}\begin{array}{c}
    \displaystyle{
  \frac{f'(\lambda)}{f(\lambda)}\sim
    \frac{1}{4}\log{(-\lambda)}+
    \frac{1}{4}
    \sum_{k=0}^\infty c_k(\nu)\,(-\lambda)^{-k}+ }\\    \\
    \displaystyle{
    +\sum_{N=1}^\infty \sum_{n=0}^\infty
    C_{N,n}(\nu,\theta)\,(-\lambda)^{-N\nu-2n-1}}\,,
\end{array}
\end{eqnarray}
donde los coeficientes $c_k(\nu)$ son polinomios en $\nu$ cuya
forma expl{\'\i}cita no presentamos, en tanto que los coeficientes
$C_{N,n}(\nu,\theta)$ est{\'a}n dados por,
\begin{equation}\label{CNn}\begin{array}{c}
  C_{N,n}(\nu,\theta)
  \displaystyle{
  =-\left( 4^{\nu}\frac{\Gamma(\nu)}{\Gamma(-\nu)}\,\theta \right)^N \left(
  {\nu+\frac{2n}{N} }\right) b_{n}(\nu,N)}\,.
\end{array}
\end{equation}
Los coeficientes $b_{n}(\nu,N)$ se definen en la ecuaci{\'o}n
(\ref{bs-as}).

\bs

Como puede verse de la ecuaci{\'o}n (\ref{te}), el desarrollo
asint{\'o}tico de $f'(\lambda)/f(\lambda)$ contiene el t{\'e}rmino
logar{\'\i}tmico $\frac 1 4 \log(-\lambda)$ y una serie de
potencias enteras negativas de $\lambda$ relacionadas con la
funci{\'o}n poligamma del {\'u}ltimo t{\'e}rmino del miembro
derecho de la ecuaci{\'o}n (\ref{int666}). Existe tambi{\'e}n una
serie de potencias de $\lambda$ que dependen de $\nu$ provenientes
del primer t{\'e}rmino del miembro derecho de (\ref{int666}).

\bigskip


Reemplazando el t{\'e}rmino logar{\'\i}tmico dominante del
desarrollo (\ref{te}) en la ecuaci{\'o}n (\ref{zetint666})
obtenemos,
\begin{equation}\label{dominant}\begin{array}{c}
    \displaystyle{
  -\frac{1}{8 \pi }\int_{1}^{\infty}\left[e^{-i\frac{\pi s}{2}}
        \log{(e^{-i\frac{\pi}{2}}\mu)}+e^{i\frac{\pi s}{2}}
        \log{(e^{i\frac{\pi}{2}}\mu)}\
        \right]\mu^{-s}
        \,d\mu= }\\ \\
        \displaystyle{
  = \frac{\sin (\frac{\pi \,s}{2})}{8\,\left(s -1  \right) } -
  \frac{\cos (\frac{\pi \,s}{2})}
  {4\,\pi \,{\left( s  -1 \right) }^2}
  =\frac{1}{4} \frac{1}{(s-1)}+ h_2(s)}\,,
\end{array}
\end{equation}
donde $h_2(s)$ es una funci{\'o}n entera. En consecuencia, la
extensi{\'o}n anal{\'\i}tica de este t{\'e}rmino presenta un
{\'u}nico polo simple en,
\begin{equation}\label{1/4}
 s=1\,,
\end{equation}
con un residuo igual a $1/4$.

\bigskip

Los t{\'e}rminos restantes en la expresi{\'o}n asint{\'o}tica de
$f'(\lambda)/f(\lambda)$ tienen la forma $A_j$ $(-\lambda)^{-j}$,
para alg{\'u}n $j\geq 0$  (ver (\ref{te}).) Reemplazando estas
potencias en la ecuaci{\'o}n (\ref{zetint666}) obtenemos,
\begin{equation}\label{next-terms}\begin{array}{c}
    \displaystyle{
  -\frac{A_j}{2\pi }\int_{1}^{\infty}\left[
        e^{-i\frac{\pi }{2}(s-j)}
        +e^{i\frac{\pi }{2}(s-j)}
        \right]\, \mu^{-s-j}
        \,d\mu= }\\ \\
        \displaystyle{
  =-\frac{A_j}{\pi }
  \cos\left(\frac{\pi}{2} (s-j)\right) \ \frac{1}{s-(1-j)}= }\\ \\
  \displaystyle{
  =-\frac{A_j\, \sin(\pi j)}{\pi}\,
  \frac{1}{s-(1-j)}+ h_3(s)}\,,
\end{array}
\end{equation}
donde $h_3(s)$ es una funci{\'o}n entera. Por lo tanto, cada
potencia en el desarrollo asint{\'o}tico $f'(\lambda)/f(\lambda)$
de la forma $A_j\,(-\lambda)^{-j}$ contribuye con un polo simple
en $s=1-j$, con re\-si\-duo $\displaystyle{- (A_j/\pi)\, \sin(\pi
j)}$, al conjunto de singularidades de la funci{\'o}n
$\zeta_A^\theta(s)$.

\bigskip

Debe observarse que este residuo se anula para valores enteros de
$j$. Esto sucede para las contribuciones provenientes del
desarrollo asint{\'o}tico de $\psi(\frac{\nu}{2}+\frac{1}{2} -
\lambda/4)$ en el {\'u}ltimo t{\'e}rmino del miembro derecho de la
ecuaci{\'o}n (\ref{int666}), con excepci{\'o}n del t{\'e}rmino
logar{\'\i}tmico. De hecho, la contribuci{\'o}n de este
t{\'e}rmino origina la {\'u}nica singularidad presente en los
casos $\theta=\infty$ y $\theta=0$ (ver (\ref{int666}).)

\bs

No obstante, para una extensi{\'o}n autoadjunta general, existen
tambi{\'e}n, para $\frac{1}{2}\leq \nu<1$, polos en posiciones
dependientes de $\nu$ provenientes de las potencias de la serie
del {\'u}ltimo t{\'e}rmino del miembro derecho (\ref{te}).

\bigskip

En conclusi{\'o}n, aparte del polo en $s=1$ con residuo $1/4$, la
funci{\'o}n $\zeta_A^\theta(s)$ de la extensi{\'o}n autoadjunta
$A^\theta$ presenta, para cada par de enteros,
\begin{equation}\label{Nn}
  (N,n)\quad {\rm con}\ N=1,2,3,\ldots\ {\rm y}\
  n=0,1,2,\ldots
\end{equation}
un polo simple en,
\begin{equation}\label{pop666}
    s_{N,n}=-N\nu-2n\in(-N-2n,- N/ 2 -2n]\,,
\end{equation}
con residuo,
\begin{equation}\label{res666}
  \left. {\rm Res}\,\{ \zeta_A^\theta(s)\} \right|_{s=-N\nu-2n} =
    \, C_{N,n}(\nu,\theta)\, \frac{\sin(\pi N\nu) }{\pi} \,,
\end{equation}
N{\'o}tese que el residuo en el polo $s_{N,n}$ es proporcional a
$\theta^N$ (ver (\ref{CNn}).)

\bs

En rigor, si $\nu$ es un n{\'u}mero racional, existe un n{\'u}mero
finito de pares $(N,n)$ que contribuyen al mismo polo y el residuo
debe obtenerse sumando las contribuciones co\-rres\-pon\-dien\-tes
a cada uno de estos pares. Por el contrario, si $\nu$ es
irracional, los polos correspondientes a distintos pares $(N,n)$
no coinciden y el residuo est{\'a} dado por la expresi{\'o}n
(\ref{res666}).

\bs

Los polos dados por (\ref{pop666}) est{\'a}n distribuidos en
sucesiones caracterizadas por un entero $N=1,2,\ldots$. En cada
sucesi{\'o}n, polos contiguos difieren en $-2$. Por ejemplo, los
polos $s_{1,n}$ correspondientes a los pares $(N=1,n)$, con $n
=0,1,2,\ldots$, est{\'a}n ubicados, de acuerdo con (\ref{pop666}),
en los puntos del plano complejo,
\begin{equation}
   -1 -2n<s_{1,n}=-\nu-2n \leq -\frac 1 2 -2n\,,
\end{equation}
y tienen residuos,
\begin{equation}
  \left.{\rm Res}\,(\cdot)\right|_{s=-\nu-2n}=\frac{C_{1,n}
  (\nu,\theta)}{\pi}
  \,  \sin(\pi\nu)\,.
\end{equation}
Para $n=0$, {\it e.g.}, obtenemos un polo en $s=-\nu$ con residuo,
\begin{equation}
    \left.{\rm Res}\,(\cdot)\right|_{s=-\nu}=\frac{C_{1,0}
  (\nu,\theta)}{\pi}
  \,  \sin(\pi\nu)=\frac{4^{\nu}}{\Gamma^2(-\nu)}\,\theta \,,
\end{equation}
que coincide con el valor calculado en (\ref{alfin}). Se confirman
entonces, para el caso del potencial $V(x)=x^2$, los resultados de
la secci{\'o}n \ref{nocom}; v{\'e}ase, en particular la
ecuaci{\'o}n (\ref{toconf}).

\bigskip

Este resultado ilustra la idea central de esta Tesis: la
funci{\'o}n-$\zeta$ correspondiente a las extensiones autoadjuntas
de $A$ posee polos simples en posiciones dependientes de $\nu$,
que, en general, no son enteros negativos. Los residuos dependen,
por su parte, de la extensi{\'o}n autoadjunta considerada.

\bigskip

Destacamos, finalmente, que un polo de $\zeta_A^\theta(s)$ en un
valor no entero $s=-N\nu-2n$ implica que el desarrollo
asint{\'o}tico a peque\~nos valores de $t$ de la traza del
heat-kernel $\displaystyle{{\rm Tr}\left\{e^{-t\,
A_{(\beta)}}\right\}}$ presenta un t{\'e}rmino de la forma,
\begin{equation}\label{heato}
     {b_{N,5n}(A)}\,\theta^N\,t^{\nu\,N+2n}\,,
\end{equation}
cuyo coeficiente est{\'a} dado por,
\begin{equation}\label{heat-coef}
    {b_{N,5n}(A)}\,\theta^N=
    \Gamma(-N\nu-2n)\,
    \left. {\rm Res}\{ \zeta_A^\theta(s)\}
    \right|_{s=-N\nu-2n}\,.
\end{equation}
Esto est{\'a} de acuerdo con el desarrollo asint{\'o}tico
(\ref{asin}).

\subsection{Comportamiento asint{\'o}tico de los autovalores}\label{coasau}

La singular estructura de polos de la funci{\'o}n
$\zeta_A^\theta(s)$ dada por la ecuaci{\'o}n (\ref{pop666}) puede
obtenerse tambi{\'e}n determinando, a partir de (\ref{spe666}), el
desarrollo asint{\'o}tico de los au\-to\-va\-lo\-res $\lambda_{n}$
para $n\gg 1$. En efecto, dado el Ansatz,
\begin{equation}\label{ansatz}
    \frac{\lambda_{n}}{4}=\frac 1 2 -\frac \nu 2+n+\varepsilon\,,
\end{equation}
podemos determinar $\varepsilon$ mediante sucesivas correcciones.
Los primeros t{\'e}rminos resultan,
\begin{equation}\label{asymp-eigen}\begin{array}{c}
    \displaystyle{
   \frac{\lambda_{n}}{4}\sim \frac 1 2 -\frac \nu 2+n -
   \frac{\Gamma(\nu)}{\Gamma(-\nu)}\,
   \frac{\theta }
  {\pi }\,\sin (\pi\nu )\, n^{-\nu } -
   } \\ \\
  \displaystyle{ -\frac{\Gamma(\nu)}{\Gamma(-\nu)}\,
  \frac{\theta }{
    2\pi } \,\nu\left(\nu-1
      \right) \,\sin (\pi\nu )\, n^{-\nu-1} - }\\ \\
  \displaystyle{-\frac{\Gamma^2(\nu)}{\Gamma^2(-\nu)}\,\frac{ {\theta }^2 }
  {2\,\pi }\,
      \sin (2\,\pi\nu )  \, n^{-2\nu}+O(n^{-2})}\,.
\end{array}
\end{equation}
Esto permite escribir la funci{\'o}n $\zeta_A^\theta(s)$ de la
forma,
\begin{equation}\label{estres}\begin{array}{c}
    \displaystyle{
  \zeta_A^\theta(s)
  \sim 4^{-s}\, \zeta_R(s)+s\, 4^{-s}\, \left(\frac \nu 2 -\frac 1 2\right)\,
  \zeta_R( s+1)\, +} \\ \\
  \displaystyle{+  s\,
    \left(s + 1 \right)\, 4^{-s}
  \frac{{\left(\frac \nu 2 -\frac 1 2\right)}^2\, }{2}\,\zeta_R( s+2 ) + }\\ \\
  \displaystyle{
  +s\,4^{-s}\, \frac{\Gamma(\nu)}{\Gamma(-\nu)}\,\frac{\theta }{\pi }\,
      \sin \left(  \pi\nu\right) \,
      \zeta_R(s+\nu+1 )+ }\\ \\
      \displaystyle{ +s\,
      \left( s + \nu+1  \right)4^{-s}\,\frac{\Gamma(\nu)}{\Gamma(-\nu)}\,
      \frac{\theta }{
      \pi } \,\left(\frac \nu 2 -\frac 1 2\right)  \,
      \sin (\pi \nu )\,
      \zeta_R(2 + s + \nu)\, + }\\ \\
      \displaystyle{ + s\, 4^{-s}\,\frac{\Gamma^2(\nu)}{\Gamma^2(-\nu)}\,
      \frac{{\theta }^2}{2\,\pi } \,
    \sin (2\pi \nu )\,
    \zeta_R( s+1  + 2\nu ) \, +  \ldots}
\end{array}
\end{equation}
donde $\zeta_R(z)$ es la funci{\'o}n-$\zeta$ de Riemann, que posee
un {\'u}nico polo simple en $z=1$, con residuo igual a $1$. Los
polos que se derivan de la expresi{\'o}n (\ref{estres}) coinciden
con los primeros polos dados por las ecuaciones (\ref{1/4}) y
(\ref{pop666}).

\subsection{Casos particulares} \label{particular}

En esta secci{\'o}n mostraremos que para las extensiones
autoadjuntas cuyos dominios son invariantes de escala,
caracterizadas por $\theta=0$ y $\theta=\infty$, la
funci{\'o}n-$\zeta$ presenta un {\'u}nico polo simple. Veremos
luego que nuestros resultados se reducen a los usuales para el
caso $\nu=0$, en el que el potencial no es singular.

\subsubsection{Las extensiones autoadjuntas invariantes de escala}

La funciones $\zeta_A^0$ y $\zeta_A^{\infty}$ pueden calcularse en
forma exacta pues conocemos sus espectros expl{\'\i}citamente
(v{\'e}anse las ecuaciones (\ref{beta=0}) y (\ref{beta=-infty})),
\begin{equation}\label{zetas}
  \begin{array}{c}
  \displaystyle{
    \zeta_A^0(s)=4^{-s} \sum_{n=0}^\infty
    (n+\frac 1 2 -\frac \nu 2)^{-s}=4^{-s}\zeta_H(s,(1-\nu)/ 2)\,,}\\ \\
  \displaystyle{
    \zeta_A^{\infty}(s)=4^{-s} \sum_{n=0}^\infty
    (n+\frac 1 2 \frac \nu 2)^{-s}=4^{-s}\zeta_H(s,(1+\nu)/ 2)\,,}
  \end{array}
\end{equation}
donde $\zeta_H(z,q)$ es a funci{\'o}n-$\zeta$ de Hurwitz cuya
extensi{\'o}n anal{\'\i}tica tiene un {\'u}nico polo simple en
$z=1$, con residuo Res\,$\zeta(z,q)|_{z=1}=1$. Esto implica que
tanto $\zeta_A^0$ como $\zeta_A^{\infty}$ presentan un {\'u}nico
polo simple en $s=1$, con residuo $1/4$, en coincidencia con la
ecuaci{\'o}n (\ref{1/4}).

\bs

Efectivamente, a partir de las ecuaciones (\ref{res666}) y
(\ref{CNn}) es evidente que todos los re\-si\-duos
correspondientes a los polos negativos se anulan para $\theta=0$.
Por otro lado, si $\theta=\infty$, $f'(\lambda)/f(\lambda)$ se
reduce a $1/4\ \psi(\nu/2+1/2-\lambda/4)$ (v{\'e}ase la
ecuaci{\'o}n (\ref{int666})) por lo que el {\'u}nico t{\'e}rmino
que contribuye a las singularidades de la funci{\'o}n-$\zeta$ es
el logaritmo del desarrollo asint{\'o}tico (\ref{te}), que conduce
a un polo en $s=1$ con residuo 1/4 (v{\'e}ase la ecuaci{\'o}n
(\ref{1/4}).)

\subsubsection{El oscilador arm{\'o}nico en la semirrecta}

Para el oscilador arm{\'o}nico en la semirrecta, correspondiente
al caso $\nu=1/2$, existe un polo simple en $s=1$, con residuo
$1/4$. Como ya hemos se\~nalado, esta es la {\'u}nica singularidad
presente si se imponen condiciones de contorno del tipo de
Dirichlet o de Neumann.

\bs

Para condiciones de contorno del tipo de Robin las singularidades
restantes se encuentran en los puntos (v{\'e}ase la ecuaci{\'o}n
(\ref{pop666})),
\begin{equation}
    s=-\frac{N}{2}-2n\,,\qquad {\rm con}\ \ N=1,2,3,\ldots\qquad\ n=0,1,2,\ldots
\end{equation}
con residuos dados por (v{\'e}ase la ecuaci{\'o}n (\ref{res666})),
\begin{equation}\begin{array}{c}
    \displaystyle{
  \left. {\rm Res}\,( \cdot )\right|_{s=-\frac{N}{2}-2n} = }\\ \\
  \displaystyle{
  =\frac{(-1)^{N}}{\pi}\, \,C_{N,n}\left(\nu =
    1/2,\beta \right) \, \sin\left(\frac{3\,\pi }{2}N\right) }\,,
\end{array}
\end{equation}
que se anulan para $N$ par.

\bigskip

Por lo tanto, los polos de la funci{\'o}n-$\zeta$, a excepci{\'o}n
del primero, en $s=1$, son semienteros negativos,
\begin{equation}
  s=-k-1/2,\ k=0,1,2.\ldots
\end{equation}
en acuerdo con el resultado (\ref{resu}).

\bs

Por otra parte, para todos aquellos pares $(N,n)$ que verifiquen
$N+4n=2k+1$, los polos correspondientes coinciden. Por
consiguiente, el residuo de $\zeta_A^\theta(s)$ en $s=-k-1/2$ es
la suma de estas contribuciones, caracterizadas por $N=2(k-2n)+1$,
con $n=0,1,2,\dots,[k/2]$. Obtenemos, entonces,
\begin{equation}\label{suma-residuos}\begin{array}{c}
  \displaystyle{
  \left. {\rm Res}\,( \zeta_A^\theta(s) )\right|_{s= -k - \frac 1 2} }
  \displaystyle{ =
   \frac{(-1)^{k+1
   }}{\pi}\,\sum_{n=0}^{[k/2]}  \,C_{[2(k-2n)+1],n}\left(
    1/2,\theta \right)  }\,.
\end{array}
\end{equation}
Existe, por ejemplo, un polo en $s=-1/2$ cuyo residuo est{\'a}
dado por,
\begin{equation}
  \left. {\rm Res}\,( \zeta_A^\theta(s) )\right|_{s= - \frac 1 2} =
  - \frac{1}{\pi} \,C_{1,0}\left(
    1/2,\theta \right)=-\frac{\theta }{2\pi}\,,
\end{equation}
en tanto que el residuo en $s=-3/2$ est{\'a} dado por,
\begin{equation}
  \left. {\rm Res}\,( \zeta_A^\theta(s) )\right|_{s= - \frac 3 2} =
  \frac{1}{\pi} \,C_{3,0}\left(
    1/2,\theta \right)=\frac{\theta^3}{2\pi}\,.
\end{equation}

\section{{Un operador de Schr\"odinger en una variedad de base compacta}}
\label{Solari}

Estudiaremos a continuaci{\'o}n el operador de Schr\"odigner dado
por (\ref{1}) co\-rres\-pon\-dien\-te al caso $V(x)=0$ sobre la
variedad de base unidimensional y compacta $[0,1]\in\mathbb{R}$.
De acuerdo con los resultados de la secci{\'o}n \ref{sec1} el
operador diferencial admite una familia de extensiones
autoadjuntas caracterizada por dos par{\'a}metros $\theta$ y
$\beta$; el primero de ellos caracteriza la condici{\'o}n de
contorno en la singularidad y el segundo est{\'a} relacionado con
condiciones de contorno del tipo Robin en $x=1$. En la presente
secci{\'o}n nos limitaremos, por simplicidad, al caso
$\beta=\infty$, que define condiciones de contorno de Dirichlet en
$x=1$.

\bs

En primer lugar, verificaremos el Teorema \ref{elthmenun} dando
una nueva interpretaci{\'o}n del factor $K(\lambda)$ (v{\'e}ase la
ecuaci{\'o}n (\ref{elthm}).) Como hemos ya demostrado, existen dos
extensiones autoadjuntas definidas por condiciones de contorno
sobre la singularidad invariantes de escala. Veremos adem{\'a}s
que cualquier extensi{\'o}n autoadjunta puede escribirse como una
combinaci{\'o}n lineal de estas dos extensiones particulares. El
factor $K(\lambda)$ determina los coeficientes de esta
combinaci{\'o}n lineal. El valor de $K(\lambda)$ que obtendremos
confirma el resultado (\ref{cha}), para el caso $V(x)=0$ y
$\beta=\infty$.

\subsection{El operador y sus extensiones autoadjuntas}
\label{the-operator}

Consideremos el operador,
\begin{equation}\label{D}
  A=-\partial_x^2+\frac{\nu^2-1/4}{x^2}\, ,
\end{equation}
con $\nu\in\mathbb{R}$ definido sobre el conjunto $\mathcal{D}(A)=
\mathcal{C}_0^\infty(0,1)$, sobre el cual $A$ es sim{\'e}trico.

\bs

El operador adjunto $A^\dagger$, que es la extensi{\'o}n maximal
de $A$, est{\'a} definido sobre el conjunto
$\mathcal{D}(A^\dagger)$ de funciones $\phi(x)\in
\mathbf{L_2}(0,1)$, que tienen una derivada segunda localmente
sumable y tales que,
\begin{equation}\label{DPhi}
    A^\dagger\phi(x)=-\phi''(x)+ \frac{\nu^2-1/4}{x^2}\,
    \phi(x)=\tilde{\phi}(x)\in \mathbf{L_2}(0,1)\, .
\end{equation}

El siguiente lema describe el comportamiento en el origen de las
funciones pertenecientes a $\mathcal{D}(A^\dagger)$.

\begin{lem} \label{Lema1-1}
Si $\phi(x)\in \mathcal{D}(A^\dagger)$ y $0<\nu<1$,
entonces\,\footnote{El caso $\nu=0$ ser{\'a} considerado
separadamente en la secci{\'o}n \ref{nu=0}.} existen constantes
$C_1[\phi]$ y $C_2[\phi]$ tales que,
\begin{equation}\label{lemaI-1}
    \left|\,\phi(x)-\frac{C_1[\phi]\, x^{\nu+\frac 1 2} +
  C_2[\phi]\, x^{-\nu+\frac 1 2}}{\sqrt{2\nu}}\,\right|\,
  \leq \frac{\|A\phi\|}{(1 -\nu)\sqrt{2\nu+\frac 3 2}} \  x^{3/2}\,,
\end{equation}
y
\begin{equation}\label{lemaI-2}
      \left|\,\phi'(x)-\frac{(\nu+\frac 1 2)\, C_1[\phi]\, x^{\nu-\frac 1 2} +
  (-\nu+\frac 1 2)\, C_2[\phi]\, x^{-\nu-\frac 1 2}}{\sqrt{2\nu}}\,\right|\,
    \leq
  \frac{3/2 \, \|A\phi\|}{(1-\nu)\sqrt{2\nu+2}} \  x^{1/2}\,,
\end{equation}
donde $\| \cdot\|$ es la norma usual en $\mathbf{L_2}([0,1])$.
\end{lem}

\noindent {\bf Demostraci{\'o}n:} Definimos $u(x):= x^{-\nu-\frac
1 2} \phi(x)$. La ecuaci{\'o}n (\ref{DPhi}) implica entonces,
\begin{equation}\label{phi-chi-en0}
    \begin{array}{c} \displaystyle{
      u'(x)= K_2 \, x^{-2\nu-1} - x^{-2\nu-1} \, \int_0^x y^{\nu+\frac 1 2}\,
      \tilde{\phi}(y)\, dy
     \,  ,} \\ \\ \displaystyle{
     u(x)=K_1- \frac{K_2}{2\nu} \, x^{-2\nu} -
      \int_0^x y^{-2\nu-1} \, \int_0^y z^{\nu+\frac 1 2}\, \tilde{\phi}(z)\,
      dz \ dy
     \, ,}
    \end{array}
\end{equation}
para algunas constantes $K_1$ y $K_2$. A partir de las
desigualdades,
\begin{equation}\label{schwarz}
    \begin{array}{c}\displaystyle{
      \left|\int_0^x y^{\nu+\frac 1 2} \, \tilde{\phi}(y)\, dy\right|\leq
      \frac{x^{\nu+1}}{\sqrt{2\nu+2}}\, \|\tilde{\phi}\| }\, ,
      \\ \\ \displaystyle{
      \left|\int_0^x y^{-2\nu-1} \int_0^y z^{\nu+\frac 1 2}\,
      \tilde{\phi}(z)\, dz\, dy\right|\leq
      \frac{x^{1-\nu}}{(1-\nu)\sqrt{2\nu+2}}\, \|\tilde{\phi}\| }\,  ,
    \end{array}
\end{equation}
obtenemos las ecuaciones (\ref{lemaI-1}) y (\ref{lemaI-2}).\fin
\begin{cor}
Sean $\phi(x),\psi(x)\in \mathcal{D}(A^\dagger)$ y\ $0<\nu<1$.
Entonces
\begin{equation}\label{DDstar}\begin{array}{c}
    \left(A^\dagger \psi, \phi\right) -
  \left(\psi, A^\dagger \phi\right) = \\ \\
  = \Big\{
  C_1[\psi]^* C_2[\phi] - C_2[\psi]^* C_1[\phi]\Big\}+
  \Big\{\psi(1)^*\, \phi'(1)
  - \psi'(1)^*\, \phi(1) \Big\}\, ,
\end{array}
\end{equation}
donde las constantes $C_{1,2}[\cdot]$ son las definidas por el
Lema (\ref{Lema1-1}).
\end{cor}

\noindent {\bf Demostraci{\'o}n:} N{\'o}tese que el producto
antisim{\'e}trico del segundo miembro de la ecuaci{\'o}n
(\ref{DDstar}) define los mapeos $K_1,K_2$ referidos en el Teorema
(\ref{k0}). La demostraci{\'o}n de este corolario se obtiene
utilizando el Lema \ref{Lema1-1} para evaluar los t{\'e}rminos de
borde en la expresi{\'o}n,
\begin{equation}\label{DDstar2}
      \left(A^\dagger \psi, \phi\right) -
  \left(\psi, A^\dagger \phi\right)
    =\displaystyle{
    \lim_{\varepsilon \rightarrow 0^+}\int_\varepsilon^1
    \partial_x \Big\{\psi(x)^*\, \phi'(x)
    - \psi'(x)^* \,\phi(x)
    \Big\} \, dx}\, .
\end{equation}
\fin

\bigskip

De acuerdo con el Lema (\ref{Lema1-1}), podemos definir el mapeo,
\begin{equation}\begin{array}{c}
    K:\mathcal{D}(A^\dagger)\rightarrow \mathbb{C}^{4}\,,\\ \\
    \phi\rightarrow \left( C_1[\phi],C_2[\phi], \phi(1), \phi'(1) \right)\,.
\end{array}
\end{equation}
El dominio $\mathcal{D}(\overline{A})$ de la clausura
$\overline{A}=(A^{\dagger})^\dagger$ del operador $A$ es el
n{\'u}cleo de $K$, en tanto que las extensiones sim{\'e}tricas de
$A$ est{\'a}n definidas sobre la preimagen de los
sub\-es\-pa\-cios de $\mathbb{C}^4$ bajo el mapeo $K$. Las
extensiones autoadjuntas, por su parte, est{\'a}n identificadas
con los subespacios lagrangianos $S\subset \mathbb{C}^4$, esto es,
tales que $S=S^{\perp}$, donde el complemento ortogonal se define
de acuerdo con la forma simpl{\'e}ctica del miembro derecho de la
ecuaci{\'o}n (\ref{DDstar}).

\bigskip

En adelante, estudiaremos las extensiones autoadjuntas que
satisfagan la condici{\'o}n de contorno local,
 \begin{equation}\label{BC1}
  \phi(1)=0\, ,
\end{equation}
correspondiente a $\beta=\infty$ en la ecuaci{\'o}n (\ref{beta}).
Cada una de estas extensiones, que denotaremos por $A_{\gamma}$
,est{\'a} definida por una condici{\'o}n de la forma,
\begin{equation}\label{BC21}
    \cos\gamma\cdot C_1[\Phi] + \sin\gamma\cdot C_2[\Phi] = 0\, ,
\end{equation}
con $\gamma \in [0,\pi)$.

\subsubsection{El espectro} \label{the-spectrum1}

Para determinar el espectro de las extensiones autoadjuntas
$A_\gamma$, para el caso $0<\nu<1$, estudiaremos las soluciones
de,
\begin{equation}\label{Ec-hom}
  (A^\dagger-\lambda)\phi_\lambda(x)=0 \, ,
\end{equation}
que satisfagan las condiciones de contorno establecidas en las
ecuaciones (\ref{BC1}) y (\ref{BC21}).

\bigskip

La soluci{\'o}n general de la ecuaci{\'o}n (\ref{Ec-hom}) para
$\lambda=0$ est{\'a} dada por,
\begin{equation}\label{lambda0}
  \phi_0(x)=\frac{1}{\sqrt{2\nu}} \left(
  A\, x^{\nu+\frac 1 2} +
  B\, x^{-\nu+\frac 1 2}\right)\, ,
\end{equation}
para dos constantes arbitrarias $A,B\in\mathbb{C}$. Sin embargo,
las condiciones de contorno dadas por las ecuaciones (\ref{BC1}) y
(\ref{BC21}) implican,
\begin{equation}\label{BBBCCC}
  A+B=0\, , \quad \cos\gamma\cdot A + \sin\gamma\cdot B=0 \, .
\end{equation}
Consecuentemente, s{\'o}lo existen modos cero para la
extensi{\'o}n autoadjunta caracterizada por $\gamma=\pi/4$.

\bigskip

Por su parte, las soluciones de la ecuaci{\'o}n (\ref{Ec-hom})
para $\lambda\neq 0$ tienen la forma,
\begin{equation}\label{sol-hom-1}\begin{array}{c}
  \displaystyle{  \phi(x) =
  \frac{A}{{\sqrt{2\nu}}}\,
    \frac{
       \Gamma(1+\nu)}{
       2^{-\nu}\,{\mu }^
        {\nu}}\,{\sqrt{x}}\,
       J_{\nu}(\mu\,x )
       }
  \displaystyle{+\frac{B}{{\sqrt{2\nu}}}\,\frac{
       \Gamma(1-\nu)}{
       2^{\nu}\,{\mu }^
        { -\nu}}\,{\sqrt{x}} \,
       J_{-\nu}(\mu\,x )\, ,}
\end{array}
\end{equation}
donde $\mu = +\sqrt{\lambda}$. Las constantes $A,B\in\mathbb{C}$
est{\'a}n relacionadas en virtud de la condici{\'o}n de contorno
en el origen. En efecto, teniendo en cuenta,
\begin{equation}\label{J-Bessel}
  J_\nu(z)= z^{\nu }\,\left\{ \frac{1}
     {2^{\nu }\,
       \Gamma(1 + \nu )}
     + O(z^2) \right\}\, ,
\end{equation}
obtenemos a partir de las ecuaciones (\ref{lemaI-1}) y
(\ref{BC21}),
\begin{equation}\label{ec-spectrum}
    \cos\gamma\cdot A + \sin\gamma\cdot B=0\, .
\end{equation}
La condici{\'o}n de contorno en $x=1$, dada por la ecuaci{\'o}n
(\ref{BC1}), se escribe,
\begin{equation}\label{sol-hom-1-en1}\begin{array}{c}
  \displaystyle{  \phi(1) =
  \frac{A}{{\sqrt{2\nu}}}\,
    \frac{
       \Gamma(1+\nu)}{
       2^{1-\nu}\,{\mu }^
        {\nu }}\,
       J_{\nu}(\mu)
       }
  \displaystyle{+\frac{B}{{\sqrt{2\nu}}}\,\frac{
       \Gamma(1-\nu)}{
       2^{\nu}\,{\mu }^
        { -\nu }}\,
       J_{-\nu}(\mu) =0\, .}
\end{array}
\end{equation}
Esta expresi{\'o}n, junto con la ecuaci{\'o}n (\ref{ec-spectrum}),
determina el espectro de la extensi{\'o}n $A_\gamma$, que
describimos a continuaci{\'o}n.

\begin{itemize}
\item Si $\gamma=\pi/2$, la ecuaci{\'o}n (\ref{ec-spectrum})
implica que $B= 0$. Por consiguiente, $\phi(1)=0
 \Rightarrow  J_{\nu}(\mu)=0$. En consecuencia, el espectro de la
extensi{\'o}n autoadjunta $A_{\pi/2}$ es positivo y est{\'a} dado
por,
\begin{equation}\label{alpha0}
  \lambda_{n} = j_{\nu,n}^2\, ,
   \quad n=1,2,\dots\, ,
\end{equation}
donde $j_{\nu,n}$ es el $n$-{\'e}simo cero
positivo\,\footnote{\label{zeroesnote} Recordemos que los ceros de
$J_{\nu}(\lambda)$ admiten el desarrollo asint{\'o}tico
\begin{equation}\label{large-zeroes}
  j_{\nu,n}\simeq \gamma -\frac{4 \nu^2-1}{8 \gamma}+O\left(\frac 1
  \gamma\right)^3\, ,
\end{equation}
con $\gamma=\left( n+\frac{\nu}{2}-\frac{1}{4} \right) \pi$.} de
la funci{\'o}n de Bessel $J_{\nu}(z)$.

\item Si $\gamma=0$, la ecuaci{\'o}n (\ref{ec-spectrum}) implica
que $A= 0$. El espectro de $A_0$ es tambi{\'e}n positivo y
est{\'a} dado por,
\begin{equation}\label{eigen-beta0}
  \lambda_{n} =  j_{-\nu,n}^2\, , \ n=1,2,\dots\, ,
\end{equation}
donde $j_{-\nu,n}^2$ son los ceros positivos de $J_{-\nu}(\mu )$.

\item Para $\gamma\neq 0,\pi/2$ obtenemos, a partir de las
ecuaciones (\ref{ec-spectrum}) y (\ref{sol-hom-1-en1}), la
siguiente ecuaci{\'o}n trascendental para los autovalores de
$A_{\gamma}$,
\begin{equation}\label{spectrumo}
    F(\mu):={\mu }^{2\nu}\,\frac{
    J_{-\nu}(\mu )}
    {J_{\nu}(\mu )}=4^\nu\frac{\Gamma(\nu)}{\Gamma(-\nu)}\,\theta\,,
\end{equation}
donde hemos definido,
\begin{equation}\label{rho}
  \theta:= -\tan\gamma\, .
\end{equation}
De acuerdo con esta definici{\'o}n, designaremos en adelante a la
extensi{\'o}n $A_\gamma$ por $A^\theta$; n{\'o}tese que el
par{\'a}metro $\theta$ coincide con el definido en la
expresi{\'o}n (\ref{saesing}).

\bs

Para los autovalores positivos, ${\lambda}=\mu^2$, ambos miembros
de la ecuaci{\'o}n (\ref{spectrumo}) han sido representados en la
Figura \ref{figure} para valores particulares de $\theta$ y $\nu$.

\begin{figure}
\center
    \epsffile{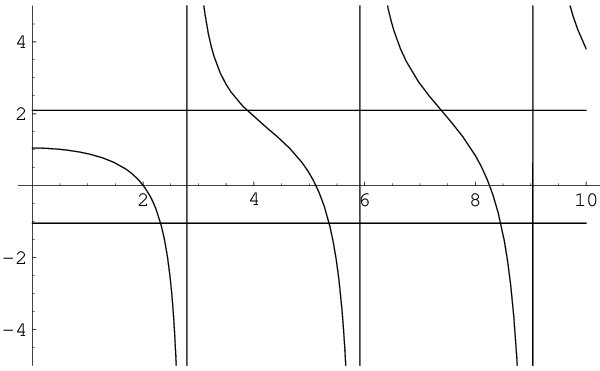} \caption{{\small Gr{\'a}fica de
    $F(\mu)$ para $\nu=1/4$. Las intersecciones con las rectas
    ho\-ri\-zon\-ta\-les representan los autovalores
    correspondientes a las extensiones dadas por $\theta=1$ y
    $\theta=-2$.}} \label{figure}
\end{figure}

\bs

Con respecto a la presencia de autovalores negativos, se puede
probar que si $\theta<-1$ entonces la extensi{\'o}n $A^{\theta}$
tiene un autovalor negativo $\lambda_0$. En efecto, si $\lambda_0
= (i \mu_0)^2 <0$, entonces,
\begin{equation}\label{Fdeimu}\begin{array}{c}
  \displaystyle{F(i\,\mu_0)={\mu_0 }^{ 2\nu}\,
  \frac{
    I_{-\nu}(\mu_0 )}
    {I_{\nu}(\mu_0 )}}
  \displaystyle{ =-4^{\nu}\,
    \frac{
     \Gamma(\nu)}{\Gamma(-\nu)}
     \left\{
     1 + \frac{\nu \,{\mu_0 }^2}
   {2\left( 1 - \nu^2 \right)}
   +{O}(\mu_0^4 )\right\} }\, ,
\end{array}
\end{equation}
donde $I_\nu(\mu)$ es la funci{\'o}n de Bessel
modificada\,\footnote{Se puede probar que este autovalor negativo
tiende a $-\infty$ a medida que $\theta \rightarrow -\infty$, en
tanto que la correspondiente autofunci{\'o}n se concentra en la
singularidad en $x=0$, como los estados de borde estudiados en la
secci{\'o}n (\ref{esta}).}. De modo que $F(i\mu_0)$ satisface la
ecuaci{\'o}n (\ref{spectrumo}) si $\theta<-1$ (v{\'e}ase la Figura
\ref{lambda-neg}.)
\begin{figure}
\center
    \epsffile{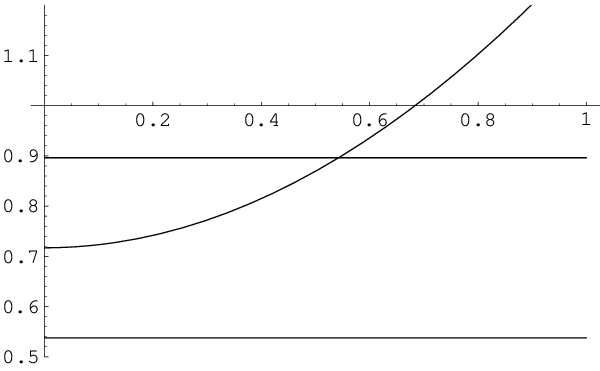} \caption{Gr{\'a}fica de $F(i\,\mu_0)$ en
    funci{\'o}n de $\mu_0$ y del miembro derecho de la ecuaci{\'o}n
    (\ref{spectrumo}) para $\nu=3/4$ y $\theta=3/4,5/4$. Esta
    ecuaci{\'o}n no tiene soluci{\'o}n para $\mu^2\in\mathbb{R}^-$ en el
    caso $\theta=3/4$ y admite una {\'u}nica soluci{\'o}n si
    $\theta=5/4$.} \label{lambda-neg}
\end{figure}

\bigskip

Observemos, por {\'u}ltimo, que el espectro es siempre no
degenerado, y que existe un autovalor positivo entre cada par de
cuadrados de ceros consecutivos de $J_{\nu}(\lambda)$. Por
consiguiente, a partir de la ecuaci{\'o}n (\ref{large-zeroes}),
obtenemos el siguiente comportamiento asint{\'o}tico para los
autovalores,
\begin{equation}\label{503}
     \lambda_n = \pi^2 \, n^2 + O(n)\,.
\end{equation}
\end{itemize}

\subsection{La resolvente} \label{the-resolvent1}

En esta secci{\'o}n construiremos la resolvente
$(A^\theta-\lambda)^{-1}$ correspondiente a las di\-fe\-ren\-tes
extensiones autoadjuntas $A^\theta$ de $A$, para el caso
$0<\nu<1$. Primeramente, con\-si\-de\-ra\-re\-mos las extensiones
autoadjuntas definidas por condiciones de contorno invariantes de
escala, caracterizadas por $\theta=\infty$ y $\theta=0$. La
resolvente para una extensi{\'o}n autoadjunta general ser{\'a}
luego obtenida como combinaci{\'o}n lineal de estos dos casos
especiales.

\bigskip

Para determinar el n{\'u}cleo $G(x,x';\lambda=\mu^2)$ de la
resolvente, que satisface la ecuaci{\'o}n,
\begin{equation}\label{G}
   (A-\mu^2)\, G(x,x'; \mu^2) =
\delta(x-x')\, ,
\end{equation}
con $-\pi/ 2 < {\rm arg} (\mu) \leq \pi/2$, necesitamos, de
acuerdo con la expresi{\'o}n (\ref{solres}), dos soluciones
particulares de la ecuaci{\'o}n (\ref{Ec-hom}). Es conveniente
entonces definir,
\begin{equation}\label{soluciones1}
    \left\{
  \begin{array}{l}
    \displaystyle{L^\infty(x,\mu)= {\sqrt{x}}\,
       J_{\nu}(\mu\,x )\,  ,}\\ \\
    \displaystyle{L^0(x,\mu)=
    {\sqrt{x}}\,
       J_{-\nu}(\mu\,x )\,  ,}\\ \\
    \displaystyle{R(x,\mu)=
    {\sqrt{x}}\,\left( J_{-\nu}(\mu)\,
     J_{\nu}(\mu\,x )
     - J_{\nu}(\mu)
     \,J_{-\nu}(\mu \,x )
     \right)
    \, .}
        \end{array}
    \right.
\end{equation}
N{\'o}tese que $L^{\infty}(x,\mu),L^{0}(x,\mu)$ presentan el mismo
comportamiento en el origen que las funciones de los dominios de
$A^\infty,A^0$, respectivamente; $R(1,\mu)$, por su parte, se
anula en $x=1$.

\bs

Definimos tambi{\'e}n los wronskianos,
\begin{eqnarray}\label{W-D1}
  W\left[ L^\infty, R\right](\mu) := L^\infty(x,\mu)\partial_x R(x,\mu)-
  \partial_x L^\infty(x,\mu)R(x,\mu)=\nonumber\\=
  -\frac{2\,\sin (\pi\nu)}
    {\pi } \,J_{\nu}(\mu )
   \, ,\\
  W\left[ L^0, R \right](\mu) := L^0(x,\mu)\partial_x R(x,\mu)-
  \partial_x L^0(x,\mu)R(x,\mu)=\nonumber\\=
  -\frac{2\,\sin (\pi\nu )}{\pi }\,J_{-\nu}(\mu )
   \, .
\end{eqnarray}

\subsubsection{La resolvente para la extensi{\'o}n $\theta=\infty$}

De acuerdo con la ecuaci{\'o}n (\ref{solres}), el n{\'u}cleo de la
resolvente para el caso $\theta=\infty$ est{\'a} dado por,
\begin{equation}\label{GD111}
  G_\infty (x,x';\mu^2)= \frac 1 {W\left[ L^\infty, R\right](\mu)}
  \times  \left\{
  \begin{array}{c}
    L^\infty(x,\mu)\, R(x', \mu),\ {\rm para}\ x\leq x'\, , \\ \\
    R(x,\mu)\, L^\infty(x',\mu), \ {\rm para}\ x \geq x'\, .
  \end{array}\right.
\end{equation}
Por lo tanto, la soluci{\'o}n $\phi(x)$ de la ecuaci{\'o}n
$(A^\infty-\mu^2)\, \phi(x)=f(x)$, dada por,
\begin{equation}\label{func-inhom1}
  \phi(x) = \int_0^1 G_\infty(x,x'; \mu^2)\, f(x')\, dx'\,,
\end{equation}
satisface $\phi(1)=0$ y $C_2[\phi]=0$, para cualquier funci{\'o}n
$f(x)\in \mathbf{L_2}(0,1)$. En efecto, de las ecuaciones
(\ref{func-inhom1}), (\ref{GD111}), (\ref{soluciones1}) y
(\ref{W-D1}), obtenemos,
\begin{equation}\label{near0-D}
    \phi(x)=
    \frac{C_1^\infty[\phi]}{\sqrt{2\nu}} \, x^{\nu+\frac 1 2} + O({x}^{3/2})\, ,
\end{equation}
con,
\begin{equation}\label{C+1}
  C_1^\infty[\phi] = -  \frac{\pi\, \mu^{\nu}\, \sqrt{2\nu} }
    {2^{1+\nu} \sin(\pi\nu)
    J_{\nu}({\mu}) \,
    \Gamma\left(1+\nu\right)}
    \ \int_0^1 R(x', \mu) f(x') \, dx' \, ,
\end{equation}
siendo $\mu$ distinto de todo cero de $J_{\nu}({\mu})$. N{\'o}tese
que $C_1^\infty[\phi]\neq 0$ si la integral en el miembro derecho
de (\ref{C+1}) no se anula.

\subsubsection{La resolvente para la extensi{\'o}n $\theta=0$}

En este caso, el n{\'u}cleo de la resolvente est{\'a} dado, de
acuerdo con (\ref{solres}), por,
\begin{equation}\label{GN111}
  G_0 (x,x';\mu^2)= \frac 1 {W\left[ L^0, R \right](\mu)} \times  \left\{
  \begin{array}{c}
    L^0(x,\mu)\, R(x',\mu),\ {\rm para}\ x\leq x'\, , \\ \\
    R(x,\mu)\, L^0(x',\mu), \ {\rm para}\ x \geq x'\, .
  \end{array}\right.
\end{equation}
La soluci{\'o}n de la ecuaci{\'o}n $(A^0-\mu^2)\,\phi(x)=f(x)$
est{\'a} entonces dada por la funci{\'o}n,
\begin{equation}\label{func-inhom-N1}
  \phi(x) = \int_0^1 G_0( x,  x'; \mu^2)\, f(x')\, dx'\,,
\end{equation}
que en este caso satisface $\phi(1)=0$ y $C_1[\phi]=0$, para
cualquier funci{\'o}n $f(x)\in \mathbf{L_2}(0,1)$. Efectivamente,
a partir de (\ref{func-inhom-N1}), (\ref{GN111}),
(\ref{soluciones1}) y (\ref{W-D1}), obtenemos,
\begin{equation}\label{near0-N}
    \phi(x)= \frac{C_2^0[\phi]}{\sqrt{2\nu}}
    \, x^{-\nu+\frac 1 2} + O({x}^{3/2})\, ,
\end{equation}
con,
\begin{equation}\label{C-1}\begin{array}{c}
  \displaystyle{C_2^0[\phi] =  - \frac{\pi\, \mu^{-\nu}
  \, {\sqrt{2\nu}}}
    {2^{1-\nu} \sin(\pi\nu)
    J_{-\nu}({\mu}) \,
    \Gamma\left(1-\nu\right)}
    \int_0^1 R(x',\mu) f(x')
    \, dx'\, ,}
\end{array}
\end{equation}
siendo $\mu$ distinto de todo cero de $J_{-\nu}({\mu})$. Por su
parte, $C_2^0[\phi]\neq 0$ si la integral en el miembro derecho de
la ecuaci{\'o}n (\ref{C-1}) (que coincide con la integral que
figura en la ecuaci{\'o}n (\ref{C+1}), correspondiente a la
extensi{\'o}n $\theta=\infty$), no se anula.

\subsubsection{La resolvente para una extensi{\'o}n autoadjunta general $A^{\theta}$}

Para una extensi{\'o}n autoadjunta arbitraria, imponemos sobre las
funciones,
\begin{equation}\label{func-inhom-gen}
  \phi(x) = \int_0^1 G_\theta( x,  x';\lambda)\, f(x')\, dx'\, ,
\end{equation}
las condiciones de contorno,

\begin{equation}\label{BC-general}
  \phi(1)=0\, ,\quad \cos\gamma\, C_1[\phi] + \sin\gamma\, C_2[\phi] =
  0\, ,\ \gamma\neq 0,\pi/2 \,.
\end{equation}
La funci{\'o}n $f(x)$ representa cualquier funci{\'o}n de $\in
\mathbf{L_2}(0,1)$. Para ello, podr{\'\i}amos repetir el
procedimiento que utilizamos para los casos $\theta=\infty,0$. En
su lugar, con\-si\-de\-ra\-re\-mos una combinaci{\'o}n lineal de
los n{\'u}cleos de las resolventes de estos casos,
\begin{equation}\label{linear-comb1}
  G_\theta(x,x'; \lambda)= \left[1- \tau(\lambda)\right]
  G_\infty(x,x'; \lambda) +
  \tau(\lambda)\, G_0(x,x';\lambda)\,
\end{equation}
y determinaremos la funci{\'o}n $\tau(\lambda)$. N{\'o}tese que,
de acuerdo con la ecuaci{\'o}n (\ref{mira}), $\tau(\lambda)$
est{\'a} relacionado con el factor $K(\lambda)$,
\begin{equation}\label{tauka}
    \tau(\lambda)=\left[1+\theta\,K(\lambda)\right]^{-1}\,.
\end{equation}

La funci{\'o}n $G_\theta(x,x'; \lambda)$ satisface la ecuaci{\'o}n
(\ref{G}) pues, de acuerdo con (\ref{linear-comb1}), es una
combinaci{\'o}n lineal de dos funciones que a su vez la
satisfacen. Adem{\'a}s, $G_\theta(x,x'; \lambda)$ satisface la
condici{\'o}n de contorno apropiada en $x=1$ pues tambi{\'e}n lo
hacen las funciones $G_\infty(x,x'; \lambda),G_0(x,x'; \lambda)$.
En consecuencia, $\tau(\lambda)$ y, consecuentemente, el factor
$K(\lambda)$ se determinan a partir de la condici{\'o}n de
contorno en $x=0$ (v{\'e}ase la ecuaci{\'o}n (\ref{BC21})) que
resulta de (\ref{near0-D}), (\ref{near0-N}) y
(\ref{linear-comb1}),
\begin{equation}\label{ec-tau1}
  \cos\gamma \cdot\left[1- \tau(\lambda)\right] C_1^\infty[\phi]
  + \sin\gamma\cdot \tau(\lambda)\, C_2^0[\phi] =0\, .
\end{equation}


Por lo tanto, a partir de las ecuaciones (\ref{C+1}), (\ref{C-1})
y (\ref{ec-tau1}), obtenemos el factor $\tau(\lambda)$,
\begin{equation}\label{taudelambda666}\begin{array}{c}
    \tau(\mu^2) = \displaystyle{\frac{\cos\gamma \, C_1^\infty[\phi]}
  {\cos\gamma \, C_1^\infty[\phi]-\sin\gamma\,  C_2^0[\phi]}} =
  \displaystyle{
  \frac{1}{1 - \theta\,4^{\nu}
  \displaystyle{\frac{\Gamma(\nu)}{\Gamma(-\nu)}\,
  \frac{
       J_{\nu}(\mu )}
       {J_{-\nu}(\mu )}\,{\mu }^{-2\nu}} }   }\, ,
\end{array}
\end{equation}
siendo $\mu^2$ distinto de todo autovalor de $A^\theta$.
Comparando las expresiones (\ref{taudelambda666}) y (\ref{tauka})
obtenemos el factor $K(\mu^2)$,
\begin{equation}
    K(\mu^2)=4^{\nu}
  \displaystyle{\frac{\Gamma(\nu)}{\Gamma(-\nu)}\,
  \frac{
       J_{\nu}(\mu )}
       {J_{-\nu}(\mu )}\,{\mu }^{-2\nu}}\,.
\end{equation}
Esta ecuaci{\'o}n confirma el desarrollo asint{\'o}tico de
$K(\lambda)$ calculado en (\ref{chau}) (v{\'e}ase la expresi{\'o}n
(\ref{JsobreJup}).)

\subsubsection{La traza de la resolvente} \label{trace-resolvent}

La ecuaci{\'o}n (\ref{linear-comb1}) implica que la resolvente de
una extensi{\'o}n autoadjunta $A^{\theta}$ puede expresarse como
una combinaci{\'o}n lineal de las resolventes correspondientes a
las extensiones correspondientes a $\theta=\infty$ y $\theta=0$.
Adem{\'a}s, dado que los autovalores de cualquier extensi{\'o}n
crecen como $\lambda_n\sim n^2$ (v{\'e}ase la ecuaci{\'o}n
\ref{503}), las resolventes son operadores tipo traza.

\bs

Podemos entonces escribir,
\begin{equation}\label{trazaG}
  {\rm Tr}(A^{\theta}-\lambda)^{-1}-
    {\rm Tr}(A^{\infty}-\lambda)^{-1}=
   \tau(\lambda)\left[
     {\rm Tr}(A^{0}-\lambda)^{-1}-
     {\rm Tr}(A^{\infty}-\lambda)^{-1}\right]\,,
\end{equation}
A partir de las ecuaciones (\ref{GD111}) y (\ref{GN111}) obtenemos
(v{\'e}anse los detalles en la secci{\'o}n \ref{integrals}),
\begin{equation}\label{traza-GD}
        {\rm Tr}\left(A^{\infty}-\mu^2\right)^{-1} =
      \int_0^1
      G_\infty(x,x; \mu^2)\}\, dx
  =\frac{\nu}{2\,\mu^2}-
  \frac{J'_{ \nu }(\mu )}
  {2\,\mu \,J_{\nu}(\mu )} \, ,
\end{equation}
y,
\begin{equation}\label{traza-GN}
        {\rm Tr}\left(A^{0}-\mu^2\right)^{-1}  =
      \int_0^1
      G_0(x,x; \mu^2)\}\, dx
  =-\frac{\nu}{2\,\mu^2}-
  \frac{J'_{-\nu}(\mu )}
  {2\,\mu \,J_{-\nu}(\mu )} \, .
\end{equation}
En consecuencia, la traza de la resolvente de una extensi{\'o}n
autoadjunta general puede obtenerse expl{\'\i}citamente y est{\'a}
dada por,
\begin{eqnarray}\label{TrG21}
  {\rm Tr}\left(A^{\theta}-\mu^2\right)^{-1} =
  \left[\frac{\nu}{2\,\mu^2}-
  \frac{J'_{ \nu }(\mu )}
  {2\,\mu \,J_{\nu}(\mu )}\right]
  \mbox{}+\tau(\mu^2) \left[
  -\frac{\nu}{\mu^2}+\frac{1}{2\,\mu}
  \left(
  \frac{J'_{\nu}(\mu )}
  {J_{\nu}(\mu )}
  - \frac{J'_{-\nu}(\mu )}
  {J_{-\nu}(\mu )}
  \right) \right] \, .\nonumber\\
\end{eqnarray}

\subsubsection{Desarrollo asint{\'o}tico de la traza de la
 resolvente}\label{Asymptotic-expansion}

Utilizando el desarrollo asint{\'o}tico de Hankel para las
funciones de Bessel \cite{A-S} (v{\'e}ase la secci{\'o}n
\ref{Hankel}), se puede obtener el desarrollo asint{\'o}tico del
primer t{\'e}rmino del miembro derecho de la ecuaci{\'o}n
(\ref{TrG21}),
\begin{eqnarray}\label{asymp-trGD-upp1}
    {\rm Tr}(A^{\infty}-\mu^2)^{-1}\sim
  \sum_{k=1}^\infty \frac{A_k(\nu,\sigma)}{\mu^k}\sim\nonumber\\
  \sim\displaystyle{
  \frac{i \,\sigma }{2\,\mu } +
  \frac{\nu+\frac 1 2}{2\,{\mu }^2} -
  \frac{i\,{\sigma } \left(\nu^2-\frac 1 4\right)
     }{{4\,\mu }^3} +
  \frac{\left(\nu^2-\frac 1 4\right)}{4\,{\mu }^4} +
  {{O}({\mu^{-5} })} }\, ,
\end{eqnarray}
donde $\sigma = 1$ para $\Im(\mu)>0$ y $\sigma = -1$ para
$\Im(\mu)<0$. Los coeficientes $A_k(\nu,\sigma)$ en esta serie
pueden ser directamente determinados a partir de las ecuaciones
(\ref{P+Q}) y (\ref{T}). N{\'o}tese que
$A_k(\nu,-1)=A_k(\nu,1)^*$, pues $A_{2k}(\nu,1)$ es real y
$A_{2k+1}(\nu,1)$ es imaginario puro.

\bs

Por su parte, el desarrollo asint{\'o}tico del segundo t{\'e}rmino
del miembro derecho de la ecuaci{\'o}n (\ref{TrG21}) se obtiene de
la expresi{\'o}n,
\begin{equation}\label{trdifasymp1}
    {\rm Tr}\left(A^{\infty}-\mu^2\right)^{-1} -
     {\rm Tr}\left(A^{0}-\mu^2\right)^{-1}
      \sim \frac{\nu}{\mu^2} \, ,
\end{equation}
obtenida a partir de la ecuaci{\'o}n (\ref{JprimasobreJasymp}) y
de,
\begin{equation}\label{tau-asymp1}
   \tau(\mu^2) \sim \displaystyle{\frac{1}
   {1 - \displaystyle{{e^
         {\sigma \,i\,
           \pi \nu }
           \,4^\nu\,\frac{\Gamma(\nu)}{\Gamma(-\nu)}\,{\theta}\,
           {\mu }^{-2\nu}}} }
        \sim }
      \displaystyle{
    \sum_{k=0}^\infty \left(
    {e^{\sigma\, i\, \pi\,\nu}\,4^\nu\,\frac{\Gamma(\nu)}
    {\Gamma(-\nu)}\,{\theta}\,
    \mu^{-2\nu}}\right)^k \, ,}
\end{equation}
donde $\sigma =1$ ($\sigma =-1$) corresponde a $\Im(\mu)>0$
($\Im(\mu)<0$) (v{\'e}ase la ecuaci{\'o}n (\ref{JsobreJup}).)
N{\'o}tese la aparici{\'o}n, en este desarrollo asint{\'o}tico, de
potencias de $\mu$ dependientes del par{\'a}metro $\nu$.

\subsection{La funci{\'o}n-$\zeta$ y la traza del heat-kernel}
\label{spectral-functions}

La funci{\'o}n $\zeta^\theta(s)$ correspondiente a una
extensi{\'o}n autoadjunta arbitraria $A^{\theta}$ satisface, para
$\Re(s)>1/2$ (v{\'e}ase la ecuaci{\'o}n (\ref{zetres})),
\begin{equation}\label{zeta}
  \zeta^{\theta}(s)=- \frac{1}{2\pi i} \oint_{\mathcal{C}}
  {\lambda^{-s}} \, {\rm Tr}\left(A^{\theta}-\lambda\right)^{-1}
  \, d\lambda\, ,
\end{equation}
donde la curva $\mathcal{C}$ encierra el espectro del operador en
sentido antihorario pero no encierra al origen. De acuerdo con la
ecuaci{\'o}n (\ref{trazaG}),
\begin{equation}\label{zeta1}
  \zeta^{\theta}(s)= \zeta^\infty(s)+\oint_{\mathcal{C}}
  {\lambda^{-s}} \,  \tau(\lambda)\,
  {\rm Tr}\left\{\left(A^{\infty}-\lambda\right)^{-1}-\left(A^{0}-
  \lambda\right)^{-1}\right\}
  \,\frac{d\lambda}{2\pi i}  \, ,
\end{equation}
donde $\zeta^\infty(s)$ es la funci{\'o}n-$\zeta$ de la
extensi{\'o}n $\theta=\infty$.

\bs

Como, de acuerdo con la discusi{\'o}n de la secci{\'o}n
\ref{the-spectrum1}, $A^\infty$ tiene un espectro positivo y las
extensiones autoadjuntas $A^{\theta}$ tienen a lo sumo un
{\'u}nico autovalor negativo $\lambda_0$, podemos escribir,
\begin{equation}\label{zeta+pri}\begin{array}{c}
  \displaystyle{  \zeta^{\theta}(s)=
  \zeta^{\infty}(s) +\theta(-\lambda_{0}) \,\lambda_{0}^{-s} -
    } \\ \\
    \displaystyle{
    - \int_{-i\,\infty+0}^{i\,\infty+0}
  {\lambda^{-s}} \,
   \tau(\lambda)\,
  {\rm Tr}\left\{\left(A^{\infty}-\lambda\right)^{-1}-\left(A^{0}-
  \lambda\right)^{-1}\right\}
  \,\frac{d\lambda}{2\pi i}
    } \, .
\end{array}
\end{equation}

Ser{\'a} tambi{\'e}n conveniente tener en cuenta,
\begin{equation}\label{zeta+1pri}\begin{array}{c}
  \displaystyle{\zeta^{\theta}(s)=
    \frac{e^{-i\,\frac{\pi}{2}\,s}}{\pi}
     \int_{1}^{\infty}
  {\mu^{1-2s}} \,
  {\rm Tr}\left(A^{\theta}-(e^{i\,\frac{\pi}{4}}\, \mu)^2\right)^{-1}
  \, d\mu\,  +
  } \\ \\
  \displaystyle{
    +\mbox{}\frac{e^{ i\,\frac{\pi}{2}\,s}}{\pi} \int_{1}^{\infty}
  {\mu^{1-2s}} \,
  {\rm Tr}\left(A^{\theta}-(e^{-i\,\frac{\pi}{4}}\, \mu)^2\right)^{-1}
  \, d\mu + {h_1(s)} \, ,
  }
\end{array}
\end{equation}
donde $h_1(s)$ es una funci{\'o}n entera.

\bs

Para determinar los polos de la funci{\'o}n $\zeta^{\theta}(s)$,
sumamos y restamos en los integrandos del miembro derecho de la
ecuaci{\'o}n (\ref{zeta+1pri}) una suma parcial del desarrollo
asint{\'o}tico de ${\rm Tr}\left(A^{\theta}-\lambda\right)^{-1}$,
que ha sido obtenido en la secci{\'o}n anterior.

\bs

En particular, para la extensi{\'o}n $\theta=\infty$ y para
$s>1/2$ obtenemos,
\begin{equation}\label{zetaD+11}\begin{array}{c}
  \displaystyle{ \zeta^{\infty}(s) }
  \displaystyle{ =
    \frac{1}{\pi}\sum_{\sigma=\pm 1} \int_{1}^{\infty}
    e^{-i\,\sigma\,\frac{\pi}{2}\,s}\,
  {\mu^{1-2 s}} \,
  \left\{ \sum_{k=1}^{N} e^{-i\,\sigma\,\frac{\pi}{4}\,k}\,
  A_k(\nu,\sigma) \, \mu^{-k} \right\}
  \, d\mu\,  +h_2(s)=
  } \\ \\
  \displaystyle{
  = \frac{1}{\pi}\,\sum_{k=1}^{N}\frac{
  \Re \left\{
  e^{-i\,\frac{\pi}{2}\,(s+k/2)}\,A_k(\nu,1) \right\}
  }{s-(1-k/2)}
  + {h_2(s)}\, ,
  }
\end{array}
\end{equation}
donde $h_2(s)$ es anal{\'\i}tica en el semiplano $\Re(s)>(1-N)/2$
(v{\'e}anse las ecuaciones (\ref{zeta+1pri}) y
(\ref{asymp-trGD-upp1}).) Consecuentemente, la extensi{\'o}n
meromorfa de la funci{\'o}n $\zeta^{\infty}(s)$
 presenta polos simples en,
\begin{equation}\label{poles-D}
    s=\frac 1 2-n\, ,\quad {\rm con}\  n=0,1,\dots,
\end{equation}
con residuos,
\begin{equation}\label{otros-residuos}
  \left. {\rm Res}\,\left\{\zeta^{\infty}(s)\right\} \right|_{s=1/2-n}
  =- \frac{1}{\pi}\,\Re\left\{i\, A_{2n+1}(\nu,1)\right\}  \, ,
\end{equation}
donde los coeficientes $A_k(\nu,1)$ est{\'a}n dados por la
ecuaci{\'o}n (\ref{asymp-trGD-upp1}). N{\'o}tese que la parte
imaginaria de estos coeficientes se anula para $k$ par, por lo que
$\zeta^\infty(s)$ no tiene polos en valores enteros de $s$. Estas
singularidades obedecen a la ecuaci{\'o}n (\ref{resu}), que es
v{\'a}lida para operadores regulares.

\bs

El residuo en $s=1/2$, {\it e.g.}, est{\'a} dado por,
\begin{equation}\label{otros-residuos-s12}
  \left. {\rm Res}\,\zeta^{\infty}(s) \right|_{s=1/2}
  =- \frac{1}{\pi}\,\Re\left\{i\, A_1(\nu,1)\right\}=
  \frac{1}{2\, \pi}  \, .
\end{equation}
Este es, por otra parte, el {\'u}nico polo de la funci{\'o}n
$\zeta^{\infty}(s)$ en el l{\'\i}mite regular $\nu\rightarrow
1/2$.

\bigskip

Para una extensi{\'o}n autoadjunta general $A^{\theta}$ debemos
tambi{\'e}n considerar, de acuerdo con la ecuaci{\'o}n
(\ref{trazaG}), las singularidades que provienen del desarrollo de
$\tau(\lambda)$ $\times$ ${\rm
Tr}\left\{\left(A^{\infty}-\lambda\right)^{-1}-\left(A^{0}-
\lambda\right)^{-1}\right\}$, dado por las ecuaciones
(\ref{trdifasymp1}) y (\ref{tau-asymp1}).

\bigskip

De la ecuaci{\'o}n (\ref{zeta+pri}), y teniendo en cuenta
(\ref{zeta+1pri}), obtenemos,
\begin{eqnarray}\label{zetadif}
    \zeta^{\theta}(s)-\zeta^{\infty}(s)=
    {h_3(s)} \, - \frac{ \nu}{\pi}\, \times\nonumber\\
  \times
     \sum_{\sigma=\pm 1}
          e^{-i\,\sigma\,\frac{\pi}{2}\,(s-k\,\nu+1)}\,
 \int_{1}^{\infty}
  \sum_{k=0}^{N}\left(
  4^\nu\frac{\Gamma(\nu)}{\Gamma(-\nu)}\,\theta\right)^k {\mu^{-2k\,\nu-2s-1}}
  \, d\mu =
   \nonumber\\=
  - \frac{\nu}{\,\pi} \sum_{k=0}^{N}\,
  \frac{1}{s+\nu k}\
  \sin\left[
  {\frac{\pi}{2} \left(k\,\nu-s\right)}\right]
  {\left[4^\nu\frac{\Gamma(\nu)}{\Gamma(-\nu)}\,\theta\right]}^k
  + \displaystyle{{h_3(s)}}\, ,\nonumber\\
\end{eqnarray}
donde $h_3(s)$ es anal{\'\i}tica para $\Re(s)>-\nu(N+1)$.

\bs

En consecuencia, $\zeta^{\theta}(s)-\zeta^{\infty}(s)$ tiene una
extensi{\'o}n meromorfa con polos simples en posiciones $s_k$
dependientes de $\nu$,
\begin{equation}\label{polos-zeta}
    s=-\nu k\, ,
     \quad {\rm con}\  k=1,2,\dots\, ,
\end{equation}
cuyos residuos dependen de la extensi{\'o}n autoadjunta
considerada y est'an dados por,
\begin{equation}\label{res-g-dep}
      \left. {\rm Res}\,\left\{\zeta^{\theta}(s)-
  \zeta^{\infty}(s)\right\} \right|_{s=-\nu k} =
    -\frac{\nu}{\pi}
    \left[4^\nu\frac{\Gamma(\nu)}{\Gamma(-\nu)}\,\theta\right]^k
    \ {\sin\left(\pi\nu k \right)}
  \, .
\end{equation}
N{\'o}tese que estos polos son irracionales para valores
irracionales de $\nu$. Adem{\'a}s, los re\-si\-duos se anulan para
la extensi{\'o}n $\theta=0$ y no es dif{\'\i}cil probar que estos
polos tampoco est{\'a}n presentes en el caso $\theta=\infty$.

\bigskip

Por otra parte, estos polos son semienteros negativos para el caso
regular $\nu=1/2$, de modo que responden al resultado (\ref{resu})
en este l{\'\i}mite.

\bigskip

Es interesante destacar que los polos en la ecuaci{\'o}n
(\ref{polos-zeta}) son tambi{\'e}n polos de la funci{\'o}n-$\zeta$
de la correspondiente extensi{\'o}n autoadjunta del operador
$-\partial_x^2+\frac{\nu^2-1/4}{x^2}+x^2$ en
$\mathbf{L}_2(\mathbb{R}^+)$, considerado en la secci{\'o}n
\ref{Wipf} (v{\'e}anse ecuaciones (\ref{pop666}) y \ref{res666}),
con exactamente los mismos residuos, como puede verificarse
f{\'a}cilmente.

\bigskip


{Realizaremos una breve consideraci{\'o}n con respecto a las
propiedades ante transformaciones de escala de las condiciones de
contorno y de la funci{\'o}n-$\zeta$. En primer lugar, n{\'o}tese
que, excepto para el caso $\theta = \infty$, el residuo de
$\zeta^{\theta}(s)$ en $s=-\nu, k$ es proporcional a $\theta^{k}$.
Esto es consistente con el comportamiento de $A$ bajo
transformaciones de escala. En efecto, consideremos la
transformaci{\'o}n $T$,
 \begin{equation}\label{escala}
    \phi_c(x):= T\cdot \phi(x)=c^{1/2}\,\phi(c\,x)
 \end{equation}
bajo la cual $\mathbf{L_2}([0,1])\rightarrow
\mathbf{L_2}([0,1/c])$. La extensi{\'o}n $A^{\theta}$ es
equivalente por una transformaci{\'o}n unitaria  al operador
$c^{-2}\,A_c^{\theta_c}$ definido an{\'a}logamente en
$\mathbf{L_2}([0,1/c])$, con $\theta_c := c^{2\nu}\, \theta$,
\begin{equation}\label{isometry}
  T\,A^{\theta} = \frac 1 {c^2} \, A_c^{\theta_c}\, T\, .
\end{equation}
S{\'o}lo para las extensiones dadas por $\theta=0,\infty$ la
condici{\'o}n de contorno en la sin\-gu\-la\-ri\-dad $x=0$, dada
por la ecuaci{\'o}n (\ref{BC21}), es invariante de escala.

\bigskip

La funci{\'o}n-$\zeta$, por su parte, cambia ante una
transformaci{\'o}n de escala de la siguiente manera,
\begin{equation}\label{zetas-isometry}
  \zeta_c^{\theta_c}(s)=
   c^{-2\, s}\,\zeta^{\theta}(s)\, ,
\end{equation}
y sus residuos
\begin{equation}\label{zetas-isometry-residues}
   \left. {\rm Res}\,\left\{\zeta_c^{\theta_c}(s)
   \right\} \right|_{s=-\nu k} = c^{2\nu k}
   \left. {\rm Res}\,\left\{{\zeta}^{\theta}(s)
   \right\} \right|_{s=-\nu k}\, .
\end{equation}
En consecuencia, el factor $c^{2\nu k}$ cancela exactamente el
efecto que el cambio de la condici{\'o}n de contorno en la
singularidad tiene sobre el factor $\theta^k$ en la ecuaci{\'o}n
(\ref{res-g-dep}),
\begin{equation}\label{change-in-rho}
  \theta^{k}
  =c^{-2\nu k}\,\theta_c^{k}\, .
\end{equation}
Por consiguiente, la diferencia entre los intervalos $[0,1]$ y
$[0,1/c]$ no tiene efecto en la estructura de estos residuos, que
conjeturamos entonces est{\'a}n
 determinados por propiedades locales en la vecindad de $x=0$. }

\bigskip

Concluimos entonces que la pre\-sen\-cia de polos de la
funci{\'o}n-$\zeta$ en posiciones dependientes de $\nu$ es
consecuencia del comportamiento singular del t{\'e}rmino de orden
cero de $A$ cerca del origen y de una condici{\'o}n de contorno
que no es invariante de escala.

\bigskip

Finalmente, a partir de las la relaciones (\ref{desaheat}),
(\ref{polres}) y (\ref{desazeta}), obtenemos el siguiente
desarrollo asint{\'o}tico,
\begin{equation}\label{heat-asymp}\begin{array}{c}
  \displaystyle{  {\rm Tr}\left\{e^{-t\,A^{\theta}}-
  e^{-t\,A^{\infty}}\right\}
  \sim \nu }
  \displaystyle{\,-\, \sum_{k=1}^\infty}
  \left[\Gamma\left(-\nu k\right)
  \frac{\nu}{\pi}\,\theta^k
    \ {\sin\left(
    \pi\nu k \right)}\right]
    t^{\nu k}\, .
\end{array}
\end{equation}
El primer t{\'e}rmino del miembro derecho, que proviene de la
ecuaci{\'o}n (\ref{trdifasymp1}) y del primer t{\'e}rmino del
desarrollo asint{\'o}tico de $\tau(\lambda)$ en la ecuaci{\'o}n
(\ref{tau-asymp1}), coinciden con el resultado de \cite{Mooers}.

\bigskip

N{\'o}tese la dependencia en $\nu$ de las potencias de $t$
presentes en el desarrollo asint{\'o}tico del miembro derecho de
la ecuaci{\'o}n (\ref{heat-asymp}) para una extensi{\'o}n
autoadjunta dada por $\theta\neq 0,\infty$. En particular, el
primero de estos t{\'e}rminos est{\'a} dado por,
\begin{equation}\label{first-term}
   \frac{\theta
    }{\Gamma\left(\nu\right)}\,t^{\nu}\,.
\end{equation}
Aunque la potencia de $t$ coincide con el resultado de
\cite{Mooers}, obtenemos aqu\'\i\ un resultado distinto para el
coeficiente.

\subsection{El caso $\nu=0$ \label{nu=0}}

El operador diferencial $A$,
\begin{equation}\label{g1/2}
    A=-\partial_x^2-\frac{1}{4\, x^2}\, ,
\end{equation}
correspondiente al valor $\nu=0$ en la expresi{\'o}n (\ref{D})
exige una consideraci{\'o}n aparte que desarrollaremos brevemente
en esta secci{\'o}n.

\bigskip

Repitiendo el procedimiento descrito en la demostraci{\'o}n del
Lema \ref{Lema1-1} se puede probar que si $\phi(x)\in
\mathcal{D}(A^\dagger)$, entonces existen constantes
$C_1[\phi],C_2[\phi]\in\mathbb{C}$ tales que,
\begin{equation}\label{l1C-1}
    \displaystyle{\left|\,\phi(x)-\left(C_1[\phi]\, \sqrt{x} +
  C_2[\phi]\, \sqrt{x}\, \log{x}\right)\right|
  \leq \frac{\|A\phi(x)\|}{\sqrt{2}} \  x^{3/2}}\,,
\end{equation}
y,
\begin{equation}\label{l1C-2}\begin{array}{c}
  \displaystyle{
      \left|\,\phi'(x)-\left[\frac{1}{2}C_1[\phi]\,x^{-1/2}+
      C_2[\phi]\left(x^{-1/2}+\frac{1}{2}x^{1/2}\log{x}
      \right)\right]\right|}
  \displaystyle{\leq
  \frac{3\|A\phi(x)\|}{2\sqrt{2}}\, x^{1/2}}\,,
\end{array}
\end{equation}
donde $\| \cdot\|$ se refiere a la norma en $\mathbf{L_2}$.

\bs

Se puede probar f{\'a}cilmente que la ecuaci{\'o}n (\ref{DDstar})
es v{\'a}lida tambi{\'e}n para el caso $\nu=0$ y que las
extensiones autoadjuntas de $A$ tambi{\'e}n est{\'a}n en
correspondencia con los subes\-pa\-cios lagrangianos $S\subset
\mathbb{C}^4$, dados por $S=S^{\perp}$ donde el complemento
ortogonal se define en t{\'e}rminos de la forma simpl{\'e}ctica
del miembro derecho de la ecuaci{\'o}n (\ref{DDstar}). Si
elegimos, adem{\'a}s, la condici{\'o}n de Dirichlet en $x=1$,
$\phi(1)=0$, entonces $A$ admite una familia de extensiones
autoadjuntas $A^{\theta}$ caracterizadas por un par{\'a}metro real
$\theta$ mediante la ecuaci{\'o}n (\ref{BC21}).

\bigskip

Existe, por otra parte, una extensi{\'o}n autoadjunta particular
$A^\infty$ definida por $\theta=\infty$, o bien, $C_2[\phi]=0$.
Las funciones de su dominio de definici{\'o}n se comportan en el
origen de acuerdo con,
\begin{equation}
    \phi(x)=C_1[\phi]\,\sqrt{x}+O(x^{3/2})\,.
\end{equation}
Las autofunciones de $A^\infty$ de autovalor $\lambda$ est{\'a}n
dadas por,
\begin{equation}
    \phi(x)=C_1[\phi]\sqrt{x}\, J_0(\mu x)\,,
\end{equation}
donde $\lambda=\mu^2$ y $\mu$ es un cero (positivo) de $J_0(\mu)$.

\bigskip

Para una extensi{\'o}n autoadjunta arbitraria $A^\theta$, con
$\theta\neq \infty$, las autofunciones co\-rres\-pon\-dien\-tes al
autovalor $\lambda=\mu^2$ est{\'a}n dadas por,
\begin{equation}\begin{array}{c}
  \phi(x)=\left\{C_1[\phi]-C_2[\phi](\log{\mu}-\log{2}+\gamma_E)
    \right\}\,\sqrt{x}\,J_0(\mu x)\,  \displaystyle{ +
  \frac{\pi}{2}\, C_2[\phi]\,\sqrt{x}\,N_0(\mu x) }\, ,
\end{array}
\end{equation}
donde las cantidades $C_1[\phi],C_2[\phi]$ est{\'a}n relacionadas
con $\theta$ mediante la ecuaci{\'o}n (\ref{BC21}).

\bs

La condici{\'o}n $\phi(1)=0$ conduce a la ecuaci{\'o}n,
\begin{equation}
    (\theta+\log{2}-\gamma_E-\log{\mu})J_0(\mu)+
    \frac{\pi}{2}N_0(\mu)=0\,
\end{equation}
que determina el espectro de $A^{\theta}$. N{\'o}tese que no
existen autovalores negativos.

\bigskip

Para determinar los n{\'u}cleos de las resolventes
$(A^{\infty}-\mu^2)^{-1}$ y $(A^{\theta}-\mu^2)^{-1}$
de\-fi\-ni\-mos las funciones,
\begin{equation}\label{LyR} \left\{
\begin{array}{l}
  L^{\infty}(x;\mu)=\sqrt{x}\, J_0(\mu x) \, ,\\ \\
  \displaystyle{
  L^{\theta}(x;\mu)=\sqrt{x}\,
  \left\{(\theta+\log{2}-\gamma_E-\log{\mu})J_0(\mu x)+
    \frac{\pi}{2}N_0(\mu x)\right\}\, ,}\\ \\
    R(x;\mu)=\sqrt{x}\, \left\{N_0(\mu)\,J_0(\mu
    x)-J_0(\mu)\,N_0(\mu x)\right\}\, ,
\end{array} \right.
\end{equation}
y obtenemos,
\begin{equation}\begin{array}{c}
  G^{\infty}(x,x';\mu^2)=
  \displaystyle{
  =\frac{1}
    {W[L^{\infty},R](\mu)}
    \left\{\begin{array}{lr}
    L^{\infty}(x;\mu)\,{R}(x';\mu)\, , & x\leq x' \, ,
    \\ \\
    L^{\infty}(x';\mu)\,{R}(x;\mu)\, , & x\geq x' \, ,
    \end{array}\right.}
\end{array}
\end{equation}
y,
\begin{equation}\begin{array}{c}
  G^{\theta}(x,x';\mu^2)=
  \displaystyle{
  =\frac{1}
    {W[{L}^{\beta},{R}](\mu)}
    \left\{\begin{array}{lr}
    {L}^{\theta}(x;\mu)\,{R}(x';\mu)\, , & x\leq x'\, ,
    \\ \\
    {L}^{\theta}(x';\mu)\,{R}(x;\mu)\, , & x\geq x'\,
    ,
    \end{array}\right.}
\end{array}
\end{equation}
donde los wronskianos se calculan a partir de (\ref{LyR}),
\begin{equation}\begin{array}{c}
    \displaystyle{
  W[{L}^{\infty},{R}](\mu)=
    \frac{2}{\pi}\, J_0(\mu) }\, ,\\ \\
    \displaystyle{
  W[{L}^{\theta},
    {R}](\mu)=\frac{2}{\pi}\,
    (\theta+\log{2}-\gamma_E-\log{\mu})J_0(\mu)+
    N_0(\mu)}\, .
\end{array}
\end{equation}
A partir de la ecuaci{\'o}n (\ref{large-zeroes}), puede verse que
tanto
 $\left(A^\infty-\lambda\right)^{-1}$ como
$\left(A^\theta-\lambda\right)^{-1}$ son operadores tipo traza.

\bigskip

Ahora bien, teniendo en cuenta que \cite{A-S,Mathematica},
\begin{equation}\label{primi-0}\begin{array}{c}
  \displaystyle{\int x\, Z_1(0,x)\, Z_2(0, x)\, dx }
  \displaystyle{=\frac{x^2 }{2} \left\{Z_1(0,x)\, Z_2(0, x)
    +Z_1(1,x)\, Z_2(1, x)
    \right\}\, ,  }
\end{array}
\end{equation}
para $Z_{1,2}(\nu,x)= J_\nu(x)$ o $N_\nu(x)$, las trazas de las
resolventes resultan,
\begin{equation}\label{traces-0}\begin{array}{c}
    \displaystyle{
  {\rm Tr}\left( A^{\infty}-\mu^2\right)=\int_0^1
    {G}^{\infty}(x,x;\mu^2)\,dx
    =\frac{1}{2\mu}\frac{J_1(\mu)}{J_0(\mu)} }\, ,\\ \\
    \displaystyle{
  {\rm Tr}\left( A^{\theta}-\mu^2\right)=\int_0^1
    {G}^{\theta}(x,x;\mu^2)\,dx =}\\ \\
    \displaystyle{
    = \frac{1}{2\mu}\, \frac{(\theta+\log{2}-\gamma_E-\log{\mu})J_1(\mu)+
    \frac{\pi}{2}\,N_1(\mu)}
    {(\theta+\log{2}-\gamma_E-\log{\mu})J_0(\mu)+\frac{\pi}{2}\,N_0(\mu)} }\, .
\end{array}
\end{equation}

\bigskip

Las ecuaciones (\ref{Jupper}) y (\ref{N-sigma}) proveen el mismo
desarrollo asint{\'o}tico para ambas,
\begin{equation}\label{tr-gd-0}
    \begin{array}{c}
    \displaystyle{
      Tr\left( A^{\infty}-\mu^2\right)\sim
    \frac{e^{i\sigma \frac \pi 2}}{2\mu} \left(
    \frac{P(1,\mu)- i \sigma \, Q(1,\mu)}
    {P(0,\mu)- i \sigma \, Q(0,\mu)}\right)
    \sim
    Tr\left( A^{\beta}-\mu^2\right)\sim }\\ \\
    \displaystyle{\sim \sum_{k=1}^\infty \frac{A_k(1/2,\sigma)}{\mu^k}
      = {\frac{i\,\sigma }{2\mu } } +
  \frac{1}{4\,{\mu }^2} +
  {\frac{i \,\sigma}{16{\mu }^3} } -
  \frac{1}{16\,{\mu }^4} +
  {{O}({\mu^{-5} })} }\, ,
    \end{array}
\end{equation}
donde $\sigma = +1$ ($-1$) para $\Im(\mu)>0$ ($\Im(\mu)<0$.)

\bs

N{\'o}tese que el desarrollo asint{\'o}tico de la ecuaci{\'o}n
(\ref{tr-gd-0}) coincide con el miembro derecho de la ecuaci{\'o}n
(\ref{asymp-trGD-upp1}) evaluada en $\nu=0$. Por lo tanto, de la
ecuaci{\'o}n (\ref{zetaD+11}) se concluye que las singularidades
de la funci{\'o}n $\zeta^{\theta}(s)$, para $\nu=0$, consisten en
polos simples en los puntos $s_n$,
\begin{equation}
     s_n=\frac 1 2-n\quad {\rm con}\quad n=0,1,2,\dots
\end{equation}
cuyos residuos est{\'a}n dados por,
\begin{equation}\label{residuos-D-0}
  \left. {\rm Res}\,\zeta^{\theta}(s) \right|_{s=1/2-n}
  =- \frac{1}{\pi}\,\Re\left\{i\, A_{2n+1}(1/2,1)\right\}\,.
\end{equation}
A diferencia del caso $\nu\neq 0$, esta estructura de polos es
com{\'u}n a todas las extensiones autoadjuntas de $A$.

\bigskip

De modo que la estructura de polos de las funciones-$\zeta$
correspondientes a las extensiones autoadjuntas del operador dado
por (\ref{g1/2}) obedece tambi{\'e}n a la ecuaci{\'o}n
(\ref{resu}), v{\'a}lida para operadores regulares.


\part{Operadores de Dirac}\label{dirope}

\vspace{5mm}\begin{flushright}{\it O God! I could be bounded in a
nutshell,
and count myself\\king of infinite space - were it not that I have bad dreams.\\
(Hamlet.)}
\end{flushright}

\vspace{25mm}

En el cap{\'\i}tulo \ref{opesing} demostramos que la posici{\'o}n
de los polos de la funci{\'o}n-$\zeta$ co\-rres\-pon\-dien\-te a
un operador de Schr\"odinger en una dimensi{\'o}n con un
coeficiente singular proporcional a $x^{-2}$ puede depender de la
intensidad de la singularidad. Este resultado ha sido ilustrado
con dos ejemplos en el cap{\'\i}tulo \ref{apli}.

\bs

En este cap{\'\i}tulo, consideraremos dos operadores de Dirac con
un coeficiente singular proporcional a $x^{-1}$ y mostraremos que
la posici{\'o}n de los polos de las correspondientes
funciones-$\zeta$ tampoco est{\'a}n determinadas por el orden del
operador y la dimensi{\'o}n de la variedad sino que dependen de la
intensidad del coeficiente del t{\'e}rmino singular.

\bs

En la secci{\'o}n \ref{Seeley} resolveremos un operador de Dirac
con una singularidad de la forma $x^{-1}$ sobre la variedad
unidimensional compacta $[0,1]$ \cite{FMPS}. Encontraremos la
re\-so\-lu\-ci{\'o}n espectral del operador, obteniendo una
ecuaci{\'o}n trascendente para los autovalores y una forma
expl{\'\i}cita para las autofunciones. Esta resoluci{\'o}n
permitir{\'a} calcular las singularidades de la
funci{\'o}n-$\zeta$.

\bs

Finalmente, en la secci{\'o}n \ref{tubos}, estudiamos una
part{\'\i}cula sin masa, con carga y con spin, en $2+1$
dimensiones, en presencia de un campo magn{\'e}tico uniforme y de
un flujo singular de Aharonov-Bohm. Mostraremos que la
posici{\'o}n de los polos de la funci{\'o}n-$\zeta$
correspondiente al hamiltoniano de Dirac depende del valor del
flujo magn{\'e}tico singular.

\section{{Un operador de primer orden}}\label{Seeley}

En esta secci{\'o}n consideraremos un operador de Dirac con un
coeficiente sin\-gu\-lar definido sobre spinores de dos
componentes en una variedad de base unidimensional. Aunque el
estudio de las funciones espectrales desarrollado en el
cap{\'\i}tulo \ref{opesing} se refiere a operadores de
Schr\"odigner, la ecuaci{\'o}n de autovalores del operador de
primer orden que estudiaremos en la presente secci{\'o}n est{\'a}
relacionada con un operador de la forma (\ref{1}).

\bs

De manera que, en estrecha analog{\'\i}a con el operador de
supercarga estudiado en la secci{\'o}n \ref{rupturaSUSY}, las
funciones espectrales que estudiaremos en esta secci{\'o}n no
obedecen al comportamiento de las correspondientes a un operador
regular. En particular, la posici{\'o}n de los polos de las
funciones-$\zeta$ y $\eta$ dependen del coeficiente del
t{\'e}rmino singular en el operador diferencial. Una vez m{\'a}s,
este resultado esta condicionado por la existencia de un conjunto
infinito de extensiones autoadjuntas.

\subsection{El operador y sus extensiones autoadjuntas}
\label{the-operator1}

Consideremos el operador diferencial de Dirac,
\begin{equation}\label{Dpri}
  D=\left(\begin{array}{cc}
    0  & A^\dagger \\
    A & 0 \
  \end{array}\right)\, ,
\end{equation}
donde,
\begin{equation}\label{AA}
  A:= -\partial_x + \frac{\alpha}{x}=-
  x^{\alpha} \, \partial_x\, x^{-\alpha},
  \quad {A}^\dagger := \partial_x + \frac{\alpha}{x}=
  x^{-\alpha}\,\partial_x \, x^{\alpha}\, ,
\end{equation}
y $\alpha\in \mathbb{R}$, definido sobre
$\mathcal{D}(D)=\mathbb{C}^2\otimes \mathcal{C}_0^\infty(0,1)$. Se
puede probar sin dificultad que $D$ es sim{\'e}trico sobre este
dominio de definici{\'o}n.

\bs

El operador adjunto $D^\dagger$, que es la extensi{\'o}n maximal
de $D$, est{\'a} definido en el dominio $\mathcal{D}(D^\dagger)$
de funciones $\Phi(x)\in\mathbb{C}^2\otimes \mathbf{L_2}(0,1)$,
cuyas componentes $\phi_1(x),\phi_2(x)$ tienen una derivada
localmente sumable y que satisfacen,
\begin{equation}\label{DPhipri}
    D\Phi(x)=\left( \begin{array}{c}
      {A}^\dagger \phi_2(x) \\
      A \phi_1(x)
    \end{array} \right) = \left(\begin{array}{c}
      \tilde{\phi}_1(x) \\
      \tilde{\phi}_2(x)
    \end{array}\right)\in \mathbb{C}^2\otimes\mathbf{L_2}(0,1)\, .
\end{equation}

\bigskip

\begin{lem} \label{Lema1-1pri}
Si $\Phi(x)\in \mathcal{D}(D^\dagger)$ y $-\frac{1}{2}< \alpha<
\frac{1}{2}$, entonces existen constantes complejas $C_1[\Phi]$ y
$C_2[\Phi]$ tales que,
\begin{equation}\label{lemaI}
  \big|\,\phi_1(x)-C_1[\Phi]\, x^\alpha\big| +
  \big|\,\phi_2(x)-C_2[\Phi]\, x^{-\alpha}\big|
  \leq K_{\alpha}\, \|D\Phi(x)\| \, x^{1/2}\, ,
\end{equation}
donde $\| \cdot\|$ es la norma en
$\mathbb{C}^2\otimes\mathbf{L_2}$.
\end{lem}

\noindent{\bf Demostraci{\'o}n:} La ecuaci{\'o}n (\ref{DPhipri})
implica,
\begin{equation}\label{phi-chi-en0pri}
    \begin{array}{c}
      \phi_1(x)= C_1[\Phi]\, x^\alpha - x^\alpha \, \int_0^x y^{-\alpha}\,
      \tilde{\phi}_2(y)\, dy
     \,  ,\\ \\
     \phi_2(x)= C_2[\Phi]\, x^{-\alpha} + x^{-\alpha} \, \int_0^x y^{\alpha}\,
     \tilde{\phi}_1(y)\, dy
     \, ,
    \end{array}
\end{equation}
y teniendo en cuenta,
\begin{equation}\label{schwarzpri}
    \begin{array}{c}\displaystyle{
      \left|\int_0^x y^{\alpha} \, \tilde{\phi}_1(y)\, dy\right|\leq
      \frac{x^{\alpha+1/2}}{\sqrt{1+2\alpha}}\, \|\tilde{\phi}_1\| }\, ,
      \\ \\ \displaystyle{
      \left|\int_0^x y^{-\alpha} \, \tilde{\phi}_2(y)\, dy\right|\leq
      \frac{x^{-\alpha+1/2}}{\sqrt{1-2\alpha}}\, \|\tilde{\phi}_2\| }\,  ,
    \end{array}
\end{equation}
obtenemos la ecuaci{\'o}n (\ref{lemaI}) con $K_\alpha
=(1-2\alpha)^{-1/2}+(1+2\alpha)^{-1/2}$.\begin{flushright}$\Box$\end{flushright}

\begin{cor}\label{corola}
Sea $\Phi(x)=\left(\begin{array}{c}
  \phi_1(x) \\
  \phi_2(x)
\end{array}\right),\Psi(x)=\left(\begin{array}{c}
  \psi_1(x) \\
  \psi_2(x)
\end{array}\right) \in \mathcal{D}(D^\dagger)$. Entonces,
\begin{equation}\label{DDstarpri}\begin{array}{c}
    \left(D^\dagger \Psi, \Phi\right) -
  \left(\Psi, D^\dagger \Phi\right) = \\ \\
  = \Big\{
  C_1[\Psi]^* C_2[\Phi] - C_2[\Psi]^* C_1[\Phi]\Big\}+
  \Big\{\psi_2(1)^*\, \phi_1(1)
  - \psi_1(1)^*\, \phi_2(1) \Big\}\, .
\end{array}
\end{equation}
\end{cor}

\noindent{\bf Demostraci{\'o}n:} A partir de las ecuaciones
(\ref{AA}) se obtiene,
\begin{equation}\label{DDstar2pri}\begin{array}{c}
      \left(D^\dagger \Psi, \Phi\right) -
  \left(\Psi, D^\dagger \Phi\right) = \\ \\
    =\displaystyle{
    \lim_{\varepsilon \rightarrow 0^+}\int_\varepsilon^1
    \partial_x \Big\{x^\alpha\,\psi_2(x)^*\, x^{-\alpha}\,\phi_1(x)
    - x^{-\alpha}\,\psi_1(x)^* \, x^\alpha \, \phi_2(x)
    \Big\} dx}\, ,
\end{array}
\end{equation}
a partir de la cual, teniendo en cuenta los resultados del Lema
\ref{Lema1-1pri}, se deduce la ecuaci{\'o}n
(\ref{DDstarpri}).\begin{flushright}$\Box$\end{flushright}

De acuerdo con el Corolario (\ref{corola}), podemos definir el
mapeo,
\begin{equation}\begin{array}{c}
    K:\mathcal{D}(D^\dagger)\rightarrow \mathbb{C}^{4}\,,\\ \\
    \phi\rightarrow \left( C_1[\phi],C_2[\phi], \phi_1(1), \phi_2(1) \right)\,.
\end{array}
\end{equation}
El dominio $\mathcal{D}(\overline{D})$ de la clausura
$\overline{D}=(D^{\dagger})^\dagger$ del operador $D$ es el
n{\'u}cleo de $K$, en tanto que las extensiones sim{\'e}tricas de
$D$ est{\'a}n definidas sobre la preimagen de los
sub\-es\-pa\-cios de $\mathbb{C}^4$ bajo el mapeo $K$. Las
extensiones autoadjuntas, por su parte, est{\'a}n identificadas
con los subespacios lagrangianos $S\subset \mathbb{C}^4$, esto es,
tales que $S=S^{\perp}$, donde el complemento ortogonal se define
de acuerdo con la forma simpl{\'e}ctica del miembro derecho de la
ecuaci{\'o}n (\ref{DDstarpri}).

\bigskip

En adelante, consideraremos adem{\'a}s extensiones autoadjuntas
que satisfagan la condici{\'o}n de contorno local,
\begin{equation}\label{BC1pri}
  \phi_1(1)=0\, .
\end{equation}
De modo que cada una de estas extensiones est{\'a} determinada por
una condici{\'o}n de la forma,
\begin{equation}\label{BC2}
    \cos\gamma\, C_1[\Phi] + \sin\gamma\, C_2[\Phi] = 0\, ,
\end{equation}
con $\gamma \in [0,\pi)$. Denotaremos esta extensi{\'o}n por
$D_{\gamma}$.

\subsubsection{El espectro} \label{the-spectrum}

Para determinar el espectro de las extensiones autoadjuntas
$D_\gamma$ de $D$ debemos encontrar las soluciones de,
\begin{equation}\label{Ec-hompri}
  (D-\lambda)\Phi(x)=0 \Rightarrow \left\{
  \begin{array}{c}
    {A}^\dagger \phi_2(x) = \lambda \phi_1(x)\, ,\\ \\
    A \phi_1(x) = \lambda \phi_2(x)\, ,
  \end{array}\right.
\end{equation}
que satisfagan las condiciones de contorno dadas por las
ecuaciones (\ref{BC1pri}) y
 (\ref{BC2}).

\bigskip

La soluci{\'o}n de la ecuaci{\'o}n (\ref{Ec-hompri}) para
$\lambda=0$ est{\'a} dada por,
\begin{equation}\label{lambda0pri}
  \Phi(x)=
\begin{pmatrix}
  C_1 \, x^\alpha\\
  C_2 \, x^{-\alpha}
\end{pmatrix}\, ,
\end{equation}
pero las condiciones de contorno (\ref{BC1pri}) y (\ref{BC2})
implican que $C_1 =
 0$ y $C_2=0$, a excep\-ci{\'o}n del caso $\gamma =0$. En consecuencia, s{\'o}lo existe
 un modo cero para la extensi{\'o}n autoadjunta $D_{0}$.

\bigskip

Para resolver el sistema (\ref{Ec-hompri}) en el caso $\lambda\neq
0$, aplicamos el operador ${A}^\dagger$ a la segunda de sus
ecuaciones y ,utilizando la primera, obtenemos,
\begin{equation}\label{hom-sec}
    \left\{-\partial_x^2 + \frac{\nu^2-1/4}{x^2}
    - \lambda^2\right\} \phi_1(x)=0\,,
\end{equation}
donde $0<\nu:=1/2-\alpha<1$. Las soluciones de (\ref{hom-sec})
tienen la forma,
\begin{equation}\label{sol-hom-1pri}
  \phi_1(x) = B_1\, \sqrt{y}\, J_{\nu}(y)
  + B_2 \, \sqrt{y} \, J_{-\nu}(y)\, ,
\end{equation}
siendo $y:=|\lambda|\, x$ y $B_1, B_2$ son constantes complejas.
Esto implica, junto con la segunda ecuaci{\'o}n del sistema
(\ref{Ec-hompri}), que la componente inferior de $\Phi(x)$
est{\'a} dada por,
\begin{equation}\label{hom-2}
  \phi_2(x)= \sigma
  \left\{-B_1 \, \sqrt{y}\, J_{-1+\nu}(y)
  + B_2\, \sqrt{y} \, J_{1-\nu}(y)\right\}\, ,
\end{equation}
donde $\sigma=|\lambda|/{\lambda}$.

\bs

La condici{\'o}n de contorno (\ref{BC2}) en el origen determina
una relaci{\'o}n entre las cons\-tan\-tes $B_1$ y $B_2$. Para
ello, tenemos en cuenta que,
\begin{equation}\label{J-Besselpri}
  J_\nu(y)= y^{\nu }\,\left\{ \frac{1}
     {2^{\nu }\,
       \Gamma(1 + \nu )}
     + O(y^2) \right\}\, ,
\end{equation}
y obtenemos,
\begin{equation}\label{ec-spectrumpri}\begin{array}{c}
    \cos\gamma\, C_1[\Phi] + \sin\gamma\, C_2[\Phi]
    \displaystyle{ =
    \frac{\cos\gamma\, B_2\, |\lambda|^{1/2-\nu}}
    {2^{-\nu}
  \Gamma\left(1-\nu\right)}
  - \sigma
  \frac{\sin\gamma\, B_1\,
  |\lambda|^{-1/2+\nu}}{2^{-1+\nu}
  \Gamma\left(\nu\right)}=0}\,.
\end{array}
\end{equation}
Esta ecuaci{\'o}n conduce a la relaci{\'o}n,
\begin{equation}\label{spectrum}
  \frac{B_2}{B_1} = \sigma\,
  |\lambda|^{-1+2\nu} \left[
  \frac{4^{1/2-\nu}\, \Gamma\left(1-\nu\right)}
  {\Gamma\left(\nu\right)}\right]
  \tan\gamma\, .
\end{equation}
La condici{\'o}n de contorno (\ref{BC1pri}) en $x=1$ puede
escribirse, pos su parte, de la forma,
\begin{equation}\label{sol-hom-1pripri}
  \frac{B_2}{B_1}=-\frac{J_{\nu}(|\lambda|)}{J_{-\nu}(|\lambda|)}\,.
\end{equation}
Las ecuaciones (\ref{spectrum}) y (\ref{sol-hom-1pripri})
determinan el espectro de cada extensi{\'o}n autoadjunta.

\begin{itemize}
\item Si $\gamma = \pi/2$, la ecuaci{\'o}n (\ref{ec-spectrumpri})
implica que $B_1 = 0$. Por consiguiente, $\phi_1(1)=0 \Rightarrow
J_{-\nu}(|\lambda|)=0$. El espectro de la extensi{\'o}n
$D_{\pi/2}$ es sim{\'e}trico con respecto al origen y sus
autovalores est{\'a}n dados por,
\begin{equation}\label{alpha0pri}
  \lambda_{\pm,n} = \pm j_{-\nu,n}\, ,
   \quad n=1,2,\dots\, ,
\end{equation}
donde $j_{-\nu,n}$ es el $n$-{\'e}simo cero positivo de la
funci{\'o}n de Bessel $J_{-\nu}(z)$.

\item Si $\gamma = 0$, la ecuaci{\'o}n (\ref{ec-spectrumpri})
implica que $B_2 = 0$. Por consiguiente, $\phi_1(1)=0 \Rightarrow
J_{\nu}(|\lambda|)=0$. El espectro de la extensi{\'o}n $D_{0}$ es
entonces sim{\'e}trico con respecto al origen y sus autovalores
est{\'a}n dados por,
\begin{equation}\label{eigen-beta0pri}
  \lambda_0 =0\, , \quad
  \lambda_{\pm,n} = \pm j_{\nu,n}\, , \ n=1,2,\dots
\end{equation}
donde $j_{\nu,n}$ es el $n$-{\'e}simo cero positivo de la
funci{\'o}n de Bessel $J_{\nu}(z)$.

\item En general, si $\gamma\neq 0,\pi/2$, los autovalores de la
extensi{\'o}n $D_\gamma$ est{\'a}n determinados por la siguiente
ecuaci{\'o}n trascendente que se deriva de las ecuaciones
(\ref{spectrum}) y (\ref{sol-hom-1pripri}),
\begin{equation}\label{eigenvalues}
  |\lambda|^{1-2\nu}\,
  \frac{J_{\nu}(|\lambda|)}
  {J_{-\nu}(|\lambda|)}
  =\sigma \, \beta\, ,
\end{equation}
donde hemos definido,
\begin{equation}\label{rhopri}
  \beta:= 4^{1/2-\nu}\,\nu\,
   \frac{\Gamma\left(-\nu\right)}
  {\Gamma\left(\nu\right)}\,
  \tan\gamma\, .
\end{equation}
De acuerdo con esta definici{\'o}n designaremos, en adelante, a
las extensiones autoadjuntas de $D$ por $D^\beta$.

\bs

Para los autovalores positivos $\sigma =1$ y la ecuaci{\'o}n
(\ref{eigenvalues}) se reduce a,
\begin{equation}\label{eigenvalues-pos}
  F(\lambda):=\lambda^{1-2\nu}\,
  \frac{J_{\nu}(\lambda)}{J_{-\nu}(\lambda)}
  =\beta\,.
\end{equation}
Esta relaci{\'o}n es representada en la Figura \ref{figurepri}
para valores particulares de $\beta,\nu$.

\begin{figure}
\center
    \epsffile{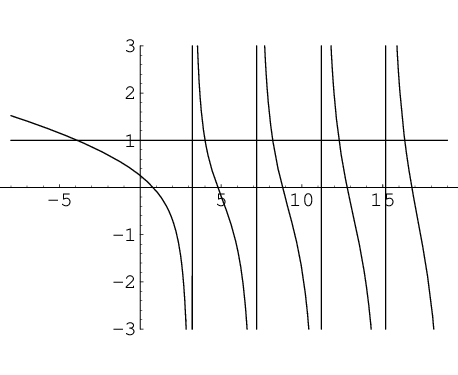}
    \caption{{\small Gr{\'a}fica de $F(\lambda):= \displaystyle{
    \lambda^{1-2\nu}\, {J_{\nu}(\lambda)}/{J_{-\nu}(\lambda)}
  }$, con $\nu=1/6$, y $\beta=1$.}} \label{figurepri}
\end{figure}

\bs

Por el contrario, para los autovalores negativos $\sigma = e^{-i\,
\pi}$ y la ecuaci{\'o}n (\ref{eigenvalues}) toma entonces la
forma,
\begin{equation}\label{eigenvalues-neg}
  F(|\lambda|)= -\beta\, .
\end{equation}
En consecuencia, los autovalores negativos de $D^{\beta}$ son los
opuestos de los au\-to\-va\-lo\-res positivos de $D^{-\beta}$.
\end{itemize}

N{\'o}tese que el espectro es no degenerado y que existe un
autovalor positivo entre cada par consecutivo de ceros de
$J_{-\nu}(\lambda)$. Adem{\'a}s, el espectro es sim{\'e}trico
respecto del origen s{\'o}lo para las extensiones  $\gamma=\pi/2$
($\beta=\infty$) y $\gamma=0$ ($\beta=0$).

\subsection{La resolvente} \label{the-resolvent}

En esta secci{\'o}n construiremos la resolvente
$(D^\beta-\lambda)^{-1}$ para cada extensi{\'o}n autoadjunta
$D^\beta$ de $D$. Primeramente, consideraremos las condiciones de
contorno invariantes de escala correspondientes a las extensiones
$\beta=\infty$ y $\beta=0$. La resolvente para una extensi{\'o}n
autoadjunta general ser{\'a} posteriormente expresada como una
combinaci{\'o}n lineal de las resolventes de estos dos casos
particulares.

\bigskip

El n{\'u}cleo de la resolvente,
\begin{equation}\label{resolv}
  G^\beta(x,x'; \lambda)=\left(\begin{array}{cc}
    G^\beta_{11}(x,x'; \lambda) & G^\beta_{12}(x,x'; \lambda) \\
    G^\beta_{21}(x,x'; \lambda) & G^\beta_{22}(x,x'; \lambda)
  \end{array}\right)\, ,
\end{equation}
verifica,
\begin{equation}\label{Gpri}
   (D-\lambda)\, G^\beta(x,x'; \lambda) =
\delta(x,x')\, \mathbf{1}\, .
\end{equation}
A partir de esta ecuaci{\'o}n obtenemos para los elementos de la
diagonal,
\begin{equation}\label{ec-dif-g-diag}
  \begin{array}{c}
  \displaystyle{
    \left\{ -\partial_x^2 + \frac{\nu^2-1/4}{x^2}
   - \lambda^2 \right\}  G^\beta_{11}(x,x'; \lambda) = \lambda\,
     \delta(x,x')\, ,} \\\\
       \displaystyle{
       \left\{ -\partial_x^2 + \frac{\nu^2-1/4}{x^2}
    - \lambda^2 \right\}  G^\beta_{22}(x,x'; \lambda) = \lambda\,
     \delta(x,x') }\, ,
  \end{array}
\end{equation}
en tanto que los elementos no diagonales est{\'a}n dados por,
\begin{equation}\label{ec-dif-g-nodiag}
  \begin{array}{c}
      \displaystyle{
    G^\beta_{21}(x,x'; \lambda)= \frac{1}{\lambda}
    \left\{- \partial_x + \frac \alpha x \right\}
    G^\beta_{11}(x,x'; \lambda)\, , }\\ \\
      \displaystyle{
    G^\beta_{12}(x,x'; \lambda)= \frac{1}{\lambda}
    \left\{ \partial_x + \frac \alpha x \right\}
    G^\beta_{22}(x,x'; \lambda)\, ,}
  \end{array}
\end{equation}
para $\lambda\neq 0$. Como la resolvente es anal{\'\i}tica en
$\lambda$, ser{\'a} suficiente evaluarla en el semiplano
$\mathcal{R}(\lambda)>0$.

\bigskip

Para construir el n{\'u}cleo de la resolvente mediante la
expresi{\'o}n (\ref{solres}) utilizaremos las soluciones de las
ecuaciones (\ref{Ec-hompri}). Definimos entonces,
\begin{equation}\label{soluciones}
    \left\{
  \begin{array}{l}
    L_1^\infty(y)= \sqrt{y}\, J_{-\nu}(y)\,  ,\\ \\
        L_2^\infty(y)= \sqrt{y}\, J_{1-\nu}(y)\,  ,\\ \\
        L_1^0(y)= \sqrt{y}\, J_{\nu}(y)\,  ,\\ \\
        L_2^0(y)= \sqrt{y}\, J_{-1+\nu}(y)\, , \\ \\
    R_1(y;\lambda)=\sqrt{y} \left[J_{-\nu}({\lambda})
     J_{\nu}(y)
    - J_{\nu}({\lambda}) J_{-\nu}(y)\right]\, ,
    \\ \\
    R_2(y;\lambda)=\sqrt{y} \left[J_{-\nu}({\lambda})
    J_{-1+\nu}(y)
    + J_{\nu}({\lambda}) J_{1-\nu}(y)\right]\, .
  \end{array}
  \right.
\end{equation}
N{\'o}tese que $R_1({\lambda};\lambda)=0$ y
$\left.{A^\dagger}\,R_2({\lambda}\, x;\lambda )\right|_{x=1}=0$.
Los wronskianos correspondientes est{\'a}n dados por,
\begin{equation}\label{W-D}
\left\{
\begin{array}{c}
  \displaystyle{
  W\left[ L_1^\infty, R_1\right](\lambda) =
   - \frac 2 \pi \, \sin(\pi\nu) \,
  J_{-\nu}({\lambda})\, ,
    } \\ \\
  \displaystyle{
  W\left[ L_2^\infty, R_2 \right](\lambda) =
   \frac 2 \pi \, \sin(\pi\nu) \,
  J_{-\nu}({\lambda})\, ,
  } \\ \\
    \displaystyle{
  W\left[ L_1^0, R_1 \right](\lambda) =
   - \frac 2 \pi \, \sin(\pi\nu) \,
  J_{\nu}({\lambda})\, ,
    } \\ \\
  \displaystyle{
  W\left[ L_2^0, R_2 \right] (\lambda)=
   - \frac 2 \pi \, \sin(\pi\nu) \,
  J_{\nu}({\lambda})
  }\, ,
  \end{array}
  \right.
\end{equation}
cuyos ceros coinciden con los ceros de $J_{-\nu}({\lambda})$ y
$J_{\nu}({\lambda})$, respectivamente.

\subsubsection{La resolvente para la extensi{\'o}n $\beta=\infty$}

Las condiciones de contorno que determinan, junto con las
ecuaciones (\ref{ec-dif-g-diag}), los elementos diagonales del
n{\'u}cleo de la resolvente se obtienen imponiendo que la
funci{\'o}n de dos componentes $\Phi(x)$,
\begin{equation}\label{func-inhom}
  \Phi(x) =\left(\begin{array}{c}\phi_1(x)\\ \phi_2(x)
  \end{array}\right)=
   \int_0^1 G^\infty( x, x'; \lambda)
  \left( \begin{array}{c}
    f_1(x') \\
    f_2(x')
  \end{array} \right) dx'
\end{equation}
satisfaga,
\begin{equation}
\phi_1(1)=0\quad{\rm y}\quad C_2[\Phi]=0\,,
\end{equation}
para cualquier par de funciones $f_1(x),f_2(x)\in
\mathbf{L_2}(0,1)$. Esto implica que,
\begin{eqnarray}
  G_{11}^\infty (x,x'; \lambda)= -\frac{1}{W\left[ L_1^\infty, R_1
  \right](\lambda) } \times  \left\{
  \begin{array}{c}
    L_1^\infty(y)\, R_1(y'; \lambda)\quad {\rm si}\ x\leq x'\, , \\ \\
    R_1(y; \lambda)\, L_1^\infty(y') \quad {\rm si}\ x \geq x'\, ,
  \end{array}\right.\label{GD11666}\\\nonumber\\
  G_{22}^\infty (x,x'; \lambda)= -\frac{1}{W\left[ L_1^\infty, R_1
  \right](\lambda) }\times  \left\{
  \begin{array}{c}
    L_2^\infty(y)\, R_2(y'; \lambda)\quad {\rm si}\ x\leq x'\, , \\ \\
    R_2(y; \lambda)\, L_2^D(y')\quad {\rm si}\ x \geq x'\, .
  \end{array}\right.\label{GD22}
\end{eqnarray}
Las componentes $G_{12}^\infty(x,x'; \lambda)$ y
$G_{21}^\infty(x,x'; \lambda)$, por su parte, est{\'a}n dadas por
la ecuaci{\'o}n (\ref{ec-dif-g-nodiag}). Las
 ecuaciones (\ref{soluciones}) y (\ref{W-D}) permiten verificar tanto las
 condiciones de contorno como $(D-\lambda)\, \Phi(x)=\begin{pmatrix}
  f_1(x) \\
  f_2(x)
\end{pmatrix}$.
En efecto, a partir de las ecuaciones (\ref{ec-dif-g-nodiag}),
(\ref{GD11666}) y (\ref{GD22}) se obtiene,
\begin{equation}\label{near0-Dpri}
    \phi_1(x)=
    C_1^\infty[\Phi] \, x^{\frac 1 2 -\nu} + O(\sqrt{x})\, ,\quad
    \phi_2(x)= O(\sqrt{x})\,  ,
\end{equation}
con,
\begin{equation}\label{C+}\begin{array}{c}
  \displaystyle{C_1^\infty[\Phi] =   \frac{-\, \pi\, \lambda^{\frac 1 2 -\nu}}
    {2^{1-\nu} \sin(\pi\nu)
    J_{-\nu}({\lambda}) \,
    \Gamma\left(1-\nu\right)}}
  \displaystyle{  \cdot
    \int_0^1 \Big[R_1(y; \lambda) f_1(y)
    - R_2(y; \lambda) f_2(y)\Big] \,dy}\, ,
\end{array}
\end{equation}
para $\lambda$ distinto de todo cero de $J_{-\nu}({\lambda})$.
N{\'o}tese que $C_1^\infty[\Phi]\neq 0$ si la integral del miembro
derecho de la
 ecuaci{\'o}n (\ref{C+}) no se anula.

\subsubsection{La resolvente de la extensi{\'o}n $\beta=0$}

En este caso, la funci{\'o}n,
\begin{equation}\label{func-inhom-N}
  \Phi(x)=\left(\begin{array}{c}\phi_1(x)\\ \phi_2(x)
  \end{array}\right) = \int_0^1 G^0(x, x'; \lambda)
  \left( \begin{array}{c}
    f_1(x') \\
    f_2(x')
  \end{array} \right) \,dx'
\end{equation}
debe verificar,
\begin{equation}
    \phi_1(1)=0\quad{\rm y}\quad C_1[\Phi]=0\,,
\end{equation}
para cualquier par de funciones $f_1(x),f_2(x)\in
\mathbf{L_2}(0,1)$. Esto implica que,
\begin{eqnarray}
  G_{11}^0 (x,x'; \lambda)= -\frac{1}{W\left[ L_1^0, R_1\right](\lambda)}
  \times  \left\{
  \begin{array}{c}
    L_1^0(y)\, R_1(y'; \lambda)\quad {\rm si}\ x\leq x'\, , \\ \\
    R_1(y; \lambda)\, L_1^0(y')\quad {\rm si}\ x \geq x'\, ,
  \end{array}\right.\label{GN11}\\\nonumber\\
  G_{22}^0 (x,y; \lambda)=-\frac{1}{W\left[ L_1^0, R_1 \right](\lambda)}
  \times  \left\{
  \begin{array}{c}
    L_2^0(y)\, R_2(y'; \lambda),\ {\rm si}\ x\leq x'\, , \\ \\
    R_2(y; \lambda)\, L_2^0(y'), \ {\rm si}\ x \geq x'.
  \end{array}\right.\label{GN22}
\end{eqnarray}
Las componentes $G_{12}^0(x,x'; \lambda)$ y $G_{21}^0(x,x';
\lambda)$ est{\'a}n dadas por la ecuaci{\'o}n
(\ref{ec-dif-g-nodiag}). Las ecuaciones (\ref{soluciones}) y
(\ref{W-D}) permiten verificar tanto las condiciones de contorno
como la
 relaci{\'o}n $(D-\lambda)\, \Phi(x)=\begin{pmatrix}
  f_1(x) \\
  f_2(x)
\end{pmatrix}$. En efecto, las ecuaciones (\ref{ec-dif-g-nodiag}),
(\ref{GN11}) y (\ref{GN22}) permiten obtener,
\begin{equation}\label{near0-Npri}
    \phi_1(x)=
    O(\sqrt{x})\, ,\quad
    \phi_2(x)= C_2^N[\Phi] \, x^{-\frac 1 2 +\nu} + O(\sqrt{x})\, ,
\end{equation}
con,
\begin{equation}\label{C-}\begin{array}{c}
  \displaystyle{C_2^N[\Phi] =   \frac{\pi\, \lambda^{\frac 1 2+\nu}}
    {2^{\nu} \sin(\pi\nu)
    J_{\nu}({\lambda}) \,
    \Gamma\left(\nu\right)} \times} \\ \\
  \displaystyle{ \phantom{C_2[\Phi] = } \times
    \int_0^1 \Big[R_1(\lambda\, x'; \lambda) f_1(x')
    - R_2(\lambda\, x'; \lambda) f_2(x')\Big] dx'\, ,}
\end{array}
\end{equation}
para $\lambda$ distinto de todo cero de $J_{\nu}({\lambda})$.

\bigskip

N{\'o}tese que $C_2^N[\Phi]\neq 0$ si la integral del miembro
derecho de la ecuaci{\'o}n (\ref{C-}) (la misma que para la
extensi{\'o}n $\beta=\infty$, dada por la ecuaci{\'o}n (\ref{C+}))
no se anula.

\subsubsection{La resolvente para una extensi{\'o}n autoadjunta general}

Para el caso general imponemos la condici{\'o}n de contorno,
\begin{equation}\label{BC-generalpri}
  \phi_1(1)=0\, ,\quad \cos\gamma\, C_1[\Phi] + \sin\gamma\, C_2[\Phi] =
  0\, \, ,
\end{equation}
para,
\begin{equation}\label{func-inhom-genpri}
  \Phi(x)=\left(\begin{array}{c}\phi_1(x)\\ \phi_2(x)
  \end{array}\right) = \int_0^1 G^\beta( x,  x'; \lambda)
  \left( \begin{array}{c}
    f_1(x') \\
    f_2(x')
  \end{array} \right) \,dx'\, ,
\end{equation}
y cualquier par de funciones $f_1(x),f_2(x)\in \mathbf{L_2}(0,1)$.
Para ello, consideramos una combinaci{\'o}n lineal de las
resolventes de los dos casos particulares $\beta=\infty$ y
$\beta=0$,
\begin{equation}\label{linear-comb}
  G^\beta(x,x'; \lambda)= \left[1- \tau(\lambda)\right] G^\infty(x,x';
  \lambda) +
  \tau(\lambda)\, G^0(x,x'; \lambda)\, .
\end{equation}

Dado que la condici{\'o}n de contorno en $x=1$ se satisface
autom{\'a}ticamente, el factor $\tau (\lambda)$ queda determinado
por la condici{\'o}n,
\begin{equation}\label{ec-tau}
  \cos\gamma \left[1- \tau(\lambda)\right] C_1^\infty[\Phi]
  + \sin\gamma\, \tau(\lambda)\, C_2^0[\Phi] =0\, .
\end{equation}
A partir de la ecuaci{\'o}n (\ref{ec-tau}) ob\-te\-ne\-mos,
\begin{equation}\label{taudelambda}\begin{array}{c}
    \tau(\lambda) = \displaystyle{\frac{\cos\gamma \, C_1^\infty[\Phi]}
  {\cos\gamma \, C_1^\infty[\Phi]-\sin\gamma\,  C_2^0[\Phi]}}=
    1-\displaystyle{ \frac{1}
  {1 - \displaystyle{\beta^{-1}\,{\lambda }^{1-2\nu}\,
  \frac{ J_{\nu}(\lambda )}{
       J_{-\nu}(\lambda )} }} \, ,          }
\end{array}
\end{equation}
para $\lambda$ distinto de todo cero de ${\beta}J_{-\nu}(\lambda
)-{\lambda^{1-2\nu}}J_{\nu}(\lambda )$.

\subsubsection{La traza de la resolvente} \label{trace-resolvent-1}

La ecuaci{\'o}n (\ref{linear-comb}) indica que la resolvente
$\left(D^\beta-\lambda\right)^{-1}$ de una extensi{\'o}n
autoadjunta arbitraria $D^\beta$ puede expresarse en t{\'e}rminos
de las resolventes de las dos extensiones particulares
$\left(D^\infty-\lambda\right)^{-1}$ y
$\left(D^0-\lambda\right)^{-1}$. Adem{\'a}s, dado que los
autovalores de cualquier extensi{\'o}n crecen linealmente con $n$
(v{\'e}ase la secci{\'o}n \ref{the-spectrum}), estas resolventes
son o\-pe\-ra\-do\-res de Hilbert-Schmidt y sus derivadas con
respecto a $\lambda$ son operadores tipo traza.

\bs

El cuadrado $\left(D^\beta-\lambda\right)^{-2}$ de la resolvente
puede escribirse entonces de la siguiente manera,
\begin{equation}\label{derivG}\begin{array}{c}
    \left(D^\beta-\lambda\right)^{-2}=\partial_\lambda \,
    \left(D^\beta-\lambda\right)^{-1} = \\ \\=
    \partial_\lambda\,\left(D^\infty-\lambda\right)^{-1}
      \mbox{}+ \partial_\lambda\tau(\lambda) \left[\left(D^0-
      \lambda\right)^{-1}-\left(D^\infty-\lambda\right)^{-1}\right]+\\ \\
      \mbox{}+ \tau(\lambda)
      \left[\partial_\lambda \,\left(D^0-\lambda\right)^{-1}
    -\partial_\lambda\,\left(D^\infty-\lambda\right)^{-1}\right]\, .
\end{array}
\end{equation}
La ecuaci{\'o}n (\ref{derivG}) indica que la diferencia
$\left(D^0-\lambda\right)^{-1}-\left(D^\infty-\lambda\right)^{-1}$
es un operador fuertemente anal{\'\i}tico de $\lambda$, excepto en
los ceros de $\partial_\lambda\tau(\lambda)$, que toma valores en
el ideal de operadores tipo traza.

\bigskip

A partir de las expresiones para los elementos de la diagonal de
$\left(D^\infty-\lambda\right)^{-1}$ y
\linebreak$\left(D^0-\lambda\right)^{-1}$ obtenidas en la
secciones anteriores (v{\'e}anse ecuaciones (\ref{GD11666}),
(\ref{GD22}), (\ref{GN11}) y (\ref{GN22})),
obtenemos\,\footnote{Los detalles pueden encontrarse en la
secci{\'o}n \ref{integrals2}},
\begin{equation}\label{traza-derivGD}\begin{array}{c}
        {\rm Tr}\,\partial_\lambda \,\left(D^\infty-\lambda\right)^{-1} =
        \displaystyle{
      \int_0^1
      {\rm tr}\{
      \partial_\lambda\, G^\infty( x, x; \lambda)\}\, dx} =
    \\ \\
    = \displaystyle{  \partial_\lambda
      \left\{\frac{J_{1-\nu}(\lambda)}
      {J_{-\nu}(\lambda)}\right\} =
      1-\frac{1-2\nu}{\lambda}\,
    \frac{J_{1-\nu}(\lambda)}{J_{-\nu}(\lambda)}
    +\frac{J_{1-\nu }^2(\lambda)}{J_{-\nu}^2(\lambda)}
    =} \\ \\
    \displaystyle{
    =1 - \frac{\left(\frac 1 2 -\nu\right)^2}{{\lambda }^2} +
  {\left( \frac{1}{2\,\lambda } +
      \frac{{J'_{-\nu}}( \lambda )}
      {J_{-\nu}(\lambda )} \right) }^2 \, .}
\end{array}
\end{equation}
En esta expresi{\'o}n, ``${\rm tr}$'' representa la traza
matricial, en oposici{\'o}n a la traza del operador ``${\rm
Tr}$.'' An{\'a}logamente,
\begin{equation}\label{TrGD-GN}\begin{array}{c}
      {\rm Tr\,}\{\left(D^0-\lambda\right)^{-1}-
    \left(D^\infty-\lambda\right)^{-1}\}
    =\displaystyle{
    \frac{2\nu-1}{\lambda} -
        \frac{J_{1-\nu }(\lambda)}{J_{-\nu}(\lambda)} -
    \frac{J_{-1+\nu}(\lambda)}
    {J_{\nu}(\lambda)}=
    } \\ \\
  \displaystyle{= \frac{1-2\nu}{\lambda } +
  \frac{{J'_{\nu}}(\lambda )}
  {J_{\nu}(\lambda )} -
  \frac{{J'_{-\nu}}(\lambda )}
  {J_{-\nu}(\lambda )}}\, .
\end{array}
\end{equation}
Por su parte, dado que,
\begin{eqnarray}\label{derTr}
  \partial_\lambda {\rm Tr}\,\{\left(D^0-\lambda\right)^{-1}
    -\left(D^\infty-\lambda\right)^{-1}\}
    ={\rm Tr}\,\{\partial_\lambda \,\left(D^0-\lambda\right)^{-1}-
    \partial_\lambda\,\left(D^\infty-\lambda\right)^{-1}\}\, ,
\end{eqnarray}
obtenemos,
 \begin{equation}\label{trazas}
  \begin{array}{c}
    {\rm Tr}\,\{\partial_\lambda\,\left(D^0-\lambda\right)^{-1} -
    \partial_\lambda\,\left(D^\infty-\lambda\right)^{-1}\}
    =\\ \\
    \displaystyle{=
    \frac{1-2\nu}{\lambda^2}+
    \frac{1-2\nu}{\lambda}\left[
    \frac{J_{1-\nu}(\lambda)}{J_{-\nu}(\lambda)}+
    \frac{J_{-1+\nu}(\lambda)}{J_{\nu}(\lambda)}
    \right] - \left[
    \frac{J_{1-\nu}^2(\lambda)}{J_{-\nu}^2(\lambda)}-
    \frac{J_{-1+\nu}^2(\lambda)}{J_{\nu}^2(\lambda)}
    \right]=
    } \\ \\
    \displaystyle{=
    \frac{2\nu-1}{{\lambda }^2} -
  {\left( \frac{1}{2\,\lambda } +
      \frac{{J'_{\nu}}(
         \lambda )}{J_{\nu}(\lambda )}
      \right) }^2 + {\left( \frac{1}
       {2\,\lambda } +
      \frac{{J'_{-\nu}}( \lambda )}
      {J_{-\nu}( \lambda )} \right) }^2}\, .
  \end{array}
 \end{equation}

\bs

Estas expresiones permiten calcular la traza del cuadrado de la
resolvente de una extensi{\'o}n autoadjunta general,
\begin{eqnarray}\label{TrG2}
  {\rm Tr}\left(D^\beta-\lambda\right)^{-2} =  {\rm Tr}\,\partial_\lambda\,
    \left(D^\infty-\lambda\right)^{-1} +\nonumber\\+
  \partial_\lambda\, \Big[ \tau(\lambda)\,
  {\rm Tr}\,\{\left(D^0-\lambda\right)^{-1}-\left(D^\infty-
  \lambda\right)^{-1}\}\Big]\, .
\end{eqnarray}

\subsubsection{Desarrollo asint{\'o}tico de la traza de la
 resolvente}\label{Asymptotic-expansion-1}

Utilizando el desarrollo asint{\'o}tico de Hankel para las
funciones de Bessel (v{\'e}ase la secci{\'o}n \ref{Hankel})
obtenemos para el primer t{\'e}rmino del miembro derecho de la
ecuaci{\'o}n (\ref{TrG2}),
\begin{equation}\label{asymp-trGD-upp}\begin{array}{c}
    \displaystyle{{\rm Tr}\,\partial_\lambda\,\left(D^\infty-
    \lambda\right)^{-1} \sim
  \sum_{k=2}^\infty \frac{A_k(\nu,\sigma)}{\lambda^k}
  = }\\ \\ =\displaystyle{
   - \frac{\frac 1 2-\nu }
   {{\lambda }^2} + i\,\sigma\,
  \frac{\nu^2 -\frac 1 4}{{\lambda }^3} -
  \frac{3}{2}\, \frac{
      \nu^2 -\frac 1 4}{{\lambda }^4} \,  +
   {{O}\left(\frac{1}
      {\lambda }\right)}^6
    }\, ,
\end{array}
\end{equation}
donde $\sigma = 1$ para $\Im(\lambda)>0$ y $\sigma = -1$ para
$\Im(\lambda)<0$. Los coeficientes de esta serie puede ser
evaluados a partir de las ecuaciones (\ref{P+Q}) y (\ref{T}).
N{\'o}tese que $A_k(\nu,-1)=A_k(\nu,1)^*$, pues $A_{2k}(\nu,1)$ es
real y $A_{2k+1}(\nu,1)$ es imaginario puro.

\bs

An{\'a}logamente, de las ecuaciones (\ref{TrGD-GN}),
(\ref{trazas}) y (\ref{JprimasobreJasymp}) obtenemos,
\begin{equation}\label{trdifasymp}
   {\rm Tr}\,\{\left(D^0-\lambda\right)^{-1}-
    \left(D^\infty-\lambda\right)^{-1}\} \sim \frac{1-2\nu}{\lambda}\,,
\end{equation}
y,
\begin{equation}\label{trdifderasymp}
   {\rm Tr}\,\{\partial_\lambda \left(D^0-\lambda\right)^{-1} -
    \partial_\lambda\left(D^\infty-\lambda\right)^{-1}\} \sim
    -\frac{1-2\nu}{\lambda^2}\, .
\end{equation}

\bigskip

Por otra parte, teniendo en cuenta la ecuaci{\'o}n
(\ref{JsobreJup}),
\begin{equation}\label{tau-asymp}\begin{array}{c}
   \tau(\lambda) \sim \displaystyle{1 -
\left(1 - \displaystyle{\frac{e^
         {\sigma \,i
           \pi \nu }
           \,{\lambda }^{1-2\nu}}{\beta        }}\right)^{-1}
        \sim }\\ \\  \sim   \left\{
  \begin{array}{l}
  \displaystyle{
    -\sum_{k=1}^\infty \left( \frac{e^{\sigma\, i \pi \nu}
    \lambda^{1-2\nu}}{\beta}\right)^k  \quad {\rm si}\quad
    \frac 1 2 < \nu < 1\, ,} \\ \\
      \displaystyle{
    \sum_{k=0}^\infty \left(\beta\,
    {e^{-\sigma\, i \pi \nu} \,
    \lambda^{-1+2\nu}}\right)^k  \quad {\rm si}\quad
    0 < \nu < 1/2\, ,}
  \end{array}
  \right.
\end{array}
\end{equation}
donde $\sigma =1$ ($\sigma =-1$) corresponde a $\Im(\lambda)>0$
($\Im(\lambda)<0$.) N{\'o}tese la presencia de potencias de
$\lambda$ no enteras, dependientes de $\nu$, en este desarrollo.

\bs

De manera an{\'a}loga obtenemos,
\begin{equation}\label{tauprima}\begin{array}{c}
    \partial_\lambda\tau(\lambda) \sim \displaystyle{-
    \frac{e^
     {\sigma\, i \pi  \nu }{\lambda }^{ -2\nu }}
    {\beta }\,
  {\left[ 1 - (1-2\nu)\frac{e^
         {\sigma\, i \pi   \nu }\,{\lambda }^{1-2\nu}}{\beta
        } \right] }^{-2}
         \sim } \\ \\
   \sim \displaystyle{
   \left\{\begin{array}{l}
     \displaystyle{
    - \frac{1-2\nu}{\lambda}\,\sum_{k=1}^\infty k
    \left( \frac{e^{\sigma\, i \pi \nu}\,
    \lambda^{1-2\nu}}{\beta}\right)^k  \quad {\rm si}\quad
    \frac 1 2 < \nu < 1\, ,} \\ \\
           \displaystyle{
    -\frac{1-2\nu}{\lambda}\,
    \sum_{k=1}^\infty k  \left(\beta\,
    {e^{-\sigma\, i \pi \nu} \,
    \lambda^{-1+2\nu}}\right)^k  \quad {\rm si}\quad
    0 < \nu < \frac 1 2\,  ,}
   \end{array}
   \right. }
\end{array}
\end{equation}
que, como se puede ver, corresponden a las derivadas de los
correspondientes desarrollos asint{\'o}ticos de la expresiones
(\ref{tau-asymp}).

\bs

Por lo tanto,
\begin{eqnarray}\label{des-asymp-deriv-tau-Tr}\begin{array}{c}
   \partial_\lambda \Big[ \tau(\lambda)\,
  {\rm Tr}\,\{\left(D^0-\lambda\right)^{-1} -\left(D^\infty-
  \lambda\right)^{-1} \}\Big]
  \sim (2\nu-1)\times\\ \\
   \times \left\{
  \begin{array}{l}
  \displaystyle{
     \sum_{k=1}^\infty
    \left( \frac{e^{\sigma\, i \pi \nu}
    }{\beta}\right)^k
    \left[(1-2\nu) \, k -1\right] \lambda^{ (1-2\nu) \, k -2}
    \quad         {\rm si}\quad
    \frac 1 2 < \nu < 1\, ,} \\ \\
      \displaystyle{
    \sum_{k=0}^\infty \left(\beta\,
    {e^{-\sigma\, i \pi \nu}
    }\right)^k
    \left[(1-2\nu) k + 1\right] \lambda^{-(1-2\nu) k -2}
     \quad {\rm si}\quad
    0 < \nu < \frac 1 2\, .}
  \end{array}
  \right.
\end{array}\nonumber\\
\end{eqnarray}
Debe observarse, nuevamente, la presencia de potencias de
$\lambda$ dependientes de $\nu$ en este desarrollo.

\subsection{Las funciones $\zeta(s)$ y $\eta(s)$} \label{spectral-functions11}

Como los autovalores negativos de la extensi{\'o}n autoadjunta
$D^{\beta}$ son opuestos a los autovalores positivos de la
extensi{\'o}n autoadjunta $D^{-\beta}$ ser{\'a} suficiente
considerar la funci{\'o}n-$\zeta$ parcial $\zeta_+^{\beta}(s)$ que
se define como la suma,
\begin{equation}\label{parc}
    \zeta_+^\beta(s)=\sum_{\lambda_n>0}\lambda_n^{-s}\,,
\end{equation}
donde $\lambda_n$ representan los autovalores positivos de
$D^\beta$. Por consiguiente, $\zeta_+^\beta$ satisface
\cite{Seeley1}, para $\Re(s)>1$,
\begin{equation}\label{zeta-1}
  \zeta_+^\beta(s)=- \frac{1}{2\,\pi\,i} \oint_{\mathcal{C}}
  \frac{\lambda^{1-s}}{s-1} \, {\rm Tr}\left(D^\beta-\lambda\right)^{-2}
  \, d\lambda\, ,
\end{equation}
donde la curva $\mathcal{C}$ encierra la parte positiva del
espectro del operador en sentido antihorario, manteni{\'e}ndose a
la derecha del origen.

 \bs

De acuerdo con la ecuaci{\'o}n (\ref{TrG2}),
\begin{eqnarray}\label{zeta1-1}
  \zeta_+^\beta(s)=
  \oint_\mathcal{C}
  \frac{\lambda^{1-s}}{s-1} \, Tr\left(D^\beta-\lambda\right)^{-2}
  \, \frac{d\lambda}{2\pi i}  = \nonumber\\\nonumber\\ =\zeta_+^\infty(s)-
   \oint_{\mathcal{C}}
  \frac{\lambda^{1-s}}{s-1} \, \partial_\lambda \Big[ \tau(\lambda)\,
  {\rm Tr}\,\left\{\left(D^0-\lambda\right)^{-1}-
    \left(D^\infty-\lambda\right)^{-1}\right\}\Big]
  \, \frac{d\lambda}{2\pi i} =\nonumber\\\nonumber\\=
  \zeta_+^\infty(s)-
  \int_{-i\,\infty+0}^{i\,\infty+0}
  \frac{\lambda^{1-s}}{s-1} \, \partial_\lambda \Big[ \tau(\lambda)\,
  {\rm Tr}\,\left\{\left(D^0-\lambda\right)^{-1}-
    \left(D^\infty-\lambda\right)^{-1}\right\}\Big]
  \, \frac{d\lambda}{2\pi i} \, ,\nonumber\\\nonumber\\
\end{eqnarray}
donde $\zeta_+^\infty(s)$ es la funci{\'o}n-$\zeta$ parcial
(v{\'e}ase la ecuaci{\'o}n (\ref{parc})) de la extensi{\'o}n
$\beta=\infty$.

\bs

Adem{\'a}s, para una extensi{\'o}n autoadjunta general, podemos
escribir,
\begin{equation}\label{zeta+1}\begin{array}{c}
  \displaystyle{\zeta_+^{\beta}(s)=
    \frac{1}{2\,\pi} \int_{1}^{\infty}
    e^{i\,\frac{\pi}{2}\,(1-s)}\,
  \frac{\mu^{1-s}}{s-1} \,
  {\rm Tr}\,\left(D^\beta-e^{i\,\frac{\pi}{2}}\, \mu\right)^{-2}
  \, \frac{d\mu}{2\pi} \,  +
  } \\ \\
  \displaystyle{
    + \int_{1}^{\infty}
    e^{- i\,\frac{\pi}{2}\,(1-s)}\,
  \frac{\mu^{1-s}}{s-1} \,
  {\rm Tr}\,\left(D^\beta-e^{-i\,\frac{\pi}{2}}\, \mu\right)^{-2}
  \, \frac{d\mu}{2\pi}+ \frac{h_1(s)}{s-1}\, ,
  }
\end{array}
\end{equation}
donde $h_1(s)$ es una funci{\'o}n entera. Podemos entonces
determinar los polos de $\zeta_+^{\beta}(s)$, sumando y
substrayendo en los integrandos del miembro derecho de la
ecuaci{\'o}n (\ref{zeta+1}) una suma parcial del desarrollo
asint{\'o}tico de la traza del cuadrado de la resolvente obtenido
en la secci{\'o}n anterior.

\bs

En particular, los polos de la funci{\'o}n $\zeta_+^\infty(s)$
est{\'a}n dados por,
\begin{equation}\label{zetaD+1}\begin{array}{c}
  \displaystyle{ \zeta_+^{\infty}(s) =
    \frac{1}{s-1} \int_{1}^{\infty}
    e^{i\,\frac{\pi}{2}\,(1-s)}\,
  {\mu^{1-s}} \,
  \left\{ \sum_{k=2}^{N} e^{-i\,\frac{\pi}{2}\,k}\,
  A_k(\nu,1) \, \mu^{-k} \right\}
  \, \frac{d\mu}{2\pi}\,  +
  } \\ \\
  \displaystyle{
    \mbox{}+\frac{1}{s-1} \int_{1}^{\infty}
    e^{-i\,\frac{\pi}{2}\,(1-s)}\,
  {\mu^{1-s}} \,
  \left\{ \sum_{k=2}^{N} e^{i\,\frac{\pi}{2}\,k}\,
  A_k(\nu,1)^* \, \mu^{-k} \right\}
  \, \frac{d\mu}{2\pi} + \frac{h_2(s)}{s-1}=
  } \\ \\
  \displaystyle{
  = \frac{1}{\pi\,(s-1)}\,\sum_{k=2}^{N}\frac{1}{s-(2-k)}\, \Re \left\{
  e^{i\,\frac{\pi}{2}\,(1-s-k)}\,A_k(\nu,1) \right\}
  + \frac{h_2(s)}{s-1}\, ,
  }
\end{array}
\end{equation}
donde $h_2(s)$ es una funci{\'o}n anal{\'\i}tica en el semiplano
$\Re(s)> 1-N$. Por consiguiente, la extensi{\'o}n meromorfa de
$\zeta_+^{\infty}(s)$ tiene un polo simple en $s=1$ (v{\'e}ase la
ecuaci{\'o}n (\ref{zeta+1})), cuyo residuo est{\'a} dado por,
\begin{equation}\label{residuos=1}
  \left. {\rm Res}\,\zeta_+^{\infty}(s) \right|_{s=1}
  =  \frac{1}{2\, \pi \, i} \int_{-i\, \infty+0}^{i\, \infty+0}
  \lambda^0 \, \partial_\lambda \left\{
  \frac{J_{1-\nu}(\lambda)}{J_{-\nu}(\lambda)}
  \right\}\, d\lambda = \frac 1 \pi\, ,
\end{equation}
donde hemos utilizado las ecuaciones (\ref{JsobreJup1}) y
(\ref{asymp-cociente}).

\bs

La funci{\'o}n $\zeta_+^{\infty}(s)$ presenta tambi{\'e}n polos
simples en $s=1-2n$, para $n=1,2,\dots$, con residuos,
\begin{equation}\label{otros-residuos-1}
  \left. {\rm Res}\,\zeta_+^{\infty}(s) \right|_{s=1-2n}
  = \frac{\Re\left\{i\, A_{2n+1}(\nu,1)\right\}}{(k-1)\, \pi}  \, ,
\end{equation}
donde los coeficientes $A_k(\nu,1)$ est{\'a}n dados por
(\ref{asymp-trGD-upp}).

\bs

Para una extensi{\'o}n autoadjunta general $D^{\beta}$, debemos
tambi{\'e}n considerar las singularidades que provienen del
desarrollo asint{\'o}tico de $\partial_\lambda [ \tau(\lambda)$
$Tr\{\left(D^\infty-\lambda\right)^{-1}-\left(D^0-\lambda\right)^{-1}\}]$
en la ecuaci{\'o}n (\ref{des-asymp-deriv-tau-Tr}). Resolveremos en
detalle solamente el caso $\frac 1 2 < \nu <1$, en tanto que el
caso $0<\nu<\frac1  2$ conduce a resultados similares.

\bs

A partir de la ecuaci{\'o}n (\ref{zeta+1}), y teniendo en cuenta
la ecuaci{\'o}n (\ref{des-asymp-deriv-tau-Tr}), si $s>1$,
\begin{equation}\label{zetadif-1
}\begin{array}{c}
    \zeta_+^{\beta}(s)-\zeta_+^{\infty}(s)=
    \displaystyle{\frac{h_3(s)}{s-1}}\,-
  \displaystyle{
     \frac{\nu- \frac 1 2}{\pi\,(s-1)}}\,\times\\ \\
     \times\left\{\displaystyle{\int_{1}^{\infty}
    e^{i\,\frac{\pi}{2}\,(-s-1)}\,
  {\mu^{1-s}} \,
  \left[ \sum_{k=1}^{N}\frac{ e^{i\,k\,\frac{\pi}{2}}}
  {\beta^k}\,
  [(2\nu-1)k+1] \, \mu^{-(2\nu-1)k-2} \right]
  \, d\mu-
  } \right.\\ \\ \left.
     -\displaystyle{\int_{1}^{\infty}
    e^{-i\,\frac{\pi}{2}\,(-s-1)}\,
  {\mu^{1-s}} \,
  \left[ \sum_{k=1}^{N}\frac{ e^{-i\,k\,\frac{\pi}{2}}}
  {\beta^k}\,
  [(2\nu-1)k+1] \, \mu^{-(2\nu-1)k-2} \right]
  \, d\mu
  }\right\}= \\ \\
  \displaystyle{
  = -\frac{2\nu-1}{\pi\,(s-1)}\, \sum_{k=1}^{N}\,
  \left[\frac{1+(2\nu-1)k}{s+(2\nu-1)k}\right]\, \Re \left\{
  \frac{e^{i\,\frac{\pi}{2} (k-s-1)}}{\beta^k}\right\}
  + \displaystyle{\frac{h_3(s)}{s-1}}\, ,
  }
\end{array}
\end{equation}
donde $h_3(s)$ es anal{\'\i}tica para $\Re(s)>-(2\nu-1)\,(N+1)$.

\bs

En consecuencia, el residuo del polo de la extensi{\'o}n meromorfa
de $\zeta_+^{\beta}(s)-\zeta_+^{\infty}(s)$ en $s=1$ se anula,
\begin{equation}\label{res-dif-s=1}\begin{array}{c}
    \left. {\rm Res}\,\left(\zeta_+^{\beta}(s)-
  \zeta_+^{\infty}(s)\right) \right|_{s=1}
  = \\ \\ \displaystyle{
  = \int_{-i\,\infty+0}^{i\,\infty+0}
  {\lambda^{0}}\,
  \partial_\lambda \Big[ \tau(\lambda)\,
  Tr\{\left(D^{0}-\lambda\right)^{-1}-\left(D^\infty-\lambda\right)^{-1}\}\Big]
  \, \frac{d\lambda}{2\pi i}  =0}\, ,
\end{array}
\end{equation}
como se desprende de las ecuaciones (\ref{trdifasymp}) y
(\ref{tau-asymp}). Existen, sin embargo, polos simples en valores
no enteros dependientes de $\nu$,
\begin{equation}
  s=(1-2\nu)k=- |1-2\nu|\,k\quad{\rm con}\quad k=1,2,\dots
\end{equation}
cuyos residuos dependen de la extensi{\'o}n autoadjunta
considerada,
\begin{equation}\label{res-g-dep-1}\begin{array}{c}
      \left. {\rm Res}\,\left\{\zeta_+^{\beta}(s)-
  \zeta_+^{\infty}(s)\right\} \right|_{s=(1-2\nu)k} =
    \displaystyle{ \frac{2\nu-1}{\pi\,\beta^k}
    \ {\sin\left[\pi\nu k\right]}
  }\, .
\end{array}
\end{equation}

\bigskip

Ahora bien, de acuerdo con lo comentado a continuaci{\'o}n de la
ecuaci{\'o}n (\ref{eigenvalues-neg}), la funci{\'o}n
$\zeta^\beta(s)$ est{\'a} dada por,
\begin{equation}\label{zeta-completa}
  \zeta^{\beta}(s) = \zeta_+^{\beta}(s)+
  e^{-i\,\pi\,s} \, \zeta_+^{-\beta}(s)\, .
\end{equation}

\bs

{En particular, para la extensi{\'o}n $\beta=\infty$,
\begin{equation}\label{zeta-alpha0}
  \zeta^\infty(s) = \left(1 + e^{-i\,\pi\,s}\right)
  \zeta_+^{\infty}(s),
\end{equation}
puesto que el espectro de $D^{\infty}$ es sim{\'e}trico respecto
del origen (v{\'e}ase la ecuaci{\'o}n (\ref{alpha0pri}).) Se
concluye entonces que $\zeta^\infty(s)$ es entera. En efecto, a
partir de la ecuaci{\'o}n (\ref{otros-residuos-1}), el residuo en
$s=1-2n$ est{\'a} dado por,
\begin{equation}\label{residuos-nulos}
   \left. {\rm Res}\,\left\{\zeta^{\infty}(s)\right\}
   \right|_{s=1-2n} =
   \left(1 + e^{-i\,\pi\left(1-2n\right)}\right)
   \left. {\rm Res}\,\left\{\zeta_+^{\infty}(s)\right\}
   \right|_{s=1-2n} =0\, .
\end{equation}}

Por otra parte, para una extensi{\'o}n autoadjunta arbitraria las
singularidades de la funci{\'o}n $\zeta^{\beta}(s)$ consisten en
polos simples en los puntos $s_k$ dados por,
\begin{equation}
s_k=-(2\nu-1)k<0\quad {\rm con}\quad k=1,2,\dots
\end{equation}
con residuos,
\begin{equation}\label{residues-zeta-completa}\begin{array}{c}
 \displaystyle{
  \left. {\rm Res}\,\left\{\zeta^{\beta}(s)-
  \zeta^{\infty}(s)\right\} \right|_{s=-(2\nu-1)k} =
  }\\ \\
  = \displaystyle{\left. {\rm Res}\,\left\{
  \left[\zeta_+^{\beta}(s)-
  \zeta_+^{\infty}(s)\right] + e^{-i\,\pi\,s} \,
  \left[\zeta_+^{-\beta}(s)-
  \zeta_+^{\infty}(s)\right]
  \right\} \right|_{s=-(2\nu-1)k} } =\\ \\
  \displaystyle{=
  \frac{2\nu-1}{\pi} \
  \frac{\sin(2\pi\nu k)}{\beta^k}}\,
  e^{i\pi\nu k}\, .
\end{array}
\end{equation}

\bigskip

An{\'a}logamente, la asimetr{\'\i}a espectral (v{\'e}ase la
ecuaci{\'o}n (\ref{eta})) satisface,
\begin{equation}\label{eta-completa}
  \eta^{\beta}(s) = \zeta_+^{\beta}(s)-
  \zeta_+^{-\beta}(s)\, .
\end{equation}

En particular, $\eta^{\infty}(s) = \eta^{0}(s)= 0$, pues los
espectros de las extensiones $D^\infty,D^0$ son sim{\'e}tricos
(v{\'e}anse las ecuaciones (\ref{alpha0pri}) y
(\ref{eigen-beta0pri})).

\bs

Por el contrario, para una extensi{\'o}n autoadjunta general, si
$1/2 < \nu <1$, la funci{\'o}n $\eta^{\beta}(s)$ no se anula
id{\'e}nticamente y tiene polos simples en los puntos $s_k$ dados
por,
\begin{equation}
    s_k=-(2\nu-1)(2k+1) \quad{\rm con}\quad k=1,2,\dots
\end{equation}
con residuos,
\begin{equation}\label{residues-eta}
\displaystyle{
  \left. {\rm Res}\,\left\{\eta^{\beta}(s)\right\}
  \right|_{s=-(2\nu-1)k} = 2\frac{2\nu-1}{\pi} \,
  \frac{\sin\left[(2k+1)\pi\nu\right]}{\beta^{2k+1}}\,.}
\end{equation}

\bigskip

Para el caso $0<\nu<1/2$, un c{\'a}lculo completamente similar
muestra que $\zeta_+^{\beta}(s)-\zeta_+^{\infty}(s)$ admite una
extensi{\'o}n meromorfa con polos simples en puntos $s_k$
dependientes de $\nu$,
\begin{equation}
    s=-(1-2\nu)k
\end{equation}
para $k=1,2,\dots$, cuyos residuos dependen de la extensi{\'o}n
autoadjunta y est{\'a}n dados por,
\begin{equation}\label{res-gpos-dep}\begin{array}{c}
      \left. {\rm Res}\,\left\{\zeta_+^{\beta}(s)-
  \zeta_+^{\infty}(s)\right\} \right|_{s=-(1-2\nu)k} =
 \displaystyle{ -\,\frac{1-2\nu}{\pi}\,\beta^k
    \ {\sin\left(\pi\nu k\right)}
  }\, .
\end{array}
\end{equation}
A partir de este resultado, es inmediato obtener los residuos de
las funciones $\zeta^\beta(s)$ y $\eta^\beta(s)$. De hecho, se
verifica que el resultado se obtiene substituyendo en las
ecuaciones (\ref{residues-zeta-completa}) y (\ref{residues-eta})
$\beta$ y $e^{i\pi\nu k}$ por sus inversos.

\bigskip

{Es interesante notar que si $\beta\neq 0,\infty$, los residuos de
la funci{\'o}n $\zeta_+^{\beta}(s)$ en los puntos del plano
complejo $s_k =-|1-2\nu|k$ son proporcionales a $\beta^{\pm k}$.
Esto es consistente con el comportamiento del operador $D$ ante
las transformaciones de escala (\ref{escala}) que aplican
$\mathbf{L_2}(0,1)\rightarrow
 \mathbf{L_2}(0,1/c)$.

\bs

La extensi{\'o}n $D^{\beta}$ es equivalente por una
transformaci{\'o}n unitaria al operador $(1/c){D_c}^{\beta_c}$
definido similarmente en $\mathbf{L_2}(0,1/c)$, con $\beta_c =
c^{1-2\nu} \, \beta$,
\begin{equation}\label{isometry-1}
  T\,D^{\beta} = \frac 1 c \,
  {D_c}^{\beta_c}\, T\, .
\end{equation}
S{\'o}lo para las extensiones con $\beta=0,\infty$ la
condici{\'o}n de contorno en la singularidad $x=0$, dada por la
ecuaci{\'o}n (\ref{BC2}), es invariante ante cambios de escala.

\bs

Por consiguiente, la funci{\'o}n $\zeta_+^\beta(s)$ transforma
ante un cambio de escala de la siguiente manera,
\begin{equation}\label{zetas-isometry-1}
  (\zeta_+^{\beta_c})_c(s)=
   c^{-s}\,\zeta_+^{\beta}(s)\, ,
\end{equation}
y los residuos correspondientes est{\'a}n dados por,
\begin{equation}\label{zetas-isometry-residues-1}
   \left. {\rm Res}\,\left\{(\zeta_+^{\beta_c})_c(s)
   \right\} \right|_{s=-|1-2\nu|\,k} = c^{|1-2\nu|k}
   \left. {\rm Res}\,\left\{{\zeta}_+^{\beta}(s)
   \right\} \right|_{s=-|1-2\nu|k}\, .
\end{equation}
El factor $c^{|1-2\nu|k}$ cancela exactamente el efecto que tiene
el cambio en la condici{\'o}n de contorno en el origen sobre
$\beta$,
\begin{equation}\label{change-in-rho-1}
  \beta^{\pm k}
  =c^{-|1-2\nu|\,k}\, \beta_c^{\pm k}\, .
\end{equation}
Por lo tanto, la diferencia entre los intervalos $(0,1)$ y
$(0,1/c)$ no tiene efecto alguno en la estructura de estos
residuos que, entonces conjeturamos est{\'a}n determinados por
propiedades locales en las vecindades de $x=0$. }

\bigskip

Para finalizar, se\~nalamos, que estos polos an{\'o}malos no
est{\'a}n presentes en el caso regular $\alpha =0$. En efecto, en
este caso $\nu=1/2$ y $\tau(\lambda)$ en la ecuaci{\'o}n
(\ref{taudelambda}) admite un de\-sa\-rro\-llo asint{\'o}tico
constante, mientras que ${\rm
Tr}\,\{\left(D^0-\lambda\right)^{-1}-\left(D^\infty-\lambda\right)^{-1}\}$
tiende asint{\'o}ticamente a cero (v{\'e}ase la  ecuaci{\'o}n
(\ref{trdifasymp}).) Adem{\'a}s, los residuos de los polos
provenientes de la funci{\'o}n $\zeta_+^{\infty}(s)$ son todos
nulos (v{\'e}anse las ecuaciones (\ref{otros-residuos-1}) y
(\ref{asymp-trGD-upp})), excepto aquel correspondiente al polo en
$s=1$, que tiene un residuo igual a $1/\pi$ (v{\'e}ase la
ecuaci{\'o}n (\ref{residuos=1}).)

\bigskip

En conclusi{\'o}n, la presencia de polos de las funciones
$\zeta^\beta(s)$ y $\eta^\beta(s)$ en valores no enteros es
consecuencia del comportamiento singular del t{\'e}rmino de orden
cero en $D^\beta$ y de la forma en que var{\'\i}a la condici{\'o}n
de contorno ante una transformaci{\'o}n de escala.

\bigskip

\section{{El problema de Aharonov-Bohm}}\label{tubos}

El {\'u}ltimo ejemplo que consideraremos se refiere a una
part{\'\i}cula de Dirac con carga y sin masa, en $(2+1)$
dimensiones, en presencia de un campo magn{\'e}tico uniforme y de
un tubo magn{\'e}tico singular con flujo no entero. Este problema
ha sido considerado en \cite{FP}, donde se ha demostrado que el
hamiltoniano restringido a un subespacio de momento angular {\it
cr{\'\i}tico} admite extensiones autoadjuntas no triviales cuyos
espectros satisfacen una ecuaci{\'o}n similar a (\ref{spe666})
(v{\'e}anse tambi{\'e}n
\cite{Beneventano:1998ai,Aharonov:1959fk,deSousaGerbert:1988yt,Adami:1997ib,Sitenko:1999yx,Polychronakos:1986ik}.)

\bs

En esta secci{\'o}n determinaremos el espectro de energ{\'\i}as de las
part{\'\i}culas y sus estados estacionarios en relaci{\'o}n con las
extensiones autoadjuntas del hamiltoniano. Calcularemos adem{\'a}s
la estructura de polos de la funci{\'o}n-$\zeta$ y mostraremos que
existen polos en posiciones dependientes del valor del flujo
magn{\'e}tico de Aharonov -Bohm.

\subsection{El operador y su espectro}\label{tuboelo}

Consideremos una part{\'\i}cula de Dirac con carga $e$ y masa nula
movi{\'e}ndose en el plano en presencia de un campo magn{\'e}tico
uniforme $B$ y de un tubo de flujo magn{\'e}tico singular $\Phi= 2
\pi\kappa/e$, con $0<\kappa<1$ ubicado en el origen.

\bs

La funci{\'o}n de onda de la part{\'\i}cula es un spinor $\psi$ de
dos componentes que satisface la ecuaci{\'o}n de
Dirac\,\footnote{Utilizamos unidades para las que $\hbar=c=1$},
\begin{equation}
    i\,\slash\!\!\! \!  \nabla\,\psi=0\,,
\end{equation}
donde la derivada covariante\,\footnote{\label{gammas} Utilizamos
la siguiente representaci{\'o}n del {\'a}lgebra de Clifford:
\begin{equation}
\gamma^{0}=\sigma^{3}\,, \ \gamma^{1}= -i\sigma^{2}\,, \
\gamma^{2}=i\sigma^{1}\,,
\end{equation}
donde $\sigma^{i}, \ i=1,2,3$ son las matrices de Pauli.} es $\nabla=\partial-ieA$.

\bs

El campo magn{\'e}tico uniforme y el tubo de flujo singular son
representados por el siguiente campo de gauge,
\begin{equation}
  \vec{A}=\left(\frac{\Omega r}{e} +
  \frac{\kappa}{e r} \right)\hat{e}_\theta\,,
\end{equation}
donde $\Omega:= eB/2$ y $\hat{e}_\theta$ es el vector unitario
perpendicular a la direcci{\'o}n radial. Como $\Omega$ tiene
dimensiones de $L^{-2}$ definimos las cantidades adimensionales
$x:= \Omega^{1/2}\,r$ y $D:= \Omega^{-1/2}\,H$, siendo $H$ el
hamiltoniano de Dirac asociado al problema.

\bs

Debido a la simetr{\'\i}a rotacional del problema, el operador $D$
conmuta con el generador de las rotaciones
$J=-i\partial_{\theta}+\sigma^{3}/2$, por lo que buscaremos
autoestados simult{\'a}neos de ambos operadores. Las autofunciones
del operador $J$ est{\'a}n dadas por,
\begin{equation}\label{auro}
  \psi(x,\theta)=\left(\begin{array}{c}
    e^{i l \theta}\phi{(x)} \\
    e^{i(l+1)\theta}\chi{(x)} \
  \end{array}\right)\in
  \mathbb{C}^2\otimes\mathbf{L_2}(\mathbb{{R}}^2,x\, dx\,d\theta ),
  \quad l \in \mathbb{Z}\,,
\end{equation}
siendo $l+1/2$ los correspondientes autovalores. De modo que los
subespacios generados por las funciones (\ref{auro}) son
invariantes ante la acci{\'o}n de $D$. La restricci{\'o}n $D_l$ de
$D$ a cada uno de estos subespacios est{\'a} dada por,
\begin{equation}\label{ham}
  D_{l}=
\left(  \begin{array}{cc}
    0 & {\displaystyle i\left(\partial_x+\frac{1-\alpha}{x}-x\right)} \\
    {\displaystyle i\left(\partial_x+\frac{\alpha}{x}+x\right)} & 0 \
  \end{array}\right)\,,
\end{equation}
con,
\begin{equation}
    \alpha = \kappa -l\,,
\end{equation}
que act{\'u}a sobre funciones de la coordenada $x$ de dos componentes,
\begin{equation}
  \psi(x) = \left(\begin{array}{c}
    \phi{(x)} \\
    \chi{(x)} \
  \end{array}\right)\in
  \mathbb{C}^2\otimes\mathbf{L_2}(\mathbb{{R}}^+,x\,dx)\,.
\end{equation}

\bs

Determinaremos las extensiones autoadjuntas y las propiedades
espectrales del o\-pe\-ra\-dor $D_l$ separadamente para $l>0$,
$l<0$ y $l=0$. Antes de ello, ser{\'a} {\'u}til se{\~n}alar que
las soluciones de la ecuaci{\'o}n diferencial
\begin{equation}
    (D_l-\lambda)\left(\begin{array}{c}
    \phi{(x)} \\
    \chi{(x)} \
  \end{array}\right)=0\,,
\end{equation}
de cuadrado integrable en $[1,\infty]$ est{\'a}n dadas por,
\begin{equation}\label{soltio}
    \left(\begin{array}{c}
    \phi{(x)} \\
    \chi{(x)} \
  \end{array}\right)=e^{-x^2/2}x^{-\alpha}\,
  \left(\begin{array}{c}
    {\displaystyle U(-{\lambda^2}/{4};1-\alpha;x^2)} \\
     {\displaystyle \frac{i\,\lambda}{2}\,
    x\,
    U(1-{\lambda^2}/{4};2-\alpha;x^2)}\
  \end{array}\right)\,.
\end{equation}
Los valores de $\lambda$ quedan determinados por el comportamiento
de las autofunciones en el origen.

\begin{itemize}

\item

Si $l>0$ se puede probar que $D_l$ es esencialmente autoadjunto de
modo que la {\'u}nica extensi{\'o}n autoadjunta que admite es su
clausura. Como en este caso $\alpha<0$, la condici{\'o}n de integrabilidad del cuadrado de la
componente $\chi(x)$, dada por (\ref{soltio}), determina el
espectro,
\begin{equation}\label{esp+}
    \lambda_n:=\pm 2\sqrt{n}\,,\qquad {\rm con}\ \ n=1,2,3,\ldots
\end{equation}
La funci{\'o}n $\zeta_l(s)$ del operador $D_l$ para $l>0$ resulta
entonces,
\begin{equation}
    \zeta_l(s)=2^{-s}\,\left(1+e^{-i\pi s}\right)\zeta_R(s/2)\,,
\end{equation}
siendo $\zeta_R(s)$ la funci{\'o}n-$\zeta$ de Riemann.

\item

Si $l<0$ se verifica tambi{\'e}n que $D_l$ es esencialmente
autoadjunto. En este caso, $\alpha>1$ y la condici{\'o}n de integrabilidad del cuadrado de la
componente $\phi(x)$, dada por (\ref{soltio}), determina el
espectro,
\begin{equation}\label{esp+1}
    \lambda_n:=\pm 2\sqrt{n+|l|+\kappa}\,,\qquad {\rm con}\ \ n=0,1,2,\ldots
\end{equation}
La funci{\'o}n $\zeta_l(s)$ del operador $D_l$ para $l<0$ est{\'a}
dada por,
\begin{equation}
    \zeta_l(s)=2^{-s}\,\left(1+e^{-i\pi s}\right)\zeta_H(s/2,\kappa+|l|)\,,
\end{equation}
siendo $\zeta_H(s,q)$ la funci{\'o}n-$\zeta$ de Hurwitz.

\item

Finalmente, si $l = 0$ entonces $D_l$ admite una familia de
extensiones autoadjuntas que designaremos por $D^\beta$
caracterizadas por un par{\'a}metro real $\beta$. El espectro de
la extensi{\'o}n $D^\beta$ est{\'a} dado por las soluciones de la
siguiente ecuaci{\'o}n trascendente,
\begin{eqnarray}\label{energ}
    G(\lambda):=\lambda\,\frac
    {\Gamma(\kappa-\lambda^2/4)}{\Gamma(1-\lambda^2/4)}
    = 4\beta^{-1}\,.
\end{eqnarray}
\begin{figure}\label{F1}
\center
    \epsffile{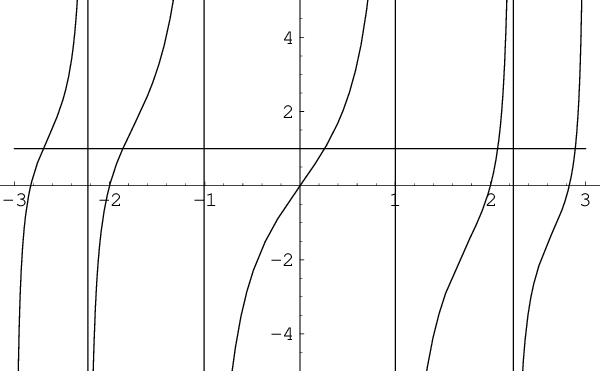} \caption{{\small Gr{\'a}fica de
    $G(\lambda)$ para $\kappa=1/4$. Las intersecciones con la
    l{\'\i}nea horizontal determinan el espectro de la extensi{\'o}n
    autoadjunta co\-rres\-pon\-dien\-te, en este caso dada por
    $\beta=4$.}}
   \end{figure}
Las soluciones $\lambda_n$ de la ecuaci{\'o}n (\ref{energ})
permiten definir la funci{\'o}n $\zeta^\beta(s)$
co\-rres\-pon\-dien\-te a este subespacio invariante,
\begin{equation}
    \zeta^\beta(s):= \sum_{\lambda_n}\lambda_n^{-s}\,.
\end{equation}
En la siguiente secci{\'o}n describiremos su estructura de polos.
\end{itemize}

\subsection{La funci{\'o}n $\zeta^\beta(s)$}\label{zetabeta}

\subsubsection{Representaci{\'o}n integral de la funci{\'o}n $\zeta^\beta(s)$}

Los espectro del operador $D^\beta$ est{\'a} dado por los ceros de la funci{\'o}n entera,
\begin{equation}\label{neoauto}
    f(\lambda):=\frac{\lambda}
    {\Gamma\left(\kappa-\frac{\lambda^2}{4}\right)}+\frac
    {\beta}{\Gamma \left(-\frac{\lambda^2}{4}\right)}\,.
\end{equation}
Como los ceros $\lambda_n$ de $f(\lambda)$ son simples, la funci{\'o}n-$\zeta$ parcial,
\begin{equation}
    \zeta_+^\beta(s):=\sum_{\lambda_n>0}\lambda_n^{-s}\,,
\end{equation}
est{\'a} dada por,
\begin{equation}\label{zir}
    \zeta_+^\beta(s)=
        \frac{1}{2\pi i}\int_\mathcal{C} \partial_\lambda\log{f(\lambda)}\
        \lambda^{-s}\,d\lambda\,,
\end{equation}
siendo $\mathcal{C}$ una curva que encierra a los ceros positivos de $f(\lambda)$ en sentido antihorario. Dado que estos crecen como $\lambda_n\sim \sqrt{n}$ la funci{\'o}n $\zeta_+^\beta$ es anal{\'\i}tica para $\mathcal{R}(s)>1$.

\bs

Para estos valores de $s$ la integral en (\ref{zir}) puede realizarse a lo largo de la curva $\mathcal{C}_+\cup \mathcal{C}_0\cup\mathcal{C}_-$, donde $\mathcal{C}_+$ representa los puntos del eje imaginario desde $i\infty$ hasta $i$, $\mathcal{C}_0$ es una curva que une $i$ con $-i$ y cuyos puntos tienen parte real positiva y menor que el primer cero de $f(\lambda)$. Finalmente, $\mathcal{C}_-$ representa los puntos del eje imaginario desde $-i$ hasta $-i\infty$. Luego de una integraci{\'o}n por partes la funci{\'o}n $\zeta^\beta_+(s)$ toma entonces la forma,
\begin{equation}\label{intrep}
    \zeta^\beta_+ (s)=
        \frac{s}{2\pi i}\int_{\mathcal{C}_+\cup \mathcal{C}_0\cup\mathcal{C}_-} \log{f(\lambda)}\
        \lambda^{-s-1}\,d\lambda\,,
\end{equation}
La estructura de polos de $\zeta_+^\beta(s)$ est{\'a} determinada por las integrales a lo largo de las curvas $\mathcal{C}_\pm$ en la ecuaci{\'o}n (\ref{intrep}):
\begin{equation}\label{intrep1}
        \frac{s}{2\pi i}\left[\int_{i\infty}^i\log{f(\lambda)}\
        \lambda^{-s-1}\,d\lambda+
        \int_{-i}^{i\infty}\log{f(\lambda)}\
        \lambda^{-s-1}\,d\lambda\right]\,.
\end{equation}
N{\'o}tese asimismo que el residuo de las integrales de la expresi{\'o}n (\ref{intrep1}) en $s=0$ es proporcional a $\zeta^\beta_+(0)$.

\bs

Estudiaremos entonces el comportamiento asint{\'o}tico de $\log{f(\lambda)}$ para $\lambda$ sobre el eje imaginario y $|\lambda|\rightarrow\infty$.

\subsubsection{Desarrollos asint{\'o}ticos}

En adelante, consideraremos solamente, por simplicidad, el caso $0<\kappa<1/2$. En primer lugar debemos determinar el comportamiento asint{\'o}tico de la funci{\'o}n,
\begin{equation}\label{logf}
    \log{f(\lambda)}=\log{\lambda}-\log{\Gamma
    \left(\kappa-{\lambda^2}/{4}\right)}
    +\log\left\{1+\beta\,\lambda^{-1}
    \frac{\Gamma\left(\kappa-\frac{\lambda^2}{4}\right)}
    {\Gamma\left(-\frac{\lambda^2}{4}\right)}\right\}\,.
\end{equation}
El desarrollo asint{\'o}tico de las funciones que intervienen en la ecuaci{\'o}n (\ref{logf}) para $\lambda=e^{\pm i\frac{\pi}{2}}x$ con $x\in\mathbb{R}^+$ y $x\rightarrow\infty$ est{\'a} dado por,
\begin{eqnarray}
    -\log{\Gamma
    \left(\kappa+{x^2}/{4}\right)}\sim
    \left(-\frac{x^2}{4}+\frac{1}{2}-\kappa\right)
    \left(2\log{x}+
    \log{\left(1+\frac{4\kappa}{x^2}\right)}-
    \log{4}\right)+\nonumber\\
    +\mbox{}\frac{x^2}{4}+\kappa-\log{\sqrt{2\pi}}-
    \sum_{m=1}^\infty\frac{B_{2m}}{2m(2m-1)}
    \left(\kappa+\frac{x^2}{4}\right)^{-2m+1}\,,\label{log1}\\
    \nonumber\\ \nonumber\\
    \frac{\Gamma\left(\kappa+\frac{x^2}{4}\right)}
    {\Gamma\left(\frac{x^2}{4}\right)}\sim
    \left(\frac{x^2}{4}\right)^\kappa
    \sum_{n=0}^\infty a_n(\kappa)x^{-2n}\,,\\
    \nonumber\\ \nonumber\\
    \log\left\{1+\beta\,e^{\mp i\frac{\pi}{2}}\,x^{-1}
    \frac{\Gamma\left(\kappa+\frac{x^2}{4}\right)}
    {\Gamma\left(\frac{x^2}{4}\right)}\right\}\sim\nonumber\\\mbox{}
    -\sum_{N=1}^\infty\sum_{n=0}^\infty
    \frac{(-1)^N}{N}\,4^{-\kappa N}\,b_{N,n}(\kappa)\,\beta^N\,
    e^{\mp i\frac{\pi}{2}N}x^{-N(1-2\kappa)-2n}\,,\label{log2}
\end{eqnarray}
donde los coeficientes $a_n(\kappa)$ y $b_{N,n}(\kappa)$ est{\'a}n definidos por,
\begin{equation}\label{aes}\begin{array}{r}
    \displaystyle{
    \sum_{n=0}^\infty a_n(\kappa)x^{-2n}:=
    \exp\left\{
    -\kappa+\left(\frac{x^2}{4}+\kappa-\frac{1}{2}\right)
    \log{\left(1+\frac{4\kappa}{x^2}\right)}
    +\right.}\\\displaystyle{
    \left.\mbox{}+
    \sum_{m=1}^\infty\frac{4^{2m-1}\,B_{2m}}{2m(2m-1)}
    \left[\left(1+\frac{4\kappa}{x^2}\right)^{-2m+1}-1\right]
    x^{-4m+2}
    \right\}}\,,\\ \\
    \sum_{n=0}^\infty b_{N,n}(\kappa)x^{-2n}:=
    \left(\sum_{n=0}^\infty a_{n}(\kappa)x^{-2n}\right)^N\,.
\end{array}
\end{equation}


\subsection{Estructura de polos de la funci{\'o}n $\zeta^\beta(s)$}

En esta secci{\'o}n reemplazaremos los desarrollos asint{\'o}ticos (\ref{log1}), (\ref{log2}) y (\ref{aes}) en la expresi{\'o}n (\ref{intrep1}) para determinar la estructura de polos de la funci{\'o}n $\zeta^\beta_+(s)$ as{\'\i} como tambi{\'e}n su valor en $s=0$.

\bs

En primer lugar, si substitu{\'\i}mos en la expresi{\'o}n (\ref{intrep1}) el primer t{\'e}rmino $\log{\lambda}$ del desarrollo (\ref{logf}) obtenemos,
\begin{eqnarray}
        \frac{s}{2\pi i}(-i)\int_1^\infty x^{-s-1}\left(e^{-i\frac{\pi}{2}(s+1)}
        \left[\log{x}+i\frac{\pi}{2}\right]
        +e^{i\frac{\pi}{2}(s+1)}
        \left[\log{x}-i\frac{\pi}{2}\right]
        \right)=\nonumber\\=
        -\frac{\cos{[\pi/2(s+1)]}}{\pi s}
        -\frac{1}{2}\sin{[\pi/2(s+1)]}\,,
\end{eqnarray}
que es una funci{\'o}n entera que se anula en $s=0$.

\bs

Consideremos ahora las potencias pares $x^{-2n}$, con $n\in\mathbb{Z}$, en el desarrollo asint{\'o}tico de $\log{f(e^{\pm i\frac{\pi}{2}}x)}$ (v{\'e}ase la ecuaci{\'o}n (\ref{log1}).) Reemplazando estas potencias en el integrando de la expresi{\'o}n (\ref{intrep1}) obtenemos,
\begin{equation}
        \frac{s}{2\pi i}\,(-i)\,2\cos{[\pi/2(s+1)]}\int_1^\infty x^{-s-1-2n}=
        -\frac{s}{\pi}\frac{\cos{[\pi/2(s+1)]}}{s+2n}\,,
\end{equation}
que es tambi{\'e}n una funci{\'o}n entera que se anula en $s=0$.

\bs

El valor de la funci{\'o}n $\zeta^\beta_+(s)$ en $s=0$ est{\'a} dado por el primer t{\'e}rmino del desarrollo (\ref{log1}),
\begin{equation}\label{log11}
    \left(-\frac{x^2}{4}+\frac{1}{2}-\kappa\right)2\log{x}\,.
\end{equation}
En efecto, si reemplazamos la expresi{\'o}n (\ref{log11}) en (\ref{intrep1}) obtenemos,
\begin{eqnarray}\label{popo1}
    \frac{s}{2\pi i}(-i)\,2\cos{[\pi/2(s+1)]}\int_{1}^{\infty}
    \left(-\frac{x^2}{4}+\frac{1}{2}-\kappa\right)2\log{x}\,
    x^{-s-1}\,dx=\nonumber\\=
    \frac{s}{2\pi}\cos{[\pi/2(s+1)]}\frac{1}{(s-2)^2}-
    (1-2\kappa)\frac{\cos{[\pi/2(s+1)]}}{\pi s}\,,
\end{eqnarray}
cuyo valor en $s=0$ es $1/2-\kappa$. Adem{\'a}s la cantidad (\ref{popo1}) presenta un polo simple en,
\begin{equation}\label{lis1}
    s=2\,,
\end{equation}
con residuo,
\begin{equation}\label{lis2}
    1/2\,.
\end{equation}

Finalmente, las contribuciones correspondientes a los t{\'e}rminos del desarrollo (\ref{log2}) a las integrales (\ref{intrep1}) est{\'a}n dadas por,
\begin{eqnarray}\label{popo2}
    \frac{s}{2\pi i}(-i)\,\beta^N\,A_{N,n}(\kappa)\,2\cos{[\pi/2(s+1+N)]}\int_{1}^{\infty}
    x^{-s-1-N-2n+2N\kappa}\,dx=\nonumber\\=
    -\frac{s}{\pi}\,\beta^N\,A_{N,n}(\kappa)
    \frac{\cos{[\pi/2(s+1+N)]}}{s+N(1-2\kappa)+2n}\,,
\end{eqnarray}
siendo
\begin{equation}
    A_{N,n}(\kappa):=-\frac{(-1)^N}{N}\,4^{-\kappa N}\,b_{N,n}(\kappa)\,.
\end{equation}

La expresi{\'o}n (\ref{popo2}) se anula en $s=0$ y presenta un polo simple en,
\begin{equation}\label{lis3}
    s_{N,n}=-N(1-2\kappa)-2n\,,\qquad N=1,2,3,\ldots\qquad n=0,1,2,\ldots
\end{equation}
con residuo,
\begin{equation}\label{lis4}
    (-1)^n\,\frac{N(1-2\kappa)+2n}{\pi}\,\beta^N\,A_{N,n}(\kappa)\sin{\pi N\kappa}\,.
\end{equation}

\subsubsection{Valor de $\zeta^\beta_+(s)$ en $s=0$}

Los c{\'a}lculos de esta secci{\'o}n indican que la {\'u}nica contribuci{\'o}n a la cantidad $\zeta^\beta_+(0)$ est{\'a} dada por el primer t{\'e}rmino de la ecuaci{\'o}n (\ref{popo1}). En consecuencia,
\begin{equation}\label{lis5}
    \zeta^\beta_+(0)=\frac{1}{2}-\kappa\,.
\end{equation}

\subsubsection{Una extensi{\'o}n autoadjunta particular}

El espectro de la extensi{\'o}n correspondiente a $\beta=0$ est{\'a} dado por (v{\'e}ase la ecuacion (\ref{neoauto})),
\begin{equation}
    \lambda_n=\pm 2\sqrt{n+\kappa}\,,\qquad n=0,1,2,\ldots\,,
\end{equation}
Por consiguiente, las funci{\'o}n-$\zeta$ parcial $\zeta^0_+(s)$ est{\'a} dada, para $\mathcal{R}(s)>2$, por,
\begin{equation}
    \sum_{n=0}^{\infty}\lambda_n^{-s}=2^{-s}\zeta_H(s/2,\kappa),
\end{equation}
donde $\zeta_H(s,q)$ es la funci{\'o}n-$\zeta$ de Hurwtiz. En consecuencia, $\zeta^0_+(s)$ presenta un {\'u}nico polo simple en $s=2$ con residuo
$1/2$ y su valor en $s=0$ est{\'a} dado por $\zeta_H(0,\kappa)=1/2-\kappa$, en acuerdo con las ecuaciones (\ref{lis1}), (\ref{lis2}), (\ref{lis3}), (\ref{lis4}) y (\ref{lis5}).

\subsection{Desarrollo asint{\'o}tico de la traza del heat-kernel de $D^2$}

Debido a la no compacidad de la variedad de base, el heat-kernel $e^{-t D^2}$ co\-rres\-pon\-dien\-te al cuadrado del hamiltoniano de Dirac $D$ no es tipo traza. En efecto, puede verse de la ecuaci{\'o}n (\ref{esp+}) que la suma de las contribuciones correspondientes a los subespacios de momento angular $l> 0$ es divergente.

\bs

En consecuencia, calcularemos la traza de la diferencia $e^{-t D^2}-e^{-t \underline{D}^2}$, donde $\underline{D}^2$ es el cuadrado del operador de Dirac $\underline{D}$ correspondiente a $ \kappa=0$.

\subsubsection{Contribuciones de los subespacios con $l\neq 0$}

Como los espectros correspondientes a $l>0$ no dependen de $\kappa$, estos subespacios no contribuyen a la traza del operador $e^{-t D^2}-e^{-t \underline{D}^2}$. Por su parte, las contribuciones de los subespacios correspondientes a $l<0$ est{\'a}n dadas por,
\begin{eqnarray}\label{hkneq}
    2\sum_{l<0}\sum_{n=0}^{\infty}
    \left(e^{-4t(n+|l|+\kappa)}-e^{-4t(n+|l|)}\right)=
    2\sum_{m=1}^{\infty}m\,\left(e^{-4t(m+\kappa)}-e^{-4tm}\right)=\nonumber\\=
    -e^{-2\kappa t}\,\frac{\sinh{2\kappa t}}{\sinh^2{2t}}\,.
\end{eqnarray}

La expresi{\'o}n (\ref{hkneq}) admite un desarrollo asint{\'o}tico para $t\rightarrow 0^+$ en potencias enteras de $t$ cuyos primeros t{\'e}rminos son,
\begin{equation}
    -\frac{\kappa}{2}\ t^{-1}+\kappa^2+O(t)\,.
\end{equation}

\subsubsection{Contribuciones del subespacio $l=0$}

De acuerdo con la ecuaci{\'o}n (\ref{neoauto}) los valores absolutos de los autovalores negativos co\-rres\-pon\-dien\-tes a la extensi{\'o}n caracterizadada por $\beta$ son los autovalores positivos co\-rres\-pon\-dien\-tes a la extensi{\'o}n caracterizada por $-\beta$. En consecuencia, la funci{\'o}n-$\zeta$ del operador $(D^\beta)^2$ est{\'a} dada por,
\begin{equation}
    \zeta^{\beta}_{(D^\beta)^2}(s)=
    \zeta^\beta_+(2s)+\zeta^{-\beta}_+(2s)\,.
\end{equation}
Por consiguiente, de acuerdo con las ecuaciones (\ref{lis1}), (\ref{lis2}), (\ref{lis3}) y (\ref{lis4}) la funci{\'o}n $\zeta^{\beta}_{(D^\beta)^2}(s)$ presenta polos simples en,
\begin{equation}\label{lisa1}
    s=1\,,
\end{equation}
con residuo,
\begin{equation}\label{lisa2}
    1/2\,,
\end{equation}
y en,
\begin{equation}\label{lisa3}
    s_{N,n}=-N(1-2\kappa)-n\,,\qquad N=1,2,3,\ldots\qquad n=0,1,2,\ldots
\end{equation}
con residuos,
\begin{equation}\label{lisa4}
    (-1)^n\,\frac{2N(1-2\kappa)+2n}{\pi}\,\beta^{2N}
    \,A_{2N,n}(\kappa)\sin{(2\pi N\kappa)}\,.
\end{equation}
Asimismo,
\begin{equation}\label{lisa5}
    \zeta^{\beta}_{-(D^\beta)^2}(0)=1-2\kappa\,.
\end{equation}

Finalmente, utilizando el desarrollo asint{\'o}tico de la expresi{\'o}n (\ref{hkneq}) y las ecuaciones (\ref{lisa1}-\ref{lisa5}) obtenemos, en virtud de las relaciones (\ref{zetcero}), (\ref{desaheat}), (\ref{polres}) y (\ref{desazeta}) el desarrollo asint{\'o}tico,
\begin{eqnarray}\label{dhkfin}
    {\rm Tr}\,\left(e^{-t D^2}-e^{t \underline{D}^2}\right)\sim
    -\frac{\kappa}{2}\ t^{-1}+\kappa(\kappa-2)
    +\sum_{N=1}^\infty\sum_{n=0}^\infty
    \beta^{2N}\,C_{N,n}(\kappa)\ t^{N(1-2\kappa)+n}+\nonumber\\\mbox{}
    +\sum_{k=0}^{\infty}C_k\ t^k\,,
\end{eqnarray}
donde,
\begin{equation}
    C_{N,n}(\kappa):=(-1)^n\frac{2}{\pi}\,[N(1-2\kappa)+n]
    \Gamma(-N(1-2\kappa)-n)\,A_{2N,n}(\kappa)
    \,\sin{(2\pi N\kappa)}\,,
\end{equation}
y las potencias enteras del {\'u}ltimo t{\'e}rmino de la expresi{\'o}n (\ref{dhkfin}) provienen del desarrollo asint{\'o}tico de (\ref{hkneq}).


\part{Conclusiones}\label{conc}

\begin{flushright}{\it{\bf Wir m\"ussen wissen. Wir werden wissen}.\\
David Hilbert.\\(Dicho en 1930 en K\"onigsberg,\\
ahora en su epitafio en G\"ottingen.)}
\end{flushright}

\vspace{25mm}

Ya hemos se{\~n}alado que en Teor{\'\i}a Cu{\'a}ntica de Campos
las primeras correcciones cu{\'a}nticas pueden describirse en
t{\'e}rminos de funciones espectrales asociadas a operadores
diferenciales que aparecen en el t{\'e}rmino cuadr{\'a}tico de las
acciones de los campos.

\bs

Por ejemplo, la informaci{\'o}n relevante para el c{\'a}lculo de
la acci{\'o}n efectiva al orden de 1-loop puede obtenerse a partir
del desarrollo asint{\'o}tico de la traza del heat-kernel para
peque{\~n}os valores de su argumento.

\bs

De ese modo, el estudio del desarrollo asint{\'o}tico del
heat-kernel permite implementar un mecanismo de regularizaci{\'o}n
de las constantes desnudas del lagrangiano, describir las
divergencias del propagador en puntos coincidentes y determinar
las funciones $\beta$ de la teor{\'\i}a, la energ{\'\i}a de
Casimir, o cantidades de car{\'a}cter topol{\'o}gico como
anomal{\'\i}as. Asimismo, mediante la determinaci{\'o}n de la
funci{\'o}n de partici{\'o}n es posible evaluar cantidades
f{\'\i}sicas asociadas a campos a temperatura finita. Algunas de
estas relaciones han sido brevemente descritas en la secci{\'o}n
\ref{fetcc}.

\bs

Cuando los operadores en consideraci{\'o}n son sim{\'e}tricos en
cierto dominio denso, interesa determinar dominios m{\'a}s amplios
donde resulten adem{\'a}s autoadjuntos. Esto conduce en general a
considerar una variedad $\mathcal{M}$ de condiciones de contorno
admisibles que permiten definir completamente el problema.

\bs

Si los coeficientes del operador diferencial $A$ de orden $d$ son
infinitamente derivables, la variedad de base $M$ es de
dimensi{\'o}n $m$ es compacta y se imponen condiciones de contorno
locales apropiadas sobre su borde, entonces la traza del
heat-kernel admite un desarrollo asint{\'o}tico en potencias de
$t$ cuyos exponentes est{\'a}n dados por $(n-m)/d$, donde
$n=0,1,2,\ldots$. De ese modo, estos exponentes est{\'a}n
determinados por el orden del ope\-ra\-dor y la dimensi{\'o}n de
la variedad. La dependencia de este desarrollo con los
coeficientes del operador diferencial, la forma de la variedad y
las condiciones de contorno (locales) se encuentra contenida en
los coeficientes de dichas potencias de $t$. Estos coeficientes
son integrales sobre $M$ y $\partial M$ de invariantes
geom{\'e}tricos locales. La derivaci{\'o}n de este resultado ha
sido presentada en el Cap{\'\i}tulo \ref{fe}.

\bs

Estas propiedades del desarrollo asint{\'o}tico de ${\rm
Tr}\,e^{-tA}$ pueden no ser v{\'a}lidas bajo otras hip{\'o}tesis.
En efecto, es sabido que si $A$ es un operador pseudodiferencial,
o si se imponen condiciones de contorno no locales del tipo de
Atiyah-Patodi-Singer \cite{APS}, entonces el de\-sa\-rro\-llo
asint{\'o}tico puede presentar logaritmos de $t$
\cite{Gr1,Gr2,GS,GS2,GS3}. Por otra parte, si la variedad de base
tiene una singularidad c{\'o}nica entonces los exponentes de las
potencias de $t$ pueden depender del {\'a}ngulo de deficiencia
\cite{Mooers}.

\bs

En esta Tesis hemos demostrado que la traza del heat-kernel
correspondiente a un operador diferencial con coeficientes
singulares  presenta en general potencias de $t$ cuyos exponentes
dependen de las caracter{\'\i}sticas de dicha singularidad.

\bs

En el Cap{\'\i}tulo \ref{opesing} hemos estudiado el operador de
Schr\"odinger en una dimensi{\'o}n,
\begin{equation}\label{opeult}
    A:=-\partial^2_x+\frac{\nu^2-1/4}{x^2}+V(x)\,,
\end{equation}
definido sobre
$\mathcal{D}(A):=\mathcal{C}_0^\infty(\mathbb{R}^+)$, donde $V(x)$
es una funci{\'o}n anal{\'\i}tica de $x\in\mathbb{R}^+$ acotada
inferiormente.

\bs

En primer lugar (v{\'e}ase la Secci{\'o}n \ref{sec1}) hemos
mostrado que, para $0 \leq \nu <1$, el operador $A$ admite una
familia de extensiones autoadjuntas $A^\theta$, donde $\theta$ es
un par{\'a}metro que caracteriza la condici{\'o}n de contorno
sobre la singularidad en $x=0$.

\bs

Hemos luego se{\~n}alado que s{\'o}lo dos de estas extensiones
autoadjuntas, que co\-rres\-pon\-den a $\theta=0,\infty$, definen
condiciones de contorno sobre la singularidad que resultan
invariantes de escala.
\bs

Los resultados de la Secci{\'o}n \ref{nocom} indican
que la traza del operador $e^{-tA^\theta}-e^{-tA^\infty}$ admite un desarrollo
asint{\'o}tico de la forma,
\begin{equation}\label{desult}
    {\rm Tr}(e^{-tA^\theta}-e^{-tA^\infty})\sim\sum_{n=1}^{\infty}a_n(A)\,t^{n/2}+
    \sum_{N=1}^\infty\sum_{n=1}^{\infty}
    b_{N,n}(A)\,\theta^N\,t^{\nu N +n/2-1/2}\,.
\end{equation}
Los coeficientes $a_n(A),b_{N,n}(A)$ dependen de la forma del
potencial $V(x)$ y de $\nu$ pero no del par{\'a}metro $\theta$ que
caracteriza la condici{\'o}n de contorno. Hemos presentado una
f{\'o}rmula de recurrencia para la determinaci{\'o}n de estos
coeficientes.

\bs

En particular, el desarrollo asint{\'o}tico (\ref{desult})
presenta potencias de $t$ cuyos exponentes dependen del
par{\'a}metro $\nu$. Este es uno de los resultados centrales de
esta Tesis cuyo {\'u}nico antecedente, a nuestro saber, se
encuentran en el trabajo de E.\ Mooers \cite{Mooers}, donde se
obtiene un t{\'e}rmino proporcional a $t^\nu$ en el desarrollo
asint{\'o}tico del heat-kernel de las extensiones autoadjuntas del
laplaciano sobre una variedad con una singularidad c{\'o}nica,
estando $\nu$ relacionado en ese caso con el {\'a}ngulo de
deficiencia de la variedad.

\bs

Por otra parte, se{\~n}alemos que este resultado implica otras
propiedades espectrales que no tienen los operadores con
coeficientes regulares: la funci{\'o}n-$\zeta$ del operador
di\-fe\-ren\-cial presenta polos simples en posiciones
dependientes de $\nu$, la traza de la resolvente
$(A^\theta-\lambda)^{-1}$ admite un desarrollo asint{\'o}tico para
grandes valores de $|\lambda|$ en potencias de $\lambda$ cuyos
exponentes dependen de $\nu$ y el comportamiento asint{\'o}tico de
los autovalores $\{\lambda_n\}_{n\in\mathbb{N}}$ para grandes
valores de $n$ presenta potencias de $n$ con exponentes
tambi{\'e}n dependientes de $\nu$ (v{\'e}ase la Secci{\'o}n
\ref{relfe}.)

\bs

Posteriormente, en la Secci{\'o}n \ref{Wipf} hemos ilustrado el
desarrollo (\ref{desult}) mediante la resoluci{\'o}n
expl{\'\i}cita del operador (\ref{opeult}) para el caso
$V(x)=x^2$. Para ello, hemos determinado la clausura del operador
diferencial, sus extensiones autoadjuntas y hemos analizado una
representaci{\'o}n integral de la funci{\'o}n-$\zeta$.

\bs

Hemos tambi{\'e}n extendido la expresi{\'o}n (\ref{desult}) al
caso del un operador de la forma (\ref{opeult}), pero definido
sobre la variedad de base compacta $[0,1]\subset\mathbb{R}^+$
(v{\'e}ase la Secci{\'o}n \ref{cccp}.) En este caso las
extensiones autoadjuntas que hemos analizado est{\'a}n
caracterizadas por dos par{\'a}metros $\theta$ y $\beta$ que
describen la condici{\'o}n de contorno sobre la
sin\-gu\-la\-ri\-dad en $x=0$ y una condici{\'o}n local en $x=1$,
respectivamente.

\bs

Los valores $\theta=0,\infty$ definen condiciones de contorno que son invariantes ante una transformaci{\'o}n de escala en las proximidades de $x=0$. Para estas extensiones la traza del heat-kernel admite un desarrollo en potencias de $t$ cuyos exponentes est{\'a}n dados por $n/2$ con $n=-1,0,1,2,\ldots$, al igual que para un opeador de orden 2 con coeficientes regulares en una variedad de base compacta de dimensi{\'o}n 1.

\bs

Sin embargo, el desarrollo asint{\'o}tico correspondiente a las restantes
extensiones autoadjuntas presenta potencias de $t$ con
exponentes dependientes de $\nu$. En la Secci{\'o}n \ref{Solari}
se ha considerando a modo de ejemplo el operador (\ref{opeult})
para el caso $V(x)=0$.

\bs

Perm{\'\i}tasenos resumir estos resultados mediante algunas
consideraciones cualitativas. N{\'o}tese que el coeficiente
singular del operador (\ref{opeult}), dominante frente a $V(x)$ en
proxi\-mi\-da\-des de $x=0$, se transforma de la misma manera ante
una transformaci{\'o}n de escala que el t{\'e}rmino de orden 2.
Por otra parte, las extensiones autoadjuntas que definen
condiciones de contorno que no son invariantes ante
transformaciones de escala introducen un par{\'a}metro $\theta$
cuya dimensi{\'o}n corresponde a una potencia de unidad de
longitud con un exponente dependiente del par{\'a}metro $\nu$.
Esto est{\'a} relacionado con la dependencia de $\nu$ de los
exponentes de las potencias de $t$ en el desarrollo asint{\'o}tico
de la traza del heat kernel.

\bs

Esto sugiere la posibilidad de obtener el mismo tipo de desarrollo
asint{\'o}tico en relaci{\'o}n con operadores de primer orden con
un coeficiente proporcional a la inversa de la distancia al punto
de la singularidad.

\bs

En efecto, en la secci{\'o}n \ref{Seeley} hemos considerado el
operador de Dirac,
\begin{equation}\label{dirult}
  D=\left(\begin{array}{cc}
    0  &  \displaystyle{\partial_x + \frac{\alpha}{x}} \\
    \displaystyle{-\partial_x + \frac{\alpha}{x}} & 0 \
  \end{array}\right)\, ,
\end{equation}
definido sobre
$\mathcal{D}(D):=\mathcal{C}_0^\infty(0,1)\otimes\mathbb{C}^2$. Si
se imponen ciertas condiciones locales fijas en $x=1$, sus
extensiones autoadjuntas para valores de $\alpha \in (-1/2,1/2)$
resultan caracterizadas por un {\'u}nico par{\'a}metro $\beta$ que
describe la condici{\'o}n de contorno sobre la singularidad en
$x=0$. Existen, en particular, dos extensiones autoadjuntas
correspondientes a $\beta=0,\infty$ que definen condiciones de
contorno invariantes de escala.

\bs

La traza de la resolvente correspondiente a las restantes
extensiones autoadjuntas, que rompen la invarianza de escala,
admite un desarrollo asint{\'o}tico en potencias de $\lambda$ con
exponentes que dependen del par{\'a}metro $\alpha$ y coeficientes
que dependen de $\beta$.

\bs

Finalmente, en la secci{\'o}n \ref{tubos}, hemos analizado en este
contexto un problema que ha sido estudiado en sus muy diversos
aspectos. Consideramos el hamiltoniano de Dirac $D$ de una
part{\'\i}cula de spin 1/2, de masa nula y con carga $e$ en $2+1$
dimensiones en presencia de un campo magn{\'e}tico homog{\'e}neo
$B$ y de un flujo de Aharonov-Bohm $\Phi$ en el origen $r=0$.

\bs

El coeficiente de $D$ proveniente del campo de gauge
correspondiente al campo mag\-n{\'e}\-ti\-co $B$ tiende a $\infty$
cuando $r\rightarrow \infty$, de modo que el operador $H$ tiene un
espectro discreto. Por su parte, el coeficiente proveniente del
campo de gauge que representa el flujo singular rompe la
simetr{\'\i}a traslacional caracter{\'\i}stica del problema de
Landau, presentando una singularidad en $r=0$.

\bs

El sistema conserva, no obstante, la invarianza rotacional, por lo
que los subespacios caracter{\'\i}sticos de momento angular
$l+1/2$, con $l\in\mathbb{Z}$, son subespacios invariantes del
hamiltoniano $D$.

\bs

Si $0<\Phi<2\pi/e$, las restricciones del hamiltoniano a los
subespacios correspondientes a $l\neq 0$ admiten una {\'u}nica
extensi{\'o}n autoadjunta, en tanto que la restricci{\'o}n al
subespacio caracterizado por $l=0$ admite una familia de
extensiones autoadjuntas caracterizadas por un par{\'a}metro que
describe el comportamiento de la funci{\'o}n de onda en las
proximidades del flujo singular.

\bs

Hemos calculado, para $0<\Phi<\pi/e$, la estructura de singularidades de la funci{\'o}n-$\zeta$ correspondiente a la restricci{\'o}n del hamiltoniano a ese subespacio particular, encontrando que presenta polos simples
en los puntos,
\begin{equation}\label{pop-11}
    s_{N,n}=-N\left(1-\frac{e\Phi}{\pi}\right)-2n\,,
    \qquad N=1,2,3,\ldots\qquad n=0,1,2,\ldots
\end{equation}
cuyos residuos dependen de la extensi{\'o}n autoadjunta
considerada.

\bs

La existencia de estos polos determina la presencia de potencias
de $t$ con exponentes dependientes de $\Phi$ en el desarrollo
asint{\'o}tico de la traza del heat-kernel correspondiente al cuadrado del hamiltoniano de Dirac.

\bs

Con respecto a las implicaciones de estos resultados, mencionemos
en relaci{\'o}n con este problema particular (pero haciendo
referencia a todos los operadores con coeficientes singulares
analizados) que esta dependencia con $\Phi$ se traslada al
comportamiento de las divergencias a 1-loop. En particular, el
propagador $H^{-1}(x,x')$ presenta un desarrollo asint{\'o}tico
para $x\rightarrow x'$ en potencias de $|x-x'|$ con exponentes
dependientes del flujo singular $\Phi$. Resulta tambi{\'e}n de
inter{\'e}s estudiar la renormalizaci{\'o}n de este tipo de
teor{\'\i}as en las que se presentan potencias del cutoff que
dependen de la cantidad $\Phi$.

\bs

Otro ejemplo de inter{\'e}s que se propone considerar en el futuro
consiste en el estudio de campos cu{\'a}nticos de distintos spines
a temperatura finita en proximidades del horizonte interno de un
agujero negro. En este problema, la energ{\'\i}a libre al orden de
1-loop se obtiene a partir de la funci{\'o}n de partici{\'o}n
calculada mediante estos m{\'e}todos. Para el caso de un agujero
negro de Schwartzschild, la variedad eucl{\'\i}dea resultante
presenta, en proximidades del horizonte interno, una singularidad
c{\'o}nica cuyo {\'a}ngulo de deficiencia est{\'a} determinado por
la temperatura.

\bs

Los resultados presentados en esta Tesis sugieren considerar las
funciones espectrales de las distintas extensiones autoadjuntas
del operador diferencial relevante, que tiene coeficientes
singulares, a fin de estudiar el comportamiento de cantidades
f{\'\i}sicas como la entrop{\'\i}a.

\bs

Para finalizar, mencionaremos otros dos resultados de esta Tesis.
En primer lugar, se{\~n}alemos que el resultado (\ref{desult}) fue
obtenido a partir de una t{\'e}cnica distinta de la utilizada para
estudiar el desarrollo asint{\'o}tico de la traza del heat-kernel
de operadores con coeficientes regulares. Una de las dificultades
en este sentido consiste en que los coeficientes del desarrollo
para el caso regular son integrales sobre la variedad y su borde
de los invariantes construidos en t{\'e}rminos de los coeficientes
del operador diferencial, el tensor de curvatura de la variedad,
la curvatura extr{\'\i}nseca del borde y los operadores que
definen las condiciones de contorno. Si los coeficientes del
operador diferencial presentan singularidades entonces estas
integrales no resultan convergentes.

\bs

Como ya hemos mencionado, las condiciones de contorno que definen
un problema en mec{\'a}nica cu{\'a}ntica sobre una variedad con
borde o en presencia de singularidades est{\'a}n determinadas por
las extensiones autoadjuntas del hamiltoniano. Esto garantiza la
unitariedad de la teor{\'\i}a. Las distintas extensiones
autoadjuntas representan las propiedades microsc{\'o}picas del
borde o de la singularidad.

\bs

Para derivar el desarrollo asint{\'o}tico (\ref{desult}) hemos
aprovechado la existencia de dos extensiones autoadjuntas,
definidas por condiciones de contorno invariantes de escala, para
las cuales los coeficientes $b_{N,n}$ en (\ref{desult}) se anulan.

\bs

El paso siguiente consisti{\'o} en establecer una relaci{\'o}n
entre las resolventes correspondientes a distintas extensiones
autoadjuntas del operador diferencial con coeficientes singulares
(v{\'e}ase la Secci{\'o}n \ref{sec2}.) Esta relaci{\'o}n
representa una generalizaci{\'o}n de la f{\'o}rmula de Krein,
v{\'a}lida para operadores con coeficientes regulares (v{\'e}ase
la Secci{\'o}n \ref{regu}.)

\bs

En el Cap{\'\i}tulo \ref{rupturaSUSY} hemos considerado un
problema en mec{\'a}nica cu{\'a}ntica supersim{\'e}trica definido
por un superpotencial singular en el cual las condiciones de
contorno cumplen un rol esencial con respecto a las simetr{\'\i}as
del hamiltoniano. Los operadores diferenciales analizados
constituyen una realizaci{\'o}n formal del {\'a}lgebra de N=2 SUSY
en 0+1 dimensiones. Sin embargo, la validez de esta
realizaci{\'o}n est{\'a} limitada por la determinaci{\'o}n de
dominios de definici{\'o}n apropiados.

\bs

Hemos estudiado las extensiones autoadjuntas del hamiltoniano y de
las supercargas y concluido que existen s{\'o}lo dos extensiones
cuyas condiciones de contorno, invariantes de escala,  definen una
realizaci{\'o}n del {\'a}lgebra de la SUSY con dos supercargas.

\bs

Para una de ellas, el estado fundamental del hamiltoniano tiene
energ{\'\i}a nula y el resto del espectro es doblemente
degenerado, de modo que la SUSY es expl{\'\i}cita. Por el
contrario, la otra extensi{\'o}n presenta un espectro doblemente
degenerado con un estado fundamental de energ{\'\i}a estrictamente
positiva y, en consecuencia, ruptura espont{\'a}nea (din{\'a}mica)
de la SUSY.

\bs

Para el resto de las extensiones autoadjuntas s{\'o}lo es posible
definir una {\'u}nica supercarga, y la SUSY est{\'a}
din{\'a}micamente rota como consecuencia de que el estado
fundamental tiene energ{\'\i}a positiva y el espectro es no
degenerado.

\bs

Este resultado resuelve la controversia originada a partir del
trabajo de A.\ Jevicki y J.P.\ Rodrigues \cite{Jevicki} en el que
proponen un mecanismo de ruptura espont{\'a}nea de la
supersimetr{\'\i}a mediante la presencia de potenciales
singulares. Esta propuesta recibi{\'o} cues\-tio\-na\-mien\-tos
basados en el estudio de una regularizaci{\'o}n que preserva la
supersimetr{\'\i}a \cite{Pernice,Das,Gangopadhyaya:2002jr}.

\bs

El problema tratado en el Cap{\'\i}tulo \ref{rupturaSUSY} muestra
que la variedad de condiciones de contorno admisibles sobre la
singularidad del superpotencial proveen un mecanismo de ruptura de
la SUSY en Mec{\'a}nica Cu{\'a}ntica Supersim{\'e}trica.


\part{Problemas de inter{\'e}s}\label{proy}

\begin{flushright}{\it El n{\'u}mero de p{\'a}ginas de este libro es exactamente infinito.\\Ninguna es la primera; ninguna, la {\'u}ltima.\\No s{\'e} por qu{\'e} est{\'a}n numeradas de ese modo arbitrario.\\Acaso para dar a entender que\\los t{\'e}rminos de una serie infinita admiten cualquier n{\'u}mero.\\
Jorge L.\ Borges (El Libro de Arena.)}
\end{flushright}

\vspace{25mm}

\section{M{\'e}todo Functorial}\label{functorial}

En este {\'u}ltimo cap{\'\i}tulo consideraremos un procedimiento
para obtener el desarrollo asint{\'o}tico de la traza del
heat-kernel de operadores diferenciales singulares distinto del
desarrollado en esta Tesis. Este procedimiento carece de una
justificaci{\'o}n completa por lo que simplemente supondremos la
validez de la generalizaci{\'o}n de un m{\'e}todo ampliamente
utilizado en el caso de operadores diferenciales regulares
\cite{Seeley1,Seeley3,Seeley4,Gilkey,Vass1}.

\bs

Consideremos primeramente un operador diferencial autoadjunto de
segundo orden $A$ positivo definido con coeficientes
$\mathcal{C}^\infty$ definido sobre secciones de un fibrado
vectorial sobre una variedad de base $M$ compacta, de
dimensi{\'o}n $m$ y con borde suave $\partial M$ y sujeto a
condiciones de contorno locales. La traza del operador $e^{-t A}$
admite un desarrollo asint{\'o}tico para peque{\~n}os valores de
$t$ dado por las ecuaciones (\ref{teohea}-\ref{coefiult}).

\bs

Los coeficientes $c_n(A),c^b_n(A)$ de estas ecuaciones son
integrales de combinaciones lineales de todos los invariantes
geom{\'e}tricos de la dimensi{\'o}n apropiada; de manera que
est{\'a}n determinados por los coeficientes del operador
diferencial $A$, el tensor de curvatura de Riemann de la variedad
$M$, el tensor curvatura extr{\'\i}nseca del borde $\partial M$ y
los o\-pe\-ra\-do\-res de borde que definen la condici{\'o}n de
contorno local.

\bs

Los invariantes relacionados con cada coeficiente
$c_n(A),c^b_n(A)$ pueden determinarse analizando las dimensiones
de las cantidades involucradas. Consideremos, por e\-jem\-plo, un
operador de Schr\"odinger $A$ caracterizado por un potencial
$V(x)$. Los coeficientes $c_0(A)$ y $c_1(A)$ tiene dimensiones
$L^{m}$ y $L^{m-1}$, respectivamente, por lo que los coeficientes
$c_0(A,x)$ y $c^b_1(A)$ son adimensionales y, por consiguiente,
constantes. De modo que los coeficientes $c_0(A)$ y $c_1(A)$ son
proporcionales a los vol{\'u}menes de la variedad $M$ y del borde
$\partial M$, respectivamente.

\bs

El coeficiente $c_2(A)$ tiene dimensi{\'o}n $L^{m-2}$, por lo que
los coeficientes $c_2(A,x)$ y $c^b_2(A,x)$ tienen dimensiones
$L^{-2}$ y $L^{-1}$, respectivamente. En consecuencia, $c_2(A,x)$
es una combinaci{\'o}n lineal con coeficientes constantes del
tensor de curvatura escalar $\mathcal{R}$ y del potencial $V(x)$.
El coeficiente $c^b_2(A,x)$, por su parte, es una combinaci{\'o}n
lineal con coeficientes constantes de la traza del tensor
curvatura extr{\'\i}nseca y del operador de borde.

\bs

De esta manera, una vez determinados los invariantes
geom{\'e}tricos en distintas dimensiones, cada cantidad $c_n(A)$
se expresa como una integral de una combinaci{\'o}n lineal de los
invariantes correspondientes cuyos coeficientes resultan ser
cantidades universales, independientes de los coeficientes del
operador, de la variedad y de las condiciones locales de contorno.
La resoluci{\'o}n de problemas particulares cobra entonces
especial relevancia en el estudio del desarrollo asint{\'o}tico de
la traza del heat-kernel de operadores regulares pues permite, en
general, determinar algunos de los coeficientes universales que
expresan las cantidades $c_n(A)$ en t{\'e}rminos de los
invariantes geom{\'e}tricos.

\bs

Esta t{\'e}cnica no puede utilizarse, {\it a priori}, en el caso
de operadores diferenciales con coeficientes singulares como los
considerados en esta Tesis. La ecuaci{\'o}n (\ref{666}) representa
a las cantidades $c_n(A)$ en t{\'e}rminos de integrales de
cantidades locales que son combinaciones lineales de los
invariantes geom{\'e}tricos, entre los que se cuentan los
coeficientes del operador diferencial. Si el operador posee
coeficientes singulares, sus potencias y derivadas no ser{\'a}n,
en general, integrables y los resultados mencionados pierden
validez.

\bs

No obstante, generalizaremos, a continuaci{\'o}n, estas ideas con
el prop{\'o}sito de es\-tu\-diar los coeficientes del desarrollo
asint{\'o}tico del heat-kernel para un operador diferencial
singular. Los resultados sugieren la posibilidad de extender la
t{\'e}cnica de determinar los coeficientes universales al caso de
operadores singulares.

\bs

Consideremos el operador,
\begin{equation}
    A=-\partial^2_x+\frac{\kappa}{x^2}+V(x)\,.
\end{equation}
Las dimensiones de las cantidades involucradas est{\'a}n dadas
por,
\begin{equation}
    \begin{array}{ccc}
    \left[A\right]=L^{-2}\,,&
    \left[\kappa\right]=[\nu]=L^{0}\,,&

    \left[V(x)\right]=L^{-2}\,,\\ \\
    \left[\theta\right]=L^{-2\nu}\,,&
    \left[z\right]=L^{-2}\,,&
    \left[t\right]=L^{2}\,.\
    \end{array}
\end{equation}
Las ecuaci{\'o}n (\ref{resuinf}) indica que la traza de la
diferencia de las resolventes correspondientes a las extensiones
$A^0$ y $A^{\infty}$ admite un desarrollo asint{\'o}tico de la
forma,
\begin{equation}
    {\rm Tr}\,\{(A^0+z)^{-1}-(A^\infty+z)^{-1}\}\sim
    \sum_{n=2}^{\infty}\alpha_n^{V}(\nu)\,z^{-n/2}\,.
\end{equation}
Esto implica, de acuerdo con las ecuaciones (\ref{desaheat}) y
(\ref{desareso}), el siguiente desarrollo asint{\'o}tico,
\begin{equation}
    {\rm Tr}\,\{e^{-t\,A^0}-e^{-t\,A^\infty}\}\sim
    \sum_{n=0}^{\infty}a_n(A)\,t^{n/2}\,,
\end{equation}
donde coeficientes $a_n(A)=\alpha_{n+2}^{V}(\nu)/\Gamma(n/2+1)$ y
sus dimensiones est{\'a}n dadas por,
\begin{equation}
    [a_n(A)]=L^{-n}\,.
\end{equation}
Si los coeficientes $a_n(A)$ s{\'o}lo dependen del par{\'a}metro
adimensional $\nu$ y de potencias de las derivadas del potencial
$V(x)$ en el borde de la variedad, esto es, en $x=0$, entonces
podemos escribir,
\begin{eqnarray}\label{ans}
    \begin{array}{lll}
    a_0(A)=C_1(\nu)\,,&&a_3(A)=C_3(\nu)\cdot\partial_xV(0)\,,\\
    \\a_1(A)=0\,,&&
    a_4(A)=C_4(\nu)\cdot V^2(0)+C_5(\nu)\cdot\partial^2_xV(0)\,,\\
    \\a_2(A)=C_2(\nu)\cdot V(0)
    \,,&&
    a_5(A)=C_6(\nu)\cdot \partial_xV^2(0)+C_7(\nu)\cdot\partial^3_xV(0)\,.
    \end{array}\nonumber\\\nonumber\\
\end{eqnarray}

\bs

Supongamos ahora la validez de la relaci{\'o}n,
\begin{equation}\label{tras}
    \left.\frac{d}{d\epsilon}{\rm Tr}\,\{e^{-t(A-\epsilon)}\}\right|_{\epsilon=0}
    =t\, {\rm Tr}\,e^{-tA}\,,
\end{equation}
que ha sido probada para operadores diferenciales regulares. De
acuerdo con la ecuaci{\'o}n (\ref{tras}),
\begin{equation}\label{trasa}
    \left.\frac{d}{d\epsilon}a_n(A-\epsilon)\right|_{\epsilon=0}=
    a_{n-2}(A)\,.
\end{equation}
La ecuaci{\'o}n (\ref{trasa}) describe el efecto de la
transformaci{\'o}n $A\rightarrow A-\epsilon$ que equivale a
reemplazar $V(0)$ por $V(0)-\epsilon$. Teniendo en cuenta las
expresiones (\ref{ans}) podemos encontrar las relaciones,
\begin{equation}\label{rela}\begin{array}{c}
    -C_2(\nu)=C_1(\nu)\,,\\
    -2C_4(\nu)=C_2(\nu)\,,\\
    -2C_6(\nu)=C_3(\nu)\,,\end{array}
\end{equation}
y se verifica, adem{\'a}s, que $a_1(A)=0$.

\bs

Aunque buscamos un m{\'e}todo para determinar los coeficientes
$a_n(A)$ distinto del provisto por la ecuaci{\'o}n (\ref{tam})
(junto con (\ref{tam2}), (\ref{fin}) y (\ref{hache})), una simple
inspecci{\'o}n del t{\'e}rmino dominante en la ecuaci{\'o}n
(\ref{tam}) permite deducir que $\alpha_2^V(\nu)=\nu$. Por
con\-si\-guien\-te,
\begin{equation}\label{ito}
    a_0(A)=\nu\,.
\end{equation}
En consecuencia, utilizando las relaciones (\ref{rela})
reescribimos las ecuaciones (\ref{ans}),
\begin{eqnarray}\label{ans2}
    \begin{array}{lll}
    a_0(A)=\nu\,,&&a_3(A)=C_3(\nu)\cdot\partial_xV(0)\,,\\ \\a_1(A)=0\,,&&
    a_4(A)={\displaystyle\frac{\nu}{2}}\cdot V^2(0)+C_5(\nu)\cdot
    \partial^2_xV(0)\,,\\ \\a_2(A)=-\nu\cdot V(0)
    \,,&&
    a_5(A)=C_3(\nu)\cdot V(0)\partial_xV(0)+C_7(\nu)\cdot\partial^3_xV(0)\,.
    \end{array}\nonumber\\
\end{eqnarray}
El prop{\'o}sito del m{\'e}todo que exponemos es determinar los
coeficientes $C_i(\nu)$ desconocidos a partir de las funciones
espectrales que se hubieren calculado para ejemplos particulares.

\bs

Consideremos el caso particular $V(x)=0$. En este caso, los
n{\'u}cleos de las resolvente de las extensiones $\theta=0$ y
$\theta=\infty$ se conocen expl{\'\i}citamente,
\begin{eqnarray}
    G_0(x,x',z)=\sqrt{xx'}\left[(\theta(x-x')\,I_{-\nu}(\sqrt{z}x)
    K_{\nu}(\sqrt{z}x')-
    \right.\nonumber\\\mbox{}-\left.
    \theta(x'-x)\,I_{-\nu}(\sqrt{z}x')K_{\nu}(\sqrt{z}x)\right]\,,
    \\\nonumber\\
    G_\infty(x,x',z)=\sqrt{xx'}\left[(\theta(x-x')\,I_{\nu}
    (\sqrt{z}x)K_{\nu}(\sqrt{z}x')-
    \right.\nonumber\\\mbox{}-\left.
    \theta(x'-x)\,I_{\nu}(\sqrt{z}x')K_{\nu}(\sqrt{z}x)\right]\,.
\end{eqnarray}
La traza de la diferencia de las resolventes est{\'a} entonces
dada por,
\begin{eqnarray}
    {\rm Tr}\,\{(A^0+z)^{-1}-(A^\infty+z)^{-1}\}=\frac{\int_0^{\infty}
    [I_{-\nu}(y)-I_\nu(y)]\,K_\nu(y)\,dy}{z}=\frac{\nu}{z}\,,\nonumber\\\label{v0}
\end{eqnarray}
que confirma el resultado (\ref{ito}). Esta soluci{\'o}n permite
escribir la soluci{\'o}n para el caso $V(x)=V_0$, siendo $V_0$ na
constante en $\mathbb{R}$. En efecto, para este caso se deduce
inmediatamente a partir de (\ref{v0}),
\begin{eqnarray}
    {\rm Tr}\,\{(A^0+z)^{-1}-(A^\infty+z)^{-1}\}=\frac{\nu}{z+V_0}\nonumber\\
    \sim \frac{\nu}{z}
    \left(1-\frac{V_0}{z}+\frac{V_0^2}{z^2}-\ldots\right)\,.
\end{eqnarray}
Los primeros t{\'e}rminos del desarrollo asint{\'o}tico para el
caso $V(x)=V_0$ est{\'a}n dados, consecuentemente, por,
\begin{equation}
    {\rm Tr}\,\{e^{-t\,A^0}-e^{-t\,A^\infty}\}\sim
    \nu-\nu V_0\,t+\frac{\nu}{2}V_0^2\,t^2-\ldots
\end{equation}
de donde verificamos las primeras dos relaciones de (\ref{rela}).

\bs

El caso $V(x)=x^2$, cuyas funciones espectrales ya hemos
estudiado, permitir{\'a} determinar el coeficiente $C_5(\nu)$. En
efecto, la funci{\'o}n (\ref{tam2}) est{\'a} dada en este caso
por,
\begin{equation}
    R(y,z)=y^{\nu+1/2}\,e^{-y^2/2z}\,U(z/4+\nu/2+1/2,1+\nu,y^2/z)\,.
\end{equation}
cuyo comportamiento en el origen es,
\begin{equation}
    R(y,z)=\frac{\Gamma(\nu)\,z^{\nu}}{\Gamma{\displaystyle\left(
    \frac z 4+\frac \nu 2+\frac 1 2\right)}}
    \,y^{-\nu+1/2}\,+
    \frac{\Gamma(-\nu)}{\Gamma{\displaystyle\left(\frac z 4-\frac
    \nu 2+\frac 1 2\right)}}
    \,y^{\nu+1/2}+\dots
\end{equation}
Teniendo en cuenta la relaci{\'o}n (\ref{venelori}) obtenemos el
valor de $H(z)$ para el potencial $V(x)=x^2$,
\begin{equation}
    H(z)=4^\nu\,\frac{\Gamma{\displaystyle\left(\frac z 4+\frac {1+\nu}{2}\right)}}
    {\Gamma{\displaystyle\left(\frac z 4+\frac {1-\nu}{2}\right)}}\,z^{-\nu}\,.
\end{equation}
Podemos entonces calcular la traza de la diferencia de las
resolventes correspondientes a las extensiones $\theta=0$ y
$\theta=\infty$ mediante la ecuaci{\'o}n (\ref{tam}),
\begin{eqnarray}\label{tamx2}
    {\rm Tr}\left\{(A^{0}+z)^{-1}-(A^{\infty}+z)^{-1}\right\}=\nonumber\\
    \nonumber\\=
    {\frac{2^{1-2\nu}\,z^{2\nu}}{\Gamma(\nu)\Gamma(1-\nu)}
    \,\frac{\Gamma{\left(\frac z 4+\frac {1-\nu}{2}\right)}}
    {\Gamma{\left(\frac z 4+\frac {1+\nu}{2}\right)}}}\,\times\nonumber\\
    \times\,
    \int_0^\infty
    x^{2\nu+1}\,e^{-x^2}\,U^2(z/4+\nu/2+1/2,1+\nu,x^2)
    \,dx=\nonumber\\\nonumber\\=
    \frac{1}{4}\left[\psi\left(\frac z 4+\frac {1+\nu}{2}\right)-
    \psi\left(\frac z 4+\frac {1-\nu}{2}\right)\right]\,.
\end{eqnarray}
Podemos entonces escribir los primeros t{\'e}rminos del desarrollo
asint{\'o}tico,
\begin{equation}
     {\rm Tr}\left\{(A^{0}+z)^{-1}-(A^{\infty}+z)^{-1}\right\}\sim
     \frac{\nu}{z}+\frac{4}{3}\nu(\nu^2-1)\frac{1}{z^3}+O(z^{-5})\,.
\end{equation}
A partir de este desarrollo obtenemos para el caso $V(x)=x^2$,
\begin{equation}
    {\rm Tr}\,\{e^{-t\,A^0}-e^{-t\,A^\infty}\}\sim
    \nu+\frac{2}{3}\nu(\nu^2-1)\,t^2+O(t^4)\,,
\end{equation}
de donde se deduce,
\begin{equation}
    C_5(\nu)=\frac{\nu(\nu^2-1)}{3}\,.
\end{equation}
De esta manera, la resoluci{\'o}n de problemas particulares
permite determinar los coeficientes $C_i(\nu)$ que proveen el
desarrollo asint{\'o}tico del la traza de la diferencia del
heat-kernel de las extensiones $\theta=0$ y $\theta=\infty$.

\bs

Utilizaremos ahora el mismo m{\'e}todo para estudiar el desarrollo
asint{\'o}tico del heat-kernel para una extensi{\'o}n autoadjunta
arbitraria. De acuerdo con la ecuaci{\'o}n (\ref{resuinf}),
\begin{equation}\label{asin-HK}
    {\displaystyle {\rm Tr}\,\{e^{-t\,A^\theta}-e^{-t\,A^\infty}\}\sim
    \sum_{n=0}^{\infty}a_n(A)\,t^{n/2}+
    \sum_{N=1}^\infty\sum_{n=0}^{\infty}
    b_{N,n}(A)\,\theta^N\,t^{\nu N +n/2-1/2}\,,}
\end{equation}
donde $b_{N,n}(A)=\beta_{N,n}/\Gamma(\nu N+n/2+1/2)$. N{\'o}tese
que $[\theta^N]=L^{-2\nu N}$ de modo que cancelan las dimensiones
dependientes de $\nu$ que provienen de las potencias $t^{\nu N}$;
las dimensiones de los coeficientes $b_{N,n}(A)$, que no dependen
de $\theta$, est{\'a}n entonces dadas por,
\begin{equation}
    [b_{N,n}(A)]=L^{1-n}\,.
\end{equation}
Podemos entonces escribir,
\begin{equation}
    \begin{array}{llll}
    b_{1,0}(A)=D_1(\nu)\,,&
    b_{2,0}(A)=D_2(\nu)\,,&
    b_{3,0}(A)=D_3(\nu)\,,&\ldots\\ \\
    b_{1,1}(A)=0\,,&
    b_{2,1}(A)=0\,,&
    b_{3,1}(A)=0\,,&\ldots\\ \\
    b_{1,2}(A)=D_4\,V(0)(\nu)\,,&
    b_{2,2}(A)=D_5(\nu)\,V(0)\,,&\ldots\\ \\
    b_{1,3}(A)=D_6(\nu)\,\partial_xV(0)\,,&\ldots
    \end{array}
\end{equation}
La propiedad de transformaci{\'o}n (\ref{tras}) implica para los
coeficientes $b_{N,n}(A)$,
\begin{equation}\label{trasa-1}
    \left.\frac{d}{d\epsilon}b_{N,n}(A-\epsilon)\right|_{\epsilon=0}=
    b_{N,n-2}(A)\,,
\end{equation}
que conduce a las relaciones,
\begin{eqnarray}
    \begin{array}{c}
    D_4(\nu)=-D_1(\nu)\,,\\
    D_5(\nu)=-D_2(\nu)\,,
    \end{array}
\end{eqnarray}
y confirman que $b_{N,1}(A)=0$.

\bs

El estudio del caso $V(x)=0$ es suficiente para establecer los
valores de los coeficientes $b_{N,0}(A)$. En efecto, es f{\'a}cil
ver que, si $V(x)=0$ entonces $H(z)=1$. Por consiguiente, de
acuerdo con la ecuaci{\'o}n (\ref{kyh}),
\begin{equation}\label{kyh2}
    K(z)=-4^{\nu}\frac{\Gamma(\nu)}{\Gamma(-\nu)}\;z^{-\nu}\,.
\end{equation}
De modo que, de acuerdo con el Teorema (\ref{elthm}), los
t{\'e}rminos dependientes de $\theta$ en el desarrollo
asint{\'o}tico de la traza de la diferencia de las resolventes de
las extensiones $\theta$ arbitrario y $\theta=\infty$ est{\'a}n
dados por,
\begin{equation}
    {\rm Tr}\,\{(A^\theta+z)^{-1}-(A^\infty+z)^{-1}\}\sim
    \frac{\nu}{z}\,\sum_{N=1}^\infty
    \left[4^\nu\frac{\Gamma(\nu)}{\Gamma(-\nu)}\right]^N
    \theta^N\,z^{-\nu N}+\ldots
\end{equation}
Los t{\'e}rminos correspondientes que se obtienen del desarrollo
asint{\'o}tico de la traza del heat-kernel,
\begin{equation}\label{asin0}
    {\rm Tr}\,\{e^{-t A^\theta}-e^{-t A^\infty}\}\sim
    \nu\cdot\sum_{N=1}^\infty
    \frac{1}{\Gamma(\nu N+1)}
    \left[4^\nu\frac{\Gamma(\nu)}{\Gamma(-\nu)}\right]^N
    \theta^N\,t^{\nu N}+\ldots
\end{equation}
Comparando las ecuaciones (\ref{asin0}) y (\ref{asin-HK})
obtenemos,
\begin{equation}
    b_{N,0}=\frac{1}{N\,\Gamma(\nu N)}
    \left[4^\nu\frac{\Gamma(\nu)}{\Gamma(-\nu)}\right]^N\,.
\end{equation}
Por consiguiente,
\begin{eqnarray}\begin{array}{lll}
    {\displaystyle D_1(\nu)=\frac{4^\nu}{\Gamma(-\nu)}}\,,&
    {\displaystyle D_2(\nu)=2^{4\nu-1}
    \frac{\Gamma^2(\nu)}{\Gamma(2\nu)\,\Gamma^2(-\nu)}}\,,\\ \\
    {\displaystyle D_3(\nu)=\frac{4^{3\nu}}{3}
    \frac{\Gamma^3(\nu)}{\Gamma(3\nu)\,\Gamma^3(-\nu)}}\,,&
    {\displaystyle D_4(\nu)=-\frac{4^\nu}{\Gamma(-\nu)}}\,,\\ \\
    {\displaystyle D_5(\nu)=-2^{4\nu-1}
    \frac{\Gamma^2(\nu)}{\Gamma(2\nu)\,\Gamma^2(-\nu)}}\,.
    \end{array}
\end{eqnarray}

\section{Otro tipo de singularidad}\label{otrassing}

En esta secci{\'o}n haremos una breve consideraci{\'o}n de un
operador de Schr\"odinger en una dimensi{\'o}n con un coeficiente
singular proporcional a $x^{-1}$. Debido a la presencia de esta
singularidad la estructura de polos de la funci{\'o}n-$\zeta$
correspondiente no responde al resultado (\ref{resu}) sino que
presenta polos dobles en el plano complejo.

\bs

Sea el operador,
\begin{equation}
    A=-\partial_x^2+\frac{\alpha}{x}
\end{equation}
definido en el subconjunto denso
$\mathcal{C}_0^\infty((0,1))\subset\mathbf{L_2}([0,1])$.

\bs

Las extensiones autoadjuntas de este operador correspondientes a
condiciones de contorno locales est{\'a}n caracterizadas por dos
par{\'a}metros que describen esas condiciones en los extremos
$x=0$ y $x=1$, respectivamente. A nuestro prop{\'o}sito de
manifestar la presencia de polos de la funci{\'o}n-$\zeta$ de
multiplicidad mayor que 1, ser{\'a} suficiente con imponer
condiciones de contorno tipo Dirichlet en $x=1$.

\bs

Las extensiones autoadjuntas resultan entonces caracterizadas por
un {\'u}nico par{\'a}metro real que describe el comportamiento de
las funciones de $\mathcal{D}(A)$ en la singularidad. Puede
probarse que la funci{\'o}n-$\zeta$ correspondiente a una
extensi{\'o}n autoadjunta general presenta polos con multiplicidad
de todo orden. Sin embargo, por simplicidad, consideraremos en
esta secci{\'o}n la extensi{\'o}n autoadjunta particular
caracterizada por condiciones de tipo Dirichlet en el origen.

\bs

Como veremos, adem{\'a}s de una sucesi{\'o}n de polos simples en
semienteros negativos, la funci{\'o}n $\zeta(s)$ correspondiente a
esta extensi{\'o}n autoadjunta posee un polo doble en $s=-1/2$.

\bs

Consideremos entonces las soluciones de la ecuaci{\'o}n de
autovalores,
\begin{equation}
    (A-\mu^2)\phi=0\,,
\end{equation}
que son combinaciones lineales de las funciones,
\begin{eqnarray}
    \phi_1(x)=x\,e^{-i\mu x}\,M(1+\alpha/2 i \mu,2,2 i\mu x)\,,\label{autocou}\\
    \phi_2(x)=x\,e^{-i\mu x}\,U(1+\alpha/2 i \mu,2,2 i\mu x)\,.
\end{eqnarray}
Sus comportamiento en el origen est{\'a}n dados por,
\begin{eqnarray}
    \phi_1(x)=x+O(x^2)\,,\\
    \phi_2(x)=\frac{1}{2i\mu\,\Gamma(1+\alpha/2 i \mu)}+O(x)\,.
\end{eqnarray}
Si imponemos condiciones de contorno tipo Dirichlet en ambos
extremos del intervalo $[0,1]$, las autofunciones del operador $A$
est{\'a}n dadas por la ecuaci{\'o}n (\ref{autocou}) y los
autovalores $\lambda_n=\mu_n$ son las soluciones de la
ecuaci{\'o}n,
\begin{equation}\label{enecou}
    M(1+\alpha/2i\mu_n,2,2i\mu_n)=0\,.
\end{equation}
La funci{\'o}n $\zeta_A(s)$ admite entonces la representaci{\'o}n
integral,
\begin{eqnarray}
    \zeta(s)=\oint_\mathcal{C}\mu^{-2s}\frac{d}{d\mu}
    \log{M(1+\alpha/2i\mu,2,2i\mu)}\,\frac{d\mu}{2\pi i}=\nonumber\\=
    s\cdot\oint_\mathcal{C}\mu^{-2s-1}
    \log{M(1+\alpha/2i\mu,2,2i\mu)}\,\frac{d\mu}{\pi i}\,,\label{intcou}
\end{eqnarray}
siendo $\mathcal{C}$ una curva que encierra las soluciones de
(\ref{enecou}) en sentido antihorario.

\bs

Las singularidades de la funci{\'o}n $\zeta(s)$ pueden obtenerse a
partir de un desarrollo asint{\'o}tico del integrando en
(\ref{intcou}). Si $|\arg{\mu}|<\pi$ entonces,
\begin{eqnarray}\label{ascou}
    M(1+\alpha/2i\mu,2,2i\mu)\sim \frac{e^{i\mu+\pi \alpha/4\mu}}{2i\mu}
    \times\nonumber\\\times
    \left[
    \frac{e^{i\mu-i \alpha\log{2\mu}/2\mu}}{\Gamma(1+\alpha/2i\mu)}
    \,_2F_0(1-\alpha/2i\mu,-\alpha/2i\mu,1/2i\mu)-\nonumber\right.\\\left.
    \mbox{}-
    \frac{e^{-i\mu+i \alpha\log{2\mu}/2\mu}}{\Gamma(1-\alpha/2i\mu)}
    \,_2F_0(1+\alpha/2i\mu,\alpha/2i\mu,-1/2i\mu)
    \right]\,.
\end{eqnarray}
Podemos cambiar el camino de integraci{\'o}n en (\ref{intcou}) de
$\mathcal{C}$ al eje imaginario. Si $\mu=\pm i \rho$ con
$\rho\in\mathbb{R^+}$ el logaritmo del desarrollo asint{\'o}tico
(\ref{ascou}) toma la forma,
\begin{eqnarray}\label{ascou2}
    \log{M(1\mp \alpha/2\rho,2,\mp 2\mu)}\sim
    (1\mp 1)\rho-\log{\rho}+\frac \alpha 2\,\frac{\log{\rho}}{\rho}-
    \nonumber\\-\log{2}
    +\frac{\alpha\log{2}}{2}\,\frac 1 \rho-\log{\Gamma(1+\alpha/2\rho)}+
    \log{_2F_0(1-\alpha/2\rho,-\alpha/2\rho,1/2\rho)}
    \,.\nonumber\\
\end{eqnarray}
Reemplazando este desarrollo asint{\'o}tico en la expresi{\'o}n
(\ref{intcou}) obtenemos las singularidades de la funci{\'o}n
$\zeta(s)$.

\bs

El primer t{\'e}rmino de (\ref{ascou2}) s{\'o}lo est{\'a} presente
cuando $\mu$ pertenece al semieje imaginario negativo y conduce a
la aparici{\'o}n de un polo simple de la funci{\'o}n $\zeta(s)$ en
\begin{equation}
    s=\frac 1 2\,,
\end{equation}
con residuo,
\begin{equation}
    \left.{\rm Res}\,\{\zeta_A(s)\}\right|_{s=1/2}=\frac{1}{2\pi}\,.
\end{equation}

\bs

Las potencias impares restantes de $\rho$ en el desarrollo
asint{\'o}tico (\ref{ascou}) originan polos simples de la
funci{\'o}n $\zeta(s)$ en los puntos,
\begin{equation}
    s_n=-\frac 1 2-n\qquad {\rm con}\qquad n=0,1,2,\ldots
\end{equation}

\bs

Las singularidades restantes provienen del t{\'e}rmino
$\alpha\log{\rho}/2\rho$. Este t{\'e}rmino contribuye en la
cantidad $-\alpha/4\pi$ al residuo del polo en,
\begin{equation}
    s=-\frac 1 2\,.
\end{equation}
Adem{\'a}s, tambi{\'e}n conduce a la presencia de un polo doble en
el mismo punto $s=-1/2$ con residuo $\alpha/8\pi$.

\subsubsection{Comportamiento asint{\'o}tico de los autovalores}

Si aproximamos las soluciones de la ecuaci{\'o}n (\ref{enecou})
utilizando la expresi{\'o}n (\ref{ascou}) obtenemos el siguiente
comportamiento asint{\'o}tico para $\mu_n$,
\begin{equation}
    \mu_n\sim \pi\,n+\frac{\alpha}{2\pi}\,\frac{\log{n}}{n}
    +\frac{\alpha}{2\pi}\left(\log{2\pi}+\gamma_E\right)\,\frac 1 n+
    O(n^{-2}\log{n})\,.
\end{equation}
Si reemplazamos este desarrollo asint{\'o}tico en la
expresi{\'o}n,
\begin{equation}
    \zeta(s)=\sum_{\mu_n}\mu_n^{-s}\,,
\end{equation}
obtenemos,
\begin{eqnarray}\label{zetar}
    \zeta(s)=\pi^{-2s}\zeta_R(2s)+\alpha\pi^{-2s-2}\,s\zeta'_R(2s+2)-
    \nonumber\\ \\-
    \alpha(\log{2\pi}+\gamma_E)\pi^{-2s-2}\,s\zeta_R(2s+2)+h(s)\,,
\end{eqnarray}
siendo $\zeta_R(s)$ la funci{\'o}n-$\zeta$ de Riemann y $h(s)$ es
una funci{\'o}n anal{\'\i}tica en $\mathcal{R}(s)>-1$. La expresi{\'o}n
(\ref{zetar}) confirma los valores de los residuos que hemos
encontrado para los polos en $s=1/2$ y $s=-1/2$. Se verifica
adem{\'a}s la presencia de un polo doble en $s=-1/2$.


\part{Ap{\'e}ndice}


\section{{Operadores regulares sobre variedades no compactas}}\label{funcpart}

El resultado (\ref{resu}) determina la posici{\'o}n de los polos
de la funci{\'o}n-$\zeta$ de un o\-pe\-ra\-dor diferencial $A$ de
orden $d$ con coeficientes infinitamente derivables definido sobre
seccio\-nes de un fibrado vectorial sobre una variedad de base
compacta $M$ de dimensi{\'o}n $m$ con borde suave $\partial M$
sobre el que se imponen condiciones de contorno locales. Sin
embargo, como en esta Tesis hemos considerado operadores
diferenciales con coeficientes singulares sobre variedades no
compactas, hemos encontrado divergencias con respecto al resultado
(\ref{resu}).

\bs

Con el objeto de distinguir cu{\'a}les de estas divergencias
provienen de la presencia de t{\'e}rminos singulares y cu{\'a}les
se originan en la no compacidad de la variedad de base,
estudiaremos en esta secci{\'o}n la validez del resultado
(\ref{resu}) en el caso de variedades de base $M$ no compactas. El
argumento que presentaremos permite determinar la posici{\'o}n del
primer polo de la funci{\'o}n-$\zeta$ del operador diferencial
para operadores de Schr\"odinger con un potencial homog{\'e}neo.

\bs

Consideremos un operador de Schr\"odinger $A$,
\begin{equation}
    A=-\Delta+V(x)\,,
\end{equation}
donde $x\in\mathbb{R}^m$ y el dominio $\mathcal{D}(A)$ del
operador es un subespacio de $\mathbf{L_2}(\mathbb{R}^m)$ sobre el
cual $A$ es autoadjunto. Designemos por $\phi_n(x)$ sus
autofunciones y por $\lambda_n$ los autovalores correspondientes.

\bs

La traza del heat-kernel puede escribirse,
\begin{equation}
    {\rm Tr}\,e^{-tA}=\sum_{n}e^{-t\lambda_n}=
    \sum_{n}\int_{\mathbb{R}^m}dx\,\phi^*_n(x)\,e^{-tA}\cdot\phi_n(x)\,.
\end{equation}
Si introducimos en la {\'u}ltima expresi{\'o}n la transformada de
Fourier (ver definici{\'o}n (\ref{fourier})) en $m$ dimensiones
obtenemos,
\begin{equation}\label{path}
    {\rm Tr}\,e^{-tA}=
    \sum_{n}\int_{\mathbb{R}^{3m}}\!\!\!
    \frac{dx\,dp\,dp'}{(2\pi)^{m}}\,
    \mathcal{F}\{\phi_n\}^*(p)\,\,e^{ipx}
    e^{-tA}\cdot\,e^{-ip'x}\mathcal{F}\{\phi_n\}(p')\,.
\end{equation}
Como estamos interesados en el orden dominante a peque{\~n}os
valores de $t$, podemos re\-a\-li\-zar la siguiente
aproximaci{\'o}n,
\begin{equation}
    \exp{\left[-t\,(-\Delta+V(x))\right]}\sim
    \exp{\left(t\,\Delta\right)}\cdot
    \exp{\left(-t\,V(x)\right)}\,,
\end{equation}
pues la diferencia entre ambos miembros es proporcional al
conmutador de $t\,\Delta$ y de $t\,V(x)$ que es orden $t^{2}$. La
ecuaci{\'o}n (\ref{path}) toma entonces la forma,
\begin{eqnarray}
   {\rm Tr}\,e^{-tA}
    \sim\nonumber\\ \sim
    \int_{\mathbb{R}^{3m}}\!\!\!
    \frac{dx\,dp\,dp'}{(2\pi)^{m}}\,
    \left[\sum_{n}
    \mathcal{F}\{\phi_n\}^*(p)\cdot
    \mathcal{F}\{\phi_n\}(p')\right]
    \,e^{i(p-p')x}
    e^{-t(p'^2+V(x))}+\ldots\sim\nonumber\\\sim
    \int_{\mathbb{R}^{2m}}
    \frac{dx\,dp}{(2\pi)^{m}}\
    e^{-t(p^2+V(x))}\ldots\nonumber\\
\end{eqnarray}
La {\'u}ltima expresi{\'o}n puede interpretarse como la
aproximaci{\'o}n cl{\'a}sica\,\footnote{N{\'o}tese que si
interpretamos a la traza del heat-kernel como la funci{\'o}n de
partici{\'o}n estad{\'\i}stica, el l{\'\i}mite de peque{\~n}os
valores de $t$ corresponde al l{\'\i}mite de altas temperaturas.}
de la funci{\'o}n de partici{\'o}n en el espacio de fases. Si
realizamos la integral en el espacio de impulsos obtenemos,
\begin{equation}\label{apro}
    {\rm Tr}\,e^{-tA}\sim
    \frac{1}{(2\sqrt{\pi})^{m}}\,t^{-m/2}\int_{\mathbb{R}^{m}}
    {dx}\,
    e^{-t\,V(x)}+\ldots
\end{equation}
Supongamos ahora que el potencial depende de la variable radial
$r$ y que satisface la condici{\'o}n de homogeneidad,
\begin{equation}
    V(c\,r)=c^{h}\,V(r)\,.
\end{equation}
La aproximaci{\'o}n (\ref{apro}) puede entonces escribirse,
\begin{eqnarray}
    {\rm Tr}\,e^{-tA}\sim
    \frac{1}{(2\sqrt{\pi})^{m}}\,t^{-m/2}\int_{S^{m-1}}d\Omega
    \int_{\mathbb{R}^{+}}
    {dr}\,r^{m-1}\,
    e^{-V(t^{1/h}\,r)}+\ldots\nonumber\\
    \sim
    \frac{1}{2^{m-1}\,\Gamma(m/2)}\,t^{-m/2-m/h}\int_{\mathbb{R}^{+}}
    {dr}\,r^{m-1}\,
    e^{-V(r)}+\ldots\label{ult}
\end{eqnarray}
Para un operador regular, el primer t{\'e}rmino del desarrollo
asint{\'o}tico de la traza del heat-kernel para una variedad
compacta es proporcional a $t^{-m/2}$. Sin embargo, como indica la
ecuaci{\'o}n (\ref{ult}), esto no es cierto para un operador de
Schr\"odinger con un potencial homog{\'e}neo sobre una variedad no
compacta sino que el exponente se modifica en la cantidad $-m/h$
siendo $h$ el grado de homogeneidad del potencial.

\bs

Si la variedad de base es unidimensional y el potencial es
homog{\'e}neo de grado $2$, entonces el primer t{\'e}rmino del
desarrollo asint{\'o}tico de la traza del heat-kernel es
proporcional a $t^{-1}$. Esto implica que el primer polo de la
funci{\'o}n-$\zeta$ del operador diferencial se encuentra en
$s=1$, en coincidencia con uno de los resultados de la secci{\'o}n
(\ref{Wipf}), que corresponde a una variedad de base
unidimensional y a un potencial cuyo t{\'e}rmino dominante en el
infinito es homog{\'e}neo de grado $2$. Este resultado se
encuentra tambi{\'e}n en \cite{A}, donde se afirma, adem{\'a}s,
que el comportamiento del potencial $V(x)$ en
$|x|\rightarrow\infty$ puede conducir, incluso, a la aparici{\'o}n
de potencias de logaritmos de $t$ en el desarrollo de la traza del
heat-kernel del operador diferencial.

\section{SUSYQM: Funciones Espectrales}
\label{spectral-functions111}

\subsection{La funci{\'o}n de partici{\'o}n graduada}\label{graded}

En esta secci{\'o}n calcularemos la funci{\'o}n de partici{\'o}n
graduada \cite{Smilga} del hamiltoniano $H^{(\gamma)}$ de la
secci{\'o}n \ref{rupturaSUSY}, definida como,
\begin{equation}
    Z_\gamma^F(t):= {\rm Tr}\,\left\{\left(-1\right)^F
    e^{-t H^{(\gamma)}}\right\}\,.
\end{equation}
Si $\{\Phi_n\}_{n\in\mathbb{N}}$ representa el conjunto de
autofunciones de componentes $\phi_{1,n},\phi_{2,n}$ del
o\-pe\-ra\-dor $Q^{(\gamma)}_+$ cuyos autovalores est{\'a}n dados
por $\{\lambda_n\}_{n\in\mathbb{N}}$, podemos definir la
funci{\'o}n $\hat{Z}_\gamma ^F(t)$,
\begin{equation}\label{grad-part}
  \hat{Z}_\gamma^F(t)=
  \displaystyle{
  \sum_{\lambda_n\neq 0} e^{-t \lambda_n^2} \frac{\left(
  \Phi_n, (-1)^F \Phi_n \right)}{\|\Phi_n\|^2} }
  =\displaystyle{
  \sum_{\lambda_n\neq 0} \frac{e^{-t \lambda_n^2}}{\lambda_n}\
  \frac{\left( Q_+^\dagger \Phi_n, (-1)^F \Phi_n
  \right)}{\|\Phi_n\|^2}}\,,
\end{equation}
donde,
\begin{equation}\label{-1F}
  (-1)^F \left( \begin{array}{c}
  \phi_1   \\  \phi_2
\end{array} \right) = \left( \begin{array}{c}
  \phi_1   \\  - \phi_2
\end{array} \right)\,.
\end{equation}
Teniendo en cuenta la ecuaci{\'o}n (\ref{Q_+-def}) es
in\-me\-dia\-to probar,
\begin{equation}\label{grad-part-1}\begin{array}{c}
  \hat{Z}_\gamma^F(t) =
  \displaystyle{ -
  \sum_{\lambda_n\neq 0} \frac{e^{-t \lambda_n^2}}
  {\sqrt{2} \lambda_n \|\Phi_n\|^2} \left[
  \phi_{n,1}(x)\phi_{n,2}(x)\right]_{x=0^+} =}\\  \\
  =  \displaystyle{ \frac{1}{2}
  \sum_{\lambda_n\neq 0} \frac{\Gamma\left( \frac{1}{2}+\alpha \right)
  \Gamma\left( \frac{1}{2}- \alpha \right)e^{-t \lambda_n^2}}
  {\Gamma\left(1- \frac{\lambda_n^2}{2}\right)
  \Gamma\left( \frac{1-\lambda_n^2}{2}-\alpha \right)
  \|\Phi_n\|^2}}\,,
\end{array}
\end{equation}
donde hemos tenido en cuenta el comportamiento de las funciones en
${\cal D}\left( H_\gamma\right)$ cerca del origen (v{\'e}ase la
ecuaci{\'o}n (\ref{eigen-asymp}).)

\bs

Dado que el espectro de $Q^{(\gamma)}_+$ depende del par{\'a}metro
$\gamma$, la funci{\'o}n de partici{\'o}n gra\-dua\-da depende de
la extensi{\'o}n autoadjunta. Se puede probar, adem{\'a}s, que,
para los casos $\gamma=0,\pi/2$, $Z_\gamma^F(t)$ es independiente
de $t$ y coincide con el {\'\i}ndice de Witten $\Delta$.

\bs

En efecto, de acuerdo con los autovalores de $Q^{(0)}_+$, dados en
la ecuaci{\'o}n (\ref{eigen-gamma-0}), cada t{\'e}rmino en la
serie en (\ref{grad-part-1}) se anula. En consecuencia, como
$Z_{\gamma=0}^F(t)$ s{\'o}lo recibe una contribuci{\'o}n no nula
proveniente del modo cero, obtenemos,
\begin{equation}\label{ZF0}
  Z_{\gamma=0}^F(t)=
  \frac{\left(
  \Phi_0, (-1)^F \Phi_0 \right)}
  {\|\Phi_0\|^2} = 1 = \Delta_{\gamma=0}\,,
\end{equation}
donde hemos utilizado la ecuaci{\'o}n (\ref{eigenfunctions-1-0}).

\bs

Por su parte, de acuerdo con los autovalores de
$P_{\gamma=\pi/2}$, dados por la ecuaci{\'o}n
(\ref{eigen-gamma-pi}), todos los t{\'e}rminos de la serie en
(\ref{grad-part-1}) se anula. Obtenemos entonces,
\begin{equation}\label{ZF-pi}
  Z_{\gamma=\pi/2}^F(t)= 0=\Delta_{\gamma=\pi/2}\,.
\end{equation}

\bs

Debe destacarse que para las extensiones autoadjuntas del
hamiltoniano correspondientes a valores de $\gamma\neq 0,\pi/2$,
la funci{\'o}n de partici{\'o}n graduada $Z_\gamma^F(t)$ depende
del par{\'a}metro $t$. El {\'\i}ndice de Witten est{\'a} dado por
el l{\'\i}mite $t\rightarrow \infty$ que es igual a cero pues
$Z_\gamma^F(t)=\hat{Z}_\gamma^F(t)$ se anula exponencialmente con
$t$ debido a la ausencia de modos cero.

\subsection{La asimetr{\'\i}a espectral de la supercarga} \label{spectral-asym}

La asimetr{\'\i}a espectral $\eta(s)$ del operador
$Q^{(\gamma)}_+$,
\begin{equation}\label{sepec-asymm}
  \eta(s):=\sum_{\lambda_{\pm,n} \neq 0} {\rm sign}
  \left( \lambda_{\pm,n}\right)\,
  \left| \lambda_{\pm,n} \right|^{-s}.
\end{equation}
est{\'a} relacionada con la degeneraci{\'o}n del espectro de
$H^{(\gamma)}$.

\bs

Dado que $\left|\lambda_{\pm,n}\right| \sim \sqrt{n}$ (v{\'e}anse
las ecuaciones (\ref{eigen-cotas})), la ecuaci{\'o}n
(\ref{sepec-asymm}) define una funci{\'o}n anal{\'\i}tica en el
semiplano $\Re(s)>2$.

\bs

Si $\alpha \in (-1/2,1/2)$, la funci{\'o}n $\eta(s)$
correspondiente a las extensiones caracterizadas por $\gamma=0$ y
$\gamma=\pi/2$ se anula id{\'e}nticamente (v{\'e}anse las
ecuaciones (\ref{beta-infty}) y (\ref{beta-0}).)

\bs

Calcularemos, a continuaci{\'o}n, el valor $\eta(0)$ para las
extensiones definidas por $\gamma\neq 0,\pi/2$. En general, la
asimetr{\'\i}a espectral puede expresarse en t{\'e}rminos de las
funciones-$\zeta$ parciales $\zeta_{\pm}(s,\beta(\gamma))$,
\begin{equation}\label{eta-zeta}
  \eta(s)= \zeta_+(s,\beta(\gamma)) - e^{i \pi s} \zeta_-(s,\beta(\gamma))\,,
\end{equation}
donde,
\begin{equation}\label{zeta+-}\begin{array}{c}
   \zeta_+(s,\beta(\gamma)):= \displaystyle{
   \sum_{\lambda_{+,n}>0} \lambda_{+,n}^{-s}, }
   \\  \\
  \zeta_-(s,\beta(\gamma)):= \displaystyle{
  \sum_{\lambda_{-,n}<0} \lambda_{-,n}^{-s}.}
\end{array}
\end{equation}

\bs

De acuerdo con la ecuaci{\'o}n (\ref{trascendental}) los
autovalores de $Q^{(\gamma)}_+$ son los ceros de la funci{\'o}n
entera,
\begin{equation}\label{holomorphic}
  F(\lambda,\beta(\gamma)) := \frac{\lambda}{\Gamma\left(
  \frac12 -\alpha-\frac{\lambda^2}{2}\right)} -
  \frac{\beta(\gamma)}{\Gamma\left( -\frac{\lambda^2}{2} \right)}\,.
\end{equation}
Como los ceros de $F(\lambda,\beta(\gamma))$ son reales y simples,
podemos utilizar la siguiente representaci{\'o}n integral para la
funci{\'o}n-$\zeta$ parcial,
\begin{equation}\label{zeta+int}\begin{array}{c}
    \zeta_+(s,\beta(\gamma))=\displaystyle{\frac{1}{2 \pi i} \oint_{{\cal C}_+}
  \lambda^{-s} \frac{F'(\lambda,\beta(\gamma))}{F(\lambda,\beta(\gamma))}
  \, d\lambda
  =} \\ \\
  \displaystyle{ = -\frac{1}{2\pi} \, e^{i\pi s/2}
  \int_{-\infty+i 0}^{\infty+i 0}
  \mu^{-s} \frac{F'(e^{-i\pi/2} \mu,\beta(\gamma))}{F(e^{-i\pi/2}
  \mu,\beta(\gamma))}
  \, d\mu\,,  }
\end{array}
\end{equation}
donde la curva ${\cal C}_+$ encierra los ceros positivos de
$Q^{(\gamma)}_+$ en sentido antihorario.

\bs

Adem{\'a}s, como $F(e^{i \pi}|\lambda|,\beta(\gamma))= e^{i
\pi}F(|\lambda|,e^{- i \pi }\beta(\gamma))$, se deduce que los
ceros negativos de $F(\lambda,\beta(\gamma))$ son opuestos a los
ceros positivos de $F(\lambda,e^{- i \pi
 }\beta(\gamma))$. En consecuencia,
\begin{equation}\label{zeta-}
  \zeta_-(s,\beta(\gamma)) = e^{- i \pi s} \zeta_+(s,e^{-i\pi}\beta(\gamma))\,.
\end{equation}

\bs

Teniendo en cuenta,
\begin{equation}\label{FpsobreF}\begin{array}{c}
  \displaystyle{\frac{F'(-i \mu,\beta(\gamma))}{F(-i \mu,\beta(\gamma))}}
    =\displaystyle{\frac{\displaystyle{

    1+\mu^2 \left[\psi\left(\mu^2/2\right)-
    \psi\left(\frac12-\alpha+\mu^2/2\right) \right]}}
    {\displaystyle{- i \mu \left[
    1-i\,\beta(\gamma)\, \frac{ \Gamma\left(\frac12-\alpha+\mu^2/2\right)}
    {\mu\, \Gamma\left(\mu^2/2\right)}\right]}}\,
  -i \mu \,\psi\left(\mu^2/2\right) =} \\ \\{\displaystyle = i\,
  \Big[ \Delta_1(\mu)
  + \Delta_2(\mu,\beta(\gamma)) \Big] +
  O\left( \mu^{-3} \right)\,,}
\end{array}
\end{equation}
siendo,
\begin{equation}\label{Delta-mu}
    \begin{array}{c}
      \Delta_1(\mu) = \displaystyle{
  - \mu \, \log \left(\frac{\mu^2}{2}\right)
  +\frac{1}{\mu} }, \\ \\
  \Delta_2(\mu,\beta(\gamma)) = \displaystyle{
   \frac{2\alpha}
  {\mu}\,\left[1-i\frac{\beta(\gamma)}{\mu}\left(
  \frac{\mu^2}{2}\right)^{-\alpha+1/2}\right]^{-1}} \,,
    \end{array}
\end{equation}
vemos que la integral en (\ref{zeta+int}) converge a una
funci{\'o}n anal{\'\i}tica en el semiplano $\Re(s)>2$.
Consideraremos la extensi{\'o}n meromorfa de
$\zeta_+(s,\beta(\gamma))$ al semiplano complementario.

\bs

Como,
\begin{equation}\label{rotando}
  \frac{F'(-i e^{i \pi}  \mu,\beta(\gamma))}
  {F(-i e^{i \pi}  \mu,\beta(\gamma))}= e^{i \pi}
  \frac{F'(-i \mu,e^{-i \pi}\beta(\gamma))}
  {F(-i\mu,e^{-i \pi} \beta(\gamma))}\,,
\end{equation}
podemos escribir,
\begin{eqnarray}\label{zeta+asympt}
   \displaystyle{-2\pi \, \zeta_+(s,\beta(\gamma))=
   -2\, \sin \left( \frac{\pi s}{2} \right)
  \int_{1}^{\infty}
  \mu^{-s}  \Delta_1(\mu)\, d\mu \, +
  }  \nonumber\\ \nonumber\\
  \displaystyle{\mbox{}+
  i \int_{1}^{\infty}
  \mu^{-s} \left\{ e^{i \pi s/2} \Delta_2(\mu,\beta(\gamma)) -
  e^{- i \pi s/2} \Delta_2(\mu,e^{- i \pi}\beta(\gamma))
   \right\} d\mu + }\nonumber\\ \nonumber\\
  \displaystyle{\mbox{}
  +\,  e^{i \pi s/2}\int_{1}^{\infty}
  \mu^{-s}
   \left\{
  \frac{F'(-i \mu,\beta(\gamma))}{F(-i
  \mu,\beta(\gamma))}
   -i\Big[ \Delta_1(\mu)
  + \Delta_2(\mu,\beta(\gamma)) \Big] \right\} d\mu -}
  \nonumber\\\nonumber \\
  \displaystyle{\mbox{}
  - e^{-i \pi s/2} \int_{1}^{\infty}
  \mu^{-s}
   \left\{
   \frac{F'(-i \mu,e^{-i \pi}\beta(\gamma))}{F(-i
  \mu,e^{-i \pi} \beta(\gamma))}
   -i\Big[ \Delta_1(\mu)
  + \Delta_2(\mu,e^{-i \pi} \beta(\gamma)) \Big] \right\} d\mu +}
 \nonumber \\\nonumber \\
   \displaystyle{\mbox{}
  + e^{i \pi s/2} \int_{e^{i \pi}}^{1}
  \mu^{-s} \, \frac{F'(-i \mu,\beta(\gamma))}{F(-i \mu,\beta(\gamma))}
   \, d\mu  }\,,
\end{eqnarray}
donde la primera integral en el miembro derecho converge para
$\Re(s)>2$, la segunda para $\Re(s)>0$, la tercera y la cuarta
para $\Re(s)>-2$, y la quinta, evaluada a lo largo de una curva en
el semiplano superior que une el punto $-1$ con el $1$, es una
funci{\'o}n entera de $s$.

\bs

La extensi{\'o}n anal{\'\i}tica $I_1(s)$ del primer t{\'e}rmino
del miembro derecho de la ecuaci{\'o}n (\ref{zeta+asympt})
est{\'a} dada por,
\begin{equation}\label{an-ext-1}
    \begin{array}{c}
     I_1(s)= \displaystyle{
   -2\, \sin \left( \frac{\pi s}{2} \right)
  \int_{1}^{\infty}
  \mu^{-s}  \Delta_1(\mu)\, d\mu }
   = \\ \\
      \displaystyle{ =
   - 2 \sin \left(\pi s/2\right) \left[\frac{1}{s} -
   \frac{2}{{\left(  s-2  \right) }^2} +
   \frac{\log (2)}
   { s-2}
   \right]\, .}\\
   \end{array}
\end{equation}
La extensi{\'o}n anal{\'\i}tica del segundo t{\'e}rmino del
miembro derecho de la ecuaci{\'o}n (\ref{zeta+asympt}) resulta,
\begin{equation}\label{an-ext-2}
      \begin{array}{c}
     I_2(s)= \displaystyle{\Re \left\{
   2 i\, e^{i \pi s/2} \int_{1}^{\infty}
   \mu^{-s} \, \Delta_2(\mu,\beta(\gamma)) \, d\mu \right\}}
   = \\ \\
      \displaystyle{ =-\Re \left\{
      2 i\, e^{i \pi s/2} \lim_{\mu\rightarrow\infty}
      \int_1^{\mu^{-2\alpha}} \frac{x^{\frac{s}{2\,\alpha}-1 }  \, dx }
    {1 - i \,2^{\alpha- \frac{1}{2}   }\,
       \beta(\gamma)\,x } \right\} \,,}
   \end{array}
\end{equation}
si $\alpha\neq 0$, en tanto que $I_2(s):= 0$ para $\alpha=0$.

\bs

De acuerdo con el valor de $\alpha$, obtenemos,
\begin{itemize}
  \item {Si $0<\alpha<1/2$,
  \begin{equation}\label{I2-gpos}
    \begin{array}{c}
      I_2(s)=\displaystyle{-\frac{4 \alpha}{s}\,
       \sin \left(\frac{\pi s}{2}\right) -
       \frac{2^{\alpha+3/2} \alpha\, \beta(\gamma)}{s+2 \alpha} \,
        \cos \left(\frac{\pi s}{2}\right)
    + } \\ \\\displaystyle{
     + 2^{2 \alpha} \beta^2(\gamma)
     \int_0^1 x^{\frac{s}{2 \alpha}+1 } \,
     \frac{\sin \left(\frac{\pi s}{2}\right)+
      2^{\alpha-1/2} \beta(\gamma)\, x\,
      \cos \left(\frac{\pi s}{2}\right)}{1 + 2^{2\alpha-1}
     \beta(\gamma)^2  x^2} \, dx }\,,
    \end{array}
\end{equation}
donde la integral converge para  $s>- 4\, \alpha$. N{\'o}tese la
presencia de un polo\,\footnote{Esta singularidad implica que la
funci{\'o}n-$\zeta$ de $Q^{(\gamma)}_+$,
\begin{equation}\label{zeta-beta}
  \zeta(s,\beta(\gamma)):=\zeta_+(s,\beta(\gamma))+\zeta_-(s,\beta(\gamma))=
  \zeta_+(s,\beta(\gamma))+e^{- i \pi s} \zeta_+(s,e^{-i \pi}\beta(\gamma))\,,
\end{equation}
presenta un polo simple en $s=-2 \alpha$,
\begin{equation}\label{pole-zeta}
  \zeta(s,\beta(\gamma)) = \frac{2^{ g+{3}/{2} }
    \left( e^{2 i \pi g }  -1\right)
       g\,\beta(\gamma) \,\cos (g\,\pi )}
    { s+ 2 g } + O(s+ 2 g)^0\,.
\end{equation}
Si el par{\'a}metro $\gamma\neq 0,\pi/2$ el residuo, que depende
de la extensi{\'o}n autoadjunta, se anula solamente para el caso
regular $\alpha=0$. Este es otro ejemplo de un operador con un
potencial singular que admite extensiones autoadjuntas cuyas
funciones-$\zeta$ asociadas presentan polos en posiciones que no
responden al resultado (\ref{resu}), v{\'a}lido para operadores
regulares, sino que dependen de las caracter{\'\i}sticas de la
singularidad.} en $s=-2 \alpha$. }

\item {Si $-1/2<\alpha<0$ y $\gamma\neq 0$,
\begin{equation}\label{I2-gneg}
\begin{array}{c}
      I_2(s)=\displaystyle{-
      \frac{2^{-\alpha+5/2} \alpha}{\beta(\gamma) (s-2 \alpha) }
      \cos \left(\frac{\pi s}{2}\right)
     + } \\ \\\displaystyle{
    +  \int_1^\infty x^{ \frac{s}{2 \alpha}-2 }\,
    \frac{2 \beta(\gamma)\, x \sin\left(\frac{\pi s}{2}\right) -
    2^{-\alpha+3/2} \cos \left(\frac{\pi s}{2}\right)}{ \beta(\gamma) \left[
    1 + 2^{2\alpha-1}  \beta(\gamma)^2  x^2\right]} \,  dx
     }\,,
\end{array}
\end{equation}
donde la integral converge para $s> 4\, \alpha=-4|\alpha|$.
N{\'o}tese la presencia de un polo simple en $s=2 \alpha = - |2
\alpha|$. }

\end{itemize}

\bs

Es importante mencionar que la funci{\'o}n
$\zeta_+(s,\beta(\gamma))$ resulta anal{\'\i}tica en una vecindad
del origen. En particular, las ecuaciones (\ref{zeta+asympt}),
(\ref{an-ext-1}), (\ref{I2-gpos}) y (\ref{I2-gneg}) permiten
obtener los primeros t{\'e}rminos del desarrollo de Taylor de la
funci{\'o}n $ \zeta_+(s,\beta(\gamma))$ alrededor de $s=0$,
\begin{equation}\label{zeta+s0}
    \begin{array}{c}
   -2\pi \, \zeta_+(s \sim 0,\beta(\gamma))= - \pi +
   \displaystyle{\left\{
   \begin{array}{cr}
     -2\pi \alpha, & \alpha>0 \\ \\
     0 , & \alpha\leq 0
   \end{array}
    \right\}
   +}\\ \\
  \displaystyle{
  +   \int_{1}^{\infty}
  \left[\frac{F'(-i \mu,\beta(\gamma))}{F(-i \mu,\beta(\gamma))}
  -\frac{F'(-i \mu,e^{-i \pi}\beta(\gamma))}{F(-i \mu,e^{-i \pi}\beta(\gamma))}
  \right]\, d\mu +}\\ \\
   \displaystyle{\mbox{}+ i\Big[\log F(-i,\beta(\gamma))-\log
   F(i,\beta(\gamma))\Big]
   + O(s)}\,,
\end{array}
\end{equation}
donde la integral puede evaluarse teniendo en cuenta que,
\begin{equation}\label{gsg}
  \displaystyle{\frac{\Gamma\left(\frac 1 2 -\alpha +
  \frac{\mu^{2}}{2}\right)}{\mu \, \Gamma\left(
  \frac{\mu^{2}}{2}\right)} =
  2^{\alpha-1/2} \mu^{-2\alpha}\left\{ 1+
  O\left({\mu^{-2}}\right) \right\}}\,.
\end{equation}

Finalmente obtenemos,
\begin{equation}\label{zeta+en0}
   \zeta_+(s = 0,\beta(\gamma))=
   \displaystyle{\left\{
   \begin{array}{cr}
     \alpha\,, & \alpha>0 \\ \\ \displaystyle{
     \frac{i}{2 \pi}\,
     \log\left(\displaystyle{
     \frac{1+ e^{i \frac{\pi}{2}} \frac{\beta(\gamma)}{\sqrt{2}}}
     {1+ e^{-i \frac{\pi}{2}} \frac{\beta(\gamma)}{\sqrt{2}}}
     }\right)  } \, , & \alpha=0 \\ \\
     \displaystyle{\frac 1 2}\, , & \alpha<0
   \end{array}
    \right\}
   }\,.
\end{equation}

De las ecuaciones (\ref{eta-zeta}),(\ref{zeta-}) y
(\ref{zeta+en0}), podemos entonces calcular la asimetr{\'\i}a
espectral en $s=0$,
\begin{equation}\label{eta+s0}
    \begin{array}{c}
   \eta(s = 0)= \left. \Big[
   \zeta_+(s ,\beta(\gamma)) - \zeta_+(s ,e^{-i\pi}\beta(\gamma))\Big]
   \right|_{s=0}
   = \\ \\
   = \displaystyle{\left\{
   \begin{array}{lr}
     0 \,, & \alpha\neq 0, \\ \\
     \displaystyle{
     -\frac{2}{\pi} \arctan
     \left(\frac{\beta(\gamma)}
     {\sqrt{2}}\right)}
     \,, & \alpha=0.
   \end{array}
    \right.}
\end{array}
\end{equation}

\section{Desarrollo asint{\'o}tico del heat-kernel en varias dimensiones}\label{hke}

En esta secci{\'o}n mostraremos que la traza del heat-kernel
$e^{-tA}$ correspondiente a un operador de segundo orden $A$ con
coeficientes regulares admite un desarrollo asint{\'o}tico en
potencias de $t$ cuyos exponentes s{\'o}lo dependen del orden del
operador y de la dimensi{\'o}n de la variedad $M$.

\bs

Consideremos, entonces, el operador,
\begin{eqnarray}
    \!&&\!A  =-\Delta+V(x)\,,\\
    \!&&\!V(x)=\mathbf{V}_{\mu}(x)\cdot\partial_{\mu}+\mathbf{V}_0(x)\,,
\end{eqnarray}
donde $x\in\mathbb{R}^m$, $\Delta=-d^\dagger d$ y
$\mathbf{V}_{\mu}(x), \mathbf{V}_0(x)$ son funciones sobre
$\mathbb{R}^m$ con valores en $\in \mathbb{C}^{k\times k}$.

{\lem El heat-kernel $e^{t\Delta}(x,x')$ definido sobre
$\mathbb{R}^m\otimes\mathbb{R}^m$ es funci{\'o}n de $x-x'$ y
satisface la propiedad de homogeneidad,
\begin{equation}\label{Lema}
    e^{-t\Delta}(x-x')=t^{-m/2}\,e^{-\Delta}(t^{-1/2}\,(x-x'))\,.
\end{equation}}

\noindent{{\bf Demostraci{\'o}n:}} El n{\'u}cleo del operador
$e^{t\Delta}$ puede calcularse en forma expl{\'\i}cita (v{\'e}ase,
{\it e.g.,} \cite{CH2}.) La propiedad (\ref{Lema}) resulta
entonces inmediata.\fin

Es conveniente formular las siguientes definiciones,
\begin{eqnarray}
    \tilde{e}^{-tA}:= e^{-t\Delta}\cdot e^{-tA}\,,\label{heaint}\\
    \tilde{V}(t):= e^{t\Delta}\cdot V(x)\cdot e^{-t\Delta}\,.\label{heaint2}
\end{eqnarray}
El operador $\tilde{e}^{-tA}$ satisface entonces la ecuaci{\'o}n
diferencial,
\begin{equation}
    \left(\frac{\partial}{\partial
    t}+\tilde{V}(t)\right)\tilde{e}^{-tA}=0\,,
\end{equation}
que puede escribirse en forma integral,
\begin{equation}
 \tilde{e}^{-tA}=\mathbf{1}-\int_0^t\tilde{V}(t_1)\cdot\tilde{e}^{-t_1A}\,dt_1\,.
\end{equation}
La soluci{\'o}n de esta ecuaci{\'o}n puede obtenerse en forma
iterativa,
\begin{eqnarray}
    \tilde{e}^{-tA}\sim \mathbf{1}-\int_0^t\tilde{V}(t_1)\,dt_1+
 \int_0^t\!\int_0^{t_1}\tilde{V}(t_1)\cdot\tilde{V}(t_2)\,dt_1\,dt_2-\ldots+
 \nonumber\\
    \mbox{}+(-1)^{n}\int_0^t\!\int_0^{t_1}\!\ldots\!\int_0^{t_{n-1}}
 \tilde{V}(t_1)\tilde{V}(t_2)\ldots\tilde{V}(t_n)\,dt_1\,dt_2\ldots\,dt_n+\ldots
\end{eqnarray}
Reemplazando en esta expresi{\'o}n las definiciones (\ref{heaint})
y (\ref{heaint2}) obtenemos,
\begin{eqnarray}
    e^{-tA}\sim e^{-t\Delta}-\int_0^t e^{-(t-t_1)\Delta}\cdot V(x)\cdot e^{-t_1
    \Delta}\,dt_1+\nonumber\\
    +\int_0^t\!\!\int_0^{t_1}e^{-(t-t_1)\Delta}\cdot V(x)\cdot
    e^{-(t_1-t_2)\Delta}
    \cdot V(x)\cdot e^{-t_2 \Delta}\,dt_1\,dt_2-\ldots+\nonumber\\\mbox{}
    +(-1)^{n}\!\int_0^t\!\!\int_0^{x}\!\!\ldots\!\int_0^{t_{n-1}}
    \!e^{-(t-t_1)\Delta}\cdot V(x)\cdot e^{-(t_1-t_2)\Delta}\cdot V(x)\times
    \ldots\times\nonumber\\\times
    \, e^{-(t_{n-1}-t_n)\Delta}\cdot V(x)\cdot e^{-t_n
    \Delta}\,dt_1\,dt_2\ldots\,dt_n+
    \ldots
\end{eqnarray}
de modo que el desarrollo asint{\'o}tico para la el heat-kernel en
la diagonal est{\'a} dado por,
\begin{eqnarray}
    e^{-tA}(x,x')\sim \sum_{n=0}^{\infty}
    (-1)^n\int_0^t\!\int_0^{t_1}\!\ldots\int_0^{t_{n-1}}\!\!
    \int_{\mathbb{R}^m}\!\ldots\int_{\mathbb{R}^m}
    e^{-(t-t_1)\Delta}(x,x_1)
     \times\nonumber\\\times\ e^{-(t_1-t_2)\Delta}(x_1,x_2)\ldots
    e^{-t_n \Delta}(x_n,x')
    \cdot V(x_1)\ldots V(x_n)
    \ dx_1\ldots\,dx_n
    \,dt_1\ldots\,dt_n\,.\nonumber\\
\end{eqnarray}
Si hacemos el cambio de variables $t_i\rightarrow t\cdot t_i$ para
$i=1,\ldots,n$ y utilizamos la propiedad de homogeneidad
(\ref{Lema}) obtenemos,
\begin{eqnarray}
    e^{-tA}(x,x')\sim \sum_{n=0}^{\infty}
    (-1)^n\,t^{n-\frac{m}{2}(n+1)}
    \int_0^1\!\!\ldots\!\!\int_0^{t_{n-1}}\!\!
    \int_{\mathbb{R}^m}\!\!\ldots\!\!\int_{\mathbb{R}^m}
    e^{-(1-t_1)\Delta}(t^{-1/2}\,(x-x_1))\cdot\nonumber\\
    \cdot\, e^{-(t_1-t_2)\Delta}(t^{-1/2}\,(x_1-x_2))\cdot
    \ldots
    \cdot
    \,e^{-t_n \Delta}(t^{-1/2}\,(x_n-x'))\cdot\nonumber\\
    \cdot\,V(x_1)\ldots V(x_n)
    \ dx_1\ldots\,dx_n
    \,dt_1\ldots\,dt_n\,.\nonumber\\
\end{eqnarray}
Finalmente, hacemos el cambio de escala en las coordenadas
$x_i\rightarrow t^{-1/2}(x-x_i)$ para $i=1,\ldots,n$,
\begin{eqnarray}
    e^{-tA}(x,x')\sim \sum_{n=0}^{\infty}
    (-1)^{n(m+1)}\,t^{n-\frac{m}{2}}
    \int_0^1\!\!\ldots\!\!\int_0^{t_{n-1}}\!\!
    \int_{\mathbb{R}^m}\!\!\ldots\!\!\int_{\mathbb{R}^m}
    e^{-(1-t_1)\Delta}(x_1)\times\nonumber\\
    \times\,e^{-(t_1-t_2)\Delta}(x_2-x_1)
    \ldots
    e^{-t_n \Delta}(t^{-1/2}\,(x-x')-x_n)\times\nonumber\\
    \times\,V(x-\sqrt{t}x_1)\cdot V(x-\sqrt{t}x_2)\ldots V(x-\sqrt{t}x_n)
    \,\,dt_1\ldots\,dt_n\,dx_1\ldots\,dx_n\,.\nonumber\\
\end{eqnarray}

El desarrollo asint{\'o}tico del heat-kernel en la diagonal
$e^{-tA}(x,x)$ para peque{\~n}os va\-lo\-res de $t$ resulta
entonces,
\begin{eqnarray}\label{dam}
    e^{-tA}(x,x)\sim t^{-m/2}\,\sum_{n=0}^{\infty}
    c_n(A,x)
    \cdot t^n\,,
\end{eqnarray}
donde los coeficientes $c_n(A)$ est{\'a}n dados por,
\begin{eqnarray}
    c_n(A,x):=\frac{(-1)^{n(m+1)}}{n!}
    \int_0^1\!\ldots\int_0^{t_{n-1}}\,dt_1\ldots\,dt_n
    \int_{\mathbb{R}^m}\!\ldots\int_{\mathbb{R}^m}\ dx_1\ldots\,dx_n\times
    \nonumber\\\times\,
    e^{-(1-t_1)\Delta}(x_1)\cdot\,e^{-(t_1-t_2)\Delta}(x_2-x_1)
    \ldots
    e^{-t_n \Delta}(-x_n)\times\nonumber\\\times
    \left.\frac{d^{n}}{d\tau^n}
    V(x-\tau x_1)\ldots V(x-\tau x_n)\right|_{\tau=0}\,.
\end{eqnarray}

A partir del desarrollo asint{\'o}tico (\ref{dam}) del heat-kernel
en la diagonal $e^{-tA}(x,x)$ se demuestra, utilizando las
ecuaciones (\ref{desaheat}) y (\ref{desareso}), que el n{\'u}cleo
de la resolvente en la diagonal $G(x,x,\lambda)$ admite un
desarrollo asint{\'o}tico dado por,
\begin{equation}\label{teor2}
    G(x,x,\lambda)\sim \sum_{k=0}^{\infty}\gamma_{2k}(A,x)\cdot
    \lambda^{\frac{m-2k}{2}-1}\,.
\end{equation}

\bs

El desarrollo (\ref{dam}) conduce a la expresi{\'o}n
(\ref{teohea}) cuando el operador diferencial $A$ es de segundo
orden y est{\'a} definido sobre una variedad de base sin borde.
Por su parte, la expresi{\'o}n (\ref{teor2}) permite obtener el
desarrollo de la traza de la resolvente en potencias de $\lambda$
dadas por la expresi{\'o}n (\ref{desaheadesareso}) en el caso
$d=2$.

\section{Desarrollos asint{\'o}ticos de la secci{\'o}n VI.1}\label{asymptotic}

En esta secci{\'o}n calcularemos el desarrollo asint{\'o}tico de
$f'(\lambda)/f(\lambda)$ que hemos presentado en la ecuaci{\'o}n
(\ref{int666}).

\bs

El desarrollo asint{\'o}tico de la funci{\'o}n poligamma que
figura en el miembro derecho de la ecuaci{\'o}n (\ref{int666})
puede ser sencillamente derivado de la f{\'o}rmula de Stirling
\cite{A-S},
\begin{equation}\label{asymp-psi}
\psi\left(\frac{\nu}{2}+\frac{1}{2}-\lambda/4\right)\sim
\log{(-\lambda)}+\sum_{i=0}^\infty c_i(\nu) (-\lambda)^{-k},
\nonumber
\end{equation}
donde los coeficientes $c_i(\nu)$ son polinomios en $\nu$ cuya
expresi{\'o}n expl{\'\i}cita no necesitamos.

\bs

Por otra parte, teniendo en cuenta la ecuaci{\'o}n
(\ref{fun3666}), podemos escribir asint{\'o}ticamente el primer
t{\'e}rmino del miembro derecho de la ecuaci{\'o}n (\ref{int666})
de la siguiente manera,
\begin{equation}\label{asymp}     \begin{array}{c}
\displaystyle{
\frac{\left[\psi\left(-\frac{\nu}{2}+\frac{1}{2}-\frac{\lambda}{4}\right)-
            \psi\left(\frac{\nu}{2}+\frac{1}{2}-\frac{\lambda}{4}\right)\right]}
            {1-\theta\,\frac{\Gamma(\nu)}{\Gamma(-\nu)}\,
            \frac{\Gamma\left(-\frac{\nu}{2}+\frac{1}{2}-\frac{\lambda}{4}\right)}
            {\Gamma\left(\frac{\nu}{2}+\frac{1}{2}-\frac{\lambda}{4}
            \right)}}\sim }\\
            \\
\displaystyle{   \sim\sum_{N=0}^{\infty}\theta^N
            \left[\frac{\Gamma(\nu)}{\Gamma(-\nu)}\,
            \frac{\Gamma\left(-\frac{\nu}{2}+\frac{1}{2}-\frac{\lambda}{4}\right)}
            {\Gamma\left(\frac{\nu}{2}+\frac{1}{2}-\frac{\lambda}{4}
            \right)}\right]^N
            \left[\psi\left(-\frac{\nu}{2}+\frac{1}{2}-\frac{\lambda}{4}\right)-
            \psi\left(\frac{\nu}{2}+\frac{1}{2}-\frac{\lambda}{4}\right)
            \right]= }\\
            \\
\displaystyle{    =\sum_{N=0}^{\infty}{\theta^N}
             \left[\frac{\Gamma(\nu)}{\Gamma(-\nu)}\,
            \frac{\Gamma\left(-\frac{\nu}{2}+\frac{1}{2}-\frac{\lambda}{4}\right)}
            {\Gamma\left(\frac{\nu}{2}+\frac{1}{2}-\frac{\lambda}{4}\right)}
            \right]^N \,
            4\,\frac{d}{d(-\lambda)}\log
            \left[\frac{\Gamma\left(-\frac{\nu}{2}+\frac{1}{2}-\frac{\lambda}
            {4}\right)}
            {\Gamma\left(\frac{\nu}{2}+\frac{1}{2}-\frac{\lambda}{4}\right)}
            \right]=}\\
            \\
\displaystyle{    =\sum_{N=0}^{\infty}\frac{\theta^N}{N}\,4\,
            \frac{d}{d(-\lambda)}
            \left[\frac{\Gamma(\nu)}{\Gamma(-\nu)}\,
            \frac{\Gamma\left(-\frac{\nu}{2}+\frac{1}{2}-\frac{\lambda}{4}
            \right)}
            {\Gamma\left(\frac{\nu}{2}+\frac{1}{2}-\frac{\lambda}{4}\right)}
            \right]^N\,.}
    \end{array}
\end{equation}

De la f{\'o}rmula de Stirling \cite{A-S} obtenemos,
\begin{equation}\label{gqg}\begin{array}{c}
    \displaystyle{
       \log\left[\frac{\Gamma\left(-\frac{\nu}{2}+\frac{1}{2}-\frac{\lambda}
       {4}\right)}
            {\Gamma\left(\frac{\nu}{2}+\frac{1}{2}-\frac{\lambda}{4}\right)}
            \right]\sim
    -\nu\log(-\frac{\lambda}{4})}
        \displaystyle{
    +{\left\{\displaystyle{\sum_{m=1}^\infty a_m(\nu)
  (-\lambda)^{-2m}
  }\right\}}},
\end{array}
\end{equation}
donde los coeficientes en la serie est{\'a}n dados por
\begin{equation}\label{am(kappa)}\begin{array}{c}
    \displaystyle{
  a_m(\nu)=\frac{2^{4m-1}}{2m+1}\left\{
  \left[ \left(-\frac{\nu}{2}+\frac{1}{2}\right)^{2m} - \left(\frac{\nu}{2}+
  \frac{1}{2}\right)^{2m} \right]+
  \left(\frac{\nu}{2m}\right) \times\right. }\\ \\
      \displaystyle{
  \times\left[ \left(-\frac{\nu}{2}+\frac{1}{2}\right)^{2m} + \left(\frac{\nu}
  {2}+\frac{1}{2}\right)^{2m}  \right]+
  (2m+1) \sum_{p=1}^m \frac{B_{2p}}{p(2p-1)} \times }
  \\ \\    \displaystyle{
  \left. \times
  \left(\begin{array}{c}
    2m-1 \\
    2p-2
  \end{array}\right) \left[ \left(\frac{\nu}{2}+\frac{1}{2}\right)^{2(m-p)+1} -
  \left(-\frac{\nu}{2}+\frac{1}{2}\right)^{2(m-p)+1} \right] \right\} }\,.
\end{array}
\end{equation}
Por consiguiente,
\begin{equation}\label{GsGN}\begin{array}{c}
    \displaystyle{
  \left[\frac{\Gamma\left(-\frac{\nu}{2}+\frac{1}{2}-\frac{\lambda}{4}\right)}
            {\Gamma\left(\frac{\nu}{2}+\frac{1}{2}-\frac{\lambda}{4}\right)}
            \right]^N }
               \displaystyle{
  \sim \left(-\frac{\lambda}{4}\right)^{-N\nu}
            \sum_{n=0}^{\infty}b_n(\nu,N)\,(-\lambda)^{-2n} }\,,
\end{array}
\end{equation}
donde los coeficientes $b_n(\nu,N)$ est{\'a}n definidos por,
\begin{equation}
  \sum_{n=0}^{\infty}b_n(\nu,N)\,z^{-2n} :=
  \exp\left\{\displaystyle{N \sum_{m=1}^\infty a_m(\nu)\,
  z^{-2m}}\right\}\,.
\end{equation}
Estos coeficientes $b_n(\nu,N)$ son polinomios en $\nu$ y $N$
dados por,
\begin{equation}\label{bs-as}\begin{array}{c}
    \displaystyle{
      b_n(\nu,N)=\sum_{r_1,r_2,\dots,r_n}N^{r_1+ r_2+\dots+
    r_n} \, }
    \displaystyle{ \cdot \,
    \frac{a_1(\nu)^{r_1}\,a_2(\nu)^{r_2}\,\dots\,
    a_n(\nu)^{r_n}}
    {r_1!\, r_2!\, \dots \, r_n!} }\,,
\end{array}
\end{equation}
donde la suma comprende el conjunto de todos los enteros positivos
$r_1,r_2,\dots,r_n$ tales que $r_1+2\, r_2+\dots+n\, r_n=n$.

\bs

Reemplazando ahora la ecuaci{\'o}n (\ref{GsGN}) en la ecuaci{\'o}n
(\ref{asymp}) obtenemos,
\begin{equation}\label{des-asymp-fin}\begin{array}{c}
\displaystyle{
\frac{\left[\psi\left(-\frac{\nu}{2}+\frac{1}{2}-\frac{\lambda}{4}\right)-
            \psi\left(\frac{\nu}{2}+\frac{1}{2}-\frac{\lambda}{4}\right)\right]}
            {4\left[1-\beta(\gamma)\frac{\Gamma\left(-\frac{\nu}{2}+\frac{1}{2}-
            \frac{\lambda}{4}\right)}
            {\Gamma\left(\frac{\nu}{2}+\frac{1}{2}-\frac{\lambda}{4}\right)}
            \right]}\sim
            - \sum_{N=1}^\infty \sum_{n=0}^\infty
  4^{N\nu}\,\times}
            \\ \\
  \displaystyle{ \times\, \beta(\gamma)^{N}\left( \nu+\frac{2n}{N}
  \right)\, b_n(\nu,N)\, \left( -\lambda \right)^{-N\nu - 2n
  -1}:=}\\ \\
  \displaystyle{:= \sum_{N=1}^\infty \sum_{n=0}^\infty C_{N,n}
  (\nu,\theta)\,
  \left( -\lambda \right)^{-N\nu - 2n -1}}\,.
\end{array}
\end{equation}
Las ecuaciones (\ref{asymp-psi}) y (\ref{des-asymp-fin}) conducen
finalmente al desarrollo asint{\'o}tico para
$f'(\lambda)/f(\lambda)$ dado en la ecuaci{\'o}n (\ref{te}).

\section{Trazas de la secci{\'o}n VI.2}\label{integrals}

En esta secci{\'o}n describimos brevemente el c{\'a}lculo de las
trazas utilizado en la secci{\'o}n \ref{the-resolvent1}.

\bs

A partir de (\ref{GD111}) obtenemos para el n{\'u}cleo de
$G_\infty(\lambda)$ en la diagonal,
\begin{equation}\label{trGDDD}
    G_\infty(x,x;\mu^2) =-\frac{\pi}{2}\,\frac{x}{\sin{\pi\nu}\,J_{\nu}(\mu)}
    \left\{ J_{-\nu}(\mu )\,
     J_{\nu}^2(\mu \, x) -
     J_{\nu}(\mu ) \,
     J_{\nu}(\mu\, x ) \, J_{-\nu}(\mu\,x )\right\}
     \, .
\end{equation}
Para evaluar la traza de la resolvente hemos utilizado las
primitivas \cite{Mathematica,GR},
\begin{equation}\label{nu-nu}
  \int x\, J_\nu^2(\mu \,x)\,dx
  =\frac{x^2}{2}\,\left\{ {J_\nu(x\,\mu  )}^2 -
      J_{\nu-1}(x\,\mu  )\,
       J_{\nu+1}(x\,\mu  ) \right\}\,,
\end{equation}
y,
\begin{equation}\label{nu-menosnu}\begin{array}{c}
 \displaystyle{ \int x\, J_\nu(\mu \,x)\,
 J_{-\nu}(\mu \,x)\,dx= }\\ \\
     =\displaystyle{\frac{ - {\nu }^2}
      {{\mu  }^2\,
      \Gamma(1 - \nu )\,
      \Gamma(1 + \nu )}\,
      \left[ \,_1 F_2
          \left(\{ - {1}/
            {2}  \} ,
         \{ -\nu ,\nu \} ,
         -x^2 \, {\mu  }^2\right)-1  \right]}\, ,
\end{array}
\end{equation}
donde,
\begin{equation}\label{1F2}\begin{array}{c}
   _1 F_2 \left(\{ - {1}/
            {2}  \} ,
         \{ -\nu ,\nu \} ,
         -x^2 \, {\mu  }^2\right)
  \displaystyle{
  =- \frac{ \pi \,x^2\,
      {\mu  }^2\,
       \,\csc (\pi \,\nu )
       }{4\,\nu }\, \times }\\ \\
      \left\{ J_{-1 - \nu }(x\,\mu  )\,
         J_{-1 + \nu }(x\,\mu  ) +
        2\,J_{-\nu }(x\,\mu  )\,
         J_{\nu }(x\,\mu  ) +
        J_{1 - \nu}(x\,\mu  )\,
         J_{1 + \nu }(x\,\mu  )
        \right\}\, .
\end{array}
\end{equation}
Estas primitivas, junto con la relaci{\'o}n,
\begin{equation}\label{prop-J-Bessel}
  J_{\nu-1}(z)+J_{\nu+1}(z)
  =\frac{2 \nu}{z} J_{\nu}(z)\, ,
\end{equation}
conducen directamente a la ecuaci{\'o}n (\ref{traza-GD}).

\bigskip

An{\'a}logamente, el n{\'u}cleo de $G_0(\lambda)$ en la diagonal
est{\'a} dado por,
\begin{equation}\label{trGD}\begin{array}{c}
    G_0(x,x;\mu^2) =-\frac{\pi}{2}\,\frac{x}{\sin{\pi\nu}\,J_{-\nu}(\mu)}
    \left\{ J_{-\nu}(\mu )\,
     J_{-\nu}(\mu \, x)\,J_{\nu}(\mu\,x) -
     J_{\nu}(\mu ) \,
     J_{-\nu}^2 (\mu\, x )\right\}
     \, .
\end{array}
\end{equation}
El mismo argumento conduce a la ecuaci{\'o}n (\ref{traza-GN}).

\bigskip

\section{Trazas de la secci{\'o}n VII.1}\label{integrals2}

En esta secci{\'o}n describiremos brevemente el c{\'a}lculo de las
trazas utilizadas en la secci{\'o}n \ref{trace-resolvent-1}.

\bs

De las ecuaciones (\ref{GN11}) y (\ref{GN22}) obtenemos la traza
matricial de $G^0(x,x',\lambda)$ en la diagonal $x=x'$,
\begin{equation}\label{trGN-matrix}\begin{array}{c}
    tr\left\{G^0(x,x;\lambda) \right\}
  =- \, \displaystyle{\frac{\pi \,x\,\lambda\,\cosec (\nu\pi ) }
    {2\,J_{\nu}(\lambda )}}\,
    \left\{ J_{-\nu}(\lambda )\,{J_{\nu-1}^2(
          x\,\lambda )}
        + \right.
            \\ \\
  \left. +
          J_{  -\nu}(\lambda )\, J_{\nu}^2(x\,\lambda )
          - J_{\nu}(\lambda )\,J_{\nu}(x\,\lambda )
           \,J_{-\nu}(x\,\lambda )  
              +
       J_{\nu}(\lambda )\,J_{\nu-1}(x\,\lambda )\,
             J_{1-\nu}(x\,\lambda )
      \right\}\, ,
\end{array}
\end{equation}
cuyo comportamiento en proximidades del origen est{\'a} dado por,
\begin{equation}\label{TRGN-matrix-en0}
  tr\left\{G^0(x,x;\lambda) \right\}
    =
  - \frac{4^{1/2-\nu}\,\pi \,\cosec (\nu\pi )\,
         J_{-\nu}(\lambda )
         }{{\lambda }^{1-2\nu}\,
         \Gamma^2(\nu)
         J_{\nu}(\lambda )}
        \, {x^{-1+2\nu}} +  {O}(x)\,,
\end{equation}
de modo que es integrable en $[0,1]$. Para calcular esta integral
hemos utilizado las primitivas \cite{Mathematica},
\begin{eqnarray}
  \int x\, J_\nu^2(\lambda\,x)\,dx
  =\frac{x^2}{2}\,\left\{ {J_\nu(x\,\lambda )}^2 -
      J_{\nu-1}(x\,\lambda )\,
       J_{\nu+1}(x\,\lambda ) \right\}\,,\label{nu-nu-1}\\
 \displaystyle{ \int x\, J_\nu(\lambda\,x)\,
 J_{-\nu}(\lambda\,x)\,dx= }\\
     =\displaystyle{\frac{ - {\nu }^2}
      {{\lambda }^2\,
      \Gamma(1 - \nu )\,
      \Gamma(1 + \nu )}\,
      \left[ \,_1 F_2
          \left(\{ - {1}/
            {2}  \} ,
         \{ -\nu ,\nu \} ,
         -x^2 \, {\lambda }^2\right)-1  \right]}\, ,
         \label{nu-menosnu-1}
\end{eqnarray}
donde,
\begin{equation}\label{1F2-1}\begin{array}{c}
   _1 F_2 \left(\{ - {1}/
            {2}  \} ,
         \{ -\nu ,\nu \} ,
         -x^2 \, {\lambda }^2\right)
  \displaystyle{
  =- \frac{ \pi \,x^2\,
      {\lambda }^2\,
       \,\csc (\pi \,\nu )
       }{4\,\nu }\, \times }\\ \\
      \left\{ J_{-1 - \nu }(x\,\lambda )\,
         J_{-1 + \nu }(x\,\lambda ) +
        2\,J_{-\nu }(x\,\lambda )\,
         J_{\nu }(x\,\lambda ) +
        J_{1 - \nu}(x\,\lambda )\,
         J_{1 + \nu }(x\,\lambda )
        \right\}\, .
\end{array}
\end{equation}
junto con la relaci{\'o}n,
\begin{equation}\label{prop-J-Bessel-1}
  J_{\nu-1}(z)+J_{\nu+1}(z)
  =\frac{2 \nu}{z} J_{\nu}(z)\,.
\end{equation}
La traza de la resolvente correspondiente a $\beta=0$ resulta
entonces,
\begin{equation}\label{TRGN}
  {\rm Tr}\,(D^0-\lambda)^{-1}=\displaystyle{ \int_0^1
  tr\left\{G^0(x,x;\lambda) \right\}\,
  dx
  =\,\frac{2\nu-1}{\lambda } -
  \frac{J_{\nu-1}(\lambda )}{J_{\nu}(\lambda
  )}\, . }
\end{equation}

An{\'a}logamente, la traza matricial de $G^\infty(x,x';\lambda)$
en la diagonal $x=x'$ resulta,
\begin{equation}\label{trGD-1}\begin{array}{c}
     tr\left\{G^\infty(x,x;\lambda) \right\}
     =     \displaystyle{\frac{\pi \,x\,\lambda \,\cosec (\nu\pi )}
    {2\,J_{-\nu}(\lambda )}}
    \,
    \left\{- J_{-\nu}(\lambda )\,J_{\nu}(x\,\lambda )
        \,
       J_{-\nu}(x\,\lambda )\, + \right.
       \\ \\\left.
       +
      J_{-\nu}(\lambda )\,J_{\nu-1}(x\,\lambda )\,
              J_{1-\nu}(x\,\lambda ) \,
        +
      J_{\nu}(\lambda )\,
       {J_{-\nu}^2(
            x\,\lambda )} +J_{\nu}(\lambda )\,
         {J_{1-\nu}^2(
            x\,\lambda )}
      \right\}\, ,
\end{array}
\end{equation}
cuyo comportamiento en el origen est{\'a} dado por,
\begin{equation}\label{trGD-en0}
  {\rm tr}\left\{G^\infty(x,x;\lambda) \right\}
  =\displaystyle{\frac{\pi \,{\lambda }^{1-2\nu}
    \,\cosec (\nu\pi )\,
       J_{\nu}(\lambda )}{4^{1/2-\nu}\,
       {\Gamma^2(1-\nu)} \,
       J_{-\nu}(\lambda )} \,x^{1-2\nu}\,+ O(x)
  \, .}
\end{equation}
Podemos entonces calcular la traza de la resolvente
correspondiente a la extensi{\'o}n $\beta=\infty$,
\begin{equation}\label{TRGD}
  {\rm Tr}\,(D^\infty-\lambda)^{-1}=
    \displaystyle{ \int_0^1
  tr\left\{G^\infty(x,x;\lambda) \right\}\,
  dx   =
  \frac{J_{ 1-\nu }(\lambda )}
      {J_{-\nu }(\lambda )}\, . }
\end{equation}
De las ecuaciones (\ref{TRGN}) y (\ref{TRGD}) obtenemos las
expresiones (\ref{traza-derivGD}) y (\ref{TrGD-GN}).

\bigskip

N{\'o}tese adem{\'a}s que,
\begin{equation}\label{ddd}\begin{array}{c}
  \partial_\lambda
  {\rm tr}\left\{G^\infty(x,x;\lambda) \right\}= \\ \\=
  \displaystyle{
 \frac{2^{2\nu}\,
       {\lambda }^{-2\nu}\,
       \left[ 1 + (1/2-\nu)\,\pi \,
          J_{\nu}(\lambda )\,
          J_{-\nu}(\lambda )\,
          \cosec (\nu\pi ) \right] }{{
          J_{-\nu}(\lambda )}^2\,
       {\Gamma(1-\nu)}^2}\,x^{1-2\nu}+O(x)} \,.
\end{array}
\end{equation}

\section{Desarrollos asint{\'o}ticos de las Secciones VI.2 y VII.1}\label{Hankel}

Para desarrollar asint{\'o}ticamente la traza de la resolvente
utilizamos el
 desarrollo de Hankel para las funciones de Bessel, que brevemente exponemos en
 esta secci{\'o}n.

\bs

Para $|z|\rightarrow \infty$, con $\nu$ fijo y $|\arg z|<\pi$, se
verifica \cite{A-S},
\begin{equation}\label{hankel}
  J_\nu(z)\sim \left(\frac{2}{\pi\,z}\right)^{\frac 1 2}
  \left\{ P(\nu,z) \cos \chi(\nu,z)
  -Q(\nu,z) \sin \chi(\nu,z) \right\}\, ,
\end{equation}
y,
\begin{equation}\label{hankel-N}
    N_\nu(z)\sim \left(\frac{2}{\pi\,z}\right)^{\frac 1 2}
  \left\{ P(\nu,z) \sin \chi(\nu,z)
  + Q(\nu,z) \cos \chi(\nu,z) \right\}\, ,
\end{equation}
donde,
\begin{equation}\label{chi}
  \chi(\nu,z) = z- \left(\frac \nu 2 + \frac 1 4\right) \pi\,,
\end{equation}
\begin{equation}\label{P}
  P(\nu,z) \sim \sum_{k=0}^\infty
  \frac{(-1)^k\,\Gamma\left(
  \frac 1 2 + \nu + 2 k\right)}{(2k)! \, \Gamma\left(
  \frac 1 2 + \nu - 2 k\right)} \,
  \frac{1}{\left(2 z\right)^{2 k}}\, ,
\end{equation}
y,
\begin{equation}\label{Q}
  Q(\nu,z) \sim \sum_{k=0}^\infty
  \frac{(-1)^k \,\Gamma\left(\frac 1 2+
   \nu + 2 k+1 \right)}{(2 k+1)! \, \Gamma\left(
   \frac 1 2+ \nu - 2 k-1 \right)} \,
    \frac{1}{\left(2 z\right)^{2 k+1}}\, .
\end{equation}

Adem{\'a}s, $P(-\nu,z)=P(\nu,z)$ y $Q(-\nu,z)=Q(\nu,z)$, pues
estas funciones dependen solamente de $\nu^2$ (v{\'e}ase
\cite{A-S}.) Por consiguiente,
\begin{equation}\label{Jupper}
  J_\nu(z)\sim
  \frac{e^{-i\sigma z}\, e^{i\sigma \pi \left(
  \frac \nu 2 + \frac 1 4 \right)}}{\sqrt{2\pi z}}
  \left\{ P(\nu,z)
  - i\sigma \,Q(\nu,z) \right\}\, ,
\end{equation}
donde $\sigma=1$ si $\mathcal{I}(z)>0$ en el semiplano superior y
$\sigma=-1$ si $\mathcal{I}(z)<0$. An{\'a}logamente,
\begin{equation}\label{N-sigma}
    N_\nu(z)\sim i \sigma\, \frac{e^{-i \sigma z}
    \, e^{i \sigma  \pi \left(
  \frac \nu 2 + \frac 1 4 \right)}}{\sqrt{2 \pi z}}\,
  \left\{ P(\nu,z)
  - i \sigma \,Q(\nu,z) \right\}\, ,
\end{equation}
donde $\sigma =1$ si $\Im(z)>0$ y $\sigma=-1$ si $\Im(z)<0$. En
estas ecuaciones,
\begin{equation}\label{P+Q}
  P(\nu,z) - i\sigma  \,Q(\nu,z)\sim
  \sum_{k=0}^\infty \langle \nu , k\rangle
  \, \left(\frac{- i \sigma}{2 z}\right)^k\, ,
\end{equation}
donde los coeficientes,
\begin{equation}\label{coef-hankel}
  \langle \nu , k\rangle=
  \frac{\Gamma\left(\frac 1 2 +\nu + k\right)}
  {k! \, \Gamma\left(\frac 1 2 +\nu - k\right)}
  =\langle -\nu , k\rangle
\end{equation}
son los s{\'\i}mbolos de Hankel.

\bigskip

El cociente de dos funciones de Bessel admite entonces el
desarrollo asint{\'o}tico,
\begin{equation}\label{JsobreJup1}
      \frac{J_{\nu_1}(z)}{J_{\nu_2}(z)}
      \sim e^{ i\sigma \frac{\pi}{2} (\nu_1 - \nu_2)}\,
      \frac{P(\nu_1 ,z)- i\sigma \,
      Q(\nu_1 ,z) }
      {P(\nu_2,z) - i\sigma \,
      Q(\nu_2,z) }\, ,
\end{equation}
donde $\sigma =1$ para $\Im(z)>0$ y $\sigma=-1$ para $\Im(z)<0$.
Los coeficientes de este desarrollo asint{\'o}tico pueden ser
f{\'a}cilmente calculados, a todo orden, a partir de la
ecuaci{\'o}n (\ref{P+Q}),
\begin{equation}\label{asymp-cociente}\begin{array}{c}
    \displaystyle{\frac{P(\nu_1 ,z)\pm i\,
      Q(\nu_1 ,z) }
      {P(\nu_2,z) \pm i\,
      Q(\nu_2,z) }} \sim
      1 + \Big( \langle \nu_1 , 1\rangle -
     \langle \nu_2 , 1\rangle \Big) \,\left(\frac{\pm i}{2 z}\right)
      +  O\left(\frac 1 {z^2}\right)\, .
\end{array}
\end{equation}

En particular,
 \begin{equation}\label{JsobreJup}
      \frac{J_{\nu}(z)}{J_{-\nu}(z)}
      \sim e^{ i\sigma \pi\, \nu}
      \frac{P(\nu,z)- i \sigma \,
      Q(\nu,z) }
      {P(-\nu,z) - i\sigma \,
      Q(-\nu,z) } = e^{ i \sigma  \pi \,\nu}\, ,
\end{equation}
pues $P(\nu,z)$ y $Q(\nu,z)$ son funciones pares en $\nu$.

\bigskip

An{\'a}logamente, las derivadas de las funciones de Bessel admiten
los siguientes de\-sa\-rro\-llos asint{\'o}ticos \cite{A-S} para
$|\arg z|<\pi$,
\begin{equation}\label{deriv-asymp}
  J'_\nu(z) \sim -\frac{2}{\sqrt{2 \pi  z}}
  \left\{
  R(\nu,z) \sin \chi(\nu,z) + S(\nu,z) \cos\chi(\nu,z)
  \right\}\, ,
\end{equation}
y,
\begin{equation}\label{deriv-asymp-N}
  N'_\nu(z) \sim \frac{2}{\sqrt{2 \pi  z}}
  \left\{
  R(\nu,z) \cos \chi(\nu,z) - S(\nu,z) \sin \chi(\nu,z)
  \right\}\, ,
\end{equation}
donde,
\begin{equation}\label{R}
  R(\nu,z) \sim \sum_{k=0}^\infty(-1)^k\,
  \frac{\nu^2 + (2k)^2-1/4}
  {\nu^2-(2k-1/2)^2}\,
  \frac{\langle\nu,2k\rangle
  }{\left(2 z\right)^{2 k}}\, ,
\end{equation}
y,
\begin{equation}\label{S}
  S(\nu,z) \sim \sum_{k=0}^\infty(-1)^k \,
  \frac{\nu^2 + (2k+1)^2-1/4}
  {\nu^2-(2k+1-1/2)^2}\,
  \frac{\langle\nu,2k+1\rangle
  }{\left(2 z\right)^{2 k+1}}\, .
\end{equation}
Por lo tanto,
\begin{equation}\label{Jprima}
  J'_\nu(z)\sim \mp i\,
  \frac{e^{\mp i z}\, e^{\pm i \pi \left(
  \frac \nu 2 + \frac 1 4 \right)}}{\sqrt{2\pi z}}
  \left\{ R(\nu,z)
  \mp i \,S(\nu,z) \right\}\, ,
\end{equation}
donde el signo superior es v{\'a}lido para $\Im(\lambda)>0$, y el
inferior para $\Im(\lambda)<0$. Se verifica tambi{\'e}n
\begin{equation}\label{RS}
  R(\nu,z) \pm i \,S(\nu,z)=
  P(\nu,z) \pm i \,Q(\nu,z)
  + T_\pm(\nu,z)\, ,
\end{equation}
con,
\begin{equation}\label{T}
  T_\pm(\nu,z)\sim
  \sum_{k=1}^\infty
  (2k-1)\langle\nu,k-1\rangle
  \left(\frac{\pm i}{2z}\right)^{k}\, .
\end{equation}

Obtenemos entonces,
\begin{equation}\label{JprimasobreJ}
    \frac{J'_\nu(z)}{J_\nu(z)}
  \sim \mp  i \left\{
  1+ \frac{T_\mp(\nu,z)}{P(\nu,z)\mp i Q(\nu,z)}\right\}\, ,
\end{equation}
donde el signo superior vale para $\Im(\lambda)>0$, y el inferior
para $\Im(\lambda)<0$. Los coeficientes del desarrollo
asint{\'o}tico del miembro derecho de la ecuaci{\'o}n
(\ref{JprimasobreJ}) pueden obtenerse f{\'a}cilmente a partir de
las ecuaciones (\ref{P+Q}) y (\ref{T}),
\begin{equation}\label{TsobrePQ}\begin{array}{c}
    \displaystyle{
   \frac{T_\pm(\nu,z)}{P(\nu,z)\pm i Q(\nu,z)}
  = \pm\frac{ i}{2 z}
  + O\left(\frac 1 {z^2}\right)
  }
\end{array}
\end{equation}

Finalmente, como los s{\'\i}mbolos de Hankel son pares en  $\nu$
(v{\'e}anse las ecuaciones (\ref{coef-hankel}), (\ref{P+Q}),
(\ref{T}) y (\ref{JprimasobreJ})) se verifica,
\begin{equation}\label{JprimasobreJasymp}
  \frac{J'_\nu(z)}{J_\nu(z)}
  \sim \frac{J'_{-\nu}(z)}{J_{-\nu}(z)}\, .
\end{equation}

\bigskip

\vspace{3cm}

\begin{equation}
\mathbf{*\qquad*\qquad*}\nonumber
\end{equation}


\part{Agradecimientos}

\begin{itemize}

\item Agradezco la experiencia de mi Director Horacio Falomir en
ense\n ar a reconocer los aspectos relevantes de un problema, su
tenaz paciencia para revelarme los e\-rro\-res que me empe\n {\'e}
en repetir y su vocaci{\'o}n para renovar mi entusiasmo luego de
mis tantos intentos fallidos. Le agradezco todo lo que he
aprendido durante la elaboraci{\'o}n de esta Tesis y el trabajo
que ello le ha demandado.

\item Agradezco a Mariel Santangelo, Mar{\'\i}a Amelia Muschietti,
Rafael Benguria y Fidel Schaposnik el haber aceptado conformar el
jurado de esta Tesis.

\item Agradezco muy especialmente al Profesor Robert Seeley por su
generosa lectura de esta tesis, que supo transitar la oscura
redacci{\'o}n de mi castellano. {\'E}l me ha se{\~n}alado cada uno de los
numerosos errores que comet{\'\i}. El lector ya no habr{\'a} percibido
aquellos que fui capaz de corregir. Aquellos que s{\'\i} ha percibido,
en cambio, son los que no fui capaz de comprender.

\item Agradezco nuevamente a Mariel Santangelo su apost{\'o}lica
dedicaci{\'o}n a la tarea de aprender, que transmite permanente
motivaci{\'o}n a los que tenemos el privilegio de trabajar junto a
ella.

\item Agradezco a Gabriela y Karin por brindarme atenci{\'o}n y
consejos siempre que los necesit{\'e}.

\item Agradezco conocer a Los Justos: Diego el Esenio, Leo el
Paciente Peregrino, Nico, Guille y Ale, que descubren con placer,
que cultivan un jard{\'\i}n, que justifican el mal que se ha hecho
y prefieren que los dem{\'a}s tengan raz{\'o}n.

\item Agradezco a mi familia el agrado que siento al mencionarlos
y que testimonien que cada cosa engendra su semejante.

\item Agradezco al Consejo Nacional de Investigaciones
Cient{\'\i}ficas y T{\'e}cnicas, a la Comisi{\'o}n de
Investigaciones Cient{\'\i}ficas de la Provincia de Buenos Aires,
a la Fundaci{\'o}n Antorchas, al Instituto de F{\'\i}sica de La
Plata y a la Universidad Nacional de La Plata por el apoyo que he
recibido y que brindan a quienes ejercen esta profesi{\'o}n.

\end{itemize}

\bigskip

\begin{equation}
    *\qquad *\qquad *\nonumber
\end{equation}



\part{Bibliograf{\'\i}a}


\end{document}